\documentclass[12pt, lof, singlespace]{puthesis}

\title{Zonal Flows and Turbulence\\ in Fluids and Plasmas}

\submitted{September 2014}  
\copyrightyear{2014}  
\author{Jeffrey Bok-Cheung Parker}
\adviser{John A.~Krommes}  
\department{Astrophysical Sciences -- Program in Plasma Physics}

\usepackage{graphicx}

\usepackage{verbatim}

\usepackage{multirow}
\usepackage{longtable}

\usepackage{booktabs}

\usepackage[sort,authoryear]{natbib}
\setcitestyle{round,comma,aysep={},yysep={,}}

\ifdefined\printmode

\usepackage{url}

\else

\ifdefined\proquestmode

\usepackage{hyperref}
\hypersetup{bookmarksnumbered}

\makeatletter
\hypersetup{pdftitle=\@title,pdfauthor=\@author}
\makeatother

\else

\usepackage{hyperref}
\hypersetup{colorlinks,bookmarksnumbered}

\makeatletter
\hypersetup{pdftitle=\@title,pdfauthor=\@author}
\makeatother

\fi 
\fi 

\ifodd 0
\renewcommand{\maketitlepage}{}
\renewcommand*{\makecopyrightpage}{}
\renewcommand*{\makeabstract}{}

\else

\abstract{
In geophysical and plasma contexts, zonal flows are well known to arise out of turbulence.  We elucidate the transition from statistically homogeneous turbulence without zonal flows to statistically inhomogeneous turbulence with steady zonal flows.  Starting from the Hasegawa--Mima equation, we employ both the quasilinear approximation and a statistical average, which retains a great deal of the qualitative behavior of the full system.  Within the resulting framework known as CE2, we extend recent understanding of the symmetry-breaking `zonostrophic instability'.  Zonostrophic instability can be understood in a very general way as the instability of some turbulent background spectrum to a zonally symmetric coherent mode.  As a special case, the background spectrum can consist of only a single mode.  We find that in this case the dispersion relation of zonostrophic instability from the CE2 formalism reduces exactly to that of the 4-mode truncation of generalized modulational instability.  We then show that zonal flows constitute pattern formation amid a turbulent bath.  Zonostrophic instability is an example of a Type I$_s$ instability of pattern-forming systems.  The broken symmetry is statistical homogeneity.  Near the bifurcation point, the slow dynamics of CE2 are governed by a well-known amplitude equation, the real Ginzburg-Landau equation.  The important features of this amplitude equation, and therefore of the CE2 system, are multiple.  First, the zonal flow wavelength is not unique.  In an idealized, infinite system, there is a continuous band of zonal flow wavelengths that allow a nonlinear equilibrium.  Second, of these wavelengths, only those within a smaller subband are stable.  Unstable wavelengths must evolve to reach a stable wavelength; this process manifests as merging jets.  These behaviors are shown numerically to hold in the CE2 system, and we calculate a stability diagram.  The stability diagram is in agreement with direct numerical simulations of the quasilinear system.  The use of statistically-averaged equations and the pattern formation methodology provide a path forward for further systematic investigations of zonal flows and their interactions with turbulence.  
}

\acknowledgements{
In the six years I have been a graduate student at Princeton University and the Princeton Plasma Physics Laboratory, I have been helped countless times by countless people.  The completion of this thesis depended on their generosity and willingness to spend their time teaching me, and their contributions deserve acknowledgment.  Of course, any errors in this work are mine alone.

First and foremost, none of this would have been possible without my thesis advisor, John Krommes.  From the moment I sat in his class Irreversible Process in Plasmas in the spring of 2010, I saw that John has a penetrating insight into physics.  The next year, I sat in his class again, and I learned even more.  John has mentored me for nearly four years, and I attribute the greater part of my growth as a physicist to his tutelage.  He taught me everything, starting with how to teach myself.  A new challenge can seem overwhelming, but there are usually individual pieces which can be understood bit by bit and then reconstructed to understand the whole problem.  Breaking a new problem down into its simplest manifestations is the most important lesson I learned from John.  I owe him a great debt for his tireless efforts in holding me and my work to the highest academic standards.

I next acknowledge my Readers, Ilya Dodin and Hong Qin.  They, in their precious spare time, agreed to look over my thesis and give critical feedback.  Ilya in particular went above and beyond the call of duty in providing comments and suggestions on every last detail.

I worked with Peter Catto in the summer of 2011 at the MIT Plasma Science and Fusion Center in fulfillment of the practicum of the DOE Fusion Energy Sciences Fellowship, and he has been a lasting influence.  Peter's intuitive style combined with mathematical talent left an indelible impression on me, though the most treasured aspect of our relationship has been his friendship and support.

Cynthia Phillips mentored me in a summer research internship at PPPL when I was an undergraduate, and she has been a friend ever since.  In that summer, I experienced my first true glimpse of independent research.  Cynthia's mentorship played a large role in me coming back to Princeton for graduate school.  I also owe a great deal to DOE's SULI, the NUF program, and the Science Education Department at PPPL that made that summer possible.  It has been a pleasure collaborating with Andrew Zwicker and Deedee Ortiz of the Science Education Department since then on science outreach.

Yevgeny Raitses advised my first-year experimental project.  Even though I knew I would be a theorist, I learned so much thanks to Yevgeny's willingness and trust to grant me responsibility in handling expensive equipment.  Never again will I fear Langmuir probes.  Roscoe White, my second-year theory project advisor, has been a fantastic teacher and a wonderful role model for a life inside and outside of physics.  Greg Hammett has been a constant source of original ideas and inspiration, and at any time I could pop into his office to ask a question.  Nat Fisch has been a supporter in so many ways that I cannot thank him enough.  Furthermore, as Chair of the Plasma Physics Graduate Committee I brought a number of concerns to Nat, and he was always receptive and willing to talk about them.  Others at PPPL I wish to acknowledge for their teaching and advice include Amitava Bhattacharjee, Sam Cohen, Hantao Ji, and Bill Tang.  Jennifer Jones was always helpful and willing to lend a hand, especially after she hit my car.

Barbara Sarfaty, the den mother of the graduate students, took care of so many little details and was always willing to go the extra mile.  It is with much wistfulness that I watch Barbara step down just before I depart.  Beth Leman has big shoes to fill.

I want to thank those who laid the intellectual groundwork upon which my thesis builds: Brian Farrell, Petros Ioannou, Brad Marston, Kaushik Srinivasan, Steve Tobias, and Bill Young.  Without their efforts, my dissertation would not exist.  They are the giants whose shoulders I stand on.  I've interacted with all and met in person all but one, and it has been a pleasure the whole way.  I especially owe Kaushik for buying my train ticket after I landed in Zurich and discovered I couldn't withdraw any cash at an ATM.  How fortuitous it was that we were on the same plane.

Several people were influential in supporting my travel to international meetings and workshops, contributing to my growth as a physicist and as a person.  The Stix Prize Committee, for granting me the support to travel to the International Centre for Theoretical Physics in Trieste for the 2009 Summer College on Plasma Physics (the first time I ever traveled internationally, I might add) and meeting many young physicists from developing countries.  Felix Parra, who provided for me to attend the Madrid Workshop on gyrokinetics and turbulence, giving me my first taste of interacting closely with the international plasma turbulence community.  Amitava Bhattacharjee, who in addition to offering advice gave me extremely generous support so I could travel to Switzerland to take part in an upcoming book on zonal flows.

I would like to acknowledge funding from a National Science Foundation Graduate Research Fellowship as well as a Department of Energy Fusion Energy Sciences Fellowship.  
These fellowships gave me the freedom to pursue knowledge down the path of my own choosing.  

On a more personal note, there are many to thank for making this journey fun and unforgettable.  Seth Davidovits, for being a great officemate and friend as well as personal trainer.  Mike Hay and Eric Shi, for numerous rides to and from Princeton Junction.  And to the others with offices around me, for providing a \emph{lively} atmosphere: Josh Burby, Ben Faber, Renaud Gueroult, Clayton Myers, John Rhoads, Filippo Scotti, Daniel Ruiz, Lei Shi, Jono Squire, and Yao Zhou.  Other graduate students and friends were no less influential: Tyler Abrams, Tarek Anous, Jessica Baumgaertel, Dennis Boyle, Dan Dennhardt, Lee Ellison, Martin Griswold, Katy Ghantous, Xiaoyin Guan, Josh Kallman, Mike Lekas, Chaney Lin, Matt Lucia, Brendan Lyons, Jake Nichols, Luc Peterson, Kelsey Tresemer, and countless others I am no doubt forgetting.  And I must give a shoutout to the Tokabats softball team and all the players and fans.  Winning the B-League Championship was a highlight I shall not forget.

Finally, with all my heart I thank my parents and my brother and sister for their support in everything I've done.  My parents gave me the opportunity to follow my dreams no matter where they might lead.  What I have achieved so far, and whatever destination those unknown future roads end up at, it is all due to them.
}

\dedication{To my parents}

\fi  

\usepackage{amsmath}
\usepackage{amssymb}
\usepackage{bm} 

\usepackage{mycommands}

\newcommand{\RB}{Rayleigh-B\'{e}nard\ }
\newcommand{\azf}{\hat{\a}_{ZF}}
\newcommand{\LD}{L_d^{-2}}
\newcommand{\nablabarsq}{\ol{\nabla}^2}
\newcommand{\ybar}{{\ol{y}}}
\newcommand{\xbar}{{\ol{x}}}
\newcommand{\pbsq}{\ol{p}^2}
\newcommand{\qbsq}{\ol{q}^2}
\newcommand{\hbpsq}{\ol{h}_+^2}
\newcommand{\hbmsq}{\ol{h}_-^2}
\newcommand{\kbsq}{\ol{k}^2}
\newcommand{\kybarU}{k_{\ybar,U}}
\newcommand{\kybarW}{k_{\ybar,W}}
\newcommand{\kbsqybarU}{\ol{k}_{\ybar,U}^2}

\newcommand{\kybardU}{k_{\ybar,\de U}}
\newcommand{\kybardW}{k_{\ybar,\de W}}
\newcommand{\kbsqybardU}{\ol{k}_{\ybar,\de U}^2}

\newcommand{\hbpsqdW}{\ol{h}_{+,\de W}^2}
\newcommand{\hbmsqdW}{\ol{h}_{-,\de W}^2}
\newcommand{\sU}[1]{\s_{#1,U}}
\newcommand{\sdU}[1]{\s_{#1,\de U}}

\newcommand{\sinc}{\operatorname{sinc}}
\newcommand{\ecrossb}{\v{E} \times \v{B}}
\newcommand{\ti}[1]{\widetilde{#1}}
\renewcommand{\O}{O}

\newcommand{\figref}[1]{Figure~\ref{#1}}	
\newcommand{\secref}[1]{Section~\ref{#1}}	
\newcommand{\chref}[1]{Chapter~\ref{#1}}
\newcommand{\appref}[1]{Appendix \ref{#1}}
\newcommand{\symbline}[3]{#1 & #2 & \eref{#3} \\}

\newcommand{\zf}{ZF}

\newcommand{\zfs}{ZFs}

\renewcommand{\cite}{\citep}

\begin{document}

\makefrontmatter


\chapter{Introduction}
\label{ch:intro}

Zonal flows are turbulence-driven sheared flows.  They are usually associated with a direction of symmetry.  In planetary atmospheres, they flow along lines of latitude, parallel to the equator, and the direction of flow alternates with latitude.  In that context, zonal flows are associated with the azimuthal symmetry.  In magnetically confined toroidal plasmas, zonal flows consist of $\ecrossb$ flows produced by toroidally and poloidally symmetric fluctuations of electric potential.  The direction of flow is along a flux surface and varies radially.

Zonal flows have taken on special significance in plasma physics because they are thought to regulate drift-wave turbulence.  In particular, evidence is mounting that turbulence driven by the ion-temperature-gradient (ITG) instability in toroidal plasmas is suppressed by zonal flow or mean shear flow.\footnote{Zonal flow refers to turbulence-driven flow, and it typically oscillates in space with finite radial wavenumber.  Mean shear flow is caused by diamagnetic effects associated with the mean pressure profile.}  Furthermore, zonal flows are thought to play a role in triggering the L--H transition.  The enhanced plasma performance of the H-mode is viewed as essential to any viable fusion reactor, and zonal flows may play an important part of the H-mode.

One mechanism by which shear flow is believed to suppress turbulence is shear-enhanced decorrelation \cite{biglari:1990,terry:2000,diamond:2005}.  The basic idea is that the flow causes a turbulent eddy to stretch and elongate, making it more likely for that eddy to break apart.  This reduces the length scale of turbulence and hence reduces the resultant turbulent transport.  Numerical simulations seem to corroborate the idea, directly implicating zonal flows in reducing the levels of turbulent fluctuations \cite{lin:1998}.  This simple, powerful idea has been incredibly influential, spawning an entire genre of inquiry, and zonal flows have been under intense study by the plasma physics community ever since.  Any means that might help in taming the beast of tokamak turbulence is pursued with vigor.

\begin{figure}
		\centering
		\includegraphics[width=4in]{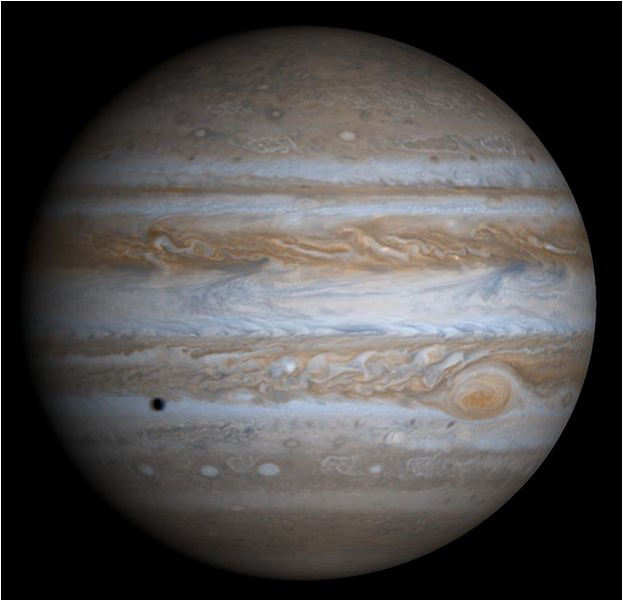}
		\caption{Jupiter, with zonal bands visible.  Image from NASA spacecraft Cassini.}
		\label{intro:fig:jupiter}
	\end{figure}

Zonal flow is also prominent in geophysical contexts.  For example, \figref{intro:fig:jupiter} shows Jupiter with visible zonal bands.  The alternating bands flow in alternating directions.\footnote{Animated images are available at \url{http://ciclops.org/view/92/Jupiter_Mosaics_and_Movies_-_Rings_Satellites_Atmosphere} and \url{http://www.nasa.gov/centers/goddard/multimedia/largest/EduVideoGallery.html}.}  All of the gas giants in our solar system have zonal flows, not just Jupiter.  Due to their visibility, the atmospheric science community has studied zonal flow for decades.

Zonal flows and zonal magnetic fields are also beginning to be observed in astrophysical simulations of accretion disc turbulence driven by the magnetorotational instability \cite{johansen:2009,kunz:2013}.  We cannot currently observe and may never be able to directly observe zonal structure of accretion discs, but our understanding of their dynamics may hinge upon the behavior of zonal fields.

Since zonal flows are driven by turbulence, any understanding of zonal flows must begin with an understanding of turbulence.  In this chapter we start by introducing some important aspects of turbulence.  Then, we turn to zonal flows and review the experimental, numerical, and theoretical literature, separated into geophysical and plasma physics sections.

\section{Turbulence in Fluids and Plasmas}
The word turbulence conjures up images of chaotic motion, of disorder.  Turbulence would seem to destroy any semblance of regularity or organization.  Typically, smooth, laminar flow such as regular pipe flow or \RB convection rolls gives way to disorder, turbulence, and a jumble of scales.  Out of this turbulence, seemingly by magic, coherent structures such as zonal flows can form, as we shall see.

Turbulence theory in fluids and plasmas has varying objectives.  In 3D, homogeneous, isotropic, incompressible Navier-Stokes turbulence, theory has been trying to understand intermittency of the inertial range.  In geophysical fluid dynamics, the goal of theory is to understand the atmospheres of not only other planets, but also our own.  The Earth's combined atmosphere-ocean system constitutes an incredibly complex dynamical system, one which determines our climate.  In fusion theory, the ultimate objective of turbulence theory is to predict and control the level of turbulent transport.  When one attempts to build a fusion reactor, out of the many, many factors that must be considered, the effect of microturbulence often boils down to a single number: the energy confinement time.  The greater the level of turbulence, the worse the confinement of heat and energy within the plasma.

In the following sections we introduce a tiny bit of basic turbulence theory.  For a comprehensive introduction, see \citet{davidson:book}.

\subsection{Cascade in 3D}
The natural place to start is with Kolmogorov's explanation of the energy cascade in 3D neutral-fluid turbulence \cite{frisch:1995}.  The famous Kolmogorov scaling is one of the most fundamental and celebrated results of neutral fluid turbulence theory.  It is one of the first quantitative, successful predictions of fully developed turbulence.  The result concerns turbulence of the incompressible Navier-Stokes equation,
	\begin{gather}
		\pd{\v{v}}{t} + \v{v}\cdot\nabla\v{v} = -\nabla p + \nu \nabla^2 \v{v}, \label{cascades:nse} \\
		\nabla\cdot\v{v} = 0,
	\end{gather}
where $p$ is the pressure divided by density and $\n$ is the viscosity.  The Kolmogorov theory makes a definite prediction for the energy spectrum in wavenumber space.  First, several assumptions are made:
	\begin{enumerate}
		\item The turbulence is statistically homogeneous and isotropic.
		\item Energy flows \emph{locally} in $k$-space.
		\item There is an \emph{inertial range} of $k$-space where the turbulence does not ``know'' about forcings at the large scale or viscosity at the small scale.
	\end{enumerate}

The average energy density\footnote{Using the density instead of the total energy prevents us from having to deal with infinite energies in an infinite fluid.} $\hat{E}$ and omnidirectional energy spectrum $E(k)$ of the flow are related by
	\begin{equation}
		\hat{E} = \left\langle \frac{1}{2} v^2 \right\rangle = \int_0^\infty dk\, E(k)
	\end{equation}
The energy density $\hat{E}$ is decomposed as a sum over wavenumbers.  The spectrum $E(k)$ depends only on the magnitude of the wavenumber, $k \defineas |\v{k}|$ (where $\defineas$ denotes a definition), due to the isotropy assumption.

Physically, one often speaks of ``eddies'' in a turbulent flow.  The typical picture is that energy is somehow injected into the system at large scales, perhaps due to mechanical stirring of the fluid, and gives rise to eddies.  These turbulent eddies interact with each other in some way, giving rise to smaller scale eddies.  Energy flows from the larger scales to the smaller scales in this scenario.  This is related to assumption 2.  Eventually, when energy reaches small enough scales, viscosity becomes important and the energy is dissipated.  The physical attributes at an intermediate scale are assumed to not depend on the precise behavior at the large or the small scales.

Let us make this more precise.  Energy is injected at a rate $\varepsilon$ at the large scales, the forcing scales, designated by wavenumber $k_f$.  Energy is assumed to flow locally through $k$-space without dissipation until it reaches the viscosity-dominated small scales, designated by wavenumber $k_\nu$.  In a statistically steady state, the energy flux through every scale must, on the average, be $\varepsilon$, until it is dissipated.  At intermediate wavenumbers, $k_f \ll k \ll k_\nu$, the locality assumption means that the turbulence cannot depend upon $k_f$ or $k_\nu$, but only on the scale $k$ and the energy flux $\varepsilon$.  There are no other local quantities it can depend on.

These ideas may be expressed as an advection equation in $k$-space.\footnote{The author first learned of this approach from G.~Hammett.}  While the Kolmogorov argument is essentially dimensional and not a quantitative calculation, the advection equation is handy for systematizing the assumptions and tracking the dimensions of all quantities.  The assumptions lead one to being able to write the advection equation
	\begin{equation}
		\pd{E(k,t)}{t} + \pd{}{k}\left(\frac{\Delta k}{\Delta t} E(k,t) \right) = \varepsilon \delta(k-k_f).
	\end{equation}
And in a statistically steady state, $E(k,t)$ will not depend on $t$.  This is a local conservation equation, where forcing but not dissipation has been built in.  This equation can be considered as part of a Fokker-Planck equation, where $\Delta k/\Delta t$ is the ``drift velocity'' through $k$-space.  We will consider this equation at some $k$ larger than $k_f$.

When speaking of a given scale $k$ or $l \sim k^{-1}$, it will be convenient to assign a width to the scale.  It is most natural to break the scales up logarithmically.  That is, starting from the largest scale $k_f$
	\begin{equation}
		k_f \longrightarrow 2^1 k_f \longrightarrow 2^2 k_f \longrightarrow 2^3k_f \longrightarrow 2^4 k_f \longrightarrow \cdots.
	\end{equation}
A given scale labeled `$k$' can be considered to contain wavenumbers from $k/2$ to $k$.  By the locality assumption, the $\Delta k$ appearing in the advection equation can only be $k$.

$\Delta t$ is what is called a nonlinear correlation time, or an ``eddy turnover time'' $\tau_k$.  An eddy turnover time is defined as the time it takes for a fluid element of speed $v_k$ to cross an eddy of size $l \sim k^{-1}$,
	\begin{equation}
		\tau_k = \frac{l}{v_k}.
	\end{equation}
Here $v_k$ is the characteristic speed of eddies of size $l$ and is quantified through
	\begin{align}
		\frac{1}{2}v_k^2 &= \int_{k/2}^k dk'\, E(k') \\
		&\sim kE(k),
	\end{align}
which gives $v_k \sim \sqrt{kE(k)}$ (ignoring constants of order unity).

Now, integrate the advection equation from $k=0$ to some $k>k_f$.  The energy spectrum $E(k)$ is assumed to vanish at $k=0$.  One finds
	\begin{align}
		\frac{\Delta k}{\Delta t} E(k) &= \varepsilon, \\
		k^2 v_l E(k) &= \varepsilon, \\
		k^{5/2} E^{3/2} &= \varepsilon.
	\end{align}
One thus obtains the Kolmogorov scaling for the inertial range,
	\begin{equation}
		E(k) = C \varepsilon^{2/3} k^{-5/3},
	\end{equation}
where $C$ is simply an order-unity constant.

\subsubsection{Kolmogorov Scaling from Pure Dimensional Analysis}
Another way to obtain the Kolmogorov scaling is through dimensional analysis without any recourse to the physics.  This approach yields less intuition than the physical picture of eddy turnover, but it is a useful demonstration of the power of dimensional analysis.  The locality hypothesis demands
	\begin{equation}
		E(k) = f(\varepsilon,k).
	\end{equation}

The dimensions of the average energy density, energy spectrum, and energy flux are given below, where $L$ is the dimension of length and $T$ is the dimension of time.
	\begin{gather}
		\hat{E} \sim \frac{\mathrm{L}^2}{ \mathrm{T}^2}, \\
		E(k) \sim \frac{\mathrm{L}^3}{\mathrm{T}^2}, \\
		\varepsilon \sim \frac{\mathrm{L}^2}{\mathrm{T}^3}.
	\end{gather}

Now, suppose the function $f \sim \varepsilon^\alpha k^\beta$.  Then the dimension of $E$ would be
	\begin{equation}
		E \sim \frac{\mathrm{L}^3}{\mathrm{T}^2} \sim \left( \frac{\mathrm{L}^2}{\mathrm{T}^3} \right) ^\alpha \frac{1}{\mathrm{L}^\beta}.
	\end{equation}
Satisfying dimensional consistency requires $\alpha = 2/3$ and $\beta = -5/3$.  Thus, $E \sim \varepsilon^{2/3}\, k^{-5/3}$ is recovered.

\subsubsection{Energy Dissipation and the Viscous Scale}
\begin{figure}
		\centering
		\includegraphics{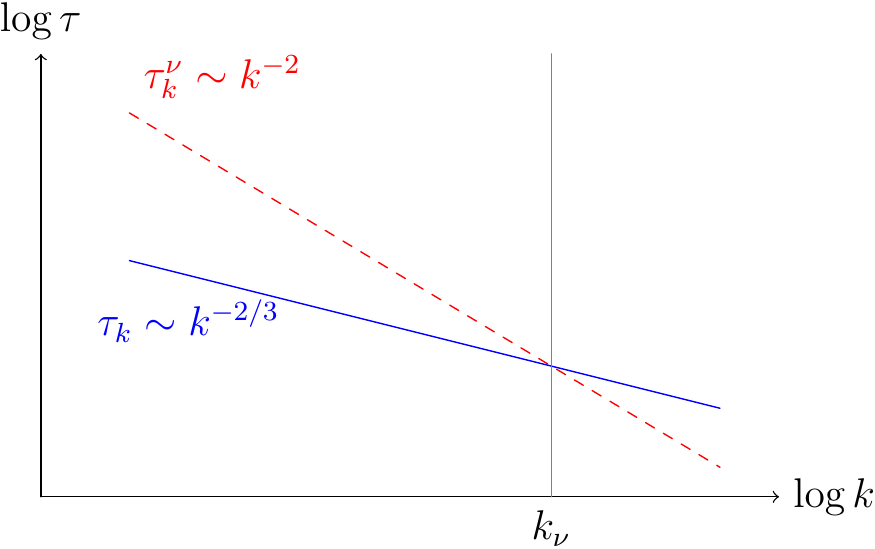}
		\caption{Time scale of inertial effects (blue, solid) and viscous effects (red, dashed) as a function of wavenumber.}
		\label{intro:fig:kolmogorov_scaling}
	\end{figure}

At a given $k$ in the inertial range, the eddy turnover time $\tau_k$ is given by
	\begin{equation}
		\tau_k \sim \frac{1}{k v_k} \sim k^{-3/2} E^{-1/2} \sim \varepsilon^{-1/3}\, k^{-2/3}.
	\end{equation}
From the Navier-Stokes equation, the timescale for viscous processes at a scale $k$ can be seen to be
	\begin{equation}
		\tau_k^\nu = \frac{1}{\nu k^2}.
	\end{equation}
The process with the \emph{shorter} time scale dominates.  Dissipation becomes important at the scale $k_\nu$ where $\tau_k^\nu = \tau_k$.  For $k<k_\nu$, inertial effects dominate, while for $k>k_\nu$, viscous effects dominate (see \figref{intro:fig:kolmogorov_scaling}).

Setting $\tau_k^\nu = \tau_k$ gives an estimate for the viscous scale, or Kolmogorov scale:
	\begin{equation}
		k_\nu = \biggl( \frac{\varepsilon}{\nu^3} \biggr)^{1/4} \qquad \text{or} \qquad l_\nu = \biggl( \frac{\nu^3}{\varepsilon}\biggr)^{1/4}.
	\end{equation}
	
The dissipation rate is $\dot{\hat{E}} = \langle \nu \v{v}\cdot \nabla^2 \v{v} \rangle$.  We have already assumed that it acts only at scales smaller than or comparable to the Kolmogorov scale $l_\nu$.  If we look \emph{at} the Kolmogorov scale, then substituting $\nabla \sim k_\nu$, we find
	\begin{align*}
		\dot{\hat{E}} &\sim \nu v_\nu^2 k_\nu^2 \\
			&\sim \nu^3 k_\nu^4 \\
			&\sim \varepsilon.
	\end{align*}
This shows that the dissipation acts primarily at the Kolmogorov scale; nothing much is happening at smaller scales.  This result is also important because it shows that dissipation is independent of the viscosity, even as $\nu \to 0$.

\subsection{Cascade in 2D}
The nature of cascades are different in two dimensions \cite{kraichnan:1967}.  Instead of just the energy being conserved, in 2D there are two quadratic quantities that are conserved by the nonlinear interaction: energy and enstrophy.  The 2D case is important because geophysical flows are quasi-2D due to atmospheric stratification \cite{pedlosky:book,vallis:2006}, and plasma flows are quasi-2D due to the magnetic field.

Again assume statistical isotropy and homogeneity.  Instead of forcing at large scales as in 3D, assume that forcing occurs at some intermediate length scale or wavenumber.  Then there is a \emph{dual cascade}, with two inertial ranges rather than just one.  Energy cascades from the forcing scale to \emph{larger} scales, whereas enstrophy cascades from the forcing scale to \emph{smaller} scales.  The energy cascade is called the \emph{inverse cascade}, while the enstrophy cascade is called the \emph{direct cascade}.  The energy spectrum in the inverse cascade range is $E(k) = \ve^{2/3} k^{-5/3}$, where $\ve$ is the energy flux through wavenumber space.  In the direct cascade range, the energy spectrum is $E(k) = \eta^{2/3} k^{-3}$, where $\eta$ is the enstrophy flux through wavenumber space.\footnote{\citet{kraichnan:1971} showed that a logarithmic correction needs to be applied in the enstrophy inertial range; see also \citet{bowman:1996}.}

The flow of energy to large scales and the flow of enstrophy to small scales can be understood as a consequence of the conservation laws \cite{fjortoft:1953,kraichnan:1967}.  There is some energy spectrum, $E(k)$, with total energy density given by
	\begin{equation}
		\hat{E} = \int dk\, E(k).
	\end{equation}
Let $Z(k)$ be the enstrophy spectrum.  It is related to the energy spectrum by $Z(k) = k^2 E(k)$, so that the total enstrophy density is
	\begin{equation}
		\hat{Z} = \int dk\, k^2 E(k).
	\end{equation}
In other words, the enstrophy is weighted by a higher power of wavenumber than the energy is.  \citet{vallis:2006} showed that if the energy spectrum spreads out under the constraint of conservation of both total energy and enstrophy, then the centroid of the energy spectrum must move to smaller $k$ (larger scales) and the centroid of the enstrophy spectrum must move to larger $k$ (smaller scales).  The tendency for energy to accumulate at large scales will be especially important for understanding the generation of zonal flows in geophysical contexts.



\subsection{Statistical Theories of Turbulence}
The statistical approach to understanding turbulence, which this thesis takes, complements other methods such as making detailed measurements of plasma fluctuations or performing direct numerical simulations (DNS).  Those methods can accumulate reams of data so vast that it can be unclear how one should go about making sense of it all.  The aim of the statistical approach is to focus on the macroscopic quantities of interest, such as transport coefficients, energy spectra, and the like.  By working with averaged quantities from the outset, one can circumvent the rapid spatiotemporal fluctuations and potentially see a clearer view of the physics.  Of course, there is no free lunch.  As a consequence of averaging a nonlinear equation, one is generally left with the average of an unknown quantity: a closure problem.  Various statistical closures, perhaps motivated by physical considerations, provide different approximations for the unknown terms.  A major difficulty with this approach is that the closures are essentially uncontrolled approximations; the nonlinearity inherent to turbulence makes it hard to know exactly what is lost.  The closure might obliterate some highly coherent or correlated phenomena.  Nevertheless, these difficulties do not invalidate the statistical approach, from which much has been learned \cite{frisch:1995,krommes:2002,kraichnan:1959,kraichnan:1964a}.  Historically, the majority of theoretical studies into turbulence that follow this approach assume \emph{homogeneous} statistics, where the statistics of turbulent quantities do not depend on position.  Consequently, most of the theoretical machinery that has been developed also applies only to homogeneous statistics, with comparatively little effort devoted to \emph{inhomogeneous} statistics.  The main line of work in this thesis involves inhomogeneous statistics.

\section{Zonal Flows}
The simplest model in which zonal flows arise naturally out of turbulence is the 2D system
\begin{equation}
		\partial_t w + \v{v} \cdot \nabla w + \b \partial_x \psi  = \ti{f} + D,
		\label{intro:vorticityeqn}
	\end{equation}
where $\ti{f}$ is a forcing term, $D$ represents dissipation, and
	\begin{equation}
		w = \nabla^2 \psi.
	\end{equation}
Here, $\psi$ is the stream function, $\v{v} = \unit{z} \times \nabla \psi$ is the velocity, and $w = \unit{z} \cdot \nabla \times \v{v}$ is the vorticity.  This equation will be discussed much more fully in Chapter \ref{ch:eom}.  The equation is often used as the simplest, most reduced description of atmospheric turbulence.  The behavior of turbulence and zonal flow even in this simple system is still studied today.  In this introductory chapter, we use this equation to highlight a few key points.

\subsection{Zonal Flows in Geophysics}
We briefly\footnote{Very briefly, since this is not the author's area of expertise.} review some of what is known about zonal flows in geophysical contexts.  For more information, see the works of \citet{vasavada:2005,vallis:2006,pedlosky:book} and references therein.

Jupiter, for example, has prominent, easily visible zonal jets.  It has roughly 30 zonal jets, and they have been remarkably stable over time.  Measurements by Voyager in 1979 and Cassini in 2000 indicate the zonal wind profile has barely changed in that time period.  Compared to Jupiter's equatorial radius of 70,000 km, we can directly observe at most only a few hundred kilometers into the atmosphere.  Little is known about the turbulence and zonal wind deeper down.  In the upper atmosphere, zonal jet speed is mainly measured by assuming that clouds are passive tracers of the zonal wind.  This is not perfect, due to for example, larger clouds averaging over an extended spatial region, but it seems to be somewhat successful.  This technique can only measure jet speed at cloud level.  The energy source of the zonal jets is hypothesized to be buoyant convection from a hot planetary interior \cite{vasavada:2005}.  On Earth, zonal flows occur can occur in both the ocean and atmosphere, but the flows tend to meander with complex dynamics, and there are not as many jets.

One idea deserves special note.  The notion of the Rhines scale has been enormously influential in the geophysical literature \cite{rhines:1975}.  The Rhines scale purportedly estimates the jet width or spacing, and is given by
	\begin{equation}
		L_R = \sqrt{\frac{U}{\b}},
	\end{equation}
where $U$ is the rms velocity and $\b$ is the northward gradient of the Coriolis parameter.  Inversely, we can express the characteristic Rhines wavenumber as
	\begin{equation}
		\label{intro:rhineswavenumber}
		k_R = (\b / U)^{1/2}.
	\end{equation}

We give a couple of ways of obtaining the Rhines scale \cite{vallis:1993,vasavada:2005}.  The first method is essentially dimensional analysis.  Let
	\begin{equation}
		\v{v} = u\unit{x} + v\unit{y}
	\end{equation}
and $w = \unit{z} \cdot \nabla \times \v{v} = \partial_x v - \partial_y u$.  Then \eref{intro:vorticityeqn}, rewritten here as
	\begin{equation}
			\partial_t w + \v{v} \cdot \nabla w + v\b = \ti{f} + D,
	\end{equation}
can be used to find the Rhines scale by heuristically balancing the magnitudes of the Rossby wave term (the $\b$ term) and the nonlinear advection term.  If we treat $u \sim v \sim U$ and $k_x \sim k_y \sim k$, then we find that the nonlinear advection term is roughly $k^2 U^2$, and the linear term is roughly $U \b$.  Where these are equal gives this Rhines scale.

A slightly more refined analysis would allow for the zonal jets to have a different magnitude and length scale than the eddies.  Let $\ol{\v{v}}$ and $\ol{\z}$ be the zonally-averaged velocity and vorticity, $U$ be the characteristic velocity of zonal flow, and $u',v'$ be the characteristic velocity of the eddies.  We suppose $U \gg u',v'$.  Then
	\begin{align}
		\v{v} \cdot \nabla \z &\approx \v{v}' \cdot \nabla \ol{\z} + \ol{\v{v}} \cdot \nabla \z' \\
			&= -v' \partial_y^2 \ol{u}(y) + \ol{u}(y) \partial_x \bigl( \partial_x v' - \partial_y u' \bigr) \label{intro:rhines1}.
	\end{align}
If we look particularly at the first term of \eref{intro:rhines1},\footnote{This is not particularly justified without further argumentation.  Based on magnitudes, one might expect the other term to dominate because the length scale of turbulence is smaller and so its spatial derivatives are larger.} then we see that the advection term goes like $v k_R^2 U$, which can be compared with $v\b$.  Equating them gives the Rhines scale.

\begin{figure}
		\centering
		\includegraphics{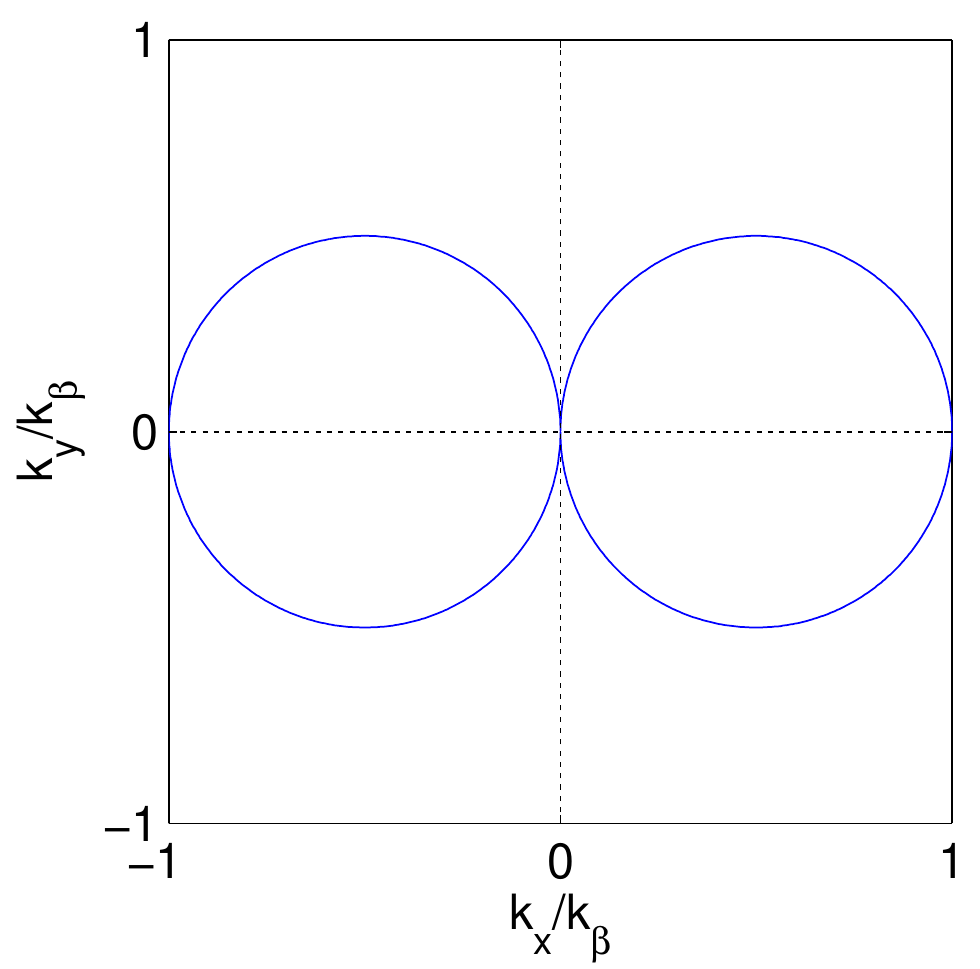}
		\caption{Anisotropic Rhines scale.  Outside the dumbbell, at high $k$, inertial turbulence has a shorter time scale and dominates.  Inside the dumbbell, Rossby wave dynamics are faster. The hypothesis is that the inverse cascade cannot penetrate into the dumbbell, so energy piles up on the $k_y$ axis where $k_x=0$.}
		\label{intro:fig:rhines_dumbbell}
	\end{figure}

A more physical argument views the Rhines scale as a transition scale between the regimes where inertial, isotropic turbulence and Rossby-wave activity dominates \cite{rhines:1975,vasavada:2005}.  Assume the turbulence is forced at small scales.  At wavenumbers greater than the Rhines scale, the eddy-turnover time scale is shorter than the time scale of Rossby waves, so standard 2D turbulence results with an inverse cascade.  Energy proceeds towards larger scales until it reaches the Rhines scale.  Then the time scale of Rossby waves becomes shorter than the eddy turnover time, so Rossby waves dominate.  The idea is that these large scale waves are inefficiently forced by the turbulence, and energy cannot easily cascade to length scales larger than the Rhines scale, so energy piles up at $k_R$ and the inverse cascade slows.  The turbulent frequency is roughly $U k$ and the Rossby wave frequency is $\w = -k_x \b / k^2$.  Let $\phi$ be the angle between east (the $\unit{x}$ direction) and the direction of wave phase propagation, so that $\cos \p = k_x/k$.  Then equating the turbulent frequency and the wave frequency leads to an anisotropic Rhines scale,
	\begin{equation}
		k_R^2 = \frac{\b}{U} |\cos \p|.
	\end{equation}
A plot of this anisotropic ``dumbbell'' shape is shown in \figref{intro:fig:rhines_dumbbell}.  The dumbbell outline is where the inverse cascade halts.  The anisotropic shape offers an explanation for why energy piles up on the $k_y$ axis where $k_x=0$, leading to the preference of zonally-symmetric structures.  This scenario appears to have been confirmed \cite{vallis:1993}, although some have called it into question by arguing that the small scales directly force the zonal flows \cite{huang:1998}.

Large-scale friction or drag is critical for getting the Jovian jets correct \cite{vasavada:2005}.  Without it, energy would slowly leak past the Rhines scale into larger scales.  If this energy is not damped somehow, then on a long enough time scale, even the large scales would isotropize and distinct zonal jets would not exist.  Large-scale friction damps the energy that would leak past the Rhines scale.  However, if friction were too large, energy would damp before it could cascade up to the Rhines scale, and so no jets would form and isotropic turbulence would result.

\subsection{Zonal Flows in Plasmas}
Our definition of zonal flow in plasma refers only to the zero-frequency flows.  We exclude geodesic acoustic modes (GAMs) \cite{winsor:1968} from our definition of zonal flow.  Some authors describe GAMs as oscillatory zonal flows, but in this thesis we do not.

\subsubsection{Theory and Simulations}
Efforts at developing a systematic theory of zonal flows have almost exclusively focused on the scenario where the zonal flows are assumed to be long wavelength compared to the scale of the turbulence \cite{krommes:2000,smolyakov:2000b,diamond:2005,connaughton:2011}.  This remains true despite the fact that in simulations and experiments, zonal flows tend to be of scale comparable to that of the turbulence.  When the long-wavelength zonal flow assumption is made, the resulting interaction of turbulence and zonal flow can be described in terms of a wave kinetic equation.  In this type of description, one might imagine a sea of drift-wave packets evolving in an weakly-inhomogeneous medium of zonal flows.  The wave-kinetic formulation has intuitive advantages because the turbulent wave action is materially conserved along phase-space trajectories.  In addition to the long-wavelength assumption, many studies make a \emph{single-harmonic} assumption where only a single Fourier mode of the zonal flows is retained \cite{connaughton:2011}.  When that is done the zonal flow wavelength cannot be found from the theory but is left as an undetermined parameter.  The state of things indicates that theory of zonal flows is still in its infancy.

Some studies have focused on a generalized modulational instability, in both the geophysics and plasma literature \cite{lorenz:1972,gill:1974,manin:1994,smolyakov:2000b,connaughton:2010,wordsworth:2009}.\footnote{Generalized in the sense that it is not restricted to the original meaning of long-wavelength modulations that vary in the same direction as the primary wave.}  In analytic studies of these instabilities, typically a single eigenmode, referred to as the primary wave, is used as the background upon which the perturbation grows.  For example, in a periodic box, a single Fourier mode is an exact solution to the nonlinear vorticity equation.  A conceptually close cousin of modulational instability is secondary instability \cite{rogers:2000,plunk:2007,pueschel:2013}.  In secondary instability, a growing linear eigenmode, the primary mode, acts as a background upon which a secondary perturbation grows.  If the secondary mode grows much faster than the primary, then the primary can be treated as stationary.

Other simulations have investigated various aspects of zonal flows.  For instance, \citet{nakata:2012} examined entropy transfer via zonal flows in gyrokinetic simulations.  Along the same lines, \citet{makwana:2012,makwana:2014} looked at how zonal flows interact with damped modes to regulate turbulence.  \citet{xanthopoulos:2011} studied the effects of the magnetic equilibrium in stellarator geometry.  \citet{waltz:2008}, after turning off drift wave--drift wave nonlinear couplings, concluded that the drift wave--zonal flow coupling accounts for most of the nonlinear saturation.  Other models, using fluid equations and simplified geometry, are also a fruitful ground with which to gain intuition and insight.  The Hasegawa--Wakatani system is one such model which has been used to study zonal flows \cite{hasegawa:1983,hasegawa:1987,pushkarev:2013}.  \citet{numata:2007} first studied the Modified Hasegawa-Wakatani system, which corrects the treatment of zonal flows when the equations are restricted to two dimensions.

\subsubsection{Experimental Observations}
It is tough to make direct measurements of zonal flows in plasmas.  First, the zonal flow involves only an electric potential and plasma flow, which are not easily observed.  In contrast, the GAM is associated with an $m=1$ poloidal density fluctuation which can be measured more readily.  Furthermore, the GAM oscillates at moderate frequency, whereas the zonal flow fluctuates at zero or low frequency, which is also difficult to measure by certain techniques.  Zonal flows are often not zero frequency in practice, but fluctuate on a much slower time scale (a few kHz) than the turbulence (tens of kHz).

The main diagnostic tools used to measure zonal flows are the Langmuir probe, the heavy ion beam probe (HIBP), beam emission spectroscopy (BES), and Doppler reflectometry \cite{fujisawa:2009,estrada:2014}.

Langmuir probes measure the ion saturation current and floating potential \cite{hutchinson:book}.  Fluctuations in floating potential are usually analyzed as fluctuations in plasma potential and fluctuations in ion saturation current  as fluctuations in plasma density.  Potential measurements in multiple locations can be used to calculate electric fields and hence $\ecrossb$ flows.  One signature of zonal flow is correlation in electric potential between positions on the same flux surface but separated toroidally or poloidally.  Density measurements are useful because finding a weak correlation in density fluctuations while detecting a strong correlation in potential fluctuations enhances one's confidence that the observed phenomenon is in fact a zonal flow.  Langmuir probes are restricted to cooler plasmas because the probes would otherwise not survive, so Langmuir probes cannot be used for measurements in the core of high-performance plasmas.  Owing to their simplicity, Langmuir probes are widely used when feasible.

The HIBP diagnostic provides a direct measurement of plasma potential, even in the plasma core \cite{crowley:1994,ido:2002}.  Heavy, singly charged ions are injected into the plasma at high energies (hundreds of keV).  Upon impact with electrons, some ions undergo ionization into a double charge state, and these so-called secondary ions are deflected more strongly in the magnetic field.  The secondary ions also gain energy at the ionization point due to the increase in potential energy from a higher charge state.  The plasma potential $\phi$ at the ionization point can be determined by measuring the difference in kinetic energy between primary and secondary ions at a detector.

The BES diagnostic measures local density fluctuations \cite{fonck:1990}.  A neutral beam is injected into the plasma and undergoes collisional fluorescence.  The emitted light is approximately proportional to the local density.  The Doppler shift of the emitted light due to the beam's velocity allows for the separation of the beam $H_\a$ or $D_\a$ emission from the bulk plasma emission.  High spatial and temporal resolution is possible with a 2D imaging system.  Flow velocity can be calculated from the motion of turbulence structures between poloidally separated channels using time-delay-estimation techniques.

Doppler reflectometry yields measurements of both flow and density fluctuations \cite{hirsch:2001}.  Unlike traditional reflectometry, Doppler reflectometry uses an angle between the incoming microwave beam and the cutoff layer.  Scanning the tilt angle allows wavenumber-resolved turbulence measurements.  Flow velocity is measured from the Doppler shift of the scattered signal.  This diagnostic can provide high temporal and spatial resolution.

The first direct observation of zonal flows in the core region of a toroidal plasma came from a dual-HIBP measurement in the CHS stellarator \cite{fujisawa:2004}.  Zonal flows were then found in the core of a tokamak plasma in DIII-D with BES \cite{gupta:2006}.  In both of these cases, the measured zonal flows exhibited a short radial length scale comparable to that of the turbulence.  Many other identifications of zonal flows can be found in the reviews of \citet{fujisawa:2009} and \citet{estrada:2014} and references therein.

\subsubsection{L--H Transition}
The transition from L-mode to H-mode has been the subject of intense interest since its discovery \cite{wagner:1982}.  H-mode is associated with a transport barrier and a reduced level of turbulence, along with a sharper plasma pressure gradient which improves performance.  A number of studies have implicated sheared $\ecrossb$ flows in the L--H transition, although there is not yet a detailed understanding of the associated physics.

Measurements of $E_r$ find that the $\ecrossb$ flow varies rapidly during the L--H transition, whereas the pressure profiles and the resulting diamagnetic flow take longer to evolve.  It also found that the increase in $E_r$ shear occurs before the decrease in turbulent fluctuations, consistent with shear flow causing turbulent suppression \cite{moyer:1995,burrell:1999,estrada:2009,meyer:2011}. 

With recently improved spatiotemporal resolution, many devices have observed between L-mode and H-mode an intermediate, transient phase, which is called I-phase \cite{colchin:2002,estrada:2010,estrada:2011,estrada:2012,xu:2011,schmitz:2012}.\footnote{The I-phase should not be confused with the I-mode regime first observed on the Alcator C-Mod tokamak \cite{whyte:2010}.}  The I-phase is characterized by oscillations in the zonal flow and turbulent fluctuations.  These oscillations often show a characteristic predator-prey behavior, with the $\ecrossb$ flow (the predator) following the density fluctuations (the prey) with a phase delay of $90^\circ$.

On the theoretical side, no first-principles simulation has reproduced the L--H transition.  Consequently, theoretical investigations have focused primarily on reduced models with various assumptions and approximations.  Initially, these studies were 0D and modeled the predator-prey interaction only between the mean shear flow and the fluctuation level \cite{diamond:1994}.  Later, \citet{kim:2003} extended the model to include suppression of turbulence by both mean flow and zonal flow.  This two-predator, one-prey model exhibits pre-transition oscillations.  In the model, the zonal flow triggers the transition, and then the steep-gradient-driven mean flow sustains the H-mode.  That work has been developed further into a 1D, radially-extended model that evolves turbulence intensity, zonal and mean flow shear, and pressure and density profiles \cite{miki:2012}.

\section{Overview of this Thesis}
The strategy of this thesis is to start at the basics and develop a systematic theory of zonal flows from the bottom up.  To this end, we use the simplest models in order to develop a sound theoretical foundation.  In \chref{ch:eom}, we introduce the Charney--Hasegawa--Mima equation, which serves as the model for almost all of the work in this thesis.  This equation neglects many of the realistic effects in plasmas and fluids, which allows for tractable analysis.  We also describe the quasilinear approximation and the CE2 statistical framework.  The quasilinear approximation denotes that the fields of interest are divided into a mean field and an eddy field, and then in the equation for the eddy field the eddy-eddy nonlinearity is neglected.  We perform all of our analysis within the context of the quasilinear approximation.  While it is clearly not a realistic approximation in all cases, numerical simulation provides convincing support that the qualitative behavior, at least of zonal flows, is similar as in the full model.  This approximation leads naturally to the CE2 statistical framework.  Because statistical theories of turbulence average over small-scale fluctuations, these theories have the advantageous feature of allowing for a steady-state, statistical description of a turbulent equilibrium.  For example, many theories, including CE2, describe turbulence in terms of a two-point correlation function.

\chref{ch:beyondzonostrophic} contains the main physics content of the thesis.  We first review the recently discovered zonostrophic instability (ZI).  In ZI, a statistically homogeneous turbulent state that is on average uniform in space becomes unstable to a inhomogeneous perturbation.  These perturbations grow into saturated zonal flows.  We draw connections between ZI and modulational instability.  Then we show that zonal flow can be interpreted as pattern formation, and we expand upon the insights that brings.  For instance, as a control parameter is varied and the homogeneous turbulent state becomes ZI unstable and a new stable inhomogeneous state appears, the bifurcation is described by a simple equation with universal behavior.  One immediate consequence, previously remarked in scattered observations but never explicitly understood mathematically, is the existence of multiple solutions to the CE2 equations with varying ZF wavelengths.  In other words, the width of the zonal jets is not unique; we derive this mathematically.  We analytically calculate the bifurcation at which zonal flows appear and verify it numerically.  In \chref{ch:numerical}, we solve CE2 numerically to find equilibria of nonlinearly interacting turbulence and zonal flows.  To do this, we use Newton's method to directly solve the steady-state CE2 equations.  This technique is common for pattern-forming systems.  Once the equilibria are found, we also calculate their stability.  In terms of the ZF wavelength, the region of stability calculated from CE2 is consistent with the results of direct numerical simulation of the QL equations.

\chref{ch:closure} proposes a simple closure for homogeneous turbulence.  We are able to examine in detail the stability of solutions to the closure, a property mostly neglected in the literature.  As it stands, this chapter is somewhat separate from the rest of the thesis.  But if extended, it could be connected back to zonal flows and provide a way for further analytic progress.

Finally, in \chref{ch:future}, we explore directions for future research.  For instance, we discuss more realistic turbulence closures than the quasilinear approximation and CE2, such as the DIA.  Such closures are needed to account in some way for the eddy-eddy nonlinearities.  These more sophisticated approaches would allow for better quantitative and qualitative accuracy.   Additionally, in this thesis we dealt with the Charney--Hasegawa--Mima equation where turbulence is forced by external drive, but most models of plasma turbulence relevant to fusion have an intrinsic instability.  One simple way to proceed along this path is to extend the closure described in \chref{ch:closure} to allow for inhomogeneity.  Then one could perform a similar bifurcation analysis to that in \chref{ch:beyondzonostrophic}.  The results of this thesis provide a theoretical foundation for those more sophisticated models, but research is needed to understand those situations in detail.  Toroidal geometry presents a challenge, especially for analytic work.  One plausible path forward is to use CE2 to describe both ZFs and geodesic acoustic modes (GAMs) together.  Just as we have gained definite insights into the behavior of ZFs, we are optimistic that similar insights are possible for GAMs.

\section{Mathematical Conventions}
A table of important mathematical symbols is given in Table \ref{table:importantsymbols}.

\begin{table}
\label{table:importantsymbols}
\caption{Important symbols, their meaning, and the equation in which they are first used.}
\vspace{.4cm}
\begin{tabular}{r|l|l}
Symbol & Meaning & Equation \\ \hline
$\defineas$ & Definition & \\
\symbline{$\n$}{Viscosity}{cascades:nse}
\symbline{$w$}{Generalized vorticity}{intro:vorticityeqn}
\symbline{$\psi$}{Stream function}{intro:vorticityeqn}
\symbline{$\b$}{Planetary vorticity gradient (or plasma density gradient)}{intro:vorticityeqn}
\symbline{$L_d$}{Deformation radius (or plasma sound radius)}{eom:equivbarotropicinversion}
\symbline{$\azf$}{Modifies relation of $w$ and $\psi$ (is either 1 or 0)}{eom:unifiedalpha}
\symbline{$\m$}{Friction}{eom:dissipationoperator}
\symbline{$h$}{Hypervisocity factor}{eom:dissipationoperator}
\symbline{$\g$}{Fundamental dimensionless paramater}{gamma}
\symbline{$U$}{Zonal flow velocity}{qlsystem}
\symbline{$W$}{Covariance of vorticity}{Wdef}
\symbline{$\Psi$}{Covariance of stream function}{CE2}
\symbline{$F$}{Covariance of random, external forcing}{CE2}
\symbline{$x,y$}{Difference coordinates of 2-point correlation function}{CE2}
\symbline{$\ybar$}{Sum coordinate of 2-point correlation function}{CE2}
\symbline{$U_\pm$}{$U(\ybar \pm \tfrac12 y)$}{CE2}
\symbline{$\nablabarsq$}{$\nabla^2 - \LD = \partial_x^2 + \partial_y^2 - \LD$}{CE2}
\symbline{$\ol{I}$}{$1 - \azf \LD \partial_\ybar^{-2}$}{CE2}
\symbline{$\kbsq$}{$k^2 + \LD$}{zi:kbsqdefinition}
\symbline{$q$}{Wavenumber of zonal flow}{zi:perturbationfields}
\symbline{$\l$}{Eigenvalue (i.e., growth rate)}{zi:perturbationfields}
\symbline{$\qbsq$}{$q^2 + \azf \LD$}{zi:Ubarpp}
\symbline{$W_{mnp}$}{Fourier coefficients of $W(x,y \mid \ybar)$}{iscalc:galerkinseries}
\symbline{$U_p$}{Fourier coefficients of $U(\ybar)$}{iscalc:galerkinseries}
\end{tabular}
\end{table}

\subsection{Coordinate Convention}
\begin{figure}
		\centering
		\includegraphics{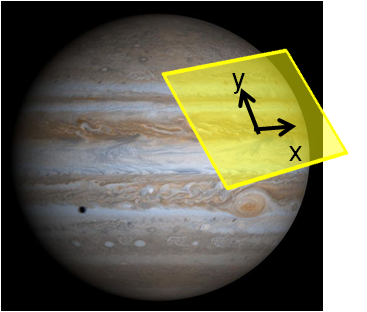}
		\caption{Local coordinate system on a sphere.}
		\label{intro:fig:planet_coords}
	\end{figure}

\begin{figure}
		\centering
		\includegraphics[scale=0.8]{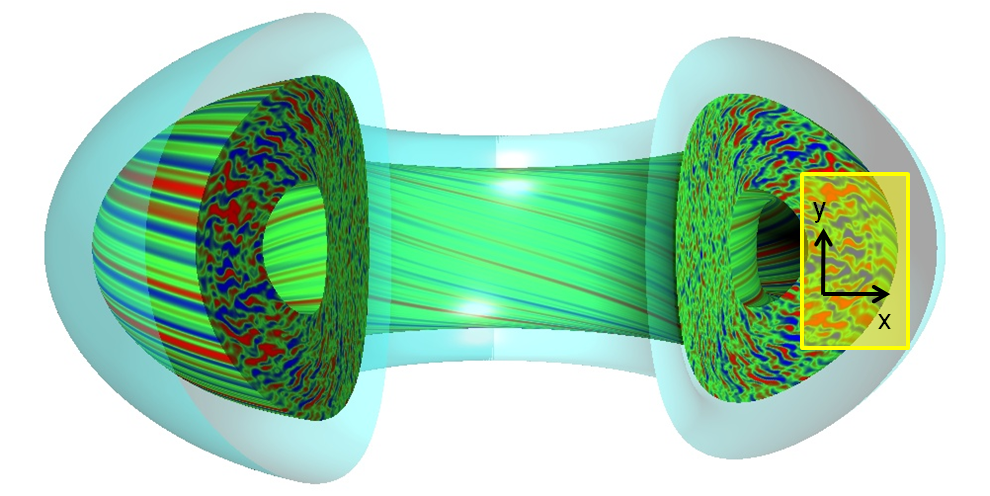}
		\caption{Local coordinate system at the outboard midplane of a tokamak.  (Image from the GYRO team.)}
		\label{intro:fig:tokamak_coords}
	\end{figure}

The geophysical communities and plasma communities use opposite coordinate conventions for the direction of inhomogeneity in two-dimensional planar models.  In planetary atmospheres, zonal flows run in the east--west ($x$) direction and vary in the north--south ($y$) direction (see \figref{intro:fig:planet_coords}).  In tokamaks, zonal flows run in the poloidal direction.  If one imagines a small box placed at the outboard midplane of a tokamak, the poloidal direction becomes the $y$ direction and the radial direction becomes the $x$ direction (see \figref{intro:fig:tokamak_coords}).  These opposite conventions for the direction of zonal flow require us to make a choice as to which will be followed.  We follow the convention used in the geophysical community.  We do this because the method of approach used in this thesis is closely related to recent works in the geophysical literature.  It is significantly easier to comprehend the literature when the same convention has been used everywhere.  Thinking in terms of the usual tokamak convention then requires flipping only a single mental switch.  If the alternative of using opposite conventions had been chosen, then the practitioner must separately assimilate each equation that is encountered.  However, as a compromise, we also give the rule to transform between conventions for key equations.  As an aside, it might also be noted that the tokamak convention could be made consistent with the geophysical convention, if one were to place the region of interest not at the outboard midplane but at a poloidal angle of $90^\circ$.

\subsection{Fourier Transform Convention}
Throughout, we use the convention
	\begin{align}
		\hat{f}(k) &= \int dx\, e^{-ikx} f(x) \\
		f(x) &= \frac{1}{2\pi} \int dk\, e^{ikx} \hat{f}(k)
	\end{align}
Often, we work only in Fourier space and drop the hat.

\chapter{Equations of Motion}
\label{ch:eom}
\section{Modified Hasegawa--Mima Equation}
We take as the starting point the Modified Hasegawa-Mima equation \cite{smolyakov:2000,krommes:2000}.  The Hasegawa--Mima equation (HME) has been studied for decades as a paradigm of electrostatic turbulence \cite{hasegawa:1978,horton:1994}.  The basic physics involve a nonuniform background density profile and the motion of charged fluid elements due to $\ecrossb$ and polarization drifts where the fluctuating electric field is self-consistently determined.  The HME was originally derived from a fluid perspective using the Braginskii equations \cite{hasegawa:1978}.  It can also be derived in a simple way using a cold ion limit of the gyrokinetic equation \cite{krommes:australia}.

The Modified Hasegawa--Mima (mHME) equation fixes a defect of the original version in its treatment of zonal flows.   Built into the HME is the assumption of adiabatic electrons.  But the adiabatic electron response relies on the fast motion of electrons along magnetic field lines.  Electrons can rapidly flow along magnetic field lines to neutralize charge imbalance, but cannot flow across field lines in the same way.  For fluctuations that are constant on a magnetic surface (i.e., with $k_\parallel=0$), the adiabatic electron response breaks down.   Therefore zonal flows, which by definition are produced by electrostatic fluctuations constant on a magnetic surface, are not treated correctly in the HME \cite{hammett:1993}.  The mHME modifies the electron response to be more physically correct.  

In a 2D formulation, the mHME is typically written as
	\begin{equation}
		\partial_t w(x,y) + \v{v} \cdot \nabla w - \k \partial_y \phi = \ti{f} + D,
	\end{equation}
where $x$ corresponds to a radial-like direction, $y$ to a poloidal-like direction, $\phi = (L_n / \r_s) e\vp / T_e$ is the normalized electrostatic potential, $L_n$ is the density gradient scale length, $\r_s$ is the sound radius, $T_e$ is the electron temperature, $w = \nabla^2 \phi - \hat{\a}\phi$ is the generalized vorticity and is related to ion gyrocenter density fluctuations $\de n_i^G$ by $w = -(L_n / \r_s) \de n_i^G / n_0$ where $n_0$ is the background density, $\hat{\a}$ is an operator that is zero when acting on zonal flows and unity when acting on drift waves, the magnetic field is in the $\unit{z}$ direction, $\v{v} = \unit{z} \times \nabla \phi$ is the $\ecrossb$ velocity, $\k$ is related to the density scale length, $\ti{f}$ is some kind of forcing that drives turbulence, and $D$ is a dissipation term.  Lengths are normalized to the sound radius $\r_s$ and times are normalized to the drift wave period $\omega_*^{-1}=(L_n / \r_s) \Omega_i^{-1}$.  These normalizations and scalings are convenient to make $w$, $\phi$, and the active length and time scales of order unity.  Additionally, they also allow us to set $\k=1$.

The terms $\ti{f}$ and $D$ produce forced, dissipative turbulence.  Many studies examine the ideal limit in which both $\ti{f}$ and $D$ are neglected \cite{horton:1994}.  The systems in such cases are Hamiltonian and conserve an infinite number of quantities.  The difference between the ideal limit and the non-ideal limit here is much the same as the difference between the neutral-fluid Euler equation and Navier-Stokes equation.  Studies of the ideal limit may yield qualitative insight regarding statistical equilibrium or cascades of conserved quantities \cite{lee:1952,zhu:2010} or may reveal the tendency of a system to form coherent structures.  In contrast, this thesis is concerned with forced, dissipative turbulence.

Since a fixed, spatially-independent profile gradient is used, the model is called a \emph{local} model, as described in \chref{ch:intro}.  One imagines the domain of the system is a small box within the much larger physical system.  The local approach tries to extract as much physics as possible from as simple a system as possible.  The local approach can often be justified in terms of the smallness of the ratio $\rho/L$, where $\rho$ is the gyroradius of the relevant species, and $L$ is the length scale of the macroscopic parameter, e.g., for density, $L_n^{-1} = |d \ln n_0/dx|$.  This approach retains the physics of the existence of a gradient in density (or other macroscopic parameter) but does not require detailed spatial profile information.  This approach can capture a great deal of the physics involved and also remove the necessity of dealing with complicated boundary conditions.  Since turbulence has a small length scale, the usual argument goes that regardless of the boundary conditions that are used in theory or simulation of the model equations, a few correlation lengths away from the boundaries the turbulence should not be affected by the boundaries.  For simplicity, many simulations use periodic boundary conditions.  Even when the local approach cannot be rigorously justified in an asymptotic ordering for some realistic situation, it is still a useful method for obtaining qualitative insight.

\section{(Equivalent) Barotropic Vorticity Equation}
It has been long known that the HME is mathematically very similar to an equation that arises in a geophysical context \cite{pedlosky:book}.  When one writes the equation of motion for an incompressible fluid on the surface of a rotating sphere, one has what is known as the quasigeostrophic (QG) equation for barotropic vorticity.  This 2D formulation on the surface of the sphere is a decent qualitative approximation, because due to the rotation the oceans and atmosphere are  stratified into horizontal layers.  In an approach which is in very much the same spirit as the local approach described above, theoretical geophysicists consider not a rotating sphere but a $\b$ plane.  A $\b$-plane approximation simplifies the model by linearizing the variation of the Coriolis term in the equation of motion.  In the full spherical geometry, the Coriolis term varies nonlinearly (that is, sinusoidally) over the latitude of the sphere.  The $\b$-plane approximation retains the fact that there is variation, but keeps only a linear variation.  With such an approach, Rossby waves (the analog of plasma drift waves) are easily analyzed.  Sometimes the geophysical literature keeps in the $\b$-plane approximation something known as the \emph{deformation radius} $L_d$, which is the length scale at which rotational effects become as important as buoyancy or gravity waves.  Even though $L_d$ is usually thought of as being comparable to turbulent length scales, many theoretical studies continue to neglect gravity-wave effects by taking infinite $L_d$ \cite{vasavada:2005,scott:2012,srinivasan:2012}.  The reason for studying a model with all these approximations is that it is more tractable to analysis and interpretation.   With $L_d = \infty$, the equation of motion is called the barotropic vorticity equation, while with finite $L_d$ it is called the equivalent barotropic vorticity equation.

The (equivalent) barotropic vorticity equation is given by
	\begin{equation}
		\partial_t w + \v{v} \cdot \nabla w + \b \partial_x \psi  = \ti{f} + D,
		\label{QGmodel}
	\end{equation}
where
	\begin{equation}
		w = \nabla^2 \psi - \LD \psi. \label{eom:equivbarotropicinversion}
	\end{equation}
Here $w$ is the vorticity, $\psi$ is the stream function, $\v{v} = \unit{z} \times \nabla \psi$ is the horizontal velocity.  The deformation radius $L_d$ plays the same role here as the plasma sound radius $\r_s$ plays in the Hasegawa--Mima equation.

\section{Unified Equation}
The mHME and the (equivalent) barotropic vorticity equation can be unified in a single equation.  In the following we also take an explicit form for the forcing $\ti{f}$ and dissipation $D$.  The unified quasigeostrophic-mHME is given by
	\begin{equation}
		\partial_t w + \v{v} \cdot \nabla w + \b \partial_x \psi  = \ti{f} + D,
		\label{unifiedEOM}
	\end{equation}
where
	\begin{align}
		w &= \nablabarsq \psi, \\
		\nablabarsq &\defineas \nabla^2 - \hat{\a} \LD. \label{eom:unifiedalpha}
	\end{align}
The nonlinear advection term $\v{v} \cdot \nabla w$ is sometimes written as the Poisson bracket $\{\psi, w\}$, where 
	\begin{equation}
		\{A,B\} = (\partial_x A) (\partial_y B) - (\partial_y A)(\partial_x B) = \unit{z} \times \nabla A \cdot \nabla B.
	\end{equation}
We take the external forcing $\ti{f}$ to be white noise forcing $\xi$ and the dissipation operator to be
	\begin{equation}
		D \defineas - \m w - \n (-1)^h \nabla^{2h} w, \label{eom:dissipationoperator}
	\end{equation}
where $\m$ is the scale-independent friction and $\n$ is the viscosity with hyperviscosity factor $h$.  To recover the QG barotropic vorticity equation, set $\hat{\a}=1$.  To recover the physics of the mHME, set $\hat{\a}=1$ for DW modes ($k_x\neq 0$) and $\hat{\a}_{ZF}=0$ for ZF modes ($k_x=0$), and set $L_d \defineas \r_s = 1$.  To get back to the plasma physics notational conventions, make the substitutions
	\begin{equation}
		(x,y,\psi,\b,v_x,v_y) \mapsto (-y,x,\phi,\k,-v_y,v_x).
		\label{eom:QLqgtoghm}
	\end{equation}
In the plasma context, $\b$ represents the density gradient and must not be confused with the ratio of plasma pressure to magnetic pressure.

In the interest of full generality, the rest of this thesis will make use of the unified framework.  As discussed in \chref{ch:intro}, the framework sticks to the conventions of the geophysical literature.  Unless specified otherwise, all figures will be presented with just the geophysical parameters in the limit $L_d = \infty$ rather than the plasma parameters ($L_d = \r_s = 1$, $\azf = 0$).  This is done for simplicity as there are many qualitative similarities in the ZF behavior in the two cases.

\subsection{Symmetries}
Neglecting the random forcing for a moment, we examine the relevant symmetries of \eref{unifiedEOM}.  We assume that any symmetries of the equation are not spoiled by boundary conditions (e.g., take an infinite system or one with periodic boundary conditions).  These symmetries are
\begin{enumerate}
	\item $x \to x+a$ \quad (translational symmetry in $x$)
	\item $y \to y+a$ \quad (translational symmetry in $y$)
	\item $\{y, w, \vp\} \to \{-y, -w, -\vp\}$ \quad (reflection symmetry in $y$)
\end{enumerate}
In other words, if $w(x,y,t)$ is a solution, then the symmetries give us other solutions:
\begin{enumerate}
	\item $\hat{w}(x,y,t) = w(x+a,y,t)$
	\item $\hat{w}(x,y,t) = w(x,y+a,t)$
	\item $\hat{w}(x,y,t) = -w(x,-y,t)$
\end{enumerate}
Note that since $v_x = \partial_y \psi$ and $v_y = -\partial_x \psi$, we see that the third symmetry implies that the solution and its symmetric partner have the same value of $v_x$, and hence, any jets take the same form.  In other words, on the $\b$-plane, eastward and westward are fundamentally distinguishable, whereas northward and southward in some sense are equivalent.  There is no requirement for jet motion to be symmetric in the eastward and westward direction.  The east-west symmetry is broken by the planetary rotation.  In the context of plasma, the analogous statement is that there is a physical difference for flow in the ion and electron diamagnetic directions.

When the random forcing is taken into consideration, exact translational symmetry is spoiled (as is the reflection symmetry).  However, we assume the forcing is statistically homogeneous in space.  Mathematically, this means that its statistics are independent of position.  Then, we can still say that \eref{unifiedEOM} satisfies the symmetries listed above statistically.  Naively, one might expect the turbulence that results from a solution to \eref{unifiedEOM} to be statistically homogeneous.  This is not always the case, as we shall see.

\subsection{Nonlinearly Conserved (Quadratic) Quantities}
The 2D equation \eref{unifiedEOM}, like the 2D Navier-Stokes equation, possesses two quadratic quantities which are conserved by nonlinear interactions.  These are the energy and enstrophy.  The average energy density is given by
	\begin{equation}
		E_a = \frac{1}{L_xL_y} \int dx\, dy\, \frac12 \left[ (\nabla \psi)^2 + \psi^2 \LD \right],
	\end{equation}
where the two terms account for kinetic and potential energy.  This can be rewritten in another form as
	\begin{equation}
		E_a = -\frac{1}{L_xL_y} \int dx\, dy\, \frac12 w \psi.
		\label{eom:energydensity}
	\end{equation}
The average enstrophy density is given by
	\begin{equation}
		W_a = \frac{1}{L_xL_y} \int dx\, dy\, \frac12 w^2
		\label{eom:enstrophydensity}
	\end{equation}

\subsection{Fundamental Dimensionless Parameter}
A fundamental dimensionless parameter controlling the zonal flow dynamics is \cite{danilov:2004}
	\begin{equation}
		\g \defineas \varepsilon^{1/4} \b^{1/2} \m^{-5/4}.
		\label{gamma}
	\end{equation}
This parameter is related to the zonostrophy index $R_\b$ by $\g \approx R_\b^5$~\cite{galperin:2010}.  Also, if one were to modify the definition of the small parameter $\alpha$ defined by \citet{bouchet:2013} such that the normalization length scale is the Rhines scale $L_{\rm R}$ rather than the size of the domain, then $\g = \a^{-1}$.  Since it has been shown by \citet{bouchet:2013} that it is $\a$ (and $\a^{1/2}$) that naturally appear in the normalized equations of motion, we opt to use $\g$ instead of $R_\b$ as the descriptive parameter.  When $L_d = \infty$, $\g$ is essentially the only independent dimensionless parameter in the problem.

\section{Quasilinear Approximation}
\subsection{Definition}
We restrict ourselves to the quasilinear (QL) approximation of this system.  Let us be very precise about what we mean by the QL approximation.  Given some kind of averaging procedure, one can decompose a field into a mean plus a fluctuation.  Then one can write down separate equations of motion for the mean and the fluctuation.  By the QL approximation, we mean that within the fluctuation equation of motion, the fluctuation self-nonlinearities (nonlinear terms involving only the fluctuation) are neglected.  Interactions between the fluctuation and the mean field are retained.  The equation of motion for the mean is not approximated.  This definition is consistent with classical usage in plasma physics \cite{vedenov:1962,drummond:1962,krall:1973}.

At the moment, we define our average to be a \emph{zonal} average.  The zonal mean of a quantity $w$ is given by
	\begin{equation}
		\ol{w} (y) = \frac{1}{L_x} \int_0^{L_x} dx\, w(x,y).
	\end{equation}
The zonal average is the conceptually simplest route, requiring the fewest number of assumptions, to the desired result.  A substantial discussion of different types of averages will be given in \secref{sec:otherclosures}.  The fluctuation, or deviation from the zonal mean, is given by $w' = w - \ol{w}$, and is referred to as an \emph{eddy} quantity.  We make this semantic distinction because a zonal mean quantity is likely to not fluctuate much if many independent correlation lengths of the turbulence have been averaged over.  We assume the eddy quantities contain the turbulent behavior.

To illustrate the QL approximation, we temporarily set forcing $\ti{f}$ and dissipation $D$ to zero.  We decompose the flow field into a zonally symmetric part (the zonal flow) and the residual (the eddies or turbulence).   Equation \eqref{unifiedEOM} can be decomposed as
	\begin{subequations}
	\begin{gather}
		\partial_t \ol{w} + \ol{ \v{v}'  \cdot \nabla w' } = 0, \\
		\partial_t w' + \ol{\v{v}} \cdot \nabla w' + \v{v}' \cdot \nabla \ol{w} + \v{v}'  \cdot \nabla w' - \ol{ \v{v}'  \cdot \nabla w' } + \b \partial_x \psi' = 0. \label{jp:eddyeqngeneral}
	\end{gather}
	\end{subequations}
No approximation has been made thus far.  The QL approximation involves neglecting the eddy-eddy nonlinearity within the eddy equation.  The QL system is
	\begin{subequations}
	\begin{gather}
		\partial_t \ol{w} + \ol{ \v{v}'  \cdot \nabla w' } = 0, \\
		\partial_t w' + \ol{\v{v}} \cdot \nabla w' + \v{v}' \cdot \nabla \ol{w} + \b \partial_x \psi' = 0. \label{qleddygeneral}
	\end{gather}
	\end{subequations}
More explicitly, and with forcing and dissipation restored, the QL system is
	\begin{subequations}
	\label{qlsystem}
	\begin{gather}
		\partial_t w' + \{U \nablabarsq + \b - [(\partial_y^2 - \LD) U]\} \partial_x \psi' = \xi - \m w' - \n(-1)^h \nabla^{2h} w', \label{QLwprime} \\
		\bigl[ \partial_t + \m + \n(-1)^h \partial_y^{2h} \bigr]  \bigl(1 - \azf \LD \partial_y^{-2} \bigr) U(y) + \partial_y \ol{ v_x' v_y' } = 0, \label{QLU}
	\end{gather}
	\end{subequations}
where 
	\begin{equation}
		U(y) \defineas -\ol{ \partial_y \psi }
	\end{equation}
is the zonal-mean zonal velocity.  We have assumed that the zonal mean of the forcing is zero.  If desired, one could easily allow for different dissipation rates on the zonal-mean quantities.

The QL approximation does not affect the conservation of the quadratic quantities by the nonlinear interactions.  This can be easily seen from the Fourier-space point of view.  Each triad interaction individually conserves these quantities, and the QL approximation amounts to removing some of these triad interactions.  

One issue to keep in mind is that the QL approximation breaks the material conservation of potential vorticity (PV).  Potential vorticity, a scalar field defined by $q = \r^{-1} \boldsymbol{\w}_a \cdot \nabla \th$, where $\r$ is the fluid density, $\boldsymbol{\w}_a$ is the absolute vorticity, and $\th$ is the potential temperature, is a critical quantity \cite{pedlosky:book}.  Many quantities of interest can be derived from the PV, a concept known as the PV invertibility principle \cite{mcintyre:2008}.  Furthermore, $q$ is conserved following the flow.  Conservation of PV relies on the combination of the eddy-eddy interactions and the eddy-mean interactions, so the neglect of eddy-eddy interactions in the QL approximation breaks PV conservation.  As a result, the QL system may lose certain physics that are based on the conservation of PV \cite{dritschel:2008}.

\subsection{Motivation for Using the QL Approximation}
\label{sec:QLmotivation}
\begin{figure}
		\centering
		\includegraphics{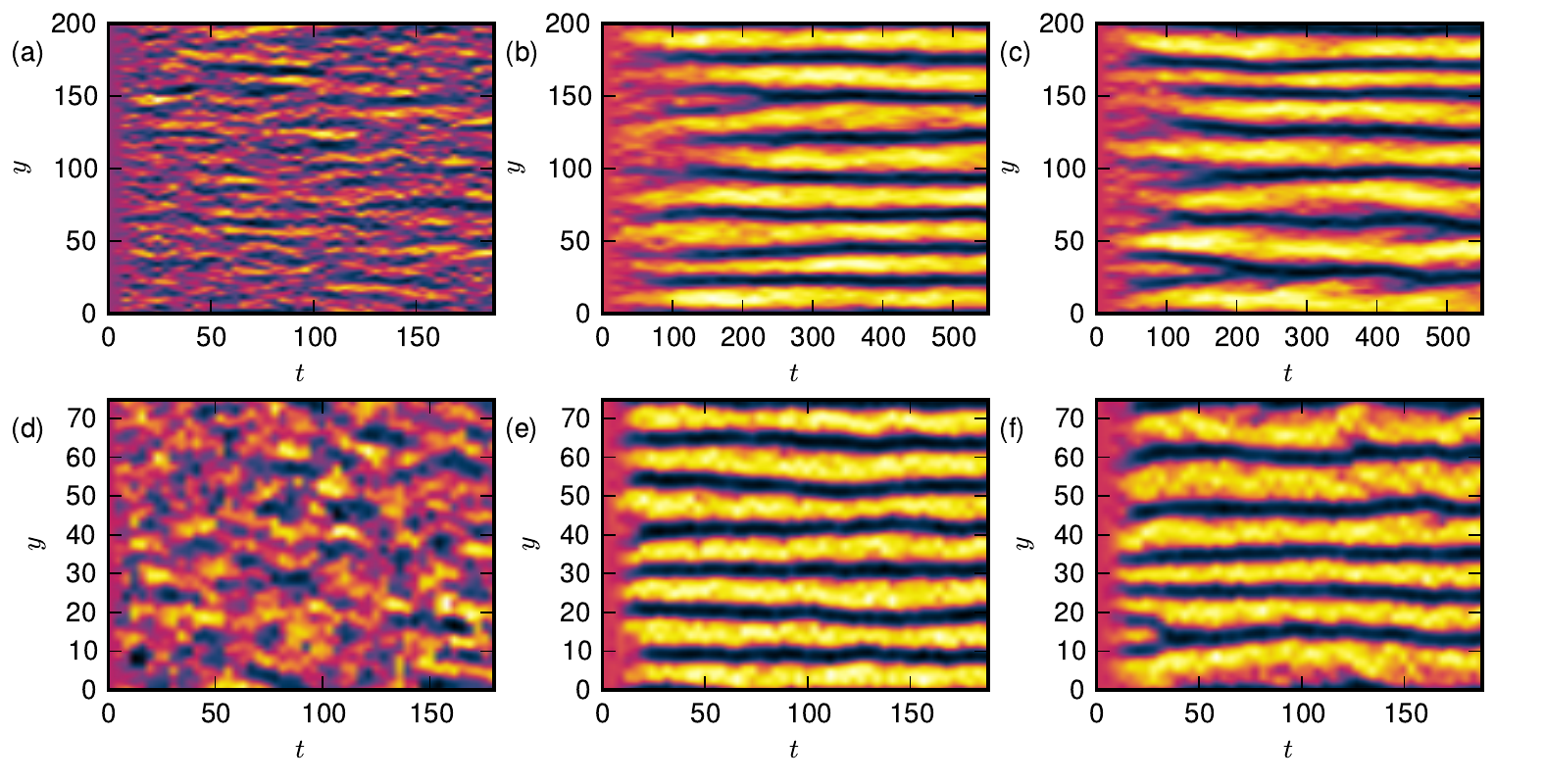}
		\caption{Space-time diagrams of zonal flow in DNS of QG.  Top: NL simulations at (a) $\m=0.08$ (no steady jets), (b) $\m=0.02$ (8 jets), and (c) $\m=0.02$ (7 jets).  Bottom: QL simulations at (d) $\m=0.29$ (no steady jets), (e) $\m=0.08$ (7 jets), and (f) $\m=0.08$ (6 jets).  The only differences between (b) and (c) and between (e) and (f) are the choice of initial conditions and the realization of the random forcing.  Merging jets can be seen in (c) at $t \approx 200$ and in (f) at $t \approx 30$.  Our numerical simulations described here and elsewhere are pseudospectral, typically using a resolution of $256 \times 256$ \cite{orszag:1969,trefethen:2000}.  We use ETDRK4 as our timestepping algorithm \cite{cox:2002,kassam:2005}.  We dealias using the $2/3$ rule \cite{orszag:1971:dealias,boyd:2001}.}
		\label{fig:zfplots}
	\end{figure}

\citet{srinivasan:2012} have shown that the QL system exhibits many of the same basic zonal jet features as the full nonlinear (NL) system, including the formation of stable jets and merging jets.  With periodic boundary conditions, the equation of motion enjoys translational symmetry in both the $x$ and $y$ directions.  As a parameter is varied, the simulations suggest a spontaneous breaking of statistical homogeneity in the $y$ direction.  At large $\m$ (small $\g$), the NL system in \figref{fig:zfplots}(a) and the QL system in \figref{fig:zfplots}(d) do not exhibit steady \zfs{}, so the behavior is statistically homogeneous.  At small $\m$ (large $\g$), both the NL system in \figref{fig:zfplots}(b) and the QL system in \figref{fig:zfplots}(e) do exhibit steady \zfs{}, implying a breaking of statistical homogeneity in the $y$ direction.  We also observe that, in both the NL and QL systems, simulations that differ only in initial conditions and realizations of the random forcing can display different numbers of jets [\figref{fig:zfplots}(b,c,e,f)].  In addition to these features, both NL and QL exhibit merging jets, evident in \figref{fig:zfplots}(c,f).

Our motivation in adopting the QL approximation is not because we believe it to be quantitatively correct, but rather because the QL system apparently retains the necessary ingredients that lead to the rich behavior of \zf{} formation.  The QL system may provide insight into the more realistic models, and the advantage, of course, is that the QL system is far more tractable.  The phenomena described above will all be explained analytically within the QL approximation.

Separate from our motivations for using the QL approximation, Bouchet et al.\ have argued that in the regime of large $\g$ the flow becomes predominantly zonal and the QL approximation becomes rigorously valid \cite{bouchet:2013}.  Our present study examines the regime in parameter space where $\g$ is not asymptotically large, for it is in this regime where \zfs{} are born at low amplitudes from turbulence.

The QL approximation has also been used by Herring in the study of thermal convection \cite{herring:1963}, where the only nonlinear interaction retained was between a horizontally-averaged temperature and the fluctuating temperature and velocity; nonlinear interactions between the fluctuating quantities were discarded.  At large Rayleigh number, this approximation was able to reproduce some of the qualitative features observed in experiments.

\section{CE2}
\label{sec:CE2}
The eddy quantity $w'$ fluctuates rapidly in space and time.  Averaging over these turbulent fluctuations enables one to work with smoothly varying functions.  Such statistical approaches provide one path to gaining physical insight.  Sometimes statistical turbulence theories strive for quantitative accuracy, which requires rather complicated methods \cite{krommes:2002}, but we eschew those methods here because they are not required for investigation of the QL system.

We consider an average of the QL system \eref{qlsystem}.  The resultant framework is called CE2, or the second-order cumulant expansion.  (If one performs a cumulant expansion of the original equations and truncates all cumulants higher than second order, one reaches the same equations.)  Derivations can be found in \citet{farrell:2003,marston:2008} though we follow \citet{srinivasan:2012} because there are advantages to that formulation.  The full derivation can be found in Appendix~\ref{app:CE2derivation}, but we give here a brief overview of the procedure.  One defines the two-point, one-time correlation function of vorticity using a zonal average as
	\begin{equation}
		W(x,y_1,y_2,t) =  \frac{1}{L_x} \int_0^{L_x} d\ol{x}|_x\, w'(x_1,y_1,t) w'(x_2,y_2,t),
		\label{Wdef}
	\end{equation}
where $L_x$ is some averaging length, the integration is over the sum coordinate $\ol{x} = \frac12 (x_1 + x_2)$, and the difference coordinate $x = x_1 - x_2$ is held fixed.  The correlation function $\Psi$ of stream function can be defined similarly as
	\begin{equation}
		\Psi(x,y_1,y_2,t) =  \frac{1}{L_x} \int_0^{L_x} d\ol{x}|_x\, \psi'(x_1,y_1,t) \psi'(x_2,y_2,t).
		\label{Wdef}
	\end{equation}
One finds an evolution equation for $W$ by taking a time derivative of \eref{Wdef}, substituting the expression for $\dot{w}'$ from \eref{QLwprime}, and performing the average.  Under an ergodic assumption, the zonal average is equivalent to a statistical ensemble average, and the stochastic forcing can be averaged to a deterministic quantity.  Then one performs a linear coordinate transform to the sum and difference variables $\ol{y} = \frac12 (y_1 + y_2)$ and $y = y_1 - y_2$.  In the \zf{} equation \eref{QLU}, the Reynolds stress term can be related to~$\Psi$.  The final equations are\footnote{To transform to the conventional plasma coordinates and notation, it follows from \eref{eom:QLqgtoghm} that one needs to make the substitution $(x,y,\ol{y},\b,U) \mapsto (-y,x,\ol{x},\k,-U)$.}
	\begin{subequations}
		\label{CE2}
	\begin{gather}
		\partial_t W(x,y\mid \ybar, t) + (U_+ - U_-) \partial_x W - \bigl(\ol{U}_+'' - \ol{U}_-''\bigr) \biggl( \nablabarsq + \frac{1}{4} \partial_\ybar^2 \biggr) \partial_x \Psi \notag \\
			\qquad \qquad - \bigl[2\b - \bigl(\ol{U}_+'' + \ol{U}_-''\bigr)\bigr] \partial_\ybar \partial_y \partial_x \Psi = F(x,y) - 2\m W - 2 \n D_h W, \label{CE2W} \\
		\bigl[ \partial_t + \m +  \n (-1)^h \partial_\ybar^{2h} \bigr] \ol{I} U(\ybar, t) + \partial_\ybar \partial_y \partial_x \Psi(0,0 \mid \ybar,t) = 0, \label{CE2U}
	\end{gather}
	\end{subequations}
where $U(\ybar,t)$ is the \zf{} velocity, and
	\begin{gather}
		U_\pm \defineas U\bigl(\ybar \pm \frac12 y, t\bigr), \\
		\ol{U}_\pm'' \defineas U_\pm'' - \azf \LD U_\pm, \\
		\nablabarsq \defineas \nabla^2 - \LD = \partial_x^2 + \partial_y^2 - \LD, \\
		\ol{I} \defineas 1 - \azf \LD \partial_\ybar^{-2},
	\end{gather}
$F(x,y)$ is the covariance of the external forcing, and $D_h$ is the hyperviscosity operator, given by
	\begin{equation}
		D_h = (-1)^h \frac{1}{2} \left\{ \left[ \partial_x^2 + \left( \partial_y + \frac12 \partial_{\ol{y}}\right)^2\right]^h + \left[ \partial_x^2 + \left( \partial_y - \frac12 \partial_{\ol{y}}\right)^2\right]^h \right\}.
	\end{equation}
In \eref{CE2U}, the notation $\partial_y \partial _x \Psi(0,0\mid\ybar,t)$ implies that the partial $x$ and $y$ derivatives are taken first, and then the result is evaluated at $x=y=0$.  It can be shown from the definitions that $W$ and $\Psi$ are related by
	\begin{equation}
		W(x,y\mid \ybar,t) = \left(\nablabarsq + \partial_y \partial_\ybar + \frac14 \partial_\ybar^2\right) \left(\nablabarsq - \partial_y \partial_\ybar + \frac14 \partial_\ybar^2\right) \Psi(x,y \mid \ybar, t).
		\label{WCrelation}
	\end{equation}
	
The use of the sum and difference coordinates $x,y,\ybar$ allows the structure of the theory and especially of the bifurcation to be more easily understood than in the original coordinates.  In the new coordinates $x$ and $y$ represent two-point separations and $\ybar$ represents the two-point average position.  If the turbulence were homogeneous, there would be no dependence on $\ybar$.  

The only assumption necessary for CE2 to be an exact description of the QL model is ergodicity in the zonal ($x$) direction, such that a zonal average is equivalent to an ensemble average.  No other assumptions are required because the QL model neglects the nonlinear eddy--eddy term that would give rise to a closure problem.  Alternatively, instead of the QL-based derivation, CE2 can be regarded as a truncated statistical closure of the NL model \cite{farrell:2003,farrell:2007,marston:2008,tobias:2011,tobias:2013}.  However, we prefer the former interpretation.

CE2, like the QL system, exhibits merging jets \cite{farrell:2007}.  Since CE2 is deterministic, if the system approaches a stable steady state then merging and branching of jets can only occur transiently.  Once the stable equilibrium is reached, the system is stuck there and no more dynamical behavior can occur.  However, if the QL system is not fully ergodic, then CE2 is not an exact description of it and dynamical behavior like merging or branching can persist even in a statistically steady state \cite{farrell:2003,bouchet:2013}.  Though ergodicity is often a useful idealization, lack of complete ergodicity is to be expected in any physical system.

Historically, CE2 was first studied by \citet{farrell:2003} under the name Stochastic Structural Stability Theory, or SSST.  Independently, \citet{marston:2008,tobias:2011} described the second-order cumulant expansion and called it CE2.  Later, \citet{srinivasan:2012} also independently derived CE2 from the quasilinear approximation and pointed out that SSST and CE2 are mathematically identical.  They opted to use the CE2 label, and we stick with the CE2 name for continuity.  Recently, the acronym for Stochastic Structural Stability Theory was rebranded from SSST to S3T \cite{constantinou:2013}.

\subsection{Symmetries of the CE2 Equation}
The CE2 equations \eref{CE2} inherit important symmetries of translation and reflection from the symmetries of the dynamical equation \eref{unifiedEOM}.  First, we note that because of the form of the expression $2\b - \bigl[\ol{U}''\bigl(\ybar + \tfrac12 y\bigr) + \ol{U}'' \bigl(\ybar - \tfrac12 y\bigr)\bigr]$, there can be no symmetry that changes the sign of $U$.  Second, we examine how the two expressions $U\bigl(\ybar + \tfrac12 y\bigr) - U \bigl(\ybar - \tfrac12 y\bigr)$ and $U\bigl(\ybar + \tfrac12 y \bigr) + U\bigl(\ybar - \tfrac12 y \bigr)$ behave under reflections.  (The same expressions with $U''$ instead of $U$ behave in the same manner.)

 If we change $\ybar \to -\ybar$ and let $\hat{U}(\ybar) = U(-\ybar)$, then for the first expression,
	\begin{align}
		\hat{U}\bigl(\ybar + \tfrac12 y\bigr) - \hat{U}\bigl(\ybar - \tfrac12 y\bigr) &= U\bigl(-\ybar - \tfrac12 y\bigr) - U\bigl(-\ybar + \tfrac12 y\bigr) \\
			&= -U\bigl(\hat{\ybar} + \tfrac12 y\bigr) + U\bigl(\hat{ \ybar} - \tfrac12 y\bigr)
	\end{align}
where in the second line we have let $\hat{\ybar} = -\ybar$.  Here, we note that the transformation has induced a minus sign.  On the other hand, for the second expression,
	\begin{align}
		\hat{U}\bigl(\ybar + \tfrac12 y\bigr) + \hat{U}\bigl(\ybar - \tfrac12 y \bigr) &= U\bigl(-\ybar - \tfrac12 y \bigr) + U \bigl(-\ybar + \tfrac12 y \bigr) \\
			&= U\bigl(\hat{\ybar} + \tfrac12 y\bigr) + U\bigl(\hat{ \ybar} - \tfrac12 y\bigr)
	\end{align}
the transformation has not induced a minus sign.

If we change $\ybar \to -\ybar, y \to -y$, still with $\hat{U}(\ybar) = U(-\ybar)$, then for the first expression,
	\begin{align}
		\hat{U}\bigl(\ybar + \tfrac12 y\bigr) - \hat{U}\bigl(\ybar - \tfrac12 y/2\bigr) &= U\bigl(-\ybar - \tfrac12 y \bigr) - U\bigl(-\ybar + \tfrac12 y\bigr) \\
			&= U \bigl(\hat{\ybar} + \tfrac12 \hat{y} \bigr) - U\bigl(\hat{ \ybar} - \tfrac12 \hat{y}\bigr)
	\end{align}
where in the second line we have let $\hat{y} = - y$.  Here, the transformation has not induced a minus sign.  Similarly, for the second expression, 
	\begin{align}
		\hat{U}\bigl(\ybar + \tfrac12 y \bigr) + \hat{U}\bigl(\ybar - \tfrac12 y \bigr) &= U\bigl(-\ybar - \tfrac12 y \bigr) + U \bigl(-\ybar + \tfrac12 y \bigr) \\
			&= U \bigl(\hat{\ybar} + \tfrac12 \hat{y} \bigr) + U \bigl(\hat{ \ybar} - \tfrac12 \hat{y} \bigr)
	\end{align}
the transformation has not induced a minus sign.

With the above relations, we can see that the equations have the following symmetries:
	\begin{subequations}
	\label{CE2symmetries}
	\begin{align}
		\ybar &\to \ybar + \de \ybar, \label{symmetry1} \\
		x, \ybar &\to -x, -\ybar, \qquad \text{assuming $F(x,y) = F(-x,y)$,}\\
		y, \ybar &\to -y, -\ybar, \qquad \text{assuming $F(x,y) = F(x,-y)$,}\\
		x, y &\to -x, -y. \label{exchangesymmetry}
	\end{align}
	\end{subequations}
In other words, if $\{W(x,y\mid \ybar,t),U(\ybar,t)\}$ is a solution, then the symmetries give us other solutions:
	\begin{subequations}
	\begin{gather}
		\{W(x,y\mid \ybar + \de \ybar, t), U(\ybar + \de \ybar, t)\}, \\
		\{W(-x,y \mid -\ybar, t), U(-\ybar, t)\}, \\
		\{W(x,-y \mid -\ybar,t), U(-\ybar, t)\},
	\end{gather}
	\end{subequations}
where $\de \ybar$ is some constant translational shift.  The symmetry \eref{exchangesymmetry}, dubbed the exchange symmetry, does not give a new solution because it is always obeyed by the correlation function such that $W(x,y \mid \ybar,t) = W(-x,-y \mid \ybar,t)$ \cite{srinivasan:2012}.  Equation \eref{CE2symmetries} gives the symmetries obeyed by the \emph{equations}.  If all of the reflection symmetries are obeyed by the \emph{solutions}, then one has that $U$ is even in $\ybar$,
	\begin{equation}
		U(\ybar, t) = U(-\ybar, t),
	\end{equation}
and three relations for $W$:
	\begin{equation}
		W(x,y \mid \ybar, t) = W(-x,-y \mid \ybar,t) = W(x,-y \mid -\ybar, t) = W(-x,y \mid -\ybar, t).
	\end{equation}

These are the symmetries in real space.  We can also state what the corresponding symmetries are in Fourier space.  It is not difficult to see that a reflection symmetry in real space corresponds to a reflection symmetry in Fourier space, which comes directly from the definition of the Fourier transform.  Suppose there is some equation in $x$, and that $f(x)$ and $\ol{f}(x)$ are both solutions, where $\ol{f}(x) = f(-x)$.  Let $\hat{f}(k) = \mathcal{F}[f(x)]$ and $\hat{\ol{f}}(k) = \mathcal{F}\bigl[\ol{f}(x) \bigr]$.  Then both $\hat{f}(k)$ and $\hat{\ol{f}}(k)$ will be solutions in Fourier space, and they will be related by
	\begin{align*}
		\hat{\ol{f}}(k) &= \mathcal{F} \bigl[\ol{f}(x) \bigr] \\
			&= \int dx\, e^{-ikx} \ol{f}(x) \\
			&= \int dx\, e^{-ikx} f(-x) \\
			&= \int dx\, e^{ikx} f(x) \\
			&= \hat{f}(-k).
	\end{align*}
Thus, the reflection symmetries correspond to
	\begin{subequations}
	\begin{align}
		k_x, \ybar &\to -k_x, -\ybar, \\
		k_y, \ybar &\to -k_y, -\ybar, \\
		k_x, k_y &\to -k_x, -k_y, 
	\end{align}
	\end{subequations}
if $\ybar$ is kept in real space, or
	\begin{subequations}
	\begin{align}
		k_x, k_\ybar &\to -k_x, -k_\ybar, \\
		k_y, k_\ybar &\to -k_y, -k_\ybar, \\
		k_x, k_y &\to -k_x, -k_y, 
	\end{align}
	\end{subequations}
if $\ybar$ is also transformed to Fourier space.

We also point out that as a result of the exchange symmetry, the mixed real space--Fourier space quantity $W(\v{k} \mid \ybar, t)$ is purely real and must satisfy $W(\v{k} \mid \ybar, t) = W(-\v{k} \mid \ybar, t)$.

\subsection{Nonlinearly Conserved Quantities}
The average energy density and enstrophy density are conserved by nonlinear interactions in the QG-mHME and its quasilinear variant.  Accordingly, they are also conserved in CE2.  We derive here the formulas for the energy and enstrophy density within the CE2 description.

First we split the average energy density, given in \eref{eom:energydensity}, into a contribution from zonal and eddy contributions:
	\begin{align}
		E_a &= -\frac{1}{2} \frac{1}{L_y} \int_0^{L_y} dy\, \ol{w}(y) \ol{\psi}(y) - \frac{1}{2} \frac{1}{L_x L_y} \int_0^{L_x} dx \int_0^{L_y} dy\, w'(x,y) \psi'(x,y) \notag \\
			&\defineas E_{a;ZF} + E_{a;\text{eddy}}.
	\end{align}
For the \zf{} contribution, recall that $U(\ybar) = - \partial_\ybar \ol{\psi}$ and $\ol{w} = -\partial_\ybar \ol{I} U$, so that
	\begin{align}
		E_{a;ZF} &= -\frac{1}{2} \frac{1}{L_y} \int_0^{L_y} d\ybar\, \ol{w}(\ybar) \ol{\psi}(\ybar) \notag \\
			&= \frac{1}{2} \frac{1}{L_y} \int_0^{L_y} d\ybar\, [\partial_\ybar \ol{I} U] \ol{\psi} \notag \\
			&= \frac{1}{2} \frac{1}{L_y} \int_0^{L_y} d\ybar\, U(\ybar) \ol{I} U(\ybar).
	\end{align}
For the eddy contribution, we first define a symmetrized quantity
	\begin{equation}
		E_s(x,y \mid \ybar) = -\frac{1}{2} \frac{1}{L_x} \int_0^{L_x} d\xbar |_x \bigl[w'(x_1,y_1) \psi'(x_2,y_2) + \psi'(x_1,y_1) w'(x_2,y_2) \bigr].
	\end{equation}
Using the same techniques as in Appendix~\ref{app:CE2derivation}, we find
	\begin{equation}
		E_s(x,y \mid \ybar) = - \biggl( \nablabarsq + \frac{1}{4} \partial_\ybar^2 \biggr) \Psi(x,y \mid \ybar).
	\end{equation}
The energy density $E_a$ is obtained by setting $x_1=x_2$ and $y_1=y_2$, i.e., taking $x=0$ and $y=0$, then integrating over $\ybar$:
	\begin{equation}
		E_{a;\text{eddy}} = \frac{1}{2} \frac{1}{L_y} \int_0^{L_y} d\ybar \, E_s(0,0 \mid \ybar).
	\end{equation}

Similarly, the enstrophy density can be split into \zf{} and eddy contributions, with
	\begin{align}
		W_a &= W_{a;ZF} + W_{a;\text{eddy}}, \\
		W_{a;ZF} &= \frac{1}{2} \frac{1}{L_y} \int_0^{L_y} d\ybar\, \bigl[\partial_\ybar \ol{I} U(\ybar) \bigr]^2, \\
		W_{a;\text{eddy}} &= \frac{1}{2} \frac{1}{L_y} \int_0^{L_y} d\ybar\, W(0,0 \mid \ybar).
	\end{align}

\subsection{Wigner--Moyal Formalism}
\label{sec:wignermoyal}
The Wigner--Moyal formalism, which has been used in studies of wave physics in inhomogeneous media \cite{hall:2002}, is basically equivalent to CE2.  The Wigner distribution function, assuming an appropriate average is used in its definition, is closely related to the CE2 correlation function $W$: they are both the two-point, one-time, second-order correlation of fluctuations.  The Wigner--Moyal equation, which describes the evolution of the distribution function, is the analog of the CE2 equation \eref{CE2W}.  The Wigner--Moyal formalism has also been used as the starting point for a few calculations involving zonal flows \cite{mendonca:2011,mendonca:2012,mendonca:2014}.  Those papers perform some analyses similar to what is in this thesis, but they made several further approximations without stating regimes of validity.  In contrast, this thesis provides a deep understanding of the theory without further approximation and also frequently compares analytic results with numerical results to ensure correct understanding.

\subsection{Wave Kinetic Equation}
\label{sec:wavekinetic}
Other previous studies of zonal flows have used a \emph{wave-kinetic} framework of inhomogeneous turbulence \cite{dyachenko:1992,smolyakov:2000b,diamond:2005,krommes:2000,manin:1994,krommes:book}.  The wave-kinetic formalism, like CE2, describes fluctuations using a second-order, two-point correlation function.  In these studies, the wave-kinetic formalism is restricted such that the length scale of the inhomogeneity must be much longer than the small scales of the turbulence.  A traditional viewpoint is that the wave kinetic equation describes turbulence as wavepackets that propagate through an inhomogeneous medium.  CE2, on the other hand, is an exact description of the QL equations and makes no approximation or restriction on length scales.

The disparate-scale asymptotic limit of CE2 recovers the wave kinetic equation.  In \eref{CE2W}, assume $\partial_\ybar \ll \partial_y$, Taylor expand the terms $U(\ybar \pm \frac12 y)$, and Fourier transform $(x,y) \to (k_x,k_y)$.  After switching to using $\mathcal{N}(\v{k} \mid \ybar) = (1 - \azf \ol{k}^{-2} \LD ) W(\v{k} \mid \ybar)$ as the dependent variable rather than $W(\v{k} \mid \ybar)$, the disparate-scale form of CE2 takes on wave-kinetic form.

\chapter{Zonostrophic Instability and Beyond}
\label{ch:beyondzonostrophic}
This chapter develops the physics at the core of this thesis.  First, we review zonostrophic instability (ZI).  In ZI, a statistically homogeneous turbulent state is unstable to coherent, zonally-symmetric perturbations.  These perturbations grow into zonal flows (ZFs).  Since ZI is an instability of a turbulent, time-dependent but statistically-steady state, analysis of it requires a statistical formalism.  CE2 provides the simplest such formalism, and indeed, the instability was discovered through the CE2 framework.  The fundamental dynamical equations are too complicated for analytic progress.

ZI has been explored numerically and calculated analytically in detail.  We provide a thorough review of the analytic calculation and explore certain limits of the dispersion relation.  Our calculation mildly generalizes previous work because we allow for finite deformation radius (or Larmor radius) $L_d$ as opposed to infinite $L_d$ and we allow for viscosity in addition to a scale-independent drag.\footnote{\citet{srinivasan:2012} have some calculations in an appendix that include viscosity, but there is an error in the way the wavevector dependence of viscosity terms is treated.}

We also draw a connection between ZI and generalized modulational instability.  By modulational instability we mean the instability of a primary mode to a secondary mode.  We find that with the ZI dispersion relation from CE2, we can recover as a special case a previously-derived dispersion relation of modulational instability.  This discovery suggests that CE2 may be useful for future investigations of modulational or secondary instabilities or generalizations thereof.

We then extend analytic understanding of ZI beyond a linear stability calculation into the regime of nonlinearly interacting zonal flows and turbulence.  We connect the generation of zonal flows to the large body of literature of pattern formation.  At a basic level, zonal flows appear in a spontaneous symmetry-breaking bifurcation where the broken symmetry is statistical homogeneity in space.  The mechanism of the symmetry breaking is ZI.

We perform a bifurcation analysis, which yields numerous insights.  We construct explicit solutions to the nonlinear CE2 equations, and we discover important and unexpected features.   First, we find that the zonal flow wavelength is not unique.  Many wavelengths allow a steady-state solution to the equations.  Second, only some of these wavelengths correspond to solutions which are stable.  Unstable wavelengths must evolve to reach a stable wavelength; this process manifests as merging jets.  Consequently, we are able to provide a theoretical basis for the well-known merging of jets along with a simple PDE that demonstrates the behavior.  Furthermore our work links the merging of jets to the large body of research of \emph{defects}, providing new avenues for research into jet dynamics.  Our results provide a substantial theoretical foundation for further understanding of turbulence and zonal flows. 

This chapter is structured as follows.  \secref{sec:zi:toymodel} introduces a phenomenological model of the bifurcation.  As a zero-dimensional model, it orients the reader before the plunge into the full problem with its complexity of spatial dependence.  A review of ZI is provided in \secref{sec:zi}.  This calculation shows that a state of turbulence without zonal flows can be unstable to zonal flow perturbations. We discuss how ZI relates to modulational instability in \secref{sec:mi}.  Then, in \secref{sec:beyondzi} (with details in Appendix~\ref{app:glderivation}), we perform a full bifurcation analysis into the regime of nonlinearly interacting eddies and zonal flows.

\section{Phenomenological Bifurcation Model of Zonostrophic Instability}
\label{sec:zi:toymodel}
A zero-dimensional phenomenological model illustrates some of the key features of ZI and the bifurcation to a state with \zfs{} \cite{parker:2014}.  The system is a variant of another treatment which models the appearance of shear flows in the L--H transition in plasmas \cite{diamond:1994}.  However, the model we present more closely mirrors the structure and behavior of the CE2 equations.  The model includes three interacting degrees of freedom: the homogeneous, or spatial average, part of the fluctuation covariance $W_h$; the inhomogeneous, or deviation from the spatial average, part of the fluctuation covariance $W_i$; and the \zf{} amplitude (not covariance) $z$.  Both $W_i$ and $z$ may be positive or negative.  The model is given by
	\begin{subequations}
	\begin{align}
		\dot{W}_h &= -\m W_h - \a W_i z + F, \\
		\dot{W}_i &= -\m W_i + \eta W_h z, \\
		\dot{z} &= -\n z + \a W_i.
	\end{align}
	\end{subequations}
	
The structure of the model reflects that of the CE2 equations \eref{CE2} in several ways.  To affect the homogeneous part of the turbulence, the \zf{} interacts only with the inhomogeneous part.  Similarly, the \zf{} interacts with the homogeneous part to affect the inhomogeneous part.  (CE2 also contains an interaction between the \zfs{} and the inhomogeneous part to affect the inhomogeneous part; this is neglected here, as in ZI analysis.)  Finally, it is the inhomogeneous part of the turbulence that is responsible for driving steady \zfs{}.  The above model neglects the eddy self-nonlinearities as in CE2.  The appearance of the same coefficient $\a$ in $\dot{W}_h$ and in $\dot{z}$ reflects the conservation by the nonlinear interactions of an energy-like quantity $W_h + \frac12 z^2$.  We take all of the coefficients $\m, \n, \a, \eta, F$ to be positive.

The model allows a homogeneous equilibrium, in which forcing $F$ is balanced by dissipation $\m$, and for which $W_i$ and $z$ are zero.  The homogeneous equilibrium is unstable if $F \eta \a / \m^2 \n > 1$.  Increasing the forcing or decreasing the dissipation tends to make the homogeneous equilibrium more zonostrophically unstable, which is characteristic of the more rigorous analysis.

When the homogeneous equilibrium goes unstable, it connects to an inhomogeneous equilibrium at $W_h = \m\n / \eta\a$, $W_i^2 = (\n/\a^2)(F - \m^2\n / \eta\a)$, $z=\a W_i / \n$.  This new equilibrium is stable (when it exists), which can be seen by constructing the eigenvalues graphically from the characteristic polynomial.  Furthermore, there are actually two symmetric solutions, with either sign of $z$ and $W_i$.  There is thus a supercritical pitchfork bifurcation; this feature is also present in the complete model, but the discrete $z \to -z$ symmetry becomes a continuous symmetry associated with translational invariance.  

The model demonstrates some of the qualitative features of ZI, although in simplifying it we have tossed out spatial dependence.  Spatial dependence makes the problem both immensely more complicated and immensely more interesting.  The CE2 equations contain the full spatial dependence.  Detailed analysis of ZI and beyond proceeds in the next few sections.
\section{Zonostrophic Instability}
\label{sec:zi}
In this section we review ZI, for which substantial understanding has been recently obtained \cite{srinivasan:2012,bakas:2011}.  To give a brief overview, ZI refers to an instability where a state of homogeneous turbulence without \zfs{} can be unstable to \zf{} perturbations.  In the regime where ZI is present, \emph{inhomogeneous turbulence} results.  The instability as well as the nonlinear growth and saturation can be handled self-consistently within the CE2 framework.  This section is devoted to the study of the instability of the homogeneous equilibrium, with later sections handling the nonlinear saturation.

We examine the homogeneous equilibrium of the CE2 equations, which has no zonal flows.  We calculate the linear response of the equilibrium to zonal perturbation.  Much analytic progress is possible, which provides substantial insight.  Although the final dispersion relation must be solved numerically, it can be reduced to a single nonlinear equation.  In some regimes of parameter space the equilibrium is unstable, and the instability has been named \emph{zonostrophic instability}.  This instability has been studied analytically in detail \cite{srinivasan:2012} and aspects of it have also been examined numerically \cite{farrell:2007, bakas:2011}.

As a control parameter $\r$ is varied, the homogeneous state becomes zonostrophically unstable \cite{srinivasan:2012,farrell:2007}.  Physically, ZI occurs when dissipation is overcome by the mutually reinforcing processes of eddy tilting by \zfs{} and production of Reynolds stress forces by tilted eddies.  The instability eigenmode consists of perturbations spatially periodic in $\ybar$ with zero real frequency \cite{srinivasan:2012}, so that zonostrophic instability arises as a Type I$_s$ instability \cite{cross:1993} of homogeneous turbulence.

\subsection{CE2 Homogeneous Equilibrium}
A homogeneous, steady-state solution of the CE2 equations always exists, arising from a simple balance between forcing and dissipation.  This solution is
	\begin{subequations}
		\label{zi:homogsoln}
	\begin{align} 
		W_H &= (2\m + 2\n D_h)^{-1} F, \label{zi:homogW} \\
		U &= 0,	\label{zi:homogU}
	\end{align}
	\end{subequations}
where the $H$ subscript denotes homogeneous.  From \eref{WCrelation} it is easy to relate $W_H$ and $\Psi_H$:
\begin{equation}
		W_H(x,y) = \ol{\nabla}^4 \Psi_H(x,y). \label{zi:WandChomog}
	\end{equation}
We can also give the result in Fourier space by applying the continuous Fourier transform
	\begin{equation}
		W(k_x,k_y) = \int dx\, dy\, e^{-ik_xx} e^{-i k_y y} W(x,y).
	\end{equation}
For the homogeneous equilibrium with $k_\ybar=0$, we have $2\n D_h \to 2 \n k^{2h}$, where $k^2 = k_x^2 + k_y^2$.  This gives for the homogeneous equilibrium
	\begin{equation}
		W_H(k_x,k_y) = \frac{F(k_x,k_y)}{2(\m + \n k^{2h})}
		\label{zi:homog_eqb}
	\end{equation}
and
	\begin{equation}
		W_H(k_x,k_y) = \ol{k}^4 \Psi_H(k_x,k_y),
	\end{equation}
where
	\begin{equation}
		\kbsq \defineas k^2 + \LD. \label{zi:kbsqdefinition}
	\end{equation}

\subsection{Linearization about the Homogeneous Equilibrium}
The homogeneous equilibrium is linearly stable in a certain regime of parameters.  To determine its stability one calculates the dispersion relation corresponding to ZI.  One considers perturbations about the equilibrium in \eref{zi:homogsoln}.  The derivation given here closely follows that given by \citet{srinivasan:2012}.  Because the equilibrium is independent of $\ybar$ and $t$, the $\ybar$ and $t$ dependence of the perturbations can be Fourier transformed.  The fields are written as
	\begin{subequations}
		\label{zi:perturbationfields}
	\begin{align}
		W(x,y \mid \ybar, t) &= W_H(x,y) + \de W(x,y) e^{\l t} e^{iq\ybar}, \\
		U(\ybar, t) &= \de U e^{\l t} e^{iq\ybar},
	\end{align}
	\end{subequations}
where $q$ is the \zf{} wavenumber and $\l$ is the eigenvalue.

We now substitute the perturbations into \eref{CE2W} and \eref{CE2U} and linearize.  We use that
	\begin{align}
		U_\pm &= U(\ybar \pm y/2) \to \de U e^{\l t} e^{iq \ybar} e^{\pm iqy/2}, \\
		U''_\pm &\to -q^2 U_\pm, \\
		\ol{U}''_\pm &\to -q^2 U_\pm - \azf \LD U_\pm = -\qbsq \de U e^{\l t} e^{iq\ybar} e^{\pm iqy/2}, \label{zi:Ubarpp}
	\end{align}
where $\qbsq \defineas q^2 + \azf \LD$.  The linearized equations are
	\begin{subequations}
	\begin{align}
		\l \de W &+ \de U \bigl( e^{iqy/2} - e^{-iqy/2} \bigr) \partial_x W_H + \qbsq \de U \bigl(e^{iqy/2} - e^{-iqy/2} \bigr) \nablabarsq \partial_x \Psi_H \notag \\
				& -i 2\b q \partial_x \partial_y \de \Psi = -2 \m \de W - 2 \n D_h \de W, \label{zi:linxyW} \\
		\bigl(\l + &\m + \n q^{2h} \bigr) \bigl(1 + \azf \LD q^{-2} \bigr)  \de U = -i q \partial_x \partial_y \de \Psi(0,0). \label{zi:linxyU}
	\end{align}
	\end{subequations}
Note that we can express $1 + \azf \LD q^{-2} = \qbsq / q^2$.  If for wavenumber $q$, $\l$ is an eigenvalue with eigenvector $(\de W, \de U)$, then for wavenumber $-q$,  $\l^*$ is an eigenvalue with eigenvector $(\de W^*, \de U^*)$.

It is convenient to Fourier transform in both $x$ and $y$ as well.  For the perturbations \eref{WCrelation} becomes, with $\partial_\ybar \to iq$, $\partial_y \to ik_y$, and $\nablabarsq \to -\kbsq$,
	\begin{align}
		\de W(k_x,k_y) &= \ol{h}_+^2 \ol{h}_-^2 \de \Psi(k_x,k_y), \\
		h_\pm^2 &= k_x^2 + (k_y \pm q/2)^2, \\
		\ol{h}_\pm^2 &= h_\pm^2 + \LD.
	\end{align}
We also use
	\begin{align}
		\partial_x \partial_y f(x,y)|_{x,y =0,0} &= \partial_x \partial_y \frac{1}{(2\pi)^2} \int dk_x\, dk_y\, e^{ik_x x} e^{ik_y y} f(k_x,k_y) |_{x,y=0,0} \\
			&= -\frac{1}{(2\pi)^2} \int dk_x\, dk_y\, k_x k_y f(k_x, k_y).
	\end{align}

Thus the ZF equation \eref{zi:linxyU} becomes
	\begin{equation}
		\frac{\qbsq}{q^2} \bigl(\l + \m + \n q^{2h} \bigr)\, \de U = i q \int dk_x\, dk_y\, \frac{k_x k_y}{(2\pi)^2} \de \Psi(k_x, k_y).
	\end{equation}

Now we transform the DW equation.  The second and third term of \eref{zi:linxyW} can be combined using \eref{zi:WandChomog} as
	\[ \de U \bigl(e^{iqy/2} - e^{-iqy/2}\bigr) \nablabarsq \bigl( \nablabarsq + \qbsq \bigr) \partial_x \Psi_H \].
Using the property that $\mathcal{F} \bigl[e^{iqy/2} f(y) \bigr] = \hat{f}(k_y - q/2)$, the Fourier transform of
	\[ e^{\pm iqy/2} \nablabarsq \bigl(\nablabarsq + \qbsq \bigr) \partial_x \Psi_H(x,y) \]
is
	\begin{gather}
		ik_x \bigl[\LD + k_x^2 + (k_y \mp q/2)^2 \bigr] \bigl[\LD + k_x^2 + (k_y \mp q/2)^2 - \qbsq \bigr] \Psi_H \bigl(k_x, k_y \mp \tfrac12 q\bigr) \notag\\
		= ik_x \ol{h}_\mp^2 \bigl(\ol{h}_\mp^2 - \qbsq \bigr) \Psi_H \bigl(k_x, k_y \mp \tfrac12 q\bigr).
	\end{gather}
Also, $D_h$ transforms to
	\begin{equation}
		D_h = \frac{1}{2} \bigl( h_+^{2h} + h_-^{2h}\bigr).
	\end{equation}

Equation \eref{zi:linxyW} then becomes
	\begin{align}
		\l \hbpsq \hbmsq \de \Psi(k_x,k_y) &+ ik_x \de U \left[ \hbmsq \bigl(\hbmsq - \qbsq \bigr) \Psi_H\bigl(k_x,k_y- \tfrac12 q \bigr) - \hbpsq (\hbpsq - \qbsq) \Psi_H\bigl(k_x,k_y+ \tfrac12 q\bigr) \right] \notag \\
			&+ i2\b q k_x k_y \de \Psi = -\bigl[2\m + \n (h_+^{2h} + h_-^{2h}) \bigr] \hbpsq \hbmsq \de \Psi.
	\end{align}
Let
	\begin{equation}
		\P_H^{\pm} \defineas \ol{h}_\pm^2 \bigl( \ol{h}_\pm^2 - \qbsq\bigr) \Psi_H \bigl(k_x, k_y \pm \tfrac12 q\bigr).
	\end{equation}
Rearranging slightly, the linearized equations about the homogeneous equilibrium are
	\begin{subequations}
	\begin{gather}
		\left[ \hbpsq \hbmsq \bigl(\l + 2\m + \n(h_+^{2h} + h_-^{2h})\bigr) + i 2\b q k_x k_y \right] \de \Psi(k_x,k_y) + i k_x \de U (\P_H^- - \P_H^+) = 0, \label{zi:linW}\\
		\frac{\qbsq}{q^2} \bigl(\l + \m + \n q^{2h} \bigr) \de U = iq \int dk_x\, dk_y\, \frac{k_x k_y}{(2\pi)^2} \de \Psi(k_x,k_y). \label{zi:linU}
	\end{gather}
	\end{subequations}
Here, \eref{zi:linW} and \eref{zi:linU} are exact equations for the eigenvectors.  However, for given parameters, only certain values of $\l$ allow eigenvectors.  Those are the eigenvalues.  We can determine the values of $\l$ which give solutions by using \eref{zi:linW} to solve for $\de \Psi(k_x,k_y)$ in term of $\de U$, then substituting into \eref{zi:linU}.  A nonlinear equation results.  Once we know the eigenvalues $\l$, we can find the eigenvectors by taking some value for $\de U$ and using \eref{zi:linW} to give the $\de W(k_x,k_y)$.
	
\subsection{Dispersion Relation}
We now obtain the dispersion relation.  First we solve for $\de \Psi$ in terms of $\de U$:
	\begin{equation}
		\frac{\qbsq}{q^2} \bigl(\l + \m + \n q^{2h} \bigr) \de U = iq \int dk_x\, dk_y \frac{k_x k_y}{(2\pi)^2} \frac{-i k_x ( \P_H^- - \P_H^+) \de U}{\hbpsq \hbmsq \bigl(\l + 2\m + \n(h_+^{2h} + h_-^{2h})\bigr) + 2i\b q k_x k_y}.
	\end{equation}
We substitute this back into the equation for $\de U$, and also rewrite $\P_H^\pm$ in terms of $W_H$.  This yields the dispersion relation
	\begin{equation}
		\frac{\qbsq}{q^2} \bigl(\l + \m + \n q^{2h}\bigr) = q\L_- - q\L_+,
		\label{zi:dispersionrelation}
	\end{equation}
where
	\begin{equation}
		\L_\pm = \int \frac{dk_x dk_y}{(2\pi)^2} \frac{k_x^2 k_y \bigl(1 - \qbsq / \ol{h}_\pm^2 \bigr) W_H\bigl(k_x, k_y \pm \tfrac12 q\bigr)} {\bigl[\l + 2\m + \n\bigl(h_+^{2h} + h_-^{2h}\bigr) \bigr] \hbpsq \hbmsq + 2i \b q k_x k_y},
		\label{zi:lambdapm}
	\end{equation}
and $h_\pm^2 = k_x^2 + \bigl(k_y \pm \tfrac12 q \bigr)^2$ and $\ol{h}_\pm^2 = h_\pm^2 + \LD$.  Some algebraic manipulation shows that $\L_+ = -\L_-$ \cite{srinivasan:2012}.   This is done by first noting that $W(k_x,k_y) = W(-k_x,-k_y)$ for any correlation function, and then letting $k_x \to -k_x$ and $k_y \to -k_y$ in the integral for $\L_+$.

This dispersion relation was also obtained by \citet{carnevale:1982}, in a form that allowed for arbitrary inhomogeneities rather than only zonally symmetric ones.  That paper did not, however, remark on the connection to the generation of zonal flows.

Equation \eref{zi:dispersionrelation} is the general dispersion relation.  Following \citet{srinivasan:2012}, we also provide the specialized results for an isotropic turbulent background spectrum.  Although a purely isotropic spectrum is unlikely to obtain in practice when the beta effect is present, such an investigation helps to get a simplified dispersion relation, gain intuitive understanding, and isolate the physical consequences of various effects.

In $\L_-$, make the transformation $k_y' = k_y - \frac12 q$.  After working through the transformation, then dropping the prime on $k_y'$, we find
	\begin{equation}
		2q\L_- = \int \frac{dk_x dk_y}{(2\pi)^2} \frac{ 2qk_x^2 \bigl(k_y + \tfrac12 q\bigr) \bigl(1 - \qbsq/\kbsq \bigr) W_H(k_x,k_y)}{ \bigl[\l + 2\m + \n(h_{++}^{2h} + k^{2h})\bigr] \ol{h}_{++}^2 \kbsq + 2i\b qk_x(k_y + q/2)},
	\end{equation}	
where $h_{++}^2 = k_x^2 + (k_y+q)^2$ and $\ol{h}_{++}^2 = h_{++}^2 + \LD$.  Also note that one can write $h_{++}^2 - k^2 = 2q\bigl(k_y+ \tfrac12 q\bigr)$.  Now rewrite the integral using polar coordinates, with $k_x = k \sin \p$ and $k_y = -k \cos \p$, and note
	\begin{align}
	 	2q\bigl(k_y + \tfrac12 q \bigr) &= k^2 \bigl(n^2 - 2n \cos \p \bigr), \\
	 	h_{++}^2 &= k^2 \bigl(1 - 2n \cos \p + n^2 \bigr), \\
	 	\kbsq &= k^2 (1+m), \\
	 	\ol{h}_{++}^2 &= k^2 \bigl(1 - 2n \cos \p + n^2 + m \bigr),
	\end{align}
where $n \defineas q/k$ and $m \defineas (kL_d)^{-2}$.  Assuming the equilibrium is isotropic, $W_H(k_x,k_y) = W_H(k)$, then after some manipulation the dispersion relation can be put into the form
	\begin{equation}
		\frac{\qbsq}{q^2} \bigl(\l + \m + \n q^{2h}\bigr) = \frac{1}{\b} \int_0^\infty \frac{dk}{2\pi} k^2 \biggl( 1 - \frac{\qbsq}{\kbsq} \biggr) W_H(k) S\Biggl(\frac{(\l + 2\m) \kbsq}{\b q}, \frac{\n k^{2h} \kbsq}{\b q}, \frac{q}{k}, (kL_d)^{-2}, h \Biggr),
		\label{zi:isotropic_disprelation}
	\end{equation}
where
	\begin{gather}
		S(\chi,\eta,n,m,h) \defineas \int_0^{2\pi} \frac{d\p}{2\pi} K, \\
		K \defineas \frac{(n - 2 \cos \p) \sin^2 \p}{ \{ \chi + \eta[1 + (1 -2n \cos \p + n^2)^h] \} (1 - 2n \cos \p + n^2 + m) + i (n - 2 \cos \p) \sin \p}.
	\end{gather}

We now specialize to thin-ring forcing, where the wavevectors excited by the external forcing are confined to a thin ring in $\v{k}$-space.  We take
	\begin{equation}
		F(k) = 4\pi \ve k_f \de(k - k_f),
	\end{equation}
where $\ve$ is, in the case of $L_d \to \infty$, the total energy (density) input.  Then, from \eref{zi:homog_eqb}, the homogeneous equilibrium is
	\begin{equation}
		W_H(k) = \frac{2\pi\ve k_f}{\m + \n k_f^{2h}} \de(k - k_f).
	\end{equation}
Substituting this into \eref{zi:isotropic_disprelation}, the integral over the delta function is trivial and we obtain
	\begin{equation}
		\frac{\qbsq}{q^2} \bigl(\l + \m + \n q^{2h}\bigr) = \frac{\ve}{\b} \frac{k_f^3 \bigl(1- \qbsq/ \kbsq_f \bigr)}{\m + \n k_f^{2h}} S\Biggl( \frac{(\l+2\m)\kbsq_f}{\b q}, \frac{\n k_f^{2h} \kbsq_f}{\b q}, \frac{q}{k_f}, (k_f L_d)^{-2}, h \Biggr),
		\label{zi:disprelation_isotropic}
	\end{equation}
where $\ol{k}_f^2 = k_f^2 + \LD$.  This nonlinear equation for $\l$ involves only one integral---the polar integral in $S$---that must be computed numerically.

\subsection{Behavior of the Dispersion Relation}
We begin by showing some examples of the dispersion relation solved numerically.  In each case we use the thin-ring forcing just described.  We do not attempt to draw any definitive conclusions from the few examples we show here, but rather use them to get a general sense of how the dispersion relation behaves.  Then, we analytically explore a few limits of the dispersion relation.

\subsubsection{Numerical Results}
\begin{figure}
		\centering
		\includegraphics{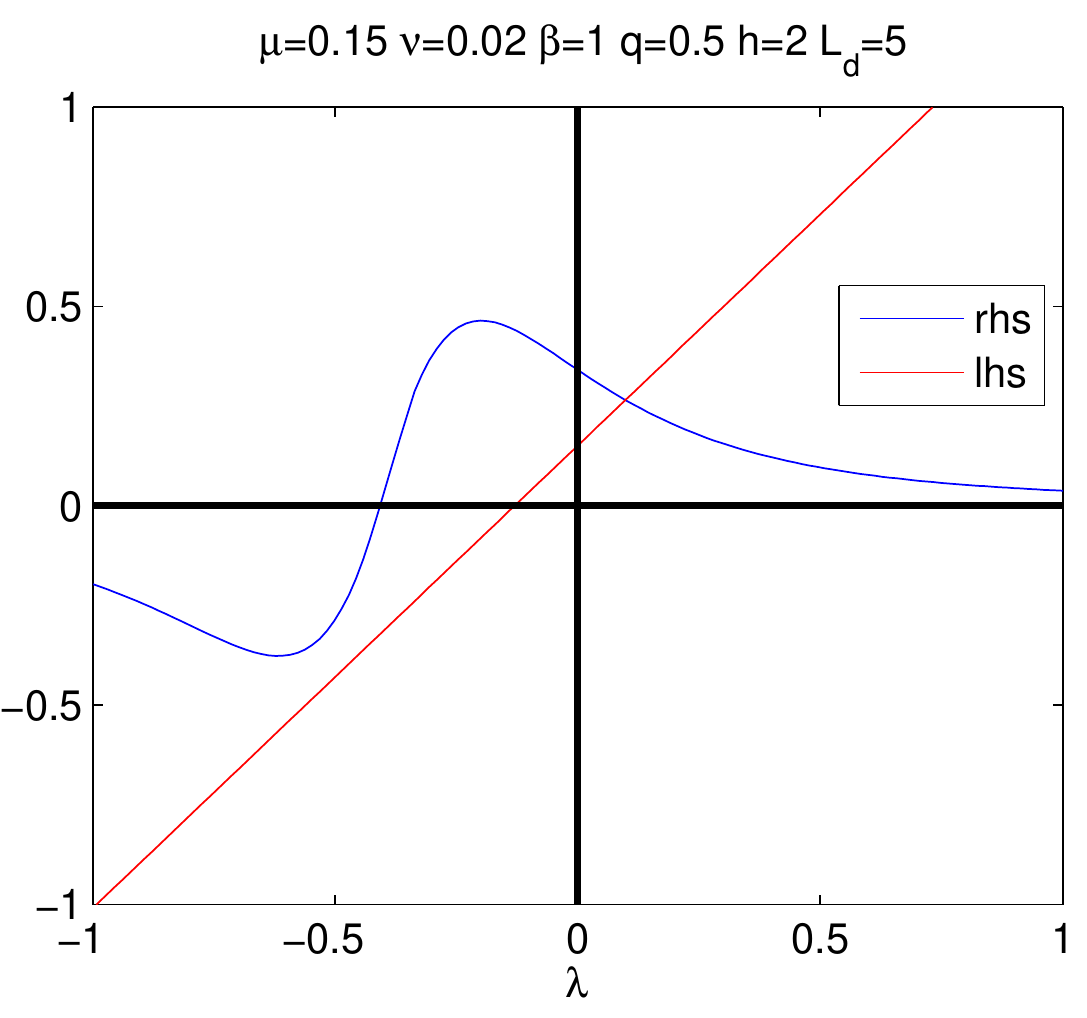}
		\caption{Plot of the RHS and LHS of \eref{zi:dispersionrelation} for real $\l$.  In this example, there is a positive eigenvalue solution, so zonostrophic instability occurs.}
		\label{fig:zi:rhs_lhs_plot}
	\end{figure}

First, in \figref{fig:zi:rhs_lhs_plot} we show the behavior of the LHS and RHS of \eref{zi:dispersionrelation} (more precisely, the isotropic version in \eref{zi:disprelation_isotropic}) as a function of $\l$.  In the example shown, there is an intersection at positive $\l$, so instability occurs.  As reported by \citet{srinivasan:2012}, numerical results indicate that all the unstable $\l$'s are pure real, and we have found the same.

\begin{figure}
		\centering
		\includegraphics{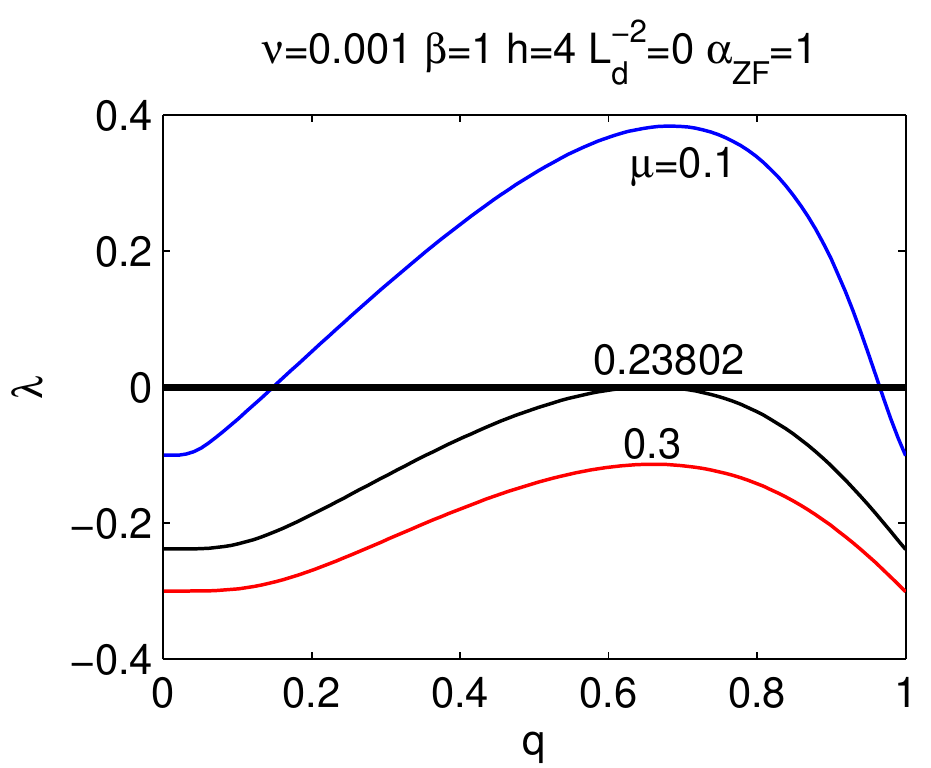}
		\caption{Plot of the dispersion relation $\l(q)$ for several values of $\m$.}
		\label{fig:zi:disprelation}
	\end{figure}

Next, we plot $\l(q)$ in \figref{fig:zi:disprelation}.  As a parameter such as $\m$ is varied, the equilibrium can go from being stable to zonal perturbations ($\l<0$ for all $q$), to having a single marginally stable mode ($\l=0$ at one $q$), to having a band of unstable modes ($\l>0$ for some $q$).

\begin{figure}
		\centering
		\includegraphics{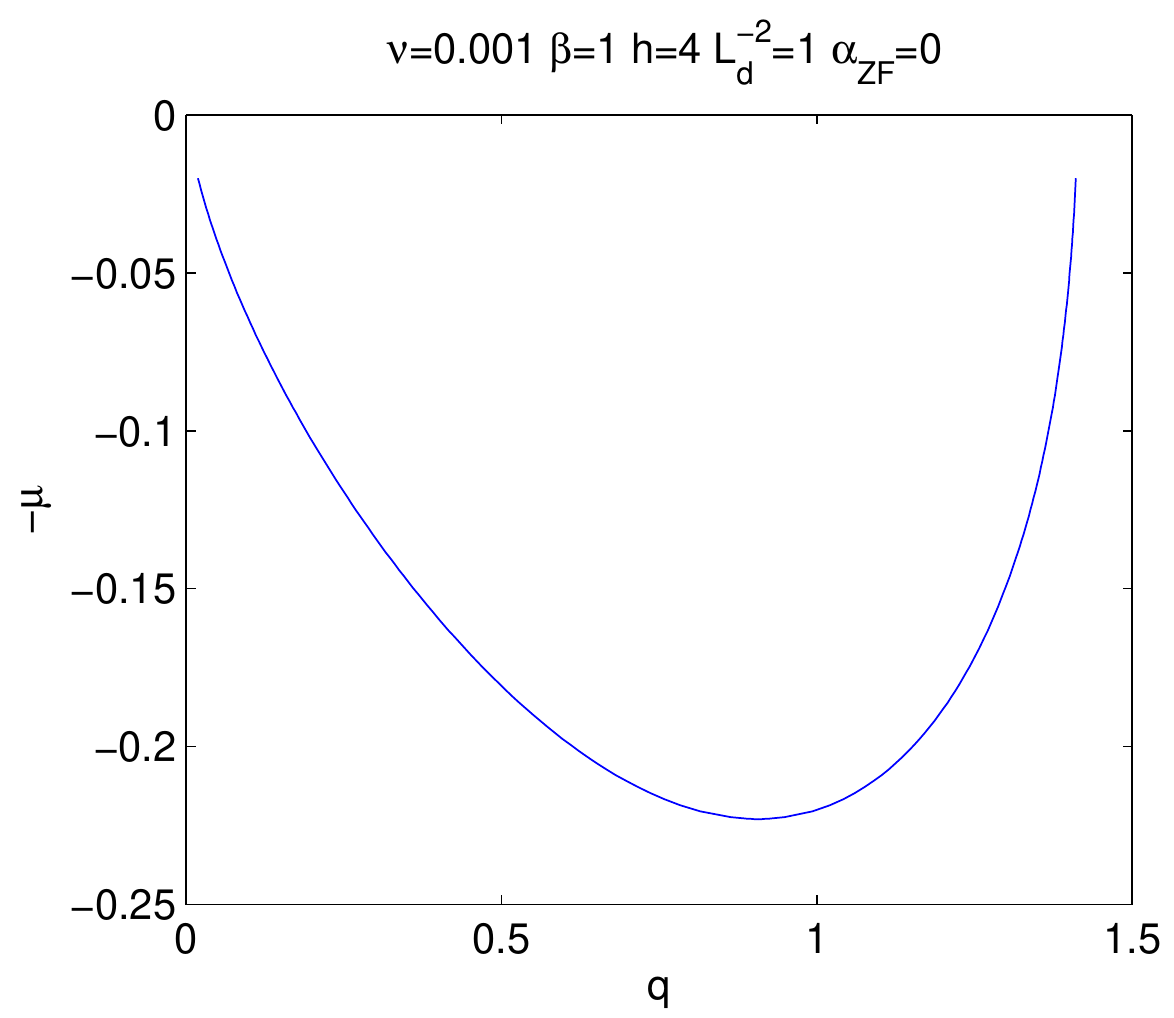}
		\caption{Neutral curve.  For values of $\m$ below the bottom of the curve, the homogeneous state is stable.  For other values of $\m$, the homogeneous state is unstable to perturbations with wavenumbers $q$ inside the neutral curve.}
		\label{fig:zi:neutralcurve}
	\end{figure}

Another result of interest is the \emph{neutral curve}.  If $\m$ is our control parameter, then the neutral curve is the curve in $(q,\m)$ space given by $\l(q,\m)=0$.  The neutral curve is the boundary between zonostrophically stable and unstable regions.  Below the bottom of the neutral curve, the homogeneous state is stable.  Above the bottom of the neutral curve, at a fixed value of $\m$, the homogeneous state is unstable to perturbations with wavenumbers $q$ inside the neutral curve.  An example of a neutral curve is shown in \figref{fig:zi:neutralcurve} (negative $\m$ is plotted because a neutral curve conventionally opens upward).  

\begin{figure}
		\centering
		\includegraphics{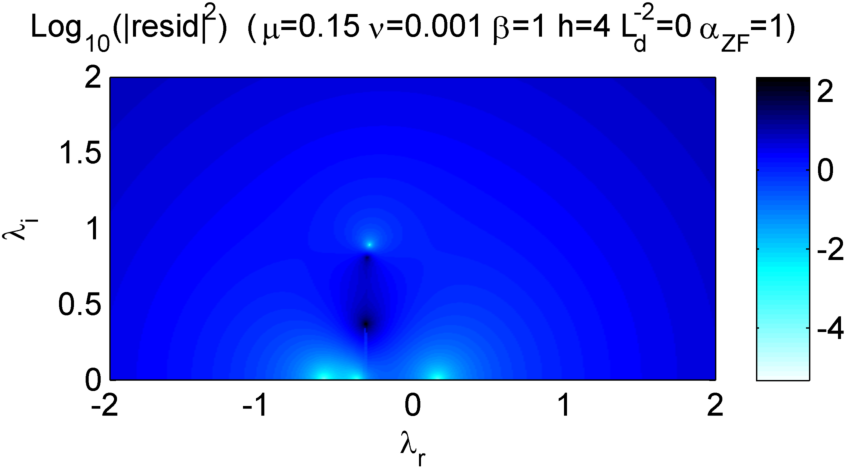}
		\caption{Magnitude squared of the residual of \eref{zi:dispersionrelation} plotted on a logarithmic scale, in the eigenvalue complex plane.  Along with three real eigenvalues, there is a single eigenvalue with nonzero imaginary part (along with its complex conjugate), which has negative real part.  This is at $q=0.5$.}
		\label{fig:zi:complexlambda}
	\end{figure}

So far we have only been concerned with real eigenvalues $\l$.  In looking to see whether there are any complex eigenvalues at all, including damped ones, we show in \figref{fig:zi:complexlambda} the residual of the dispersion relation (the difference of the RHS and LHS of \eref{zi:dispersionrelation}) as a function of $\l$ with other parameters fixed.  We find that there is a single damped, complex eigenvalue (along with its complex conjugate).

\subsubsection{Analytic Limits}
We now explore the dispersion relation analytically by examining various limits.  We explore the small $q$ (long ZF wavelength) limit and the effect of an isotropic spectrum.

First, it is easy to find the small $q$ limit of the general dispersion relation \eref{zi:dispersionrelation}.  Noting that $h_+^{2h} + h_-^{2h} = 2k^{2h} + \O(q^2)$ and $\ol{h}_+^2 \ol{h}_-^2 = \ol{k}^4 + \O(q^2)$, then keeping only to first order in $q$, we have
	\begin{equation}
		\L_\pm = \int \frac{d\v{k}}{(2\pi)^2} \biggl[c_0 \pm \frac12 q \pd{c_0}{k_y} \biggr] \biggl[W_H(k_x, k_y) \pm \frac12 q \pd{W_H}{k_y} \biggr] \frac{k_x^2 k_y}{[\l + 2\m + 2\n k^{2h}] \ol{k}^4 + 2i \b q k_x k_y},
	\end{equation}
where $c_0 \defineas 1 - \azf \LD \ol{k}^{-2}$.  Then 
	\begin{equation}
		q\L_- - q\L_+ = -q^2 \int \frac{d\v{k}}{(2\pi)^2}\, \frac{k_x^2 k_y}{[\l + 2\m + 2\n k^{2h}] \ol{k}^4 + 2i \b q k_x k_y} \pd{(c_0 W_h)}{k_y}.
	\end{equation}
If the dissipation terms $\m$ and $\n$ are not small enough to be negligible, then the $\b$ term in the denominator should be neglected as small in $q$.  However, if dissipation is negligible, then the $\b$ term cannot be ignored in general because $\l$ also turns out to be small in $q$.

Second, we examine the dispersion relation for the special case of an isotropic spectrum.  In the context of an infinite deformation radius $L_d$, the effect of an isotropic background spectrum has been studied before \cite{srinivasan:2012,bakas:2013b}.  Those studies concluded that for an isotropic background, $\b \neq 0$ is required for instability.  Additionally, they found that for an isotropic background, the eddies acted on long-wavelength zonal flows as a negative hyperviscosity instead of negative viscosity.  That is, the eddy forcing on the RHS of \eqref{zi:dispersionrelation} behaves as $q^4$ rather than $q^2$ at small $q$.  In this section, we study how these results change when finite deformation length $L_d$ is allowed.

The dispersion relation for an isotropic spectrum is given in \eref{zi:isotropic_disprelation} and involves the function $S(\chi,\eta,n,m,h)$.  For simplicity, we ignore viscosity which corresponds to setting $\eta$ to zero (in which case the hyperviscosity factor $h$ drops out also).  We explore the limit of large $\chi$, which could correspond to either small $\b$ or small $q$.  Asymptotic expansion of $S(\chi,n,m)$ for large $\chi$ reveals interesting behavior that can differ for finite vs.\ infinite $L_d$.

For infinite $L_d$ (i.e., $m=0$), $S$ behaves as\footnote{Validity of this formula requires that $1-n^2$ is not too small.} \cite{srinivasan:2012}
	\begin{equation}
		S(\chi, n, 0) =
			\begin{cases} \displaystyle \frac{n}{\chi^3} \frac{3}{8(1-n^2)} + \O(\chi^{-5}), 	&	n^2 < 1, \\[0.5cm]
				\displaystyle \frac{1}{\chi} \frac{n^2-1}{2n^3} + \O(\chi^{-3}),		&	n^2 > 1.
			\end{cases}
	\end{equation}
For small $q$, we recover $S \sim q^4$.  Additionally, we can consider the case of finite $q$ but small $\b$.  For $n^2 < 1$, the RHS of \eref{zi:isotropic_disprelation} goes as $\b^2$, which vanishes at $\b=0$.  Therefore, at $\b=0$ any thin ring of an isotropic spectrum with $k > q$ has no net effect on the zonal flow.  On the other hand, for $n^2 > 1$ the $\b$ dependence in the RHS of \eref{zi:isotropic_disprelation} vanishes.  Thus, at $\b=0$ a thin ring with $k < q$ has a net damping effect on the zonal flow.

For finite $L_d$, $S$ behaves as\footnote{Validity requires that $m \neq 0$, because for $m=0$ and $n^2 < 1$, the lowest order result vanishes.}
	\begin{align}
		S(\chi,n,m&) = (4 n^3 \chi)^{-1} \left[  -n^2(-1+m) + (1+m) \Big( -1 - m  \right. \notag  \\
						& \left. + \sqrt{[(-1+n)^2 + m][(1+n)^2 + m]} \Big)\right] +\O\left( \chi_1^{-3} \right). \label{zi:S_finiteLD}
	\end{align}
For small $\b$, the $\b$ dependence cancels out of the RHS of \eref{zi:isotropic_disprelation}.  Hence, instability is possible even with $\b=0$.  For concreteness, one might take $m=1$, for which $S$ simplifies to 
	\begin{equation}
		S(\chi,n,1) = \frac{1}{n^3 \chi} \left( -1 + \sqrt{1 + \frac{n^4}{4}} \right) +\O\left( \chi^{-3} \right).
	\end{equation}
Additionally, the small $q$ limit of \eref{zi:S_finiteLD} is
	\begin{equation}
		S(\chi,n,m) = \frac{n}{\chi} \frac{m}{2(1+m)^2} + \cdots.
	\end{equation}
Thus, for an isotropic spectrum and finite $L_d$ (and also $\m=0$), the growth rate of the zonal flows goes as $q^2$ at small $q$, rather than like $q^4$ as in the case of infinite $L_d$.

\section{Connection to Modulational Instability}
\label{sec:mi}
Zonostrophic instability can be understood in a very general way as the instability of some turbulent background spectrum to a (zonally symmetric) coherent mode.  As a special case, one can consider the background spectrum to consist of only a single mode.  \citet{parker:book} show that in this case the dispersion relation of zonostrophic instability reduces exactly to that of the 4-mode modulational instability (sometimes called parametric instability).  This correspondence was first noted by \citet{carnevale:1982}, but they did not discuss it in the context of the generation of zonal flows.

The stability of a single, primary wave $\v{p}$ to perturbations is a problem that has received attention in the past \cite{lorenz:1972,gill:1974,connaughton:2010,gallagher:2012}.  These calculations have used the fluctuating dynamical equations such as \eref{unifiedEOM} and not a statistically averaged system.  Generally one considers the unforced, undamped case, for which a single wave is an exact solution of the nonlinear dynamical equations.  Conceptually similar is the so-called secondary instability, where a growing, primary eigenmode gives rise to a secondary mode \cite{rogers:2000,plunk:2007,pueschel:2013}.  If the secondary mode grows much faster, the primary mode is treated as a stationary background.  These secondary instabilities are more complicated, since due to the toroidal geometry, the growing eigenmode has nontrivial spatial dependence.  Additionally, the eigenmode is not an exact solution of the nonlinear equations.

To calculate the stability of the primary wave using \eref{unifiedEOM}, in general one needs to retain an infinite number of coupled, perturbing modes.  However, typically one truncates the system, for example retaining a secondary mode $\v{q}$ and the sideband pair $\v{p} \pm \v{q}$.  Within this 4-mode approximation and the further assumption that the primary has $p_y=0$ such as a pure Rossby or drift wave and the secondary has $q_x=0$, the dispersion relation for 4-mode modulational instability is given by \cite{connaughton:2010}
	\begin{equation}
		\l'^3 = \l' s^4 \left( \frac{ 2M^2 (1-s^2)(1+s^2+f)(1+f)^2 - (s^2+f)}{ (1+f)^2 (1+s^2+f)^2 (s^2+f) } \right),
		\label{mi:disprelation_connaughton}
	\end{equation}
where $\l' = p \l / \b$, $s=q/p$, $f = p^{-2} L_D^{-2}$, $M = \psi_0 p^3 / \b$, and $\psi_0$ is the amplitude of the background stream function.

Some studies investigated this phenomenon by using a form of CE2 where the inhomogeneity is assumed to vary slowly in space compared to the turbulence \cite{manin:1994,dubrulle:1997,smolyakov:2000b,wordsworth:2009,trines:2010}.  With that assumption, the turbulence is described by a wave kinetic equation.  The wave kinetic equation can also be recovered from CE2 as described in \secref{sec:wavekinetic}.  While those previous studies are limited to the regime of small $q$, the CE2 framework makes no assumption about the length scale of the inhomogeneity.  Moreover, those previous studies did not draw a direct connection between the results from the statistical calculation and from the 4-mode calculation.\footnote{One reason a connection may not have been made is that the small-$q$ results in \citet{manin:1994} and \citet{smolyakov:2000b} based on the wave kinetic equation are incomplete.  Their dissipationless ($\m=0$) formulation amounts to neglecting the term $2i \b q k_x k_y$ compared to $\l$ in the denominator of \eref{zi:lambdapm}.  But this is invalid if $\l \sim q^2$ because the neglected term is larger than the retained term.  For example, when specialized to a single primary mode, both papers state that for the (unmodified) Hasegawa--Mima Equation, instability occurs when $p_x^2 + \LD - 3p_y^2 > 0$, and that $\l \sim q^2$.  When the $\b$ term is unjustifiably neglected, this result can be found from the small $q$ limit of \eref{mi:primarymode_dispersionrelation}.  Careful analysis shows this result also obtains in the $\psi_0 \to \infty$ limit.  But contrary to statements made by \citet{connaughton:2010}, the wave-kinetic formalism is \emph{not} restricted to that large-amplitude regime.  If the $\b$ term is retained, the full answer at small $q$ can be recovered from the wave-kinetic formalism.}

This dispersion relation \eref{mi:disprelation_connaughton} can be recovered from CE2 and the zonostrophic instability dispersion relation \eref{zi:dispersionrelation}.  To precisely compare, one must carefully select the background spectrum $W_H$ to correspond to a wave of stream function $\psi_0$.  If the initial background amplitude of mode $\v{p}$ is $\psi_0$, then we write
	\begin{equation}
		\psi(\v{x}) = \psi_0 \left( e^{i \v{p} \cdot \v{x} - i\w t} + e^{-i\v{p}\cdot\v{x} + i\w t} \right).
		\label{mi:backgroundpsi}
	\end{equation}
Appendix \ref{app:wave_correlation} shows that this corresponds to a one-time, two-point covariance of streamfunction
	\begin{equation}
		\Psi_H(k_x,k_y) = (2\pi)^2 \psi_0^2 \bigl[ \de(\v{k} - \v{p}) + \de(\v{k} + \v{p}) \bigr].
	\end{equation}
From \eref{WCrelation}, the corresponding covariance of vorticity is given by $W_H(k_x,k_y) = \ol{k}^4 \Psi_H(k_x,k_y)$, and thus, because of the delta functions,
	\begin{equation}
		W_H(k_x,k_y) = (2\pi)^2 A \bigl[ \de(\v{k} - \v{p}) + \de(\v{k} + \v{p}) \bigr], \label{mi:WH_from_psi0}
	\end{equation}
where we have defined $A = \psi_0^2 \bigl(p^2 + \LD \bigr)^2$.  There are two ways of achieving this background spectrum.  First, we could choose the external forcing to be $F(\v{k}) = 2\m W_H$.  Since we want the dissipation term $\m$ to disappear in the final expression, $\m$ can be chosen to be vanishingly small, in particular smaller than the eigenvalue $\l$.  Alternatively, as previously mentioned we could take the external forcing and the dissipation to be zero, in which case any arbitrary homogeneous spectrum trivially satisfies the CE2 equations.  This latter point of view is closer to the traditional stability calculations.

Substituting \eref{mi:WH_from_psi0} into \eref{zi:dispersionrelation}, we find
	\begin{equation}
		\frac{\qbsq}{q^2} \l =  2q A p_x^2 \biggl( 1 - \frac{\qbsq}{\pbsq} \biggr) \biggl( \frac{p_y + \frac12 q}{\l \pbsq_{+} \pbsq + 2i\b q p_x(p_y + \frac12 q)}  - \frac{p_y - \frac12 q}{\l \pbsq_{-} \pbsq + 2i\b q p_x(p_y - \frac12 q)} \biggr), \label{mi:primarymode_dispersionrelation}
	\end{equation}
where dissipation has been neglected, $p_{\pm}^2 = p_x^2 + (p_y \pm q)^2$, and $\pbsq_{\pm} = p_{\pm}^2 + \LD$.

When specialized to the case of a primary wave with $p_y=0$, the dispersion relation becomes
	\begin{equation}
		\frac{\qbsq}{q^2} \l = 2q A p_x^2 \left( 1 - \frac{\qbsq}{\pbsq} \right) \frac{q}{2} \frac{2 \l \pbsq_{+} \pbsq}{\l^2 \ol{p}_{+}^4 \ol{p}^4 + \b^2 q^4 p^2}.
	\end{equation}
Now, taking $\azf=1$ to specialize to quasigeostrophic physics and introducing the same normalizations as used in \eref{mi:disprelation_connaughton}, we obtain
	\begin{equation}
		\frac{s^2 + f}{s^2} \l' = \frac{ 2 A s^2 \l' (1-s^2) (1+s^2+f)}{ (\b/p)^2 [ \l'^2 (1 + s^2 + f)^2(1 + f)^2 + s^4]}.
	\end{equation}
Letting $A' = p^2 A / \b^2$, after some simplification we find
	\begin{equation}
		\l'^3 = \l' s^4 \left( \frac{ 2A' (1-s^2) (1+s^2+f) - (s^2+f) }{ (1+f)^2 (1+s^2+f)^2 (s^2+f) } \right).
		\label{mi:secondary_disprelation_zonal}
	\end{equation}
Since $A' = p^6 \psi_0^2 (1+f)^2 / \b^2 = M^2 (1+f)^2$, this exactly matches the dispersion relation given in \eref{mi:disprelation_connaughton}.

It may be at first surprising that the two dispersion relations agree exactly, but retrospectively it makes sense.  The 4-wave modulational instability contains the primary wave $\v{p}$ and the perturbations at wave vectors $\v{q}$ and $\v{p} \pm \v{q}$.  From \eref{app:W_manywaves} in Appendix \ref{app:wave_correlation} for the correlation between the primary mode $\v{k} = p \unit{x}$ and sidebands $\v{k'} = p \unit{x} \pm q \unit{y}$, we see that the spatial dependence of the correlation goes as $\cos( px \pm \tfrac12 qy \pm q \ybar)$.  Upon examining the CE2 calculations, we see that the retained modes are the zonal flow $\de U e^{\pm i q \ybar}$ (which corresponds to mode $\pm \v{q}$) and the perturbations to the spectrum $\de W(k_x,k_y) e^{\pm i q \ybar}$.  The perturbation $\de W(k_x,k_y)$ is proportional to $W_H(k_x, k_y \pm \frac12 q)$, which is nonzero at $k_x = p$ and $k_y = \pm \tfrac12 q$ for the given primary mode.  Therefore the perturbations kept within CE2 are precisely the corresponding modes kept in the 4-mode truncation.  The CE2 instability calculation neglects higher harmonics of $\v{q}$ such as $e^{2iq\ybar}$ at the linear level.  These higher harmonics are precisely what is neglected by truncation to 4 modes instead of retaining higher sidebands.

In the above calculation, we have shown that from CE2 we recover the 4-wave modulational instability in the special case of a primary wave with $p_y=0$ and a secondary wave with $q_x=0$.  We now generalize this to show that CE2 recovers the 4-wave modulational instability for an arbitrary primary wave and an arbitrary secondary wave.

The 4-wave modulational instability has the dispersion relation \cite{connaughton:2010}
	\begin{align}
		\bigl(q^2 +\LD \bigr) \l  - i\b q_x &= \psi_0^2 |\v{p} \times \v{q}|^2 \bigl(p^2 - q^2\bigr)  \biggl( \frac{p_+^2 - p^2}{\bigl(p_+^2 + \LD\bigr)(\l - i \w) - i \b (p_x + q_x)} \notag \\
		& \qquad + \frac{p_-^2 - p^2}{\bigl(p_-^2 + \LD\bigr)(\l + i \w) + i \b (p_x - q_x)} \biggr),
			\label{mi:connaughtongeneral}
	\end{align}
where $\v{p}_\pm = \v{p} \pm \v{q}$ and $\w = - \b p_x / (p^2 + \LD)$.

To allow for an arbitrary secondary wave within the CE2 formalism, we use the recent formulation of \citet{bakas:2013,bakas:2013c}.  That formulation allows for coherent structures of arbitrary spatial dependence rather than restricting to zonally symmetric structures.  (The rest of this thesis is focused on zonal flows and uses the formulation only for zonally-symmetric structure.)  Their formulation also assumed infinite deformation radius, though that could be modified.  The dispersion relation in the small forcing and small dissipation limit is \cite{bakas:2013c}\footnote{There is a seeming factor of $2\pi$ different from the formula in \citet{bakas:2013c} because of the choice of Fourier transform convention.}
	\begin{equation}
		\l q^2 - i \b q_x = \int \frac{dk_x\, dk_y}{(2\pi)^2} \frac{ N }{D} \biggl(1 - \frac{q^2}{k^2} \biggr) W_H(k_x,k_y),
		\label{mi:bakasdispersion}
	\end{equation}
where
	\begin{align}
		N &= 2 (k_x  q_y- k_y q_x) \biggl\{ q_x q_y \biggl[ \Bigl(k_x + \frac{q_x}{2}\Bigr)^2 - \Bigl(k_y + \frac{q_y}{2} \Bigr)^2 \biggr] \notag \\
		 & \qquad  \qquad \qquad \qquad  + \bigl(q_y^2 - q_x^2\bigr)\Bigl(k_x + \frac{q_x}{2} \Bigr) \Bigl(k_y + \frac{q_y}{2} \Bigr)\biggr\}, \\
		D &= \l k^2 k_+^2 - \frac12 i q_x \b \bigl[k^2 + k_+^2 \bigr] +  2 i \b \Bigl(k_x + \frac{q_x}{2}\Bigr) \Bigl[ \Bigl(k_x + \frac{q_x}{2} \Bigr) q_x + \Bigl(k_y + \frac{q_y}{2}\Bigr) q_y \Bigr],
	\end{align}
and $\v{k}_+ = \v{k} + \v{q}$.  As before, the appropriate background spectrum to correspond with that of \eref{mi:connaughtongeneral} is $W_H = (2\pi)^2 \psi_0^2 p^4  \left[ \de(\v{k} - \v{p}) + \de(\v{k} + \v{p}) \right]$.  With sufficient algebra, it is possible to show that \eref{mi:bakasdispersion} reduces exactly to the $\LD=0$ limit of \eref{mi:connaughtongeneral}.  The key is in recognizing that
	\begin{gather}
		N = (k_x q_y - k_y q_x)^2 \bigl(k_+^2 - k^2\bigr), \\
		D = k^2 \biggl[ \biggl( \l + \frac{i \b k_x}{k^2} \biggr) k_+^2 - i\b (k_x + q_x) \biggr].
	\end{gather}
	
With our finding that zonostrophic instability encompasses modulational instability (and the closely related secondary instability), we can envision future avenues for research.  For understanding how coherent structures grow, a statistical formalism like CE2 may provide a clearer window than the fundamental dynamical equations.  Indeed, a single eigenmode, which is what the calculations from the fundamental dynamical equations use, may not be unstable to coherent structures, and instead a more complete spectrum may be required.  With CE2, one could investigate how ZI depends on the background spectrum, using anywhere from a single eigenmode to a full incoherent turbulent spectrum.

In addition, future work could be done to determine how well zonostrophic instability can reproduce modulational/secondary instability when the eigenmodes are not Fourier modes, e.g., with nonperiodic boundary conditions.  The work presented here assumed a Fourier decomposition was appropriate.



\section{Beyond Zonostrophic Instability}
\label{sec:beyondzi}

\subsection{Preliminaries: Analogy Between Zonal Flows and \RB Convection Rolls}
The notion of spontaneous symmetry breaking with respect to zonal flows has been discussed before \cite{farrell:2007,srinivasan:2012}.  This section will expand on that in discussing the mechanics of the symmetry breaking, as well as specific consequences it has for the physics of zonal flows \cite{parker:2013,parker:2014}.

\begin{figure}
		\centering
		\includegraphics[width=3in]{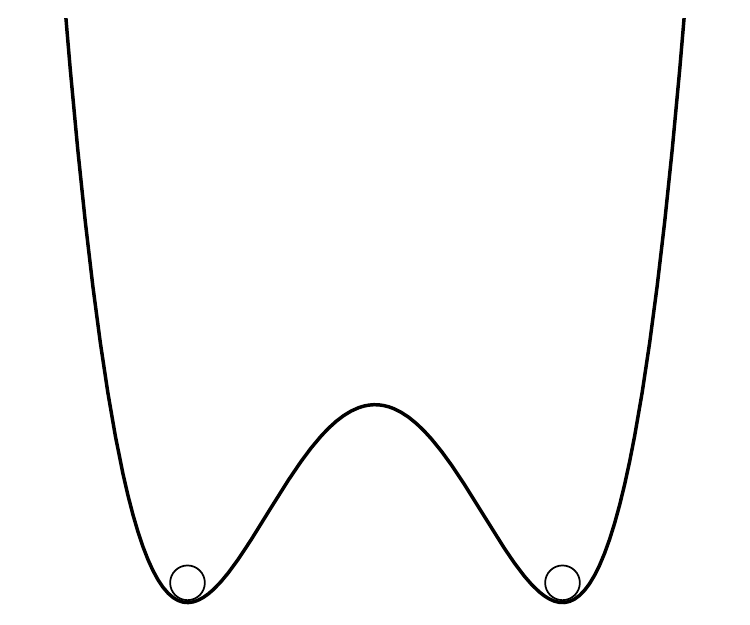}
		\caption{Discrete spontaneous symmetry breaking occurs when a ball moving in a symmetric double-well potential must, due to friction, end up in one of the wells.}
		\label{fig:gl:discretesymmetrybreaking}
	\end{figure}

An important aspect of zonostrophic instability is that it involves a spontaneous symmetry breaking.  A spontaneous symmetry breaking occurs when a situation's governing physics are invariant under a symmetry transformation but a physical realization is not invariant under the same transformation.  A simple example would be a ball moving in a symmetric double-well potential, as in \figref{fig:gl:discretesymmetrybreaking}.  The equations of motion of the ball are invariant to reflection about the center line.  But with friction the ball must eventually end up in one of the wells, a state which breaks the symmetry.

\begin{figure}
		\centering
		\includegraphics[scale=0.5]{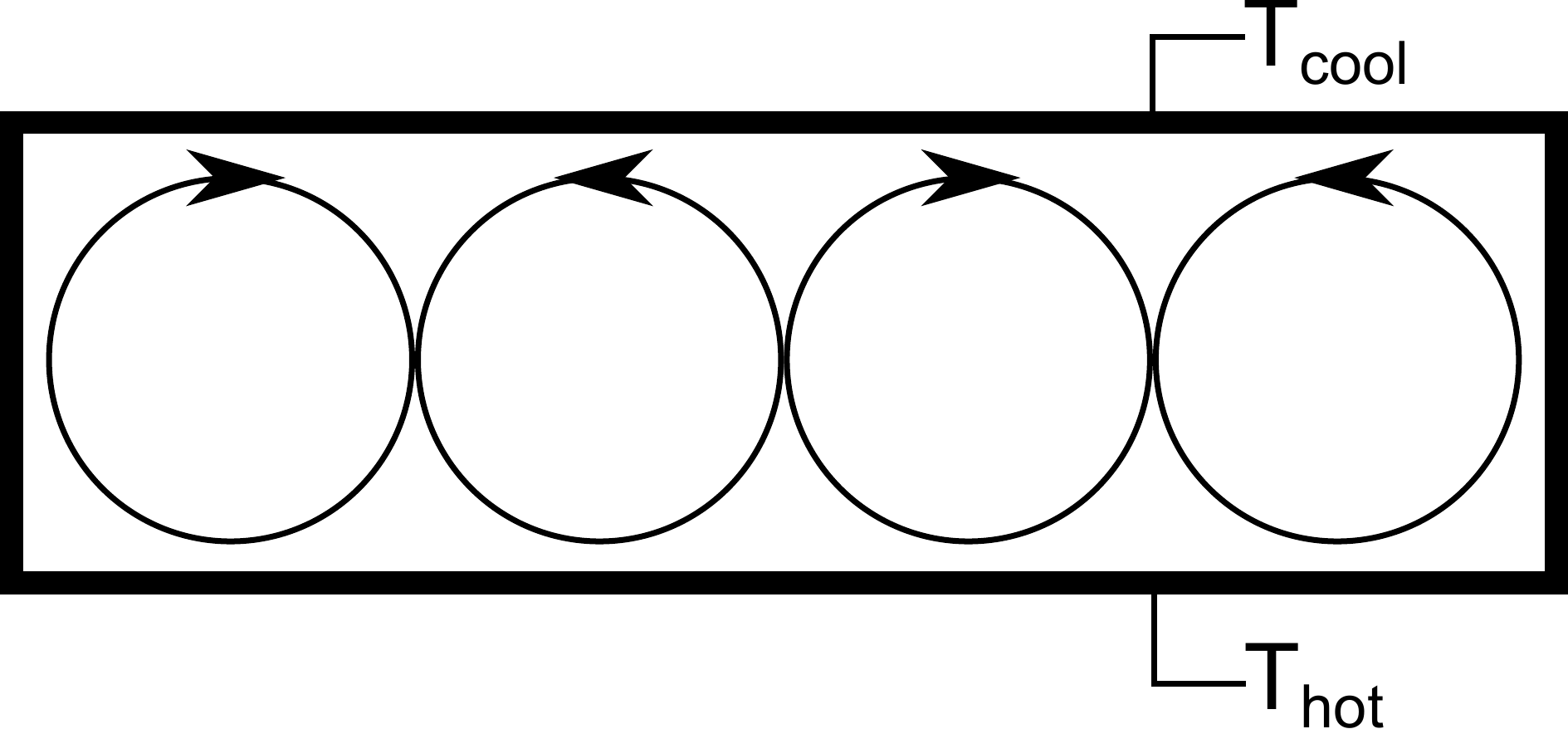}
		\caption{Convection rolls in \RB convection break the horizontal translational symmetry.}
		\label{fig:gl:schematic_convection_rolls}
	\end{figure}

\begin{figure}
		\centering
		\includegraphics[width=3.1in]{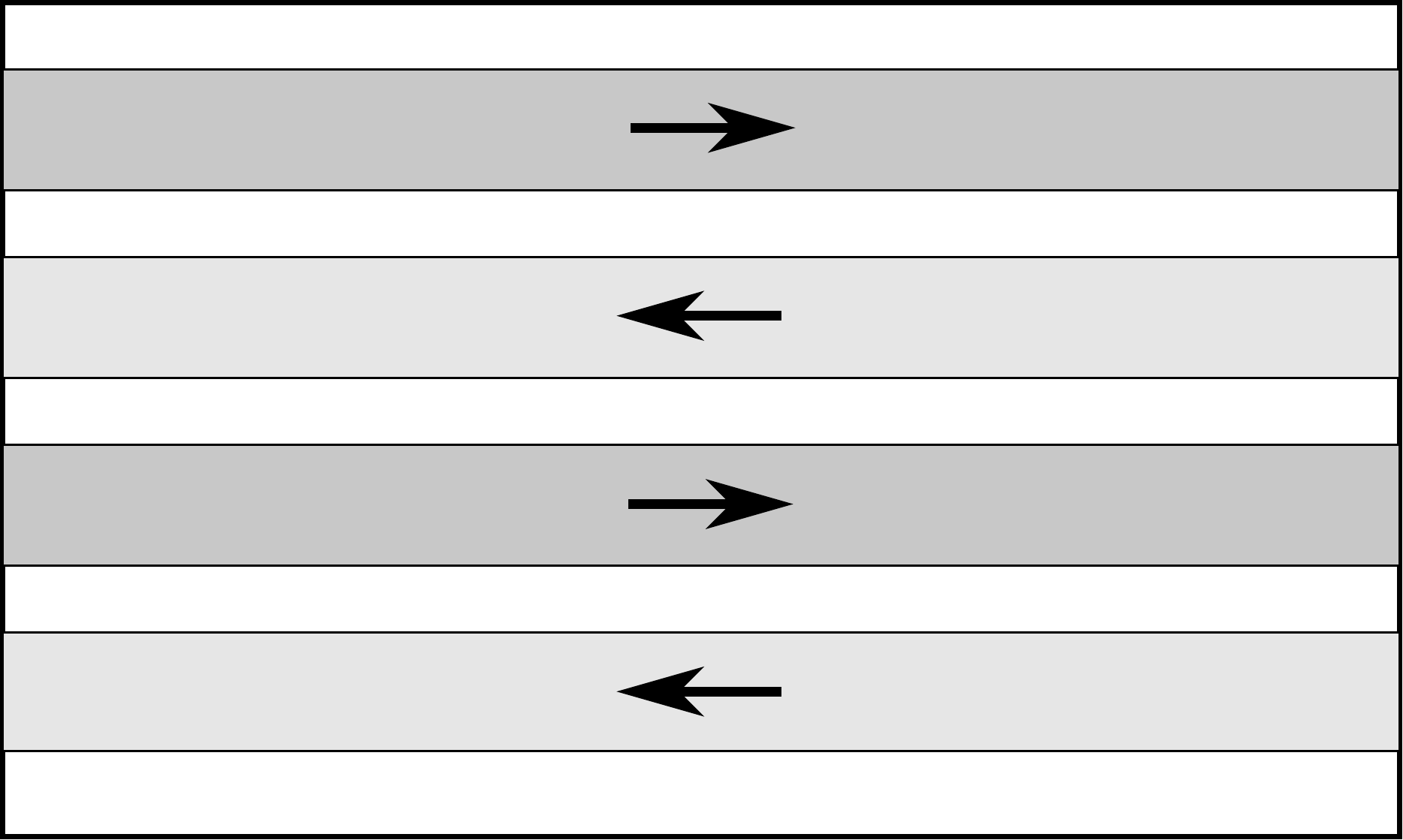}
		\caption{Zonal flows on a $\b$ plane break the north-south (statistical) translational symmetry.}
		\label{fig:gl:schematic_zonal_flows}
	\end{figure}

Another well-known example of spontaneous symmetry breaking is the formation of convection rolls in \RB convection \cite{busse:1978}.  A box of fluid, taken to be infinite in both horizontal directions and finite in vertical extent, is heated from below.  At weak heating, the heat is transferred to the cooler top surface solely by conduction, and the fluid is motionless.  But at sufficiently high heating, buoyancy forces overcome the inherent dissipation and the conduction state becomes unstable to the formation of convection rolls, as shown schematically in \figref{fig:gl:schematic_convection_rolls}.  The convection rolls are spatially periodic but steady in time.  This transition to convection is analogous to the generation of zonal flows out of homogeneous turbulence.  Like the conduction state, homogeneous turbulence is (statistically) uniform in space.  And as a drive parameter such as the strength of the forcing is varied, that uniform state becomes unstable to the formation of a periodic structure.  Born out of turbulence are spatially periodic, steady-in-time zonal flows, which are analogous to the convection rolls (see \figref{fig:gl:schematic_zonal_flows}).  More than merely descriptive, this analogy will be made mathematically precise in the following section.

\subsection{Bifurcation Analysis and the Amplitude Equation}
The existence of zonostrophic instability indicates that a homogeneous equilibrium without zonal flow is unstable.  Perturbations to this equilibrium grow exponentially, with wave number dependencies and growth rates that can be calculated.  However, the ZI calculation alone does not predict how the system saturates.

To understand the behavior in the regime of nonlinearly interacting eddies and \zfs{}, we turn to a bifurcation analysis.  Near the instability threshold, the distance from the threshold serves as a small parameter to facilitate analytic progress.  It has been demonstrated numerically that the bifurcation is supercritical \cite{farrell:2007}, which we confirm with our analytical calculations in Appendix~\ref{app:glderivation}.  Thus, only lowest-order terms in the bifurcation analysis are needed to provide saturation of the instability.

The bifurcation analysis follows a standard procedure, using a multiscale perturbation analysis, expanded around the threshold \cite{cross:1993,cross:2009}.  If the threshold occurs at some critical parameter $\r_c$, then a normalized parameter can be defined as $\e = (\r-\r_c)/\r_c$.  If we denote $u$ as the state vector relative to the homogeneous equilibrium, i.e., $u = \{W-W_H, U\}$, then the expansion proceeds as
	\begin{equation}
		u = \e^{1/2} u_1 + \e u_2 + \e^{3/2} u_3 + \cdots\, .
	\end{equation}
At first order, one finds
	\begin{equation}
		u_1 = A(\ybar,t) r + \text{c.c.},
	\end{equation}
where $\text{c.c.}$ denotes complex conjugate [analytically, we work with the quantities $W(k_x,k_y\mid \ybar)$ and $U(\ybar)$, which both must be real].  Here, $u_1$ is proportional to the eigenmode $r \sim e^{iq_c \ybar} \{ \de W, \de U\}$ that undergoes bifurcation, and $A$ is its amplitude.  The amplitude is an envelope that slowly varies in space and time.  The slow variation represents the effect of the infinity of wave numbers nearby $q_c$ that also go unstable when $\e > 0$.  The goal is to determine $A$, as then $u_1$ will be fully specified.  Here, one determines a PDE for $A$ as a solvability condition at third order in the perturbation expansion.  One eventually finds
	\begin{equation}
		c_0 \partial_t A(\ybar, t) = \e c_1 A + c_2 \partial_\ybar^2 A - c_3 |A|^2 A,
		\label{gl:amplitudeequation}
	\end{equation}
where the $c_i$ are the order unity, real, positive constants to be calculated.  If $c_3$ were negative then one would have a subcritical bifurcation.  Equation \eqref{gl:amplitudeequation} is referred to as the amplitude equation, or sometimes as the real Ginzburg-Landau equation.

It turns out that in order to understand the qualitative behavior of $A$, one does not need to carry out this calculation of the $c_i$ explicitly \cite{cross:2009}.  This is because the translation and reflection symmetries \eref{CE2symmetries} constrain the lowest-order PDE for $A$ to consist generically of the form in \eref{gl:amplitudeequation}.  For example, as a result of the translation symmetry, if $A$ is a solution then so must $A e^{i\theta}$ be for any $\th$.  This arises because the phase of $A$ determines the location of the solution in space.  This symmetry requirement demands that the lowest-order nonlinear term is uniquely determined to be $|A|^2 A$.

The behavior of \eref{gl:amplitudeequation} is universal in the sense that, as long as all of the $c_i > 0$, the qualitative behavior does not depend of the value of any of the $c_i$.  This can be seen because all parameters can be transformed to unity by simple rescaling.  The rescaling is accomplished by letting $t = t' T$, $\ybar = \ybar' L$, and $A = A' G$.  One finds that with $T = c_0 / \e c_1$, $L^2 = c_2 / \e c_1$, and $G^2 = \e c_1/c_3$, that the resulting equation for $A'$ is simply
	\begin{equation}
		\partial_{t'} A'(\ybar',t') = A' + \partial_{\ybar'}^2 A' - |A'|^2 A'.
	\end{equation}

\begin{figure}
		\centering
		\includegraphics{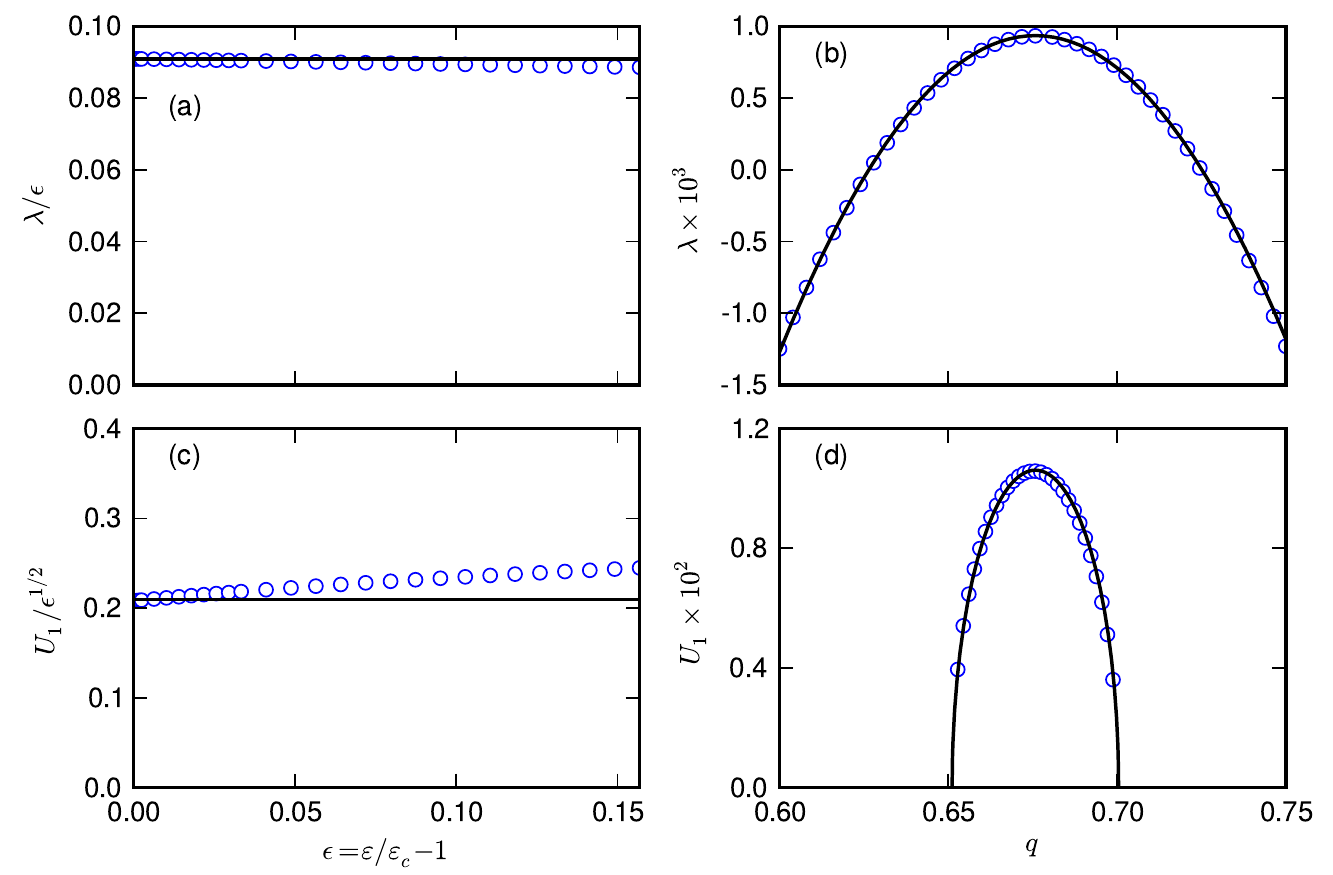}
		\caption{Comparing showing agreement between numerical solution (blue circles) and analytic solution (black line). (a)  Compensated growth rate as a function of $\e$ at $q=q_c$.  (b) Growth rate as a function of $q$ at $\e=0.01$.  (c) Compensated zonal flow amplitude as a function of $\e$ at $q=q_c$.  (d) Zonal flow amplitude as a function of $q$ at $\e=0.0025$.  In (a) and (c), the compensated growth rate and zonal flow amplitude agree with the analytic result at $\e=0$.  The deviation from the lowest order result is $\O(\e)$ for the growth rate and $\O(\e^{1/2})$ for the zonal flow amplitude.  For details, see Appendix \ref{app:glderivation}.}
		\label{fig:gl:GL_coeffs}
	\end{figure}

Even if the qualitative behavior is understood, it is still worthwhile to carry out the calculation of the coefficients $c_i$.  First, computing these and verifying the results numerically provides a concrete check on our overall understanding.  Second, the perturbation solution may be convenient for certain numerical methods where it is useful to start with a good approximation to the true solution.  In Appendix \ref{app:glderivation}, we perform the derivation of \eref{gl:amplitudeequation} and obtain expressions for the $c_i$.  This computation has also been carried out independently \cite{bakas:tbs}.  To verify our results, we compare the analytic growth rate found from \eref{gl:amplitudeequation} with that from the exact dispersion relation \eref{zi:dispersionrelation}.  Similarly, the analytic \zf{} amplitude found from \eref{gl:amplitudeequation} is compared with that from solving the ideal states numerically as in \secref{is:equilibrium}.  The results are shown in \figref{fig:gl:GL_coeffs} and are in excellent agreement.

With \eref{gl:amplitudeequation}, the analogy between the zonal flows and the convection rolls in \RB convection is complete.  The transition to convection is governed by the same class of bifurcation and subject to the amplitude equation.  The similarities between zonal flows and convection rolls alluded to in the previous section are not merely descriptive, but mathematical as well.
	
\begin{figure}
		\centering
		\includegraphics{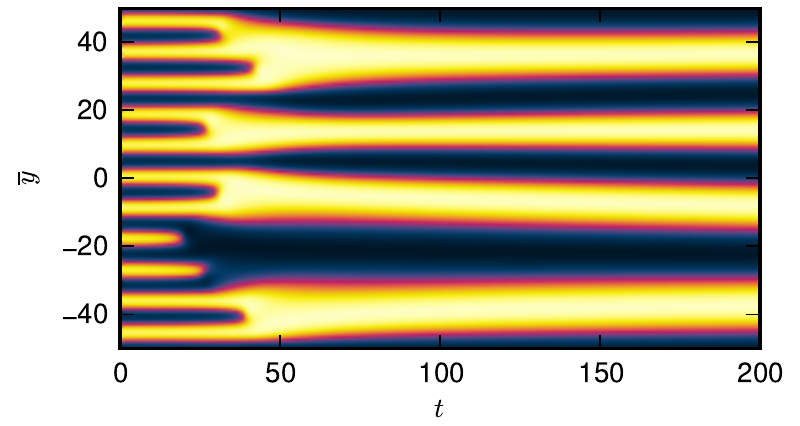}
		\caption{Merging behavior in the amplitude equation \eref{gl:amplitudeequation} [$\Re A(\ybar,t)$ is shown].}
		\label{fig:merging_jets}
	\end{figure}

The amplitude equation \eqref{gl:amplitudeequation} is well understood \cite{cross:1993,cross:2009,hoyle:2006}, and much of its qualitative behavior is seen generically in pattern formation systems.  First, with all the parameters $c_i$ and $\e$ set to unity, a steady-state solution exists for any wavenumber within the continuous band $-1 < k < 1$.  To see this, observe that $A = \a e^{ik\ybar}$ with $|\a|^2 = 1-k^2$ is a solution.  Second, only solutions with $k^2 < \frac13$ are linearly stable; those with $k^2> \frac13$ suffer the Eckhaus instability \cite{cross:2009}.  In the Eckhaus instability, long-wavelength perturbations grow atop a periodic pattern \cite{eckhaus:1965,kramer:1985,tuckerman:1990}.  This is demonstrated in \figref{fig:merging_jets}, where an unstable solution that has been slightly perturbed undergoes merging behavior until a stable wave number is reached.  Similar merging behavior was studied by \citet{manfroi:1999}.

\begin{figure}
		\centering
		\includegraphics{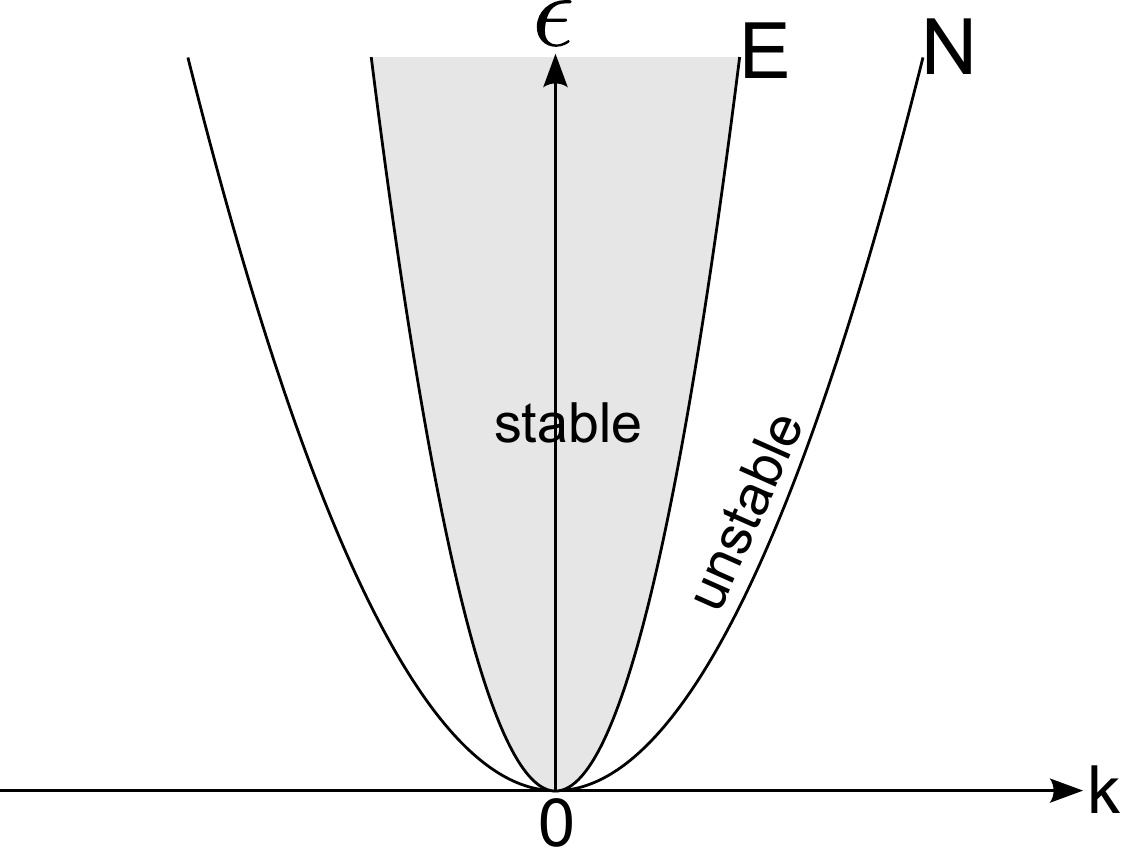}
		\caption{Stability diagram for the amplitude equation.  The labels `stable' and `unstable' refer to the nonzero-$A$ steady states.}
		\label{fig:gl:gl_stability}
	\end{figure}

The stability diagram for the amplitude equation is shown in \figref{fig:gl:gl_stability}.  The neutral curve (N) indicates marginal stability of the $A=0$ solution as a function of the wavenumber $k$ and control parameter $\e$.  The $A=0$ solution is unstable to those $k$ that are above or inside the neutral curve.  At a fixed $\e>0$, steady-state solutions with $A \neq 0$ exist at any of the $k$ inside the neutral curve.  The marginal stability of these $A \neq 0$ solutions is indicated by the Eckhaus curve (E).  Inside the E curve is a smaller band of wave numbers for which the steady-state solutions are stable.

Additionally, the amplitude equation is a gradient system, meaning that it can be written as
	\begin{equation}
		c_0 \partial_t A(\ybar) = - \fd{F[A,A^*]}{A^*(\ybar)},
	\end{equation}
where
	\begin{equation}
		F[A,A^*] = \int d\ybar \left( -\e c_1 A A^* + c_2 (\partial_\ybar A) (\partial_\ybar A^*) + \frac{1}{2} c_3 A^2 A^{*2} \right).
	\end{equation}
Along solution trajectories $F(t)$ is nonincreasing in time since
	\begin{equation}
		\d{}{t}F[A,A^*] = -\frac{2}{c_0} \int d\ybar\, \left| \pd{A}{t} \right|^2 \leq 0.
		\label{gl:dFdt}
	\end{equation}
And $F(t)$ is bounded from below because it can be rewritten in the form
	\begin{equation}
		F[A,A^*] = \int d\ybar \Biggl[ \frac{1}{2} c_3 \biggl(AA^* - \frac{\e c_1}{c_3}\biggr)^2 + c_2 |\partial_\ybar A|^2 - \frac{1}{2} \frac{\e^2 c_1^2}{c_3} \Biggr] \geq -\frac{1}{2} \frac{\e^2 c_1^2}{c_3} D,
	\end{equation}
where $D$ is the size of the integration domain, which must be selected so that boundary terms that arise in integration by parts vanish (eg., $D$ is a periodicity length).  Of all the solutions $A = \a e^{ik\ybar}$, the one with $k=0$ gives the smallest value of $F$.  Therefore, one might at first think that all initial conditions will tend towards the $k=0$ solution.  However, nonzero $k$ give legitimate steady state solutions, with $\partial A / \partial t = 0$ and hence $dF/dt=0$.  The landscape of $F$ in the function space of all possible $A$ is then such that there is a stationary value for each allowed $k$.  Around that stationary value, $F$ must be locally flat in one ``dimension'' corresponding to infinitesimal translation and locally increasing in others, but not decreasing since it is a stable equilibrium.  The $k=0$ solution gives a global minimum of $F$.  But even though $k=0$ may seem to be preferred, this does not guarantee that it is \emph{dynamically} preferred.  Simulations with periodic boundary conditions can clearly find nonzero $k$ as the steady state solution, as seen in \figref{fig:merging_jets}.  This behavior, and more generally the distribution of final wavenumbers, has been thoroughly investigated in simulations of the Swift-Hohenberg equation, which is also a gradient system \cite{schober:1986}.  However, in any realistic system, small amounts of noise are present, which perhaps has an effect in pushing a physical system towards the minimum of $F$.  It should also be noted that even though the amplitude equation is a gradient system, pattern-forming systems far from threshold are not in general gradient systems.

The CE2 system is described by this bifurcation and so near the threshold, and more generally, it exhibits solutions existing with a range of zonal flow wave numbers, with a certain stability region.  In Chapter \ref{ch:numerical}, we numerically calculate the equilibria and stability of nonlinearly interacting turbulence and zonal flows directly from the CE2 equations.

\chapter{Numerical Calculation of Ideal States}
\label{ch:numerical}

In this chapter, we study the steady-state solutions of the CE2 system \eref{CE2} numerically.    As we have learned in Chapter \ref{ch:beyondzonostrophic}, the CE2 system has the mathematical structure of pattern formation.  This means that there are multiple solutions to the equations with differing zonal flow wavelengths and there is an interesting global stability behavior of these different wavelengths.  Using established techniques from the field of pattern formation, in \secref{is:equilibrium} we compute the nonlinear steady-state solutions of CE2 to find self-consistent equilibria of interacting zonal flows and turbulence.  We follow that with a calculation of their linear stability in \secref{is:stability}.  Finally, in \secref{is:wavenumberselection} we perform some preliminary exploration into the problem of wavenumber selection of zonal jets.

In the context of an infinite domain with no boundaries, we refer to the steady-state solutions as \emph{ideal states}.  Let $q$ denote the fundamental \zf{} wavenumber of an ideal state.  For a given $q$, we solve the time-independent form of \eref{CE2} directly.  Our approach, which does not involve time evolution, differs from conventional numerical studies of turbulence.  Time-evolving simulations yield physically relevant, stable solutions.  In the vast majority of studies these are the solutions one is interested in.  But when one is interested in the nonlinear dynamics of a system as a whole, one often needs to understand the unstable solutions as well.  This approach, as well as the numerical methods we employ, was successfully used to study convection rolls in \RB convection \cite{busse:1978}.  As discussed in \secref{sec:beyondzi}, zonal flows are mathematically analogous to convection rolls.  It is therefore appropriate to use the proven techniques on our problem.

Since we are able to select $q$ and determine the \zf{} wavelength $2\pi/q$ directly, this method differs from finite-spatial-domain techniques \cite{farrell:2007,farrell:2003,tobias:2013}.  Within a finite spatial domain, the wavenumbers takes discrete values.  The dominant \zf{} mode is not preselected and is typically not the lowest mode because the system evolves self-consistently to find a solution.  Our infinite-domain technique, which allows the selection of the \zf{} wavelength, is advantageous for understanding the global dynamics.  We can solve for both stable and unstable solutions as we continuously vary $q$ and therefore can easily determine stability boundaries.

There is an important side effect to our choice of the dominant \zf{} wavenumber.  If the dominant \zf{} wavenumber does not occupy the lowest mode, subharmonics of the dominant mode can be excited.  Our method involves setting the lowest \zf{} mode to be the dominant one.  This requires fewer resolved modes.  It also excludes subharmonics (although that is not a limitation in principle).


\section{Calculation of Ideal State Equilibrium}
\label{is:equilibrium}

In this section, we compute directly the ideal states of the CE2 system \eqref{CE2}.  Using Galerkin projection, we derive a set of nonlinear algebraic equations in a form suitable for numerical implementation.  The derivation follows procedures first established for \RB convection rolls \cite{busse:1967,clever:1974,busse:1979,newell:1990,cross:2009}.

\subsection{Derivation of Formulas for Numerical Implementation}
The stationary equations for the eddies and zonal flow are
	\begin{subequations}
	\label{iscalc:CE2steadystate}
	\begin{align}
		&-(U_+ - U_-) \partial_x W + \bigl(\ol{U}_+'' - \ol{U}_-''\bigr) \biggl(\nablabarsq + \frac{1}{4} \partial_\ybar^2 \biggr) \partial_x \Psi \notag \\
			 & \qquad \qquad + \bigl[2\b - \bigl(\ol{U}_+'' + \ol{U}_-'' \bigr) \bigr] \partial_\ybar \partial_y \partial_x \Psi + F(x,y) - 2(\m  + \n D_h) W = 0, \label{iscalc:CE2W} \\
		&-\bigl[\m + \n (-1)^h \partial_\ybar^{2h}\bigr] \ol{I} U(\ybar) - \partial_\ybar \partial_y \partial_x \Psi(0,0 \mid \ybar) = 0. \label{iscalc:CE2U}
	\end{align}
	\end{subequations}
	
The Galerkin approach begins by expanding the ideal state in some suitable basis functions.  Appropriate basis functions give rapid convergence as one includes more terms, so that one does not have to keep an impractical number of terms.  We represent an ideal state using a Fourier--Galerkin series with coefficients to be determined,
	\begin{subequations}
		\label{iscalc:galerkinseries}
	\begin{align}
		W(x,y \mid \ybar) &\defineas \sum_{m=-M}^M \sum_{n=-N}^N \sum_{p=-P}^P W_{mnp} e^{imax} e^{inby} e^{ipq\ybar}, \\
		U(\ybar) &\defineas \sum_{p=-P}^P U_p e^{ipq\ybar}.
	\end{align}
	\end{subequations}
Here, $q$ is the assumed basic wavenumber of the zonal flow, giving a periodicity $2\pi/q$.  While the periodicity in $\ybar$ is desired, the periodicity in $x$ and $y$ is artificial.  The correlation function $W$ should decay smoothly to zero as $x,y \to \infty$.  Therefore, appropriate basis functions in $x,y$ would formally decay at infinity, not be periodic.  One example of such a basis set would be the Hermite functions.  However, we use the Fourier basis because of its supreme convenience.  Thus, $a$ and $b$, unlike $q$, are numerical parameters.  They represent the spectral resolution of the correlation function and should be small enough to obtain an accurate solution.  Alternatively, the box dimensions $2\pi/a$ and $2\pi/b$ must be sufficiently large.

Because the CE2 equations have translational symmetry in $\ybar$, there is an infinite number of solutions, all equivalent, corresponding to displacements along $\ybar$.  In order to obtain a well-posed numerical problem, one must restrict the set of solutions.  To this end, we again look to the symmetries \eref{CE2symmetries}.  The CE2 symmetries allow us to seek a solution for which 
	\begin{subequations}
		\label{iscalc:solnsymmetry}
	\begin{gather}
		W(x,y \mid \ybar) = W(-x,-y \mid \ybar) = W(x, -y \mid -\ybar) = W(-x, y \mid -\ybar), \\
		U(\ybar) = U(-\ybar).
	\end{gather}
	\end{subequations}
In other words, we choose the origin of $\ybar$ such that the reflection symmetries hold for the \emph{solution} itself.  We find that such solutions do exist.  It turns out that this restriction does not uniquely specify the solution, as shifting a solution by a half wavelength $\de \ybar = \pi/q$ yields a distinct but equivalent solution.  Still, this restriction is sufficient to make the problem well-posed numerically.  Put another way, the above condition acts as a way to single out solutions from a family by constraining the phase, in lieu of any other boundary conditions.  In order for the above symmetries to exist in a solution, we also require the forcing to satisfy
	\begin{equation}
		F(x,y) = F(-x,-y) = F(x,-y) = F(-x,y).
	\end{equation}

Aside from the previous statements, there is also no guarantee that there is a unique solution.  Indeed, in the zonostrophically unstable regime, once the above ansatz with a specific $q$ has been substituted, there are at least two solutions: the equilibrium with zonal flows, and the unstable homogeneous solution without zonal flows.  In some instances we also find other unstable solutions, which may be artifacts of the numerical discretization and could be unphysical.	
	
The constraints in \eref{iscalc:solnsymmetry}, along with the conditions that $U(\ybar)$ and $W(x,y\mid \ybar)$ are real, force $U_p$ to be real, and
	\begin{subequations}
	\begin{gather}
		W_{mnp} = W_{-m,n,p}^* = W_{m,-n,p}^* = W_{m,n,-p}^*, \label{iscalc:Wmnpsymmetry} \\
		U_p = U_{-p}.		
	\end{gather}
	\end{subequations}

Furthermore, we take $U_0=0$, as that would merely represent a static uniform velocity.  Equation \eqref{iscalc:Wmnpsymmetry} becomes easier to understand when separated into real and imaginary parts (which is done anyway for numerical implementation).  Let
	\begin{equation}
		W_{mnp} \defineas E_{mnp} + i F_{mnp},
	\end{equation}
where $E_{mnp}$ and $F_{mnp}$ are real.  Then the symmetries require that
	\begin{gather}
		E_{mnp} = E_{-m,n,p} = E_{m,-n,p} = E_{m,n,-p}, \\
		-F_{mnp} = F_{-m,n,p} = F_{m,-n,p} = F_{m,n,-p}.
	\end{gather}
Note that $F_{mnp}=0$ if any of $m,n,$ or $p$ are zero.  The significance of these symmetries may be even clearer when the solution is expressed with sines and cosines rather than exponentials.  One has
	\begin{subequations}
	\begin{align}
		W(x,y \mid \ybar) &= \sum_{m,n,p=0}^{MNP} \Bigl[ \hat{E}_{mnp} \cos(max) \cos(nby) \cos(pq\ybar) \notag \\
			& \qquad \qquad + \hat{F}_{mnp} \sin(max) \sin(nby) \sin(pq\ybar) \Bigr], \\
		U(\ybar) &= \sum_{p=1}^P \hat{U}_p \cos(pq\ybar).
	\end{align}
	\end{subequations}
For deriving the nonlinear algebraic equations, the exponential form is much more convenient than the sine and cosine form.

Let us count the number of independent coefficients.  For the $E_{mnp}$, there are $m=0,\ldots,M$, $n=0,\ldots,N$, and $p=0,\ldots,P$, for a total of (M+1)(N+1)(P+1) coefficients.  For the $F_{mnp}$, there are $m=1,\ldots,M$, $n=1,\ldots,N$, and $p=1,\ldots,P$, for a total of $MNP$ coefficients.  For the $U_p$, there are $P$ independent coefficients.  This gives a total of $(M+1)(N+1)(P+1) + MNP + P$ independent, real, coefficients.

Since $\Psi(x,y\mid \ybar)$ is related to $W(x,y \mid \ybar)$, we also write
	\begin{equation}
		\Psi(x,y \mid \ybar) = \sum_{mnp} C_{mnp} e^{imax} e^{inby} e^{ipq\ybar},
	\end{equation}
and let
	\begin{equation}
		C_{mnp} \defineas G_{mnp} + i H_{mnp}.
	\end{equation}
From \eref{WCrelation}, we find the $C_{mnp}$ and $W_{mnp}$ are related by
	\begin{align}
		W_{mnp} &= \biggl(\kbsq + k_yk_\ybar + \frac{1}{4} k_\ybar^2\biggr)\biggl(\kbsq - k_yk_\ybar + \frac{1}{4} k_\ybar^2\biggr) C_{mnp} \\
					&= \biggl[\LD + k_x^2 + \biggl(k_y + \frac{1}{2}k_\ybar \biggr)^2 \biggr] \biggl[\LD + k_x^2 + \biggl(k_y - \frac{1}{2}k_\ybar \biggr)^2 \biggr] C_{mnp} \\
					&= \ol{h}_+^2 \ol{h}_-^2 C_{mnp},
	\end{align}
with identical relations between the $E_{mnp}$ and $G_{mnp}$ and between the $F_{mnp}$ and $H_{mnp}$.  We have used $k_x = ma$, $k_y = nb$, $k_\ybar = pq$, and defined
	\begin{align}
		\ol{h}_+^2 &\defineas h_+^2 + \LD, \\
		\ol{h}_-^2 &\defineas h_-^2 + \LD, \\
		h_+^2 &\defineas k_x^2 + \biggl(k_y + \frac{1}{2} k_\ybar \biggr)^2, \\
		h_-^2 &\defineas k_x^2 + \biggl(k_y - \frac{1}{2} k_\ybar \biggr)^2.
	\end{align}
	
We obtain a system of nonlinear algebraic equations for the coefficients $W_{mnp}$ and $U_p$ by substituting the Galerkin series \eref{iscalc:galerkinseries} into the steady-state CE2 equations \eref{iscalc:CE2steadystate} and projecting onto the basis functions.  To demonstrate the projection for \eref{iscalc:CE2W}, let
	\begin{equation}
		\p_{mnp} \defineas e^{imax} e^{inby} e^{ipq\ybar}.
	\end{equation}
We project \eref{iscalc:CE2W} onto $\p_{rst}$ by operating with
	\begin{equation}
		\left(\frac{2\pi}{a} \frac{2\pi}{b} \frac{2\pi}{q}\right)^{-1}  \int_{-\pi/a}^{\pi/a} dx   \int_{-\pi/b}^{\pi/b} dy 	\int_{-\pi/q}^{\pi/q} d\ybar\,  \p_{rst}^*.
	\end{equation}
For instance, the term $-(U_+ - U_-) \partial_x W$ projects to $I_{rstp'mnp}^{(1)} U_{p'} W_{mnp}$, where repeated indices are summed over, $I_{rstp'mnp}^{(1)} = -i ma \de_{m,r} \de_{p'+p-t,0} (\s_+ - \s_-)$, $\s_{\pm} = \sinc ( \a_{\pm}\pi/b )$, and $\a_{\pm} = nb - sb \pm \frac12 p'q$.  The other terms of \eref{CE2W}, as well as \eref{CE2U}, are handled similarly.  In total, we generate as many equations as there are coefficients.

Appendix \ref{app:iscalc} provides the full details of the projection.  We summarize the results here.  It will be convenient to use a shorthand notation where
	\begin{align}
		k_x &\defineas ma, \label{iscalc:first_shorthand} \\
		k_y &\defineas nb, \\
		\kbsq &\defineas k^2 + \LD, \\
		\kybarU &\defineas p'q, \\
		\kybarW &\defineas pq, \\
		\kbsqybarU &\defineas \kybarU^2 + \azf \LD, \\
		h_\pm^2 &\defineas k_x^2 + (k_y \pm \tfrac{1}{2} k_\ybar)^2, \\
		\s_\pm &\defineas \sinc\left( \frac{\a_\pm \pi}{b} \right), \\
		\a_\pm &= nb - sb \pm p'q/2. \label{iscalc:last_shorthand}
	\end{align}

Using the complex coefficients $W_{mnp}$, the nonlinear algebraic equations after projection take the form:
	\begin{align}
		0 &= I_{rstp'mnp}^{(1)} U_{p'} W_{mnp} + I_{rstp'mnp}^{(2)} U_{p'} C_{mnp} + I_{rstmnp}^{(3)} C_{mnp} \notag \\
			& \qquad \qquad + I_{rstp'mnp}^{(4)} U_{p'} C_{mnp} + I^{(5)}_{rst} + I_{rstmnp}^{(6)} W_{mnp}, \\
		0 &= I_{tp'}^{(7)} U_{p'} + I_{tmnp}^{(8)} H_{mnp}.
	\end{align}

In separate real and imaginary parts, they take the form
	\begin{align}
		0 &= J_{rstp'mnp}^{(1)} U_{p'} F_{mnp} + J_{rstp'mnp}^{(2)} U_{p'} H_{mnp} + J_{rstmnp}^{(3)} H_{mnp} \notag \\
			& \qquad \qquad + J_{rstp'mnp}^{(4)} U_{p'} H_{mnp} + J^{(5)}_{rst} + J_{rstmnp}^{(6)} E_{mnp}, \\
		0 &= K_{rstp'mnp}^{(1)} U_{p'} E_{mnp} + K_{rstp'mnp}^{(2)} U_{p'} G_{mnp} + K_{rstmnp}^{(3)} G_{mnp} \notag \\
			& \qquad \qquad + K_{rstp'mnp}^{(4)} U_{p'} G_{mnp}  + K_{rstmnp}^{(6)} F_{mnp}, \\
		0 &= I_{tp'}^{(7)} U_{p'} + I_{tmnp}^{(8)} H_{mnp}.
	\end{align}

In practice, some of the sums are trivial, and a more convenient form is as follows:
	\begin{align}
		0 &= J_{rst} \defineas J_{rstp'np}^{(1)} U_{p'} F_{rnp} + J_{rstp'np}^{(2)} U_{p'} H_{rnp} + J_{rst}^{(3)} H_{rst} \notag \\
			& \qquad \qquad + J_{rstp'np}^{(4)} U_{p'} H_{rnp} + J^{(5)}_{rst} + J_{rst}^{(6)} E_{rst}, \\
		0 &= K_{rst} \defineas K_{rstp'np}^{(1)} U_{p'} E_{rnp} + K_{rstp'np}^{(2)} U_{p'} G_{rnp} + K_{rst}^{(3)} G_{rst} \notag \\
			& \qquad \qquad + K_{rstp'rnp}^{(4)} U_{p'} G_{rnp}  + K_{rst}^{(6)} F_{rst}, \\
		0 &= L_p \defineas I_{p}^{(7)} U_{p} + I_{mnp}^{(8)} H_{mnp}.
	\end{align}
In these expressions, for $J_{rst}$ and $K_{rst}$ there are implicit sums only over $p', n, p$, but no sum over $r,s,t$.  For $L_p$, there are implicit sums over $m,n$, but not over $p$.

In the above expressions,
	\begin{align}
		J_{rstp'np}^{(1)} = -K_{rstp'np}^{(1)} &= k_x (\s_+ - \s_-) \de_{p'+p-t,0}, \\
		J_{rstp'np}^{(2)} = -K_{rstp'np}^{(2)} &= -k_x \kbsqybarU \biggl(\kbsq + \frac{1}{4} \kybarW^2 \biggr) (\s_+ - \s_-) \de_{p'+p-t,0}, \\
		J_{rstp'np}^{(4)} = -K_{rstp'np}^{(4)} &= \kbsqybarU k_x k_y \kybarW (\s_+ + \s_-) \de_{p'+p-t,0},
	\end{align}
where here $k_x=ra$, $k_y=nb$, and
	\begin{align}
		J_{rst}^{(3)} = -K_{rst}^{(3)} &= 2 \b k_x k_y \kybarW, \\
		J_{rst}^{(6)} = K_{rst}^{(6)} &= - \bigl[2\m + \n \bigl(h_+^{2h} + h_-^{2h} \bigr) \bigr].
	\end{align}
where here $k_x=ra$, $k_y = sb$, $\kybarW=tq$.  We also have
	\begin{align}
		I_p^{(7)} &= - \bigl(\m + \n k_\ybar^{2h} \bigr) \kbsqybarU / \kybarU^2, \\
		I_{mnp}^{(8)} &= -k_x k_y k_\ybar.
	\end{align}
where here $k_x=ma$, $k_y=nb$, and $k_\ybar = \kybarU = pq$.

For the $J_{rst}$, we have $r=0,\ldots,M$, $s=0,\ldots,N$, $t=0,\ldots,P$.  For the $K_{rst}$, we have $r=1,\ldots,M$, $s=1,\ldots,N$, $t=1,\ldots,P$.  For the $L_p$, we have $p=1,\ldots,P$.

Schematically, we have the vector of independent coefficients
	\begin{equation}
		\v{x} = \begin{pmatrix} E_{mnp} \\ F_{mnp} \\ U_p \end{pmatrix}
	\end{equation}
and the residual vector
	\begin{equation}
		f(\v{x}) = \begin{pmatrix} J_{rst} \\ K_{rst} \\ L_p \end{pmatrix}.
		\label{iscalc:schematicresidual}
	\end{equation}
We want to solve the system of equations $f(\v{x})=0$.

The system of nonlinear algebraic equations is solved with a Newton's method \cite{kelley:2003}.  The Jacobian matrix is sparse and is easy to specify analytically, as described in the following section.  We note that because the \zf{} equation is linear, it is possible to eliminate the \zf{} degrees of freedom analytically.  This is avoided, however, because the reduction of only $P$ degrees of freedom is negligible and this step incurs the major disadvantage of making the Jacobian no longer sparse.

A Newton's method requires a good initial guess.  An accurate initial guess near the instability threshold is provided by the bifurcation calculation described in \secref{sec:beyondzi}.  To find other solutions we employ simple numerical continuation, where the solution at one value of a parameter is used as the initial guess for the solution at the next value of the parameter.

\subsection{Jacobian Matrix}
It is not too difficult to specify the Jacobian matrix.  Take the variation of the residual $f$ by varying the coordinates $U_p$, $W_{mnp}$ in \eref{iscalc:schematicresidual}:
	\begin{align}
		\de J_{rst} &= J_{rst}^{(3)} \de H_{rst} + J_{rst}^{(6)} \de E_{rst} \notag \\
				& \qquad + J_{rstp'np}^{(1)} U_{p'} \de F_{rnp} + J_{rstp'np}^{(2)} U_{p'} \de H_{rnp}  + J_{rstp'np}^{(4)} U_{p'} \de H_{rnp}   \notag \\
				& \qquad + J_{rstp'np}^{(1)} \de U_{p'} F_{rnp} + J_{rstp'np}^{(2)} \de U_{p'} H_{rnp}  + J_{rstp'np}^{(4)} \de U_{p'} H_{rnp}, \\
		\de K_{rst} &=  K_{rst}^{(3)} \de G_{rst} +  K_{rst}^{(6)} \de F_{rst} \notag \\
				& \qquad + K_{rstp'np}^{(1)} U_{p'} \de E_{rnp} + K_{rstp'np}^{(2)} U_{p'} \de G_{rnp}  + K_{rstp'np}^{(4)} U_{p'} \de G_{rnp} \notag \\
				& \qquad + K_{rstp'np}^{(1)} \de U_{p'} E_{rnp} + K_{rstp'np}^{(2)} \de U_{p'} G_{rnp}  + K_{rstp'np}^{(4)} \de U_{p'} G_{rnp}, \\
		\de L_p &= I_p^{(7)} \de U_p + I_{mnp}^{(8)} \de H_{mnp}.
	\end{align}
Then
	\begin{equation}
		\de f_{\v{x}} (\de \v{x}) = \begin{pmatrix}  \de J_{rst} \\ \de K_{rst} \\ \de L_p \end{pmatrix}.
	\end{equation}
This gives the Jacobian-vector product at a given point $\v{x}$ acting on a vector $\de \v{ x}$.  Implementing this product is virtually identical to implementing the residual vector $f$ itself.  With a little bit of work, one can easily extract the actual Jacobian matrix itself,
	\begin{equation}
		A_{ij} \defineas \pd{f_i}{x_j}.
	\end{equation}
The Jacobian matrix is sparse and should be represented as such.  When calculating the matrix coefficients, one needs to remember to convert the terms $\de G \to \de E$ and $\de H \to \de F$.

\subsection{Results}
\begin{figure}
		\centering
		\includegraphics[scale=0.9]{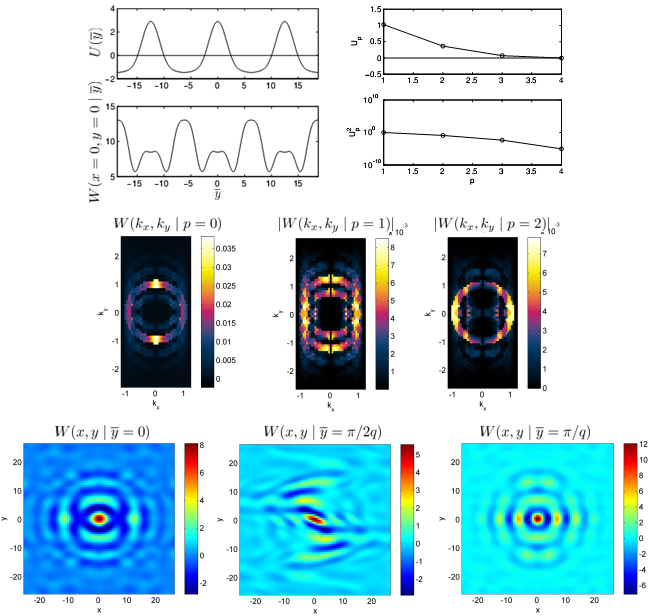}
		\caption{Ideal state equilibrium.  See text for details.}
		\label{fig:equilibrium1}
	\end{figure}

An example of an equilibrium with $\m=0.08$ and $q=0.5$ is shown in \figref{fig:equilibrium1}.  In the top left is shown the zonal flow velocity $U$ and the strength of turbulent fluctuations (measured by the local enstrophy density $W$) as a function of $\ybar$.  In the top right is a plot of the spectral content of the zonal flow, in both linear and log scale.  For these parameters, most of the \zf{} energy resides in the first two harmonics.  In the middle row is the spectral content of the correlation function $W$, for the homogeneous part $W(k_x,k_y\mid p=0)$ and the first two harmonics, $W(k_x,k_y \mid p=1)$ and $W(k_x,k_y \mid p=2)$, of the inhomogeneous part.  The external forcing is a thin ring in $\v{k}$ space around $k=1$ that drives only the $p=0$ component.  The nonlinear interactions with zonal flow act to induce a rich structure in the spectral content.  In the bottom row is $W$ as a function of the real space variables $x,y$, at several values of $\ybar$.  At $\ybar = \pi/2q$ where the \zf{} shear is strong, the correlation function in real space $W(x,y)$ is distorted compared to its more regular pattern at $\ybar=0$ and $\ybar = \pi/q$ where the shear is weak.

\begin{figure}
		\centering
		\includegraphics{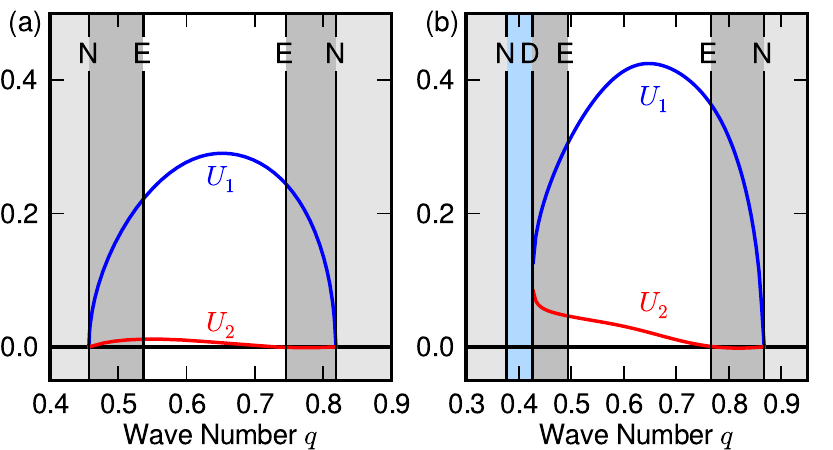}
		\caption{Zonal flow amplitude $U_1$, $U_2$ as a function of ideal state wave number $q$ at (a) $\m=0.21$ ($\g =7.03$) and (b) $\m=0.19$ ($\g=7.97$).  In the unshaded region, ideal states are stable.  The vertical lines correspond to various instabilities which separate the regions (see \figref{is:stability_balloon}).  Here, $L_d=\infty$.}
		\label{fig:zf_amplitude}
	\end{figure}

Figure \ref{fig:zf_amplitude} shows the \zf{} amplitude coefficients $U_p$ as functions of $q$ at $\m = 0.21$ and $\m = 0.19$.  Near the instability threshold, ideal states exist at all $q$ for which the homogeneous equilibrium is zonostrophically unstable [between the two lines labeled N in \figref{fig:zf_amplitude}(a)].  At fixed $\m$, as $q$ approaches the neutral curve boundary (N), the zonal flow amplitude falls to zero and the turbulence becomes homogeneous.

Farther from threshold, there is a region of $q$ where the ideal state solution disappears [between the lines N and D in \figref{fig:zf_amplitude}(b); see also \figref{is:stability_balloon}].  This latter bifurcation is not well understood, and may be a result of some other kind of instability.  We believe the feature not to be a numerical artifact.  It requires $P>1$ to exist, but adding more harmonics or refining the resolution do not alter its behavior.  Moreover, it appears robustly when using multiple variations of Newton's method as well as a distinct Levenberg-Marquardt nonlinear-least-squares algorithm.

\begin{figure}
		\centering
		\includegraphics[scale=0.9]{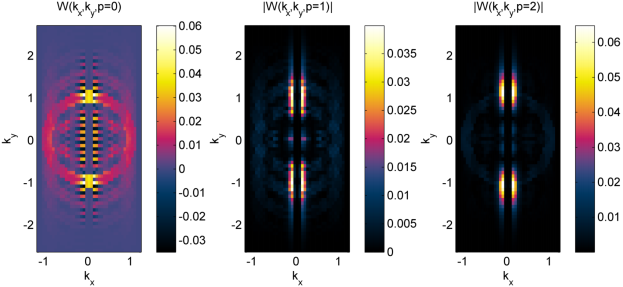}
		\caption{Unphysical solution with strange behavior found by the Newton's method in certain regimes.  See text for details.}
		\label{is:unphysicalsoln}
	\end{figure}

The computational method as described above works very well near the threshold $\m_{\rm c} = 0.237$ ($\g_{\rm c} = 6.02$).  However, far from the threshold, for $\m < 0.12$ ($\g \gtrsim14.2$), the numerical method breaks down.  This appears to be related to the existence of multiple solutions at a given parameter value, of which some are unphysical or unstable.  Far from threshold the Newton's method seems to inevitably get stuck on one of these undesirable solutions.  A plot of the spectral content of one of these solutions is shown in \figref{is:unphysicalsoln}.  At certain $k_x$ values, as $k_y$ changes there are strong oscillations in the correlation function.  This problem does not resolve with higher resolution.  However, the time-evolving simulations previously mentioned do not have these problems because the CE2 equations are statistically realizable and will only approach physical, stable solutions.

\section{Stability of Ideal States}
\label{is:stability}
With the calculations of the ideal states in hand, we now turn to calculating their stability.  Ideal-state stability, which concerns the inhomogeneous equilibria, is distinct from zonostrophic instability, which is a property of the homogeneous equilibrium.  Both types of instabilities can be described within the CE2 formalism.  

\subsection{Derivation of Formulas for Numerical Implementation}
Suppose there is an equilibrium $\{W, U\}$.  We consider perturbations about the equilibrium:
	\begin{subequations}
	\begin{align}
		W(x,y \mid \ybar, t) &= W(x,y \mid \ybar) + \de W(x,y \mid \ybar, t), \\
		U(\ybar, t) &= U(\ybar) + \de U(\ybar, t).
	\end{align}
	\end{subequations}
The CE2 equations linearized about this equilibrium are
	\begin{subequations}
	\label{is:dCE2}
	\begin{align}
		\partial_t \de W &= -(\de U_+ - \de U_-) \partial_x W - (U_+ - U_-) \partial_x \de W  \notag \\
			& \qquad +  \bigl(\de \ol{U}_+'' - \de \ol{U}_-''\bigr) \biggl(\nablabarsq + \frac14 \partial_\ybar^2 \biggr) \partial_x \Psi  + \bigl(\ol{U}_+'' - \ol{U}_-''\bigr)\biggl(\nablabarsq + \frac14 \partial_\ybar^2\biggr) \partial_x \de \Psi \notag \\
			& \qquad - \bigl(\de \ol{U}_+'' + \de \ol{U}_-''\bigr) \partial_\ybar \partial_y \partial_x \Psi - \bigl(\ol{U}_+'' + \ol{U}_-''\bigr) \partial_\ybar \partial_y \partial_x \de \Psi \notag \\
			& \qquad + 2\b \partial_\ybar \partial_y \partial_x \de \Psi - 2(\m + \n D_h) \de W, \label{is:dCE2W} \\
		\partial_t \ol{I} \de U &= -\bigl[\m + \n (-1)^h \partial_\ybar^{2h}\bigr] \ol{I} \de U - \partial_\ybar \partial_y \partial_x \de \Psi(0,0 \mid \ybar, t). \label{is:dCE2U}
	\end{align}
	\end{subequations}

With our Fourier--Galerkin solutions in \secref{is:equilibrium}, the underlying equilibrium is periodic (in every coordinate $x$, $y$, $\ybar$).  Therefore, the differential equation for the perturbations is linear with periodic coefficients.  If we had imposed periodic boundary conditions, then the perturbations would have the same periodicity.  But since we are assuming an infinite domain, more general behavior is possible.  The Bloch Theorem states that we can expand the perturbations as a Bloch state \cite{cross:2009, clever:1974}:
	\begin{subequations}
	\begin{align}
		\de W(x,y\mid \ybar, t) &= e^{\s(\v{Q},q)t} e^{i\v{Q}\cdot\v{X}} \de W_\v{Q} (x,y \mid \ybar), \\
		\de U(\ybar, t) &= e^{\s(\v{Q},q)t} e^{iQ_\ybar \ybar} \de U_\v{Q}(\ybar),
	\end{align}
	\end{subequations}
where the eigenvalue $\s(\v{Q},q)$ depends on both $\v{Q}$ and $q$, $\de W_\v{Q}$ is the Bloch function and has the same periodicity as the ideal state, and $\v{Q} = (Q_x, Q_y, Q_\ybar)^T$ and $\v{X} = (x, y, \ybar)^T$.  The Bloch wavevector $\v{Q}$ can be chosen to live in the first Brillouin zone:
	\begin{align}
		-\tfrac12 a &< Q_x \le \tfrac12a, \\
		-\tfrac12 b &< Q_y \le \tfrac12 b, \\
		-\tfrac12 q &< Q_\ybar \le \tfrac12 q.
	\end{align}
We can expand $\de W_Q$ and $\de U_Q$ in the same basis functions we used for the ideal state:
	\begin{subequations}
	\label{is:Blochfunctionexpansion}
	\begin{align}
		\de W_{\v{Q}}(x,y \mid \ybar) &= \sum_{m=-M}^M \sum_{n=-N}^N \sum_{p=-P}^P \de W_{mnp} e^{imax} e^{inby} e^{ipq\ybar}, \\
		\de U_{\v{Q}}(\ybar) &= \sum_{p=-P}^P \de U_p e^{ipq\ybar}.
	\end{align}
	\end{subequations}	
	
While the periodicity in the $\ybar$ variable is legitimate, the periodicity in $x$ and $y$ are artifacts of the use of a Fourier series.  In actuality the correlation function should decay as $x,y \to \infty$, not be periodic.  If one chooses $a$ and $b$ small enough, approximating an infinite domain better and better, one sees that $Q_x$ and $Q_y$ are restricted to lie in a smaller domain near zero.  At higher resolution, $Q_x$ and $Q_y$ would presumably get close enough to zero as to not matter.  (Nonperiodic basis functions such as Hermite functions would not lead to a Bloch wavevector.)  A separate symmetry argument also suggests taking $Q_x$ and $Q_y$ to be zero.  Due to the correlation function exchange symmetry, which we continue to enforce in the perturbations, we require $\de W(x,y \mid \ybar) = \de W(-x,-y \mid \ybar)$.  This requirement forces $Q_x$ to be either zero or $\tfrac12 a$, and $Q_y$ to be either zero or $\tfrac12 b$.  To see this, consider a function $f(x)$ of only one variable, expressed as
	\begin{equation}
		f(x) = e^{i Q x} \sum_{m=-M}^M y_m e^{imx},
	\end{equation}
where $Q$ can be chosen to lie within $-\tfrac12 < Q \le \tfrac12$.  Suppose $f$ obeys the constraint $f(x) = f(-x)$.  Then, after reindexing one of the sums with $m \to -m$, the constraint leads to
	\begin{equation}
		\sum_{m=-M}^M y_m e^{imx} = e^{-2iQx} \sum_{m=-M}^M y_{-m} e^{imx}.
	\end{equation}
This equation must be satisfied for all $x$.  It will not be satisfied unless $Q=0$ and $y_{-m} = y_m$, or $Q=\tfrac12$ and $y_m = y_{-m-1}$.  This fact, when taken with the previous argument, strongly suggests taking $Q_x$ and $Q_y$ to be zero, which is what we do.
 
Thus, we have
	\begin{subequations}
	\begin{align}
		\de W(x,y\mid \ybar, t) &= e^{\s t} e^{i Q\ybar} \de W_Q(x,y\mid \ybar), \\
		\de U(\ybar, t) &= e^{\s t} e^{iQ\ybar} \de U_Q(\ybar),
	\end{align}
	\end{subequations}
and $\de W_Q$ and $\de U_Q$ are given as in \eref{is:Blochfunctionexpansion}.  This leads to
	\begin{subequations}
	\begin{align}
		\de W(x,y\mid \ybar , t) &= e^{\s t} \sum_{mnp} \de W_{mnp} e^{i max} e^{inby} e^{i(Q+pq)\ybar}, \\
		\de U(\ybar, t) &= e^{\s t} \sum_p \de U_p e^{i(Q+pq)\ybar}.
	\end{align}
	\end{subequations}

The procedure next involves a projection and is similar to that used for the calculation of ideal states.  But several of the symmetry restrictions on the ideal states must be relaxed for the perturbations.  For instance, the reality condition no longer applies.  We are looking for a complex Bloch eigenvector.  If $\de W_1$ is an example eigenvector, real solutions are obtained from
	\begin{equation}
		\de W = A\, \de W_1 + \text{c.c.},
	\end{equation}
where $A$ is some complex amplitude.  We also cannot require $\de U(\ybar) = \de U(-\ybar)$ or $\de W(x,y\mid \ybar) = \de W(x, -y \mid -\ybar)$.  Furthermore, we should allow for nonzero $\de U_0$, as there is no reason to discard it in general (except for the case $Q=0$, in which case $\de U_0$ should be taken to vanish).  One constraint that we do retain, as mentioned previously, is the exchange symmetry, $\de W(x,y \mid \ybar) = \de W(-x, -y, \mid \ybar)$, which is a symmetry of all correlation functions.  The exchange symmetry requires that
	\begin{equation}
		\de W_{mnp} = \de W_{-m,-n,p}.
	\end{equation}

Because the eigenvectors themselves are complex, it does not seem beneficial to decompose $\de W_{mnp}$ or $\de U_p$ into real and imaginary parts, so we leave them as complex coefficients.  Let us count the number of independent coefficients.  We have the $\de W_{mnp}$, $m=-M,\ldots,M$, $n=-N,\ldots,N$, $p=-P,\ldots,P$, with the condition that $\de W_{mnp} = \de W_{-m,-n,p}$.  Therefore, at each $p$ there is a symmetry much like the reality condition of a 2D Fourier transform (but does not involve a complex conjugation).  In implementation, we choose to keep the following: for $n=0$, keep $m=0\ldots,M$, and for $n = 1,\ldots, N$, keep $m=-M,\ldots,M$.  This gives a total of $[M+1 + N(2M+1)](2P+1)$ complex coefficients from the $\de W_{mnp}$, and $2P+1$ from the $\de U_p$.  However, one could choose a different implementation such as keeping all of the $n$.  The projection of the perturbation equations \eref{is:dCE2} onto the basis functions is not affected by the particular implementation of which independent coefficients are retained.

Equation \eref{is:dCE2} is projected onto the basis functions in nearly the same way as in the ideal state calculation.  The projection results in a linear system at each $Q$ for the coefficients $\de W_{mnp}$ and $\de U_p$; this determines an eigenvalue problem for $\s$.  Appendix~\ref{app:isstability} provides the full details of the projection.  We summarize the results here.

In order to relate $\de W_{mnp}$ and $\de \Psi_{mnp}$, observe from \eref{WCrelation} that 
	\begin{align}
		\de W_{mnp} &= \left[ \LD + k_x^2 + \biggl(k_y + \frac12 \kybardW\biggr)^2\right] \left[ \LD + k_x^2 + \biggl(k_y - \frac12 \kybardW \biggr)^2 \right] \de \Psi_{mnp} \notag \\
			&= \hbpsqdW \hbmsqdW \de \Psi_{mnp},
	\end{align}
where $k_x = ma$, $k_y = nb$, $\kybardW = Q + pq$, and $\ol{h}^2_{\pm, \de W} = \LD + k_x^2 + \bigl(k_y \pm \tfrac12 \kybardW \bigr)^2$.

It will be convenient to use a shorthand notation where
	\begin{align}
		k_x &\defineas ma, \label{isstability:first_shorthand} \\
		k_y &\defineas nb, \\
		\kybarU &\defineas p'q, \\
		\kybarW &\defineas pq, \\
		\kybardU &\defineas Q+p'q, \\
		\kybardW &\defineas Q+pq, \\
		\kbsqybarU &\defineas \kybarU^2 + \azf \LD, \\
		\kbsqybardU &\defineas \kybardU^2 + \azf \LD, \\
		\kbsq &\defineas k^2 + \LD, \\
		h_{\pm,\de W}^2 &\defineas k_x^2 + \Bigl (k_y \pm \frac{1}{2} \kybardW \Bigr)^2, \\
		\s_{\pm U} &\defineas \sinc\Bigl(\frac{\a_{\pm U} \pi}{b}\Bigr), \\
		\s_{\pm \de U} &\defineas \sinc \Bigl( \frac{\a_{\pm \de U} \pi}{b} \Bigr), \\
		\a_{\pm U} &\defineas nb - sb \pm p'q/2, \\
		\a_{\pm \de U} &\defineas nb - sb \pm (Q+p'q)/2. \label{isstability:last_shorthand}
	\end{align}	

After projection, the $W$ equation takes the form
	\begin{align}
		\s \de W_{rst} &= \ti{I}_{rstp'mnp}^{(1)} \de U_{p'} W_{mnp} + \ti{I}_{rstp'mnp}^{(2)} U_{p'} \de W_{mnp} \notag\\
							 & \quad+ \ti{I}_{rstp'mnp}^{(3)} \de U_{p'} \Psi_{mnp} + \ti{I}_{rstp'mnp}^{(4)} U_{p'} \de \Psi_{mnp} \notag\\
							 & \quad+ \ti{I}_{rstp'mnp}^{(5)} \de U_{p'} \Psi_{mnp} + \ti{I}_{rstp'mnp}^{(6)} U_{p'} \de \Psi_{mnp} \notag\\
							 & \quad+ \ti{I}_{rstmnp}^{(7)} \de \Psi_{mnp} + \ti{I}_{rstmnp}^{(8)} \de W_{mnp}.
	\end{align}
As in the calculation of the ideal-state equilibrium, it is more convenient in practice to give this formula after performing some of the trivial sums.  It becomes
	\begin{align}
		\s \de W_{rst} &= \ti{I}_{rstp'np}^{(1)} \de U_{p'} W_{rnp} + \ti{I}_{rstp'np}^{(2)} U_{p'} \de W_{rnp} \notag\\
							 & \quad+ \ti{I}_{rstp'np}^{(3)} \de U_{p'} \Psi_{rnp} + \ti{I}_{rstp'np}^{(4)} U_{p'} \de \Psi_{rnp} \notag\\
							 & \quad+ \ti{I}_{rstp'np}^{(5)} \de U_{p'} \Psi_{rnp} + \ti{I}_{rstp'np}^{(6)} U_{p'} \de \Psi_{rnp} \notag\\
							 & \quad + \ti{I}_{rst}^{(7)} \de \Psi_{rst} + \ti{I}_{rst}^{(8)} \de W_{rst},
	\end{align}
where the implicit sums are only over $p', n, p$.  The zonal flow equation is written
	\begin{equation}
		\s \de U_p = \ti{I}_{mnp}^{(9)} \de \Psi_{mnp} + \ti{I}_{p}^{(10)} \de U_p,
	\end{equation}
where the implicit sums are only over $m,n$.	

In the above expressions,
	\begin{align}
		\ti{I}_{rstp'np}^{(1)} &= -ik_x (\sdU{+} - \sdU{-}) \de_{p'+p-t,0}, \\
		\ti{I}_{rstp'np}^{(2)} &= -ik_x (\sU{+} - \sU{-}) \de_{p'+p-t,0}, \\
		\ti{I}_{rstp'np}^{(3)} &= ik_x \kbsqybardU \biggl(\kbsq + \frac{1}{4} \kybarW^2 \biggr) (\sdU{+} - \sdU{-}) \de_{p'+p-t,0}, \\
		\ti{I}_{rstp'np}^{(4)} &= ik_x \kbsqybarU \biggl(\kbsq + \frac{1}{4} \kybardW^2 \biggr) (\sU{+} - \sU{-}) \de_{p'+p-t,0}, \\
		\ti{I}_{rstp'np}^{(5)} &= -i \kbsqybardU k_x k_y \kybarW (\sdU{+} + \sdU{-}) \de_{p'+p-t,0}, \\
		\ti{I}_{rstp'np}^{(6)} &= -i \kbsqybarU k_x k_y \kybardW (\sU{+} + \sU{-}) \de_{p'+p-t,0},
	\end{align}
where $k_x=ra$ here, and the other notation is as before.  Also,
	\begin{align}
		\ti{I}_{rst}^{(7)} &= -2 i \b k_x k_y \kybardW, \\
		\ti{I}_{rst}^{(8)} &= - \bigl[2\m + \n\bigl( h_{+,\de W}^{2h} + h_{-,\de W}^{2h} \bigr) \bigr], 
	\end{align}
where $k_x=ra$, $k_y=sb$, and $\kybardW = Q + tq$ here, and the other notation is as before.  Finally,
	\begin{align}
		\ti{I}_{mnp}^{(9)} &= i k_x k_y \kybardW \frac{\kybardU^2}{\kbsqybardU}, \\
		\ti{I}_{p}^{(10)}   &= - \bigl(\m + \n \kybardU^{2h} \bigr),
	\end{align}
where $\kybardU = Q + pq$ here, and the other notation is as before.

If we write the perturbation as a vector
	\begin{equation}
		\de \v{x} = \begin{pmatrix} \de W_{mnp} \\ \de U_p \end{pmatrix},
	\end{equation}
then we have an eigenvalue equation for $\s$,
	\begin{equation}
		\s \de \v{x} = A \de \v{x},
	\end{equation}
where $A$ is the linear matrix at the equilibrium point $\v{x}$.  The sums as written above give the matrix-vector product.  However, one can also extract the matrix itself without too much difficulty.  The matrix is sparse and should be represented as such.  When calculating the matrix coefficients, one needs to remember to convert the terms $\de \Psi \to \de W$.

Note that there is a different eigenvalue equation for each $Q$.  For determining stability, one must solve the eigenvalue problem for every $Q$ in $- \tfrac12 q < Q \le \tfrac12 q$.  The equilibrium is unstable if for any $Q$ there are any eigenvalues of $A$ with $\Re \s > 0$.  To calculate this efficiently, Arnoldi iterative algorithms seem to be the best approach.  Finally, we point out that these equations contain ZI as a special case, for which the equilibrium is the homogeneous one and $Q$ takes on the role of the wave number $q$.

It is possible to show two symmetries regarding the eigenvalue, which follow from the symmetry of the ideal state equilibrium.  They can be verified directly in a straightforward, if tedious, way.  First, for arbitrary $Q$, suppose $\bigl(\de W_{mnp}^{(1)}, \de U_{p'}^{(1)} \bigr)$ is an eigenvector with eigenvalue $\s$.  Then the vector $\bigl(\de W_{mnp}^{(2)}, \de U_{p'}^{(2)} \bigr)$ is also an eigenvector, with eigenvalue $\s^*$, and
	\begin{subequations}
	\begin{align}
		\de W_{mnp}^{(2)} &= \de W_{m,-n,p}^{(1)*}, \\
		\de U_{p'}^{(2)} 	&= \de U_{p'}^{(1)*}.
	\end{align}
	\end{subequations}
This guarantees that every complex eigenvalue comes in a conjugate pair.  Second, suppose at some $Q$ that $\bigl(\de W_{mnp}^{(Q)}, \de U_{p'}^{(Q)} \bigr)$ is an eigenvector with eigenvalue $\s$.  Then when the Bloch wave number is $-Q$, the vector $\bigl(\de W_{mnp}^{(-Q)}, \de U_{p'}^{(-Q)} \bigr)$ is an eigenvector with eigenvalue $\s^*$, and
	\begin{subequations}
	\begin{align}
		\de W_{mnp}^{(-Q)} &= \de W_{m,n,-p}^{(Q)*}, \\
		\de U_{p'}^{(-Q)} 	&= \de U_{-p'}^{(Q)*}.
	\end{align}
	\end{subequations}
Thus, for determining stability one actually needs to check only $0 \le Q \le \frac12 q$ because the eigenvalues for negative $Q$ are symmetric.

\subsection{Results}
\begin{figure}
		\centering
		\includegraphics{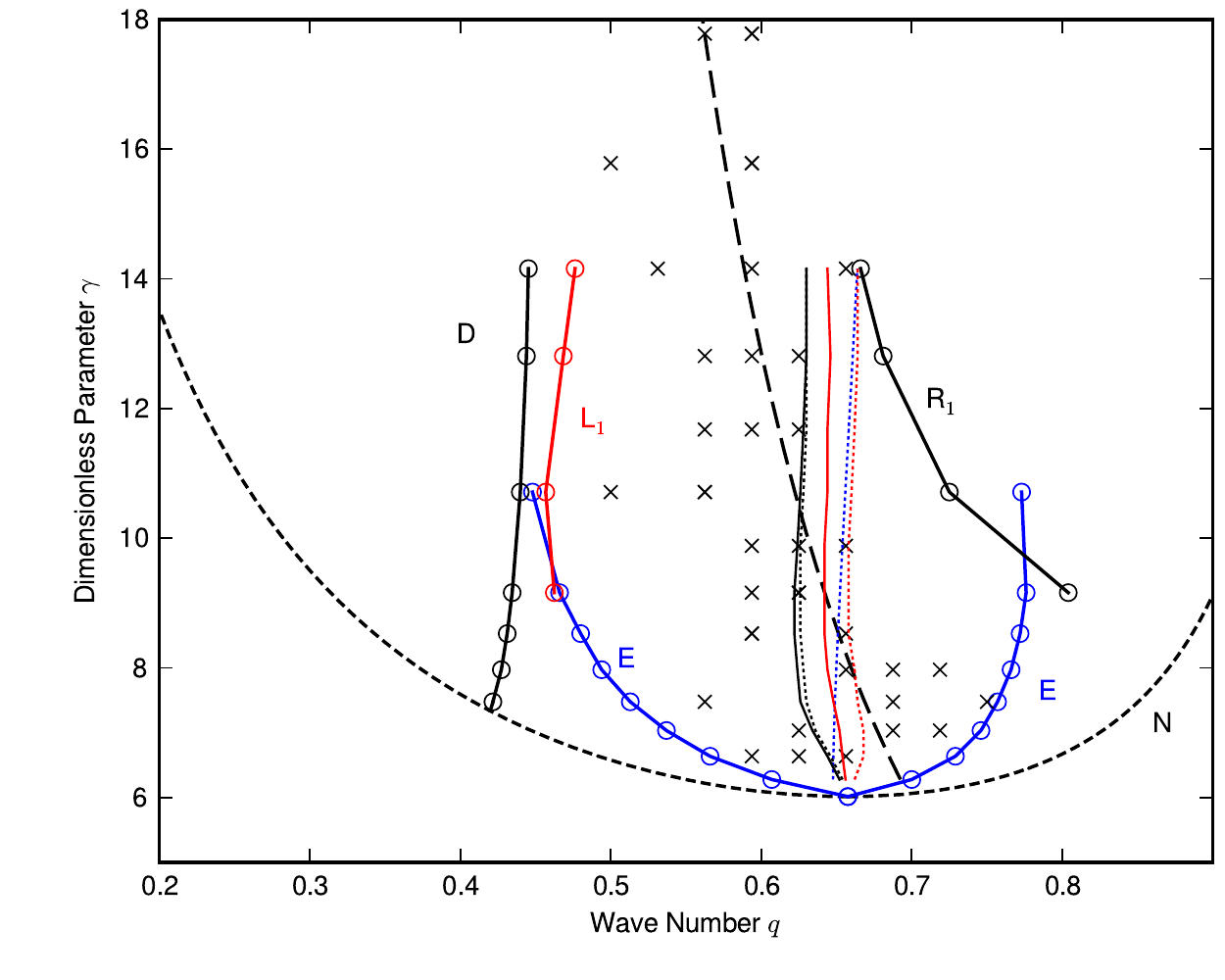}
		\caption{Stability diagram for the CE2 equations.  Above the neutral curve (N), the homogeneous turbulent state is zonostrophically unstable.  Ideal states are stable within the marginal stability curves (circles) E, L$_1$, and R$_1$.  The circle-points were computed by solving for ideal states using the method of \secref{is:equilibrium}, then finding the marginal stability boundary using the method of \secref{is:stability}.  Also shown is the dominant \zf{} wave number from independent QL simulations (crosses).  The crosses were computed by performing direct numerical simulation of the QL equations \eref{qlsystem} and determining the dominant \zf{} wave number.  The stability region calculated from CE2 is consistent with the ZFs realized in the QL simulation.  The stationary ideal states vanish to the left of curve D.  Discussed in \secref{is:wavenumberselection} are the interior curves: Rhines wave number (black dashed line), wave number of maximum growth rate for zonostrophic instability (blue dotted line), wave numbers of minimum eddy energy (red line), minimum total energy (black line), minimum eddy enstrophy (red dotted line), and minimum total enstrophy (black dotted line).  Here, $a=0.06$, $b=0.08$, $M=20$, $N=27$, $P=4$, and other parameters are given in the text.  $\g$ is varied by changing $\m$ while holding the other parameters fixed.}
		\label{is:stability_balloon}
	\end{figure}

The stability diagram is shown in \figref{is:stability_balloon}.  To vary $\gamma$, the fundamental dimensionless parameter defined in \eref{gamma}, we change $\m$ and hold other parameters fixed (at $\b=1$, $L_d=\infty$, $\n=10^{-3}$, $h=4$).  The stable ideal states exist inside of the marginal stability curves marked E, L$_1$, and R$_1$, which represent different instabilities.  The Eckhaus instability (E) is a long-wavelength universal instability, present even in the amplitude equation \eref{gl:amplitudeequation}.  The L$_1$ and R$_1$ curves represent the marginal stability boundary for novel short-wavelength instabilities.

The zonal jets are spontaneously generated by ZI for $\g > \g_{\rm c} = 6.02$.  For $6.02 < \g < 14.2$, the stability curve is consistent with the dominant \zf{} wave number observed in QL simulations.  For $\g > 14.2$, we could not calculate the stability diagram with this approach due to the aforementioned numerical issues of finding the steady state.

\begin{figure}
		\centering
		\includegraphics{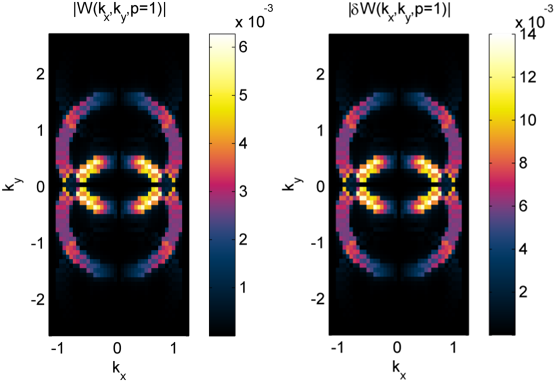}
		\caption{Left: the $W(p=1)$ component of the ideal state.  Right: the $\de W(p=1)$ component of the instability eigenvector.  The eigenvector of the Eckhaus instability is a long-wavelength modulation with otherwise the same spectral structure as the equilibrium.}
		\label{is:stability1}
	\end{figure}

Part of an unstable eigenvector for the Eckhaus instability is shown in \figref{is:stability1}.  For this figure, $q$ is just outside of the marginal stability curve, so the equilibrium is barely Eckhaus-unstable, and $Q$ is very small.  On the left is the $p=1$ spectral content of the ideal state equilibrium.  On the right is the $p=1$ spectral content of the eigenvector.  The two are proportional.  The perturbation is a long-wavelength modulation with otherwise the same spectral structure as the equilibrium.  The $L_1$ and $R_1$ instabilities have not been analyzed in detail; that could be taken up in future work.

\section{Wavenumber Selection}
\label{is:wavenumberselection}
As evident from \figref{is:stability_balloon}, we are presented with the theoretical quandary of having a wide range of allowed, stable solutions and yet a narrow preferred region where QL realizations tend to appear.  This is common to pattern-forming systems, and the problem of wavenumber selection is difficult \cite{cross:2009}.  The Rhines wavenumber \eref{intro:rhineswavenumber} can be estimated by using $\frac12 U^2 = E$ and $E = \varepsilon/2\m$ to give
	\begin{equation}
		k_{R} \approx \b^{1/2} \m^{1/4} \varepsilon^{-1/4}.
	\end{equation}
This estimate works well in giving the preferred \zf{} wavenumber.  In this section we explore what features of the equilibrium might correlate with the preferred wavenumber, in an attempt to achieve a greater understanding of what determines wavenumber selection of \zfs{} \cite{parker:2014}.

One might naturally inquire as to whether the preferred mode is the fastest growing mode in the ZI about the homogeneous equilibrium.  This does not appear to be the case away from the threshold at larger $\g$ \cite{srinivasan:2012,farrell:2007}, as seen in \figref{is:stability_balloon}.  There is, however, a plausible scenario that emerges which may explain the merging of jets often observed in the beginning stages of simulations, especially those which initialize everything at low amplitudes.  At large $\g$, it appears that the fastest growing mode may be to the right of the stability region.  In a simulation, the turbulence quickly comes to a quasi-equilibrium on a short time scale and begins to drive the \zf{}.  The growing \zf{} mode cannot stably saturate, for its wavelength is too small to coexist with the turbulence.  As the system evolves through the subsequent instability to drive the jets toward larger wavelength, a space-time visualization such as that in \figref{fig:zfplots} displays merging jets.

Another possibility is that some kind of variational principle applies.  The amplitude equation, by which CE2 is governed near threshold, is a gradient system.  The ideal states of varying wave number $q$ have varying values of the effective free energy.  However, the minimum of the effective free energy is not necessarily dynamically preferred \cite{schober:1986}.  In any case, away from threshold CE2 is not a gradient system and there is no rigorous theoretical basis for expecting variational behavior to occur.  From a different perspective, variational principles for certain 2D turbulent systems have long been discussed theoretically.  Some of these principles are based on the nonlinearly conserved quadratic quantities, the energy and the enstrophy.  For instance, in freely decaying turbulence where viscosity provides the dissipation, the enstrophy is expected to decay more quickly than the energy.  One might expect the decaying turbulence to reach a state of minimum enstrophy subject to the constraint of constant energy.  Other principles exist based on minimum dissipation or maximum entropy or entropy production \cite{majda:2006}.  Although these principles do not directly apply to the damped, driven CE2 system, they at least motivate a numerical exploration to try to discover any correlation between the preferred wave numbers and other properties.

\begin{figure}
		\centering
		\includegraphics{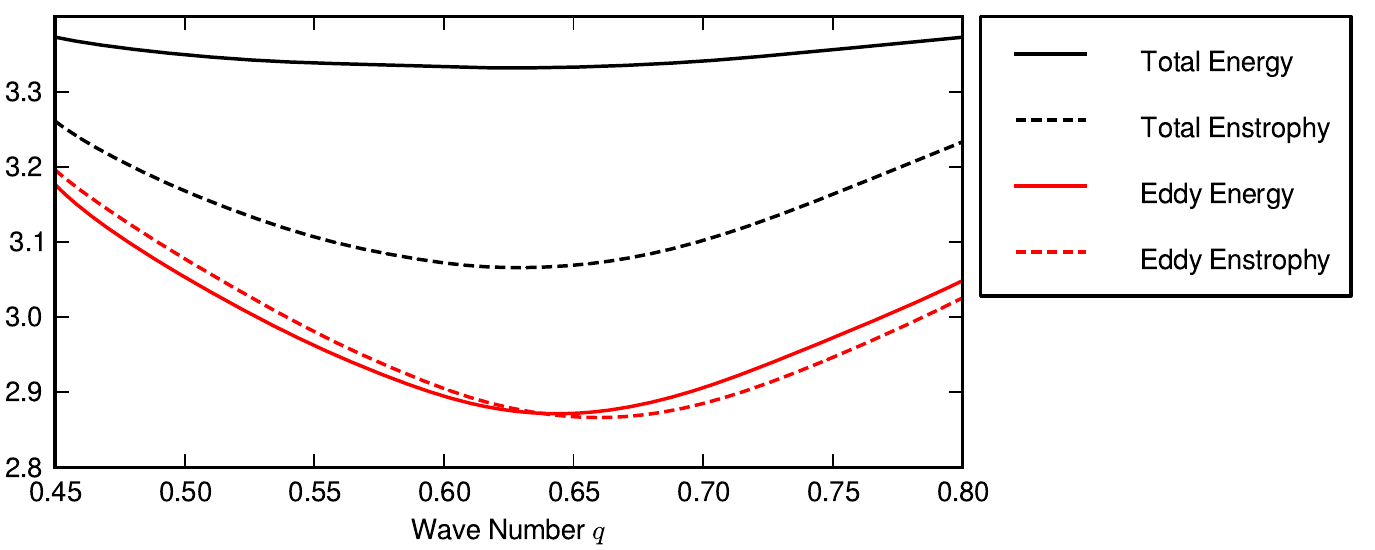}
		\caption{The energy and enstrophy, both total and that of just the eddies, for ideal states of varying wave number $q$ at $\m = 0.15$.}
		\label{ws:enstrophy}
	\end{figure}

As a simple starting point for our exploration, we examine the energy and enstrophy of the ideal states.  Plots of the energy and enstrophy, for both the total and just the eddies, are shown in \figref{ws:enstrophy} for $\m=0.15$.  For each quantity a distinct minimum is present.  We find at each $\m$ the minimum of all four quantities; the resulting curves are shown in \figref{is:stability_balloon}.  While the minima of the total energy and total enstrophy are consistent with the QL realizations, there is no clear indication that either is especially preferred.  On the other hand, the accessible regime investigated here is not too far from threshold, so this is not in the asymptotic regime of large $\g$.

There is no definitive conclusion to draw from these explorations.  Determining and understanding the length scale of zonal flow remains an important and unsolved problem in plasma physics.  Future investigations along the line discussed here may prove useful. 

\chapter{A Pedagogical Closure for Homogeneous Statistics}
\label{ch:closure}
In this chapter we propose a closure for homogeneous statistics.\footnote{This work is unpublished.}  The material here is separate and not immediately connected to the work on zonal flows in the rest of this thesis.  With some extensions described in the following paragraphs and in \chref{ch:future}, it could be connected to zonation and inhomogeneity.\footnote{This material might be more logically placed as part of \chref{ch:future}, but that would bog that chapter down.  It could be put in an appendix, but it was desired that this material not be doomed to languish in obscurity.}  But even without those extensions, it is interesting in its own right and has pedagogical value for its simplicity.  So while the work here lays the groundwork for further investigation into zonal flows, as it stands, it is simply an interesting venture into a turbulence closure.

Systematic closures for homogeneous turbulence possess important properties, including statistical equilibrium, realizability, and an H-theorem \cite{krommes:2002,carnevale:1981}.  Here, we introduce a simple closure which exhibits some of the important properties in a simple, transparent way.

Additionally, we discuss another property that has not previously received much attention, which is the stability of the closure's steady-state solutions.  From our discussion of zonostrophic instability, it is clear that stability of the homogeneous equilibrium plays a critical role.  In CE2 with external forcing and linear damping, this stability is trivial.  In the closure we introduce here, it is decidedly nontrivial, yet we are still able to prove general statements about stability.  Finally, the closure is simple enough that its equilibria can be completely characterized.  We find that there is a unique physical solution, and the necessary and sufficient conditions for its existence can be explicitly stated.  When the wavenumbers are discretized, there is a large number of nonphysical solutions.  In some special cases, analytic solutions can be found.

This ``toy'' closure is not intended to be an accurate portrayal of turbulence.  It is a stepping stone, like the works of Kraichnan and Spiegel or of Leith in the early days of analytic turbulence theory \cite{kraichnan:1962,leith:1967}, in an attempt to understand some piece of the puzzle.  Those studies used simple closures of homogeneous turbulence to try to understand the inertial range in 3D.  Here, our ultimate goal is to get at the fundamental mathematical structure that underlies the bifurcation at which zonal flows are born.

\section{Model Closure}
Consider a 1-field turbulent model,
	\begin{equation}
		\dot{\psi}_k = \frac12 \L_k \psi_k + \mathcal{N}_k[\psi_k],
	\end{equation}
where $\L_k$ represents linear terms, including drive and dissipation, and $\mathcal{N}_k$ incorporates all nonlinear terms.  We will consider the two-point correlation function, $C_k = \avg{\psi_k \psi^*_k}$.  As an equation that might represent the dynamics of $C_k$, our closure is
	\begin{equation}
		\dot{C}_k = \g_k C_k - \m_k C_k \ol{C} + f_k \ol{C}^2,
		\label{tc:Cdot}
	\end{equation}
where $\g_k = \Re{\L}_k$, and $\ol{C}$ is independent of $k$, has the same dimensions as $C_k$, and is in some way a measure of the turbulent intensity.  This form is intended to mimic the structure of turbulent damping plus nonlinear forcing, which appears in more sophisticated closures like the Direct-Interaction Approximation (DIA) or the Eddy-Damped Quasi-Normal Markovian Closure (EDQNM).  Thus, we require the $f_k$ to be positive, and we also find that we will have nice properties if all of the $\m_k$ are positive as well.  The $f_k$ and $\m_k$ are left otherwise totally unspecified.

If energy conservation among the nonlinear terms is desired, there is a unique choice for $\ol{C}$ which provides this.  Suppose the energy in a given mode is
	\begin{equation}
		E_k = \s_k C_k,
	\end{equation}
where $\s_k$ is a positive-definite weighting.  Then the total energy is $E = \sum_k \s_k C_k$, and the rate of change of energy due to nonlinear terms is
	\begin{align}
		\dot{E} \bigr|_{NL} &= \sum_k \s_k \dot{C}_k\bigr|_{NL} \notag\\
			&= \sum_k \s_k \left( -\m_k C_k \ol{C} + f_k \ol{C}^2 \right) \notag \\
			&= \ol{C} \left( \ol{C} \sum_k \s_k f_k - \sum_k \s_k \m_k C_k \right).
	\end{align}
Therefore, for $\dot{E} \bigr|_{NL} = 0$, we find
	\begin{equation}
		\ol{C} = \frac{ \sum_k \s_k \m_k C_k}{\sum_k \s_k f_k}.
	\end{equation}
Upon defining
	\begin{equation}
		F = \sum_k \s_k f_k,
	\end{equation}
we write
	\begin{equation}
		\ol{C} = \frac{1}{F} \sum_k \s_k \m_k C_k.
	\end{equation}
Thus, $\ol{C}$ is a weighted sum of the energy in each mode (weighted by $\m_k$), with a certain normalizing factor.  This closure yields only one quadratic quantity conserved by nonlinear interactions.  More sophisticated closures like the Direct-Interaction Approximation can conserve multiple such quadratic quantities \cite{kraichnan:1959}.

\section{Properties}
It turns out that this simple closure exhibits many desirable properties of statistical closures.  These properties are:
	\begin{enumerate}
		\item One can prove certain stability properties of the steady-state solutions.  In this case, every nonzero equilibrium is linearly stable.
		\item Equipartition solution is possible for statistical equilibrium with no linear terms.
		\item The nonlinear closure terms exhibit an H-Theorem for monotonic increase of entropy towards the statistical equilibrium.
		\item The system is statistically realizable.  That is, if the system initializes with all nonnegative $C_k$, they stay nonnegative.
		\item The nonlinear equilibria of the system can be completely characterized.  It can be shown that there is one, and only one, physically allowable equilibrium (with all the $C_k$ nonnegative), and a necessary and sufficient condition on the parameters for its existence can be derived.
	\end{enumerate}
There are of course deficiencies owing to the simplicity of the treatment of the nonlinear terms.  For instance, there is no real mode coupling among triads.  Only one quadratic quantity can be conserved.  The closure also does not account for the effect of linear waves.

\subsection{Linear Stability}
Clearly, $C_k = 0$ for all $k$ is one equilibrium.  However, if any of the $\g_k$ are positive, that equilibrium is unstable.  We now show that \emph{any and all} nonzero equilibria are linearly stable.  We do this by providing a positive definite functional, quadratic in the perturbation, which decays in time according to the linearized system.  That is, small deviations from the equilibrium must eventually die away.  This happens to be true even if the equilibrium is nonphysical (meaning the proof goes through even if some of the $C_k$ are negative).

To prove linear stability of any nonzero equilibrium, first note the ``steady-state condition'' of an equilibrium,
	\begin{equation}
		\g_k = \m_k \ol{C} - \frac{f_k}{C_k} \ol{C}^2.
		\label{tc:steadystatecondition}
	\end{equation}
This relation will be used to eliminate $\g_k$ later on.

\subsubsection{Linearization}
Linearize \eref{tc:Cdot} about an equilibrium $C_k$, by letting $C_k(t) = C_k + \de C_k(t)$:
	\begin{equation}
		\de \dot{C}_k = \g_k \de C_k - \m_k \de C_k \ol{C} - \m_k C_k \de \ol{C} + 2 f_k \ol{C} \de \ol{C},
	\end{equation}
where $\de \dot{C}_k \defineas \partial \de C_k / \partial t$, and $\de \ol{C} \defineas \sum_k \s_k \m_k \de C_k / F$.  Now substitute for $\g_k$ using the steady-state condition in \eref{tc:steadystatecondition} to obtain
	\begin{equation}
		\de \dot{C}_k = -\m_k C_k \de \ol{C} - \frac{f_k}{C_k} \ol{C}^2 \de C_k + 2 f_k \ol{C} \de \ol{C}.
	\end{equation}
It will be convenient to write this in terms of $w_k \defineas \de C_k(t) / C_k$:\footnote{Technically, for this to be allowed, none of the equilibrium $C_k$ can be exactly zero}
	\begin{equation}
		\dot{w}_k = -\m_k \de \ol{C} - \frac{f_k}{C_k} \ol{C}^2 w_k + \frac{2 f_k}{C_k} \ol{C} \de \ol{C},
	\end{equation}
where now $\de \ol{C} = \sum_k \s_k \m_k C_k w_k / F$.

\subsubsection{Quadratic Functional}
Consider the quadratic functional
	\begin{equation}
		W(t) = \frac{1}{2} \sum_k h_k C_k w_k(t)^2,
	\end{equation}
where $h_k$ is a positive quantity.  Observe that $W$ is positive definite.  We take
	\begin{equation}
		h_k = \frac{\s_k}{F}
	\end{equation}
as the weighting factor.  Then, the evolution of $W$ is given by
	\begin{align}
		\dot{W} &= \sum_k \frac{\s_k}{F} C_k w_k \dot{w}_k \notag \\
			&= -\sum_k \frac{\s_k}{F} \left( \m_k C_k w_k \de \ol{C} + f_k \ol{C}^2 w_k^2 - 2 f_k \ol{C} w_k \de \ol{C} \right) \notag \\
			&= -\sum_k \left( \frac{\s_k \m_k C_k w_k \de \ol{C}}{F} + \frac{\s_k f_k}{F} \ol{C}^2 w_k^2 - 2 \ol{C} \de \ol{C} \frac{\s_k f_k}{F} w_k \right) \notag \\
			&= -\left(\de \ol{C}^2 - 2 \de \ol{C} \frac{\ol{C}}{F} \sum_k \s_k f_k w_k \right) - \frac{\ol{C}^2}{F} \sum_k \s_k f_k w_k^2 \notag \\
			&= -\left[ \left( \de \ol{C} - \frac{\ol{C}}{F} \sum_k \s_k f_k w_k \right)^2 - \frac{\ol{C}^2}{F^2} \sum_{jk} \s_j \s_k f_j f_k w_j w_k \right] - \frac{\ol{C}^2}{F} \sum_k \s_k f_k w_k^2 \notag \\
			&= -\left( \de \ol{C} - \frac{\ol{C}}{F} \sum_k \s_k f_k w_k \right)^2 - \frac{\ol{C}^2}{F^2} \left( F \sum_k \s_k f_k w_k^2 - \sum_{jk} \s_j \s_k f_j f_k w_j w_k \right).
	\end{align}
In the last equality, the terms in the second set of parentheses can be written
	\begin{align}
		F \sum_k \s_k f_k w_k^2 - \sum_{jk} \s_j \s_k f_j f_k w_j w_k &= \sum_{jk} \bigl(\s_j \s_k f_j f_k w_k^2 - \s_j \s_k f_j f_k w_j w_k \bigr) \notag \\
			&= \frac{1}{2} \sum_{jk} \s_j \s_k f_j f_k \bigl(2 w_k^2 - 2 w_j w_k \bigr) \notag \\
			&=\frac{1}{2} \sum_{jk} \s_j \s_k f_j f_k \bigl(w_k^2 - 2 w_j w_k + w_j^2 \bigr) \notag \\
			&=\frac{1}{2} \sum_{jk} \s_j \s_k f_j f_k (w_j - w_k )^2,
	\end{align}
where to get to the third line, we have swapped indices $j \leftrightarrow k$ in one of the $w_k^2$ terms (with symmetric combinations in front not changing).

Thus, we find
	\begin{equation}
		\dot{W} = -\left( \de \ol{C} - \frac{\ol{C}}{F} \sum_k \s_k f_k w_k \right)^2 - \frac{\ol{C}^2}{2F^2} \sum_{jk} \s_j \s_k f_j f_k (w_j - w_k )^2,
	\end{equation}
and hence, $\dot{W} \le 0$.  And $\dot{W}=0$ only when the perturbation $w_k=0$.

Note that normalizing the perturbation $\de C_k(t)$ by the equilibrium value $C_k$ was not merely convenient.  It also served to eliminate consideration of the zero equilibrium, which is not stable if any of the $\g_k$ are positive.  What we have shown is that all nonzero equilibria are stable.

\subsubsection{External Forcing}
External forcing could be added to the system, where instead of \eref{tc:Cdot}, one might have
	\begin{equation}
		\dot{C}_k = \g_k C_k - \m_k C_k \ol{C} + f_k \ol{C}^2 + \mathcal{F}_k,
	\end{equation}
where $\mathcal{F}_k$ is some known external quantity and is nonnegative.  In this case, $C_k=0$ for all $k$ is obviously no longer a solution.  However, in a minor modification of the above proof, one still reaches the conclusion that any nonzero equilibrium is linearly stable.

\subsection{Statistical Equilibrium}
If we consider \eref{tc:Cdot} with all the $\g_k=0$ and just the nonlinear closure terms, we find that a steady state is given by
	\begin{equation}
		0 = -\m_k C_k \ol{C} + f_k \ol{C}^2,
	\end{equation}
or
	\begin{equation}
		C_k = \frac{f_k}{\m_k} A,
	\end{equation}
where $A$ is some constant and is equal to $\ol{C}(t \to \infty)$.  Note that if $f_k$ and $\m_k$ are proportional, that is, if $f_k/\m_k$ is independent of $k$, then we in fact have an equipartition statistical equilibrium.  Also, if any of the $\m_k$ are negative, then the statistical equilibrium predicts a negative $C_k$ and is unphysical.

By using the fact that energy is conserved, we can determine $A$ from the initial conditions.  We have
	\begin{align}
		E(t=0) &= \sum_k \s_k C_k(t=0), \\
		E(t\to\infty) &= \sum_k \s_k C_k(t \to \infty) = A \sum_k \frac{\s_k f_k}{\m_k},
	\end{align}
and thus, by conservation of energy,
	\begin{equation}
		A = \frac{ \sum_k \s_k C_k(t=0) }{ \sum_k \s_k f_k / \m_k }.
	\end{equation}

\subsection{Entropy and H-Theorem}
\label{sec:tc:htheorem}
\citet{carnevale:1981} have shown that in many second order closures, it is useful to think of the entropy as $S = \sum_k \ln C_k$.  Here, it turns out we need to use a slightly different definition, namely, the entropy-like quantity
	\begin{equation}
		S = \sum_k \frac{\s_k f_k}{\m_k} \ln C_k.
	\end{equation}		
(We do not prove that this is equivalent to an entropy for this model, but it has similar behavior.)  This form, like that of Carnevale et al., is scale-independent.  That is, if $C_k \to h_k C_k$, the $h_k$ terms affect only the absolute entropy but not changes in entropy because the $(\s_k f_k / \m_k) \ln h_k$ factors are simply constants.

The evolution of the entropy is given by
	\begin{align}
		\dot{S} &= \sum_k \frac{\s_k f_k}{\m_k} \frac{1}{C_k} \dot{C}_k \notag \\
			&= \sum_k \frac{\s_k f_k \g_k}{\m_k}  +  \sum_k \frac{\s_k f_k}{\m_k} \left( -\m_k \ol{C} + \frac{f_k}{C_k} \ol{C}^2 \right).
	\end{align}
The contribution from the nonlinear piece can be written
	\begin{align}
		\sum_k \frac{\s_k f_k}{\m_k} \left( -\m_k \ol{C} + \frac{f_k}{C_k} \ol{C}^2 \right)	&= \ol{C} \sum_k \left( -\s_k f_k + \frac{\s_k f_k^2}{\m_k C_k} \ol{C} \right) \notag \\
			&= \frac{\ol{C}}{F}\left( -\sum_{jk} \s_j \s_k f_j f_k + \sum_{jk} \frac{\s_k f_k^2}{\m_k C_k} \s_j \m_j C_j \right) \notag \\
			&= \frac{\ol{C}}{F} \sum_{jk} \s_j \s_k \left( \frac{\m_j f_k^2}{\m_k} \frac{C_j}{C_k} - f_j f_k \right) \notag \\
			&= \frac{\ol{C}}{2F} \sum_{jk} \frac{\s_j \s_k f_j^2 f_k^2}{\m_j \m_k C_j C_k} \left( \frac{2 \m_j^2 C_j^2}{f_j^2} - 2 \frac{\m_j \m_k}{f_j f_k} C_j C_k \right) \notag \\
			&= \frac{\ol{C}}{2F} \sum_{jk} \frac{\s_j \s_k f_j^2 f_k^2}{\m_j \m_k C_j C_k} \left( \frac{\m_j^2 C_j^2}{f_j^2} - 2 \frac{\m_j \m_k}{f_j f_k} C_j C_k + \frac{\m_k^2 C_k^2}{f_k^2} \right) \notag \\
			& = \frac{\ol{C}}{2F} \sum_{jk} \frac{\s_j \s_k f_j^2 f_k^2}{\m_j \m_k C_j C_k} \biggl( \frac{\m_j C_j}{f_j} - \frac{\m_k C_k}{f_k} \biggr)^2.
	\end{align}
And hence,
	\begin{equation}
		\dot{S} = \sum_k \frac{\s_k f_k \g_k}{\m_k} + \frac{\ol{C}}{2F} \sum_{jk} \frac{\s_j \s_k f_j^2 f_k^2}{\m_j \m_k C_j C_k} \biggl( \frac{\m_j C_j}{f_j} - \frac{\m_k C_k}{f_k} \biggr)^2.
	\end{equation}
	
First, notice that there is an H-Theorem.  If all the $\g_k$ are zero, then $\dot{S}$ is positive definite, and increases monotonically towards the statistical equilibrium state where $C_k \propto f_k / \m_k$.
	
With the $\g_k$ included, then since the contribution from the nonlinear piece is always positive, a necessary condition for a steady state to be reached is that the contribution from the linear piece be negative:
	\begin{equation}
		\sum_k \frac{\s_k f_k \g_k}{\m_k} < 0.
	\end{equation}
We will see later from a detailed analysis of the solutions that this is also a sufficient condition for an equilibrium to exist.

\subsection{Realizability}
\paragraph{Proof: Algebraic}

Assume that the ODE is solved as an initial-value problem, and all the $C_k$ are initially nonnegative (and hence $\ol{C}$ is also nonnegative).  Then, as the system evolves in time according to \eref{tc:Cdot}, the $C_k$ remain nonnegative.  For if $C_k$ were ever 0, then $\dot{C_k} = f_k \ol{C}^2 \ge 0$.  Thus, the system is realizable.

This argument does not easily generalize to models with more than one field, however, because if $C^\a_k=0$, then linear coupling to a different field $C^\b_k$ may still in principle cause $C^\a_k$ to possible become negative.

\paragraph{Proof: Langevin Equation}
Alternatively, one can prove realizability by providing a Langevin equation which has the same statistics as \eref{tc:Cdot}.  The derivation is somewhat similar to that used for the standard Markovian closures in that the terms in the Langevin equation depend on the statistics of the solution (but there is no analog to the triad interaction time here).  Consider the random equation
	\begin{equation}
		\dot{\ti{\psi}}_k = \frac{1}{2}\left[ \g_k \ti{\psi}_k - \eta_k(t) \ti{\psi}_k \right] + \ti{a}_k,
		\label{tc:langevin}
	\end{equation}
where
	\begin{gather}
		\eta_k(t) = \m_k \ol{C}(t) = \frac{\m_k}{F} \sum_k \s_k \m_k C_k, \\
		\ti{a}_k(t) = \ti{w}(t) \ol{C}(t) \sqrt{f_k}.
	\end{gather}
We assume the $C_k$ (and $\ol{C}$) in the random equation are ``known'', or at least nonrandom and independent of $\ti{\psi}_k$.  This comes from assuming that in the limit of a large ensemble, each individual $\ti{\psi}_k$ contributes only infinitesimally to the statistics to give $C_k$.  Also, $\ti{w}(t)$ is Gaussian white noise with zero mean and
	\begin{equation}
		\avg{\ti{w}(t) \ti{w}(s)} = \de(t-s).
	\end{equation}

Then, let us compute the second-order statistics for
	\begin{equation}
		C_k(t) \defineas \Bigl\langle \ti{\psi}_k(t) \ti{\psi}_k^*(t) \Bigr\rangle.
	\end{equation}
We have
	\begin{align*}
		\dot{C}_k(t) &= 2 \Re \Bigl\langle \dot{\ti{\psi}}_k(t) \psi_k^*(t) \Bigr\rangle \\
			&= \Re \bigl\langle (\g_k - \eta_k)\ti{\psi}_k \ti{\psi}^*_k \bigr\rangle  +  2 \Re \bigl\langle \ti{a}_k(t) \ti{\psi}^*_k(t) \bigr\rangle \\
			&= (\Re \g_k) C_k - (\Re \eta_k) C_k + 2 \Re \bigl\langle \ti{a}_k(t) \ti{\psi}^*_k(t) \bigr\rangle.
	\end{align*}
We assume $\g_k$ and $\eta_k$ to be real.  To compute $\bigl\langle \ti{a}_k(t) \ti{\psi}^*_k(t) \bigr\rangle$, we write down the Green's function solution of \eref{tc:langevin},
	\begin{equation}
		\ti{\psi}_k(t) = \psi_k(0) e^{\int_0^t d\t\, \ol{\g}_k (\t)} + \int_0^t ds\, e^{\int_s^t d\t\, \ol{\g}_k(\t)} \ti{a}_k(s),
	\end{equation}
where $\ol{\g}_k \defineas (\g_k - \eta_k)/2$.  We assume $\psi_k(0) = 0$.  Thus,
	\begin{gather}
		\ti{\psi}^*_k(t) = \int_0^t ds\, e^{\int_s^t d\t\, \ol{\g}_k(\t)} \ti{a}^*_k(s), \\
		\bigl\langle \ti{a}_k(t) \ti{\psi}^*_k(t) \bigr\rangle = \int_0^t ds\, e^{\int_s^t d\t\, \ol{\g}_k(\t)} \bigl\langle \ti{a}_k(t) \ti{a}^*_k(s) \bigr\rangle, \\
		\bigl\langle \ti{a}_k(t) \ti{a}^*_k(s) \bigr\rangle = \de(t-s) f_k \ol{C}^2.
	\end{gather}
So
	\begin{align*}
		\bigl\langle \ti{a}_k(t) \ti{\psi}^*_k(t) \bigr\rangle &= f_k \ol{C}^2 \int_0^t ds\, \de(t-s) e^{\int_s^t d\t\, \ol{\g}_k(\t)} \\
			&= \frac{1}{2} f_k \ol{C}^2.
	\end{align*}
Finally, we have
	\begin{equation}
		\dot{C}_k = \g_k C_k - \m_k C_k \ol{C} + f_k \ol{C}^2.
	\end{equation}
which matches \eref{tc:Cdot}.


\subsection{Nonlinear Equilibria}
Now let's actually try and solve for the equilibria of the system in \eref{tc:Cdot}.  In equilibrium, we have
	\begin{equation}
		0 = (\g_k - \m_k \ol{C}) C_k + f_k \ol{C}^2.
		\label{tc:Cdoteqb}
	\end{equation}
Since $f_k \ol{C}^2 > 0$, any equilibrium must have $(\g_k - \m_k \ol{C}) C_k < 0$.  Therefore for any physical equilibrium with all the $C_k>0$, we must have
	\begin{equation}
		\m_k \ol{C} - \g_k > 0
	\end{equation}
for every $k$.

From \eref{tc:Cdoteqb}, we may write
	\begin{equation}
		C_k\bigl(\ol{C}\bigr) = \frac{f_k \ol{C}^2}{\m_k \ol{C} - \g_k},
		\label{tc:CkofCbar}
	\end{equation}
which means that if $\ol{C}$ is known, then $C_k$ for each $k$ is known.  The method of solution is now to solve for $\ol{C}$.  Multiply \eref{tc:CkofCbar} by $\s_k \m_k$ to obtain
	\begin{equation}
		\s_k \m_k C_k = \frac{\s_k \m_k f_k}{\m_k \ol{C} - \g_k} \ol{C}^2.
	\end{equation}
Then sum over $k$ to give
	\begin{equation*}
		F \ol{C} = \ol{C}^2 \sum_k \frac{\s_k \m_k f_k}{\m_k \ol{C} - \g_k}.
	\end{equation*}
Dividing through by $\ol{C}^2$ (assuming we don't want the trivial solution with all the $C_k=0$) gives
	\begin{equation}
		\frac{F}{\ol{C}} = \sum_k \frac{\s_k f_k}{\ol{C} - \g_k/\m_k}.
		\label{tc:nonlinsolutionform1}
	\end{equation}

Substituting in the form of $F$ and combining terms gives
	\begin{equation}
		0 = \sum_k \frac{\s_k f_k \g_k / \m_k}{\ol{C} - \g_k / \m_k}.
		\label{tc:nonlinsolutionform2}
	\end{equation}

This equation completely describes the nonlinear solutions.  One finds the solution $\ol{C}$, and then computes $C_k$ from \eref{tc:CkofCbar}.  The above equation has more than one possible solution.  If there are $N$ modes in the system, then $k=1,\ldots,N$.  If one multiplies through by all the denominators to obtain a polynomial equation in $\ol{C}$, one finds a degree $N-1$ polynomial, and hence $N-1$ possible solutions to this equation (adding back in the trivial solution $\ol{C}=0$ gives a total of $N$).  To determine how many roots are real or complex, we must do something different.  Note that statistical equilibrium is retained here by noting that for $\g_k=0$, any $\ol{C}$ is an allowable solution.

\begin{figure}
		\centering
		\includegraphics[scale=0.8]{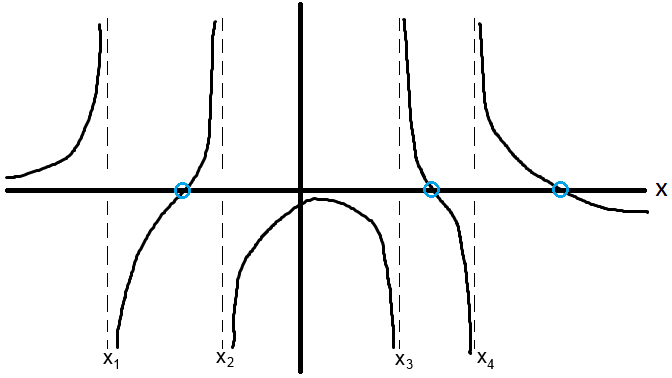}
		\caption{Schematic graph of \eref{tc:graphicalsoln} with $N=4$, with zeros highlighted.  For $\sum_k \s_k f_k x_k <0$.}
		\label{tc:fig:graphicalsolution}
	\end{figure}

A graphical approach is fruitful.  Let $x \defineas \ol{C}$, and $x_k \defineas \g_k / \m_k$.  The problem is then restated as solving for the values of $x$ which satisfy
	\begin{equation}
		\sum_{k=1}^N \frac{ \s_k f_k x_k}{x - x_k} = 0.
		\label{tc:graphicalsoln}
	\end{equation}
Graphically, this amounts to finding where the function on the LHS crosses the $x$ axis.  For simplicity, assume that all of the $x_k$ are distinct and nonzero.  Some of the $x_k$ are positive and some are negative, corresponding to positive and negative $\g_k$.  Then the graph of this function has $N$ vertical asymptotes, one at each $x_k$.  For instance, with $N=4$, the graph may look something like that depicted in \figref{tc:fig:graphicalsolution}.
Supposing that $m$ of the $x_k$ are negative and $n$ are positive, then because of the vertical asymptotes there are guaranteed to be at least $(m-1) + (n-1) = N-2$ real roots.  Since we have proven that there are only $N-1$ possible solutions, then only one other solution can exist, and so it must be real.  It is not immediately obvious whether the remaining root will occur to the left of all the $x_k$, to the right of all the $x_k$, or between the two $x_k$ asymptotes surrounding $x=0$.  We shall discover that there is never a root between the two $x_k$ asymptotes surrounding $x=0$, and that the remaining root always occurs to the right of all the $x_k$, or to the left of all the $x_k$, depending on a certain criterion.

To determine if a root exists to the right of all the $x_k$, we simply want to know if the function crosses zero.  Since the function is monotonically decaying, we can answer that question by determining whether it is positive or negative in the $x\to\infty$ limit.  Taking the large $|x|$ limit of \eref{tc:graphicalsoln}, we find the LHS goes as
	\begin{equation}
		\frac{1}{x} \sum_{k=1}^N \s_k f_k x_k.
	\end{equation}
If this is negative at large $x$, there must be a root to the right of all the $x_k$.  This condition is given by
	\begin{equation}
		\sum_{k=1}^N \s_k f_k x_k = \sum_{k=1}^N \frac{\s_k f_k \g_k}{\m_k} < 0.
		\label{tc:conditionforeqb}
	\end{equation}
On the other hand, if that quantity is negative at large negative $x$, there must be a root to the left of all the $x_k$.  This condition is given by the opposite,
	\begin{equation*}
		\sum_{k=1}^N \s_k f_k x_k = \sum_{k=1}^N \frac{\s_k f_k \g_k}{\m_k} > 0.
	\end{equation*}
Thus, assuming that sum is not exactly equal to zero, one of these two conditions must be true, and the final root occurs either to the left or to the right of all the $x_k$, and not between the two $x_k$ asymptotes surrounding $x=0$.

Now that we have a picture of where the solutions for $x$ or $\ol{C}$ are, let's determine which solutions are physically allowable.  From \eref{tc:CkofCbar}, we see that for $C_k$ to be positive, the quantity $\m_k \ol{C} - \g_k$ must be positive for all $k$, or equivalently, 
	\begin{equation}
		x - x_k > 0, \qquad \text{for all $k$}.
	\end{equation}
If $x > x_k$ for all $x_k$, then graphically this corresponds to the root of \eref{tc:graphicalsoln} being to the right of all the vertical asymptotes.  We have already shown that the condition in \eref{tc:conditionforeqb} is the necessary and sufficient condition for such a root to exist.  If it does exist, it is the unique physical solution.  The other solutions correspond to some of the $C_k$ being negative.  The number of negative $C_k$ correspond to how many of the $x_k$ are to the right of the root $x$.  Incidentally, even the unphysical equilibria are linearly stable.  If the condition in \eref{tc:conditionforeqb} is not satisfied, then the system does not saturate; it blows up.  One may think of the $\g_k$ as being too large in this case.

It is remarkable that the equilibria of \eref{tc:Cdot} can be fully characterized.  With a statistical closure, one is primarily interested in the actual steady state, not the transient evolution to the steady state.  Numerically evolving the statistical closure model, like \eref{tc:Cdot}, can be time-consuming, mainly because a small timestep is required to ensure both accuracy and stability.  It might be advantageous to write down the equation for steady state and find a way to directly compute the solutions.  However, in general certain difficulties arise when attempting this route.  In particular,
	\begin{itemize}
		\item A nonlinear solution may be unphysical (see, e.g., \secref{is:equilibrium}).  Obviously, one can discard any unphysical solution that is found, but there is still the problem of how to find a physical solution.  The time evolution method does not have this problem, so long as the model is realizable.
		\item A nonlinear solution may be (linearly) unstable.  One has no way of knowing (without further computation) whether the solution that was found is stable to small perturbations, and an unstable equilibrium has no relevance.  A separate stability calculation can be done (see, e.g., \secref{is:stability}), but this may be difficult.  The time evolution method does not have this problem, since any equilibrium found that way must be stable.
		\item Not knowing how many physical solutions may exist.  The time evolution method also has this problem---an equilibrium may be found, but it is not known if others exist.
	\end{itemize}
The closure used here is simple enough that it can be analyzed in sufficient detail to overcome all three of these difficulties.
	
In practice, it is not advisable to try to numerically solve the polynomial form of the equation for $\ol{C}$.  This is because finding the roots of high degree polynomials is an ill-conditioned problem \cite{wilkinson:1994}.  Since roundoff error is inevitable, large errors can result, including finding complex roots even though they should all be real.  Instead, using a standard nonlinear root finder on the form in \eref{tc:graphicalsoln} is preferable, especially because it is known that there is one and only one zero in the domain $(\max(x_k),\infty)$.

\subsection{Continuum}
The same results hold when using a continuum rather than discrete modes.  The results are summarized below:

	\begin{gather}
		\dot{C}(k) = \g(k) C(k) - \m(k) C(k) \ol{C} + f(k) \ol{C}^2, \\
		E(k) = \s(k) C(k), \\
		E = \int dk\, E(k), \\
		\ol{C} = \frac{1}{F} \int dk\, \s(k) \m(k) C(k), \\
		F = \int dk\, \s(k) f(k).
	\end{gather}
Then the nonlinear terms conserve the total energy $E$.  Here, $k$ can be a vector.

Equilibria are obtained from
	\begin{equation}
		0 = \bigl[ \g(k) - \m(k) \ol{C} \bigr] C(k) + f(k) \ol{C}^2.
	\end{equation}
Any physical equilibrium must have $\m(k) \ol{C} - \g(k) > 0$ for all $k$.  Divide through by $\m(k) \ol{C} - \g(k)$ to obtain
	\begin{equation}
		C(k) = \frac{f(k) \ol{C}^2}{ \m(k) \ol{C} - \g(k) }.
	\end{equation}
The above step is only valid for $\ol{C} \neq \max\bigl(\g(k)/\m(k)\bigr)$ for any $k$.  This is satisfied if $\ol{C} > \max\bigl(\g(k)/\m(k)\bigr)$.  Multiply by $\s(k) \m(k)$ and integrate over $k$ to obtain
	\begin{equation}
		F \ol{C} = \ol{C}^2 \int dk\, \frac{\s(k) \m(k) f(k)}{\m(k) \ol{C} - \g(k)}.
	\end{equation}
This can be rearranged as
	\begin{equation}
		\frac{F}{\ol{C}} = \int dk\, \frac{\s(k) f(k)}{\ol{C} - \g(k)/\m(k)},
		\label{tc:continuumnonlinform1}
	\end{equation}
or as
	\begin{equation}
		0 = \int dk\, \frac{\s(k) f(k) \g(k) / \m(k)}{\ol{C} - \g(k)/\m(k)}.
	\end{equation}
	
In the discrete case, we could use the properties of polynomial equations to prove that there was only one possible solution for $\ol{C} > \max( \g_k/\m_k)$.  Here in the continuum case, we have not found a proof that a solution is unique, though that does not mean a proof does not exist.

In the discrete case, we also found a criterion that was both necessary and sufficient for a unique solution.  In the continuous case, that proof gives only a sufficient condition for a solution.  To see this, rewrite \eref{tc:continuumnonlinform1} as
	\begin{equation}
		\frac{F}{x} = \int dk\, \frac{\s f}{x - \ti{\g}},
	\end{equation}
where $\ol{C}=x$ and $\ti{\g} = \g/\m$.  If the RHS is smaller than the LHS as $x \to \infty$, then there is guaranteed to be at least one solution.  If we expand the RHS for small $x$, we obtain
	\begin{align*}
		\int dk\, \frac{\s f}{x - \ti{\g}} &= \frac{1}{x} \int dk\, \s f + \frac{1}{x^2} \int dk\, \s f \ti{\g} + \cdots \\
			&= \frac{F}{x} + \frac{1}{x^2} \int dk\, \s f \ti{\g} + \cdots\, .
	\end{align*}
Therefore, the RHS is smaller than the LHS above as $x \to \infty$ if
	\begin{equation}
		\int dk\, \s(k) f(k) \ti{\g}(k) = \int dk\, \frac{\s(k) f(k) \g(k)}{\m(k)} < 0.
	\end{equation}
Since we have not yet proven the solution is unique, this approach does not show that the criterion is a necessary condition.  However, entropy considerations in the continuous analog to calculations in \secref{sec:tc:htheorem} do prove it is a necessary condition.  

To find an equilibrium, the nonlinear equations to solve are
	\begin{equation}
		\frac{1}{x} \int dk\, \s(k) f(k) = \int_0^\infty dk\, \frac{ \s(k) f(k) }{ x - \ti{\g}(k)},
	\end{equation}
or
	\begin{equation}
		\int dk\,  \frac{ \s(k) f(k) \ti{\g}(k)} { x - \ti{\g}(k)} = 0.
	\end{equation}

\subsubsection{A Solvable Example}
Here we provide an example that can be integrated directly and solved for $x$.  Suppose
	\begin{align}
		\s(k) &= 1, \\
		f(k) &= \frac{f_0}{1 + \b k^2}, \\
		\g(k) &= \g_0 \frac{1-\b k^2}{1 + \b k^2}, \\
		\m(k) &= \frac{\m_0}{1 + \b k^2}, \\
		\ti{\g}(k) &= \frac{\g_0}{\m_0} (1 - \b k^2),
	\end{align}
where $f_0, \m_0, \g_0, \b >0$ and let $k$ extend from 0 to $\infty$ (or from $-\infty$ to $\infty$; it doesn't change the result).  This $\ti{\g}(k)$ peaks at $k=0$ and is one-to-one.  The max of $\ti{\g}$ is $\g_0/\m_0$, so $x>\g_0/\m_0$ is required.  The nonlinear equation for a steady state is
	\begin{equation}
		\int_0^\infty dk\, \frac{f_0}{1+\b k^2} \frac{(\g_0/\m_0) (1 - \b k^2)}{x- (\g_0 / \m_0) ( 1- \b k^2)} = 0.
	\end{equation}
Let $z \defineas x \m_0/\g_0 - 1 > 0$.  Then
	\begin{equation}
		\int_0^\infty dk\, \frac{1-\b k^2}{1+\b k^2} \frac{1}{z + \b k^2} = 0.
	\end{equation}
Change integration variables to $y = \b k^2$, to obtain
	\begin{equation}
		\int_0^\infty dy\, \frac{1}{y^{1/2}} \frac{1 -y}{(y+1)(y+z)} = 0.
		\label{tc:solvableexample1}
	\end{equation}
We use the integrals
	\begin{align}
		\int_0^\infty dy\, \frac{1}{y^{1/2}(y+a)(y+b)} &= \frac{\pi}{\sqrt{ab}(\sqrt{a}+\sqrt{b})}, \\
		\int_0^\infty dy\, \frac{y^{1/2}}{(y+a)(y+b)} &= \frac{\pi}{\sqrt{a}+\sqrt{b}},
	\end{align}
valid for $a,b > 0$.  Equation \eref{tc:solvableexample1} becomes
	\begin{equation}
		\frac{\pi}{\sqrt{z}(1+\sqrt{z})} = \frac{\pi}{1+\sqrt{z}},
	\end{equation}
or
	\begin{gather}
		1 = \sqrt{z}, \\
		1 = z, \\
		1 = \frac{x \m_0}{\g_0} - 1,
	\end{gather}
so that
	\begin{equation}
		x = \ol{C} = 2 \frac{\g_0}{\m_0}.
	\end{equation}
Substituting into $C(k)$, we have
	\begin{equation}
		C(k) = 4 \frac{\g_0 f_0}{\m_0^2} \frac{1}{1 + \b k^2}.
	\end{equation}

A slight modification to the previous example with $\g(k) = \g_0(1 - \b k^2)$ is also solvable.  This $\g(k)$ gets continuously more negative at large $k$, like viscosity, instead of saturating at a constant negative value.

\subsubsection{Another Solvable Example---2D Isotropic}
It is also possible to construct an integrable example that is 2D and isotropic.  Let
	\begin{align}
		\s(k) &= 1, \\
		f(k) &= \frac{f_0}{1 + \b k^4}, \\
		\g(k) &= \g_0 \frac{1-\b k^4}{1 + \b k^4}, \\
		\m(k) &= \frac{\m_0}{1 + \b k^4}, \\
		\ti{\g}(k) &= \frac{\g_0}{\m_0} (1 - \b k^4).
	\end{align}
Note that if $f(k)$ went like $k^{-2}$ at large $k$, the integral for $F$ would not converge.  The nonlinear equation to solve becomes
	\[ \int_0^\infty dk\, \frac{k}{1 + \b k^4} \frac{1 - \b k^4}{z + \b k^4} = 0, \]
where $z \defineas x \m_0/\g_0 - 1 > 0$.  Once again the solution is $z=1$, or
	\begin{equation}
		x = \ol{C} = 2 \frac{\g_0}{\m_0}.
	\end{equation}
Substituting into $C(k)$, we have
	\begin{equation}
		C(k) = 4 \frac{\g_0 f_0}{\m_0^2} \frac{1}{1 + \b k^4}.
	\end{equation}

\section{Discussion}
As far as we are aware, the literature on statistical closures has neglected any kind of detailed examination of stability of the steady states.  We believe the proof here of linear stability is the first such result obtained.  In \appref{app:kraichnanspiegel}, we provide a similar proof of stability for the Kraichnan-Spiegel closure (allowing for linear drive), which encompasses the Leith diffusion closure.  A similar proof for EDQNM remains elusive, despite significant effort spent.  \cite{orszag:1977} clearly believes solutions to EDQNM to be stable.  His arguments are compelling, although he was only considering turbulent drive due to external forcing, whereas we want to allow the case of arbitrary linear drive.  It is not surprising that we have been unable to find a stability proof for EDQNM.  EDQNM is far more complicated and may well allow for solutions which are linearly unstable in certain situations.  A proof may not exist, or it may be beyond the limits of our imagination.

This closure for homogeneous turbulence could be extended with the appropriate terms for inhomogeneous interactions (which would mostly amount to pasting in terms from the CE2 equations).
\chapter{Suggestions for Future Research}
\label{ch:future}
Although this thesis has answered some questions, it has raised many new ones.  Our theoretical analysis has taken place in the simplest possible setting, a 2D infinite (or periodic) system driven by white-noise external forcing.  Naturally, one might wonder how to extend our analysis to more complicated, more realistic systems.  For example, what happens in a realistic, physical geometry like a tokamak?  If the system is driven by an intrinsic instability rather than external forcing, is there any qualitative difference?  Do our previous results still hold in these instances?  If not, why not, and how should the analysis be modified?

We present a few of the issues in some detail, along with some ideas on how to proceed.  Some of these issues could be studied within the CE2 formalism, while others would require more sophisticated statistical closures.  This chapter necessarily includes some speculation in order to offer possible fruitful research directions.

\section{Using CE2}
There are several directions for future research even within the CE2 framework.  First, one could perform some quantitative studies.  For example, it has yet to be determined how the \zf{} length scale depends on other scales in the problem.  For our numerical work we have done only two things.  We have taken the deformation radius to be infinite, in which case the forcing length scale is the only external scale in the problem and sets the size of the \zfs{}.  And we have taken the deformation radius to be of the same size as the forcing scale, in which case the \zf{} size must inevitably be similar to both.  Parameter scans should be performed where the deformation radius and forcing scale are varied independently to determine their affect on the \zf{} size.  Additionally, it would be interesting to study how large scale vs.~small scale dissipation affects \zf{} saturation.  For these studies, one might use DNS in addition to CE2.

Second, one could attempt to understand in detail the problem with the Newton's method used to solve for the ideal states numerically in \secref{is:equilibrium}.  Near the instability threshold there was no issue, but far from threshold multiple solutions were appearing to the equations.  The Newton's method inevitably got stuck on nonphysical solutions.  Solving this problem would be worthwhile because our direct method of solution of ideal states is otherwise limited to being near the threshold where the \zf{} is weak, and we cannot calculate the full stability diagram.  There are a couple ways we envision proceeding.  One might try using better numerical continuation methods \cite{allgower:2003}.  These might do a better job of staying on the desired branch of physical, realizable solutions than the simple continuation method we have used.  Additionally, others have used CE2 numerically with no problem \cite{farrell:2007,tobias:2013}; the difference between those methods and ours is that our method does not use a time evolution and excludes subharmonics.  Therefore, one might try some kind of hybrid method involving both time evolution and Newton's method to find the fixed point; the time-evolving method would help ensure a realizable solution.  One could also include subharmonics in our calculation.  One other difference between our method and other numerical CE2 work is the use of an alternative coordinate system: we use the sum and difference coordinates $y=y_1 - y_2$ and $\ybar = \tfrac12 (y_1 + y_2)$ instead of $y_1$ and $y_2$.  Using $y_1$ and $y_2$ has the advantage that the equations are in a form suitable for the Fast Fourier Transform, but the use of $y$ and $\ybar$ allows us to change the \zf{} wavenumber $q$ in tiny steps without changing the ``box size'' at the same time.  

The Hasegawa-Mima equation in periodic slab geometry omits a great deal of physics.  We would like to understand zonal flows in toroidal devices such as tokamaks and stellarators.  Some of the complications introduced are the magnetic geometry and linear instability.  While linear instability is a topic we discuss in \secref{sec:intrinsicinstability}, the magnetic curvature leads to the existence of geodesic acoustic modes (GAMs).  GAMs are modes with a zonally symmetric electric potential, like zonal flows, but are distinguished from zonal flows mainly in two ways: 1) GAMs oscillate at a frequency $\w \sim c_s/R$, where $c_s$ is the acoustic speed and $R$ is the major radius, and 2) GAMs are associated with a density perturbation that has $\sin \th$ dependence, where $\th$ is the poloidal angle \cite{winsor:1968,itoh:2005}.  Given how much we have learned about zonal flows using CE2, we have cautious optimism that something could be learned about GAMs as well.  Besides for toroidal plasmas, linear plasma devices such as LAPD or CSDX may also provide a testbed and a window of understanding, especially for how shear flow interacts with turbulence.  Linear devices are easier to analyze theoretically because of their simpler magnetic geometry and azimuthal symmetry.  In linear devices, shear flow is often controlled through externally-applied potentials, although sometimes spontaneous shear flow emerges \cite{carter:2009,zhou:2012,tynan:2006,holland:2006,yan:2010,yan:2010b}.

In the geophysical context, one obviously would want to know how these results extend to a rotating sphere.  The $\b$ plane we have been using is merely an approximation to the rotating sphere.  We have been emphasizing the role of symmetry breaking, but moving to the surface of a rotating sphere destroys the north-south translational symmetries associated with a $\b$ plane.  Do any of these results apply to zonal flows in spherical geometry?  Although this question should be studied in detail, we offer one possibility.  Due to the latitudinal variation of the Coriolis parameter, the turbulence is always inhomogeneous on the sphere.  A transition from homogeneous to inhomogeneous turbulence is not the right description, but perhaps some type of transition may still occur.  Besides for the development of inhomogeneity, another aspect of the bifurcation on a $\b$ plane is the spontaneous formation of a mean field, i.e., the zonal flow.  We suggest that this mean-field generation may persist for flow on a rotating sphere, and would be observable as a control parameter is varied.  The zonal flow still behaves as an order parameter in this more general type of scenario.  This idea has some support, as numerical simulations appear to have observed this behavior as the rotation rate is increased from zero \cite{nozawa:1997}.  Additionally, CE2 has been used to simulate turbulence on the rotating sphere, and \zfs{} have been observed within that framework \cite{marston:2008,tobias:2011}.  Therefore, a future line of investigation could be to use CE2 to study zonostrophic instability on the sphere.  This could be done numerically or possibly analytically by using equivariant bifurcation theory (bifurcation theory for dynamical systems with symmetry) \cite{golubitsky:book}.  Qualitative insight could be gained into the structure of the unstable eigenfunction, including the direction of the equatorial jet.

Finally, one could build upon the connection between zonostrophic instability and modulational instability described in \secref{sec:mi} to improve our understanding of both.  CE2 can be used to generalize modulational/secondary instability to more general background spectra.  CE2 offers an alternative perspective into the physics of coherent-structure formation.  It would be interesting to determine if CE2 can reproduce modulational/secondary instability when the eigenmodes are not Fourier modes (i.e., if nonperiodic boundary conditions are used).

\section{Other Statistical Formalisms and Closures}
\label{sec:otherclosures}
In the theoretical study of turbulence, one line of approach is to examine statistically averaged quantities.  That is the approach we have taken in this thesis, and it is distinct from laboratory experiments or direct numerical simulation.  In the statistical approach, one is interested often only in calculating second-order statistical quantities such as energy and transport, and so the closure problem arises for third-order terms.  Many statistical closures of this type, which approximate the third-order terms in some way, have been studied in depth, including the Direct-Interaction Approximation (DIA) and the Eddy-Damped Quasi-Normal Markovian closure \cite{kraichnan:1959,orszag:1977,bowman:1993,bowman:1997,krommes:2002}.  These closures have several important properties.  First, they conserve the same nonlinear invariants as the original dynamical equations through the same triadic mode-interaction structure.  Second, they ensure statistical realizability.  This means that statistical quantities are well-behaved under time evolution, so certain statistical constraints are guaranteed to be satisfied.  For instance, realizability prevents energy from becoming negative.  Some closures that do not respect realizability experience negative energies, an unacceptable flaw \cite{ogura:1962a,ogura:1962b}.

CE2, as previously discussed, can be categorized as a type of statistical closure.  It is a closure for the one-time, two-point correlation function and allows for inhomogeneous turbulence.  CE2 is particularly simple, since the closure technique involves nothing more than neglecting the unknown terms.  This means that CE2 totally ignores eddy self-nonlinearities, which are responsible for the traditional cascades.  Using one-time correlation functions rather than the more general two-time functions means that CE2 also lacks certain time-history information and loses some of the effects of wave propagation \cite{krommes:1987}.  To incorporate these physical effects, as well as to achieve greater quantitative accuracy, the effect of eddy self-nonlinearities and time-history information must be retained in some way through a more sophisticated closure like those described above.

Another type of approach does not focus solely on second-order or $n$th-order statistical quantities, but uses the full probability density function (or functional).  This is the approach taken by \citet{bouchet:2013}, who used it to rigorously justify the quasilinear approximation in the barotropic vorticity equation in the large $\g$ limit.

The averaging procedure to obtain the CE2 equations from the QL equations merits further discussion \cite{parker:2013}.  We used a zonal average, but for CE2 or other formalisms, other types of averages may be used.  Under appropriate assumptions, which always include some kind of ergodicity assumption, multiple choices of average will lead to the same final equations.  For instance, zonal \cite{srinivasan:2012}, short-time \cite{bakas:2011}, and coarse-graining \cite{bakas:2013} averages have been discussed.  The ergodicity assumption allows one to transform the average over the random forcing into a deterministic quantity.  One can also discuss things in terms of an ensemble average, in which case an assumption of statistical homogeneity in the zonal ($x$) direction is made, but inhomogeneity is allowed in the nonzonal ($y$) direction.  In this case, ergodicity is not required in order to derive the CE2 equations, but it becomes necessary if one wants to interpret the solutions of the equations as having anything to do with the behavior of an individual realization.  

When using the ensemble average, Kraichnan pointed out in the context of thermal convection that the definition of the statistical ensemble is somewhat subtle for the situation of spontaneous symmetry breaking \cite{kraichnan:1964a}.  Because of the translational symmetry, the zonal jets have no preferred location and are presumably equally likely to form with any particular phase.  One choice of the statistical ensemble encompasses all possible realizations consistent with the prescribed parameters, in which case the ensemble itself is statistically homogeneous and any ensemble-averaged quantity must be homogeneous also.  Therefore the average yields zero mean \zf{} (and then the \zf{} must be described as a fluctuation), despite the fact that each individual realization has a nonzero \zf{}.  This was the procedure followed in \citet{krommes:2000}.  Another possibility is that the ensemble might consist only of the realizations for which the zonal jets have a particular phase.  The latter interpretation is the one that yields the CE2 equations identical to those obtained by zonal averaging.  With the former ensemble, the ergodic assumption is invalid, since an individual realization is no longer mixing throughout the full set of realizations of this ensemble.  This is consistent with the fact that the ensemble-averaged behavior is not equivalent to the behavior of an individual realization.

\subsection{Development of Systematic Closures for Inhomogeneous Turbulence}
Historically, the majority of analytical theories of statistical turbulence assume \emph{homogeneous} statistics, where the statistics of turbulent quantities do not depend on position.  Relatively little effort has been devoted to \emph{inhomogeneous} statistics.  Progress developing inhomogeneous closures has been limited and is one area for future research.

One proposed way to go beyond CE2 is to use third-order cumulants in a CE3 framework, where fourth-order cumulants are neglected \cite{tobias:2013}.  That could be useful when eddy-eddy nonlinearities are a small perturbation.  But this approach has problems because CE3, unlike CE2, is not realizable; it must be patched up in an ad-hoc manner.

A few systematic inhomogeneous closures exist, mostly stemming from Kraichnan.  One is the full, inhomogeneous DIA \cite{kraichnan:1964a}.  Kraichnan also proposed a simpler DIA variant called the diagonalizing DIA \cite{kraichnan:1964b}.  More recently, the diagonalizing DIA has been generalized into the quasi-diagonal DIA \cite{frederiksen:1999,okane:2004}, but these ``diagonal'' DIA closures approximate the interaction between the mean field and the fluctuation.  That approximation would affect the stability properties of the \zf{} in ways currently unknown.  Additionally, an inhomogeneous Markovianized closure exists in the test-field model \cite{kraichnan:1972}, but it is not statistically realizable in the presence of waves \cite{bowman:1993,bowman:1997}.  A homogeneous realizable test-field model exists \cite{bowman:1997}, but as of yet there is no version that is both realizable and inhomogeneous.  More work along these lines needs to be done.

\subsection{Systems with Intrinsic Instability}
\label{sec:intrinsicinstability}
The statistical closures described above are intended to more faithfully represent the eddy-eddy nonlinearities than CE2 does.  This can be important for more than mere quantitative accuracy.  We can imagine at least one situation for which it is crucial to retain the eddy self-nonlinearities: a system with linear instability.  Linear instabilities in plasmas are common, such as the ion-temperature-gradient instability.  And the oceans are baroclinically unstable.  Numerical simulations of the Modified Hasegawa-Wakatani system have clearly demonstrated the symmetry-breaking bifurcation of \zf{} generation from homogeneous to inhomogeneous turbulence \cite{numata:2007}.  In order to describe this transition, a model must allow for an equilibrium of homogeneous turbulence.  But in a quasilinear (QL) or CE2 description, if no \zfs{} are present then there are no nonlinear interactions, and it is impossible for a linear instability to saturate.  With linear instability present, a QL description has no homogeneous equilibrium.  Retaining the eddy self-nonlinearities is required to allow a statistical equilibrium of homogeneous turbulence, which can then undergo zonostrophic instability to generate zonal flows.

\begin{figure}
		\centering
		\includegraphics{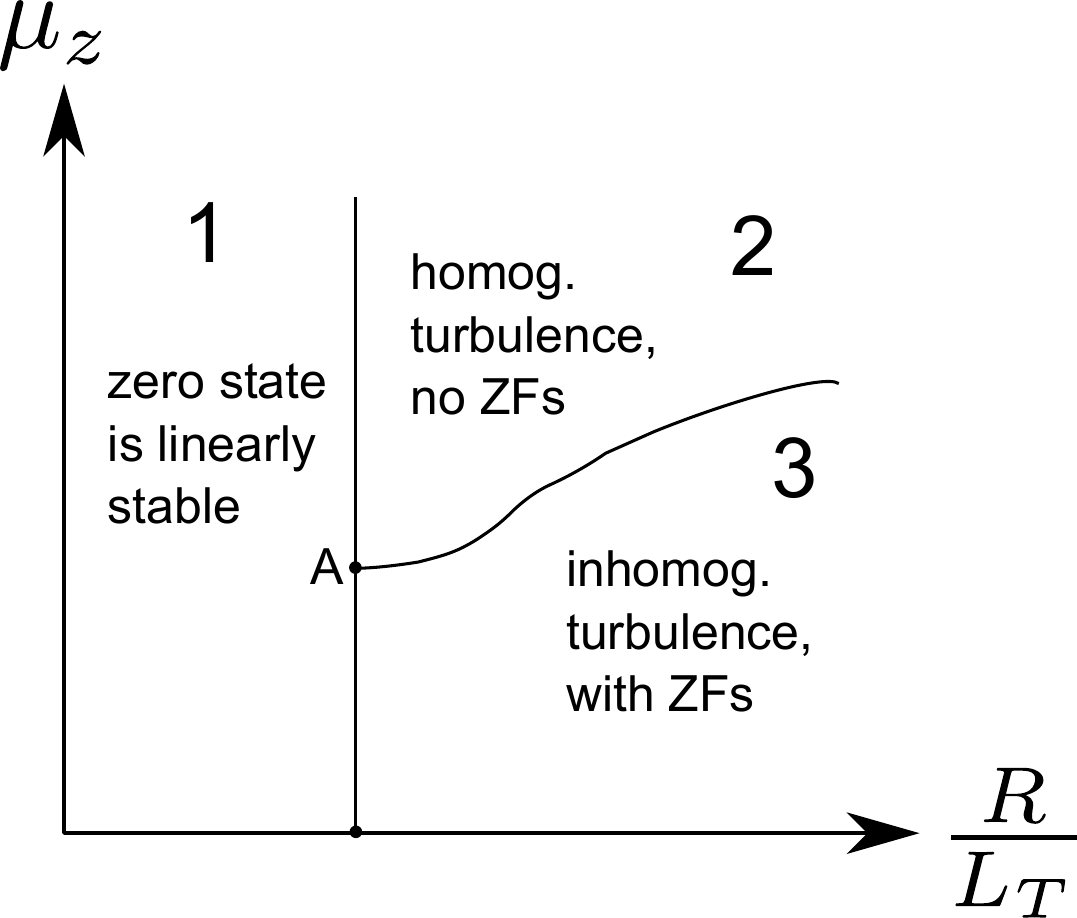}
		\caption{Hypothetical schematic of three possible parameter regimes in an ion temperature gradient (ITG) system, as a function of two parameters, the zonal flow damping rate $\m_z$ and the temperature gradient $R/L_T$.  The point A indicates the codimension-2 bifurcation point where regions 1, 2, and 3 intersect.}
		\label{future:fig:parameterspace}
	\end{figure}

A schematic of the different possible regimes as a function of parameter space is sketched in \figref{future:fig:parameterspace}.  The transition from region 1 to region 2 gives the transition to homogeneous turbulence as linear instability becomes active.  The transition from region 2 to region 3 is the zonostrophic bifurcation studied in great detail in \chref{ch:beyondzonostrophic}, where homogeneous turbulence becomes inhomogeneous as zonal flows are born.  The transition from region 1 to region 3 is not understood at this point.  The point A indicates the codimension-2 bifurcation point where regions 1, 2, and 3 intersect.  A bifurcation analysis about the point $A$ might be interesting.

However, this sketch may be too simplistic for even the least complicated plasma turbulence systems studied.  For one, we have assumed there is no subcritical turbulence\footnote{Subcritical turbulence refers to turbulence that is sustained even when the base state is linearly stable.}.  We have also let the dissipation parameter of the zonal flows, $\m_z$, be controlled independently from other parameters.  But in the Modified Hasegawa-Wakatani system, the dissipation is not so simple, and it may not be controlled directly.  Instead, much of the dissipation arises from the coupling of the zonal flows to nonzonal modes, which then suffer from resistive damping \cite{terry:2006,hatch:2011a,hatch:2011b,makwana:2011,makwana:2012}.  The dissipation is determined nonlinearly after saturation by all the mode couplings.  That kind of scenario will have to be studied in detail.  The sketch we offered in \figref{future:fig:parameterspace} is just a beginning.  And there may be other types of regimes and transitions that we are yet unaware of. 

The Dimits shift is another aspect of certain linearly unstable magnetically confined plasmas \citep{dimits:2000}.  The Dimits shift has been a phenomenon of high interest ever since it was discovered numerically, and there is as yet no experimental evidence for it.  In the Dimits shift, turbulence and turbulent heat transport are suppressed even beyond the linear marginal stability boundary.  In other words, when the ion-temperature gradient was increased to just beyond the critical value for linear instability, no turbulence was observed.  This unexpected behavior was attributed to the suppression of turbulence by nonlinearly-generated zonal flows.  As the ion-temperature gradient was increased even further, eventually turbulence and turbulent transport would develop (possibly because the zonal flows suffer their own instability and can no longer effectively suppress the turbulence).  This upshift from the linear stability boundary to some other boundary is termed the Dimits shift.

With collisionless \zfs{}, a Galerkin-truncated ITG system of just 10 modes was found to exhibit a Dimits shift  \cite{kolesnikov:2005,kolesnikov:2005b}.  However, it is unclear what exactly can be learned from that calculation, because the behavior of the system was sensitive to the number of modes retained in the truncation.  Perhaps an analysis that retains the full spatial dependence, through the inhomogeneous statistical closures we have been describing, would lead to more regular and well-behaved behavior and improved understanding of the Dimits shift.  One possibility, suggested by the pattern formation framework, is that in the Dimits shift regime, steady zonal flows exist within some stability balloon.  But at large enough profile gradients, any steady zonal flow becomes unstable, leading to rapidly fluctuating zonal flows and reduced suppression of turbulence.  This scenario would be consistent with the ideas of \citet{rogers:2000}.

The Dimits shift is not understood theoretically.  Many studies of it use collisionless \zfs{}, but not all \cite{mikkelsen:2008}.  What can be said is that the Dimits shift involves a transition that includes the generation of zonal flows.  This is a type of behavior similar to the zonostrophic bifurcation that has been successfully described by CE2.  It is possible that the statistical framework with inhomogeneous turbulence may be similarly successful in describing the Dimits shift.  To follow this route, one would want to find the simplest system that exhibited a Dimits-shift-like behavior.  For example, does the Dimits shift require the effects present in gyrokinetics, or can it be adequately captured in a fluid description?  Is toroidal magnetic geometry essential, or is there a simpler geometry that possesses sufficiently similar behavior?  A minimal model would make the analysis and physics as transparent as possible.

The zero-dimensional phenomenological bifurcation model of zonostrophic instability, presented in \secref{sec:zi:toymodel}, can be modified for the case of a linear instability by the addition of terms representing the eddy self-nonlinearity.  For instance, following the example of a nonlinear closure with quadratic terms, one might have
	\begin{subequations}
	\begin{align}
		\dot{W}_h &= \g W_h - \m W_h^2 + F W_h^2 - \a W_i z, \\
		\dot{W}_i &=  \g W_i - \m W_h W_i + \eta W_h z, \\
		\dot{z} &= -\n z + \a W_i,
	\end{align}
	\end{subequations}
where we have assumed  that the incoherent forcing term $F$ does not appear in the inhomogeneous equation.  (This model has the problem that in some circumstances $W_h$ can become negative and blow up.)  The linear instability term $\g$ might be related to the temperature gradient $R/L_T$.  We can absorb $F$ into $\m$ and thus write 
	\begin{subequations}
	\begin{align}
		\dot{W}_h &= \g W_h - \m_h W_h^2 - \a W_i z, \\
		\dot{W}_i &=  \g W_i - \m_i W_h W_i + \eta W_h z, \\
		\dot{z} &= -\m_z z + \a W_i,
	\end{align}
	\end{subequations}
where $\m_i > \m_h$.  The zero state is linearly unstable if $\g>0$.  This model has a homogeneous equilibrium at $W_h = \g / \m_h$.  Its linear stability can be checked easily.  The condition for zonostrophic instability is
	\begin{equation}
		\frac{\a\eta}{\m_h \m_z} > \frac{\m_i}{\m_h}-1.
	\end{equation}
This simple model has a structure similar to that in \figref{future:fig:parameterspace}.  When $\g>0$, the zero state is unstable, with a zonostrophic boundary depending on the value of $\m_z$ (with all other parameters fixed).  In this simple model, the zonostrophic boundary has no dependence on $\g$.  The model is also not complicated enough to have a Dimits shift.

The closure for homogeneous statistics presented in \chref{ch:closure} could be extended to inhomogeneous statistics as well.  That closure has a nonlinear damping term and so can handle intrinsic linear instabilities.  Since it is rather simple, analytic progress might even be possible, e.g., in a bifurcation analysis.

\section{Other Gaps in Knowledge}
In pattern-forming systems, the simplest theoretical approach is to eliminate boundaries and use an infinite or periodic system.  That was the initial approach taken in \RB convection, and that is the approach taken here.  However, boundaries are actually quite important.  For instance, one might expect that if a system gets very large, then far from the boundaries, the boundaries have little effect.  But in the amplitude equation \eref{gl:amplitudeequation}, a prototypical pattern-inhibiting boundary condition $A=0$ has a profound effect on the possible wavenumbers of the pattern even far from the boundary.  Instead of a band of stable, stationary solutions as in the case of infinite boundaries, an $A=0$ boundary condition in a semi-infinite system forces the pattern wavenumber to be unique and equal to the critical wavenumber \cite{cross:2009}.  Some boundaries can suppress the amplitude of patterns, and others can enhance pattern formation.

In toroidal and cylindrical plasma devices, boundaries exist and have a major influence on the plasma's behavior.  In toroidal plasmas, the magnetic geometry plays a dominant role in determining the character of the turbulence, and we should expect that the magnetic geometry and especially the separatrix influence the generation and characteristic of zonal flows.  Systematically understanding the geometry and boundary effects is a major open area for study and will rely heavily on simulations.  Some initial work has been done in terms of examining how various stellarator configurations affect \zfs{} \cite{xanthopoulos:2011}, but far more work needs to be done.

In many simulations in toroidal geometry, \zfs{} are observed to be non-steady.  This fluctuating behavior is distinct from the steady \zfs{} we have been assuming in the theoretical analysis in this thesis.  If the time scale of the \zfs{}' fluctuations are long compared to that of the turbulence, then perhaps an assumption of steady \zfs{} is an acceptable lowest-order approach.  But \zfs{} have sometimes been seen to fluctuate on the same time scale as the turbulence, in which case the theory developed here is not directly applicable.

How can we use any of this knowledge to benefit experiments, or even to talk in a language that experimentalists understand?  In the geophysical context, possibly.  Given the numerous discoveries of exoplanets and the ever-more sophisticated observational methods, we someday might encounter an exoplanet gas giant that has no zonal jets.  This would contrast with the gas giants within our solar system, which all have zonal jets.  A fundamental theoretical understanding of the zonostrophic bifurcation is key to puzzling out how various factors impact zonation.

Plasmas, on the other hand, are so messy and complex that we currently see no direct way for the theory to be directly compared with experiment.  The Hasegawa--Mima equation neglects many, many physical effects.  We discovered some general principles in the 2D slab geometry, but it is unclear if those survive in toroidal geometry.  Even the cylindrical plasma devices, with their simpler magnetic geometry, are so small that radial boundary conditions inevitably have a strong influence.

To us, the way to proceed to develop this theory for usefulness to plasma physicists is twofold.  One direction is to increment in complexity, step by step.  For instance, eddy-eddy nonlinearities can be added to handle linear instabilities.  The theory should be constructed in cylindrical geometry, then in toroidal geometry.  GAMs should be investigated.  Kinetic effects might be added.  A worthy goal would be to try to identify and understand the Dimits shift in a simple model, as explained above.  The second direction goes hand-in-hand with the first, and that is to firm up the theory with numerical simulation.  We believe that many of the principles that we have found from the QL approximation to the Hasegawa--Mima equation will apply in many other cases.  It appears generic that steady zonal flows are generated in slab geometry in plasmas.  Some gyrokinetic ITG simulations in slab geometry have seen steady zonal flows \cite{pc:hatch:2013}, in which case the pattern formation principles ought to apply.  Detailed comparisons of such simulations with theoretical predictions will undoubtedly lead to progress.  It is only by laying the groundwork that we as a community will be able to construct the elaborate theoretical towers required to understand plasma turbulence.

Finally, one area of high interest, which was originally to be one of the questions considered in this thesis but was barely touched on, is how zonal flow suppresses turbulence.  Multiple explanations have been given, but there is no firm theoretical basis for which to understand and compare them.  Since the pattern formation approach is new in the field of zonal flows, it provides a novel way to attack this problem.

\appendix 
\chapter{Derivation of CE2 in Real Space}
\label{app:CE2derivation}

Here we provide the details of the derivation of the CE2 equations \eqref{CE2}.  This procedure follows that by \citet{srinivasan:2012}.

We begin from the QL system \eqref{qlsystem}, which we restate here:
	\begin{subequations}
	\label{appCE2:qlsystem}
		\begin{gather}
		\partial_t w' + \{U \nablabarsq + \b - [(\partial_y^2 - \LD) U]\} \partial_x \psi' = \xi - \m w' - \n(-1)^h \nabla^{2h} w', \label{appCE2:QLwprime} \\
		\bigl[ \partial_t + \m + \n(-1)^h \partial_y^{2h}\bigr] \bigl(1 - \azf \LD \partial_y^{-2} \bigr) U(y) + \partial_y \ol{ v_x' v_y' } = 0, \label{appCE2:QLU}
		\end{gather}
	\end{subequations}
The covariance of the white-noise forcing $\xi$ is defined to be
	\begin{equation}
		\avg{ \xi(x_1,y_1,t_1) \xi(x_2,y_2,t_2) } = F(x_1-x_2,y_1-y_2) \de(t_1-t_2)
	\label{appCE2:forcingcovariance}
	\end{equation}
The forcing is taken to be homogeneous in space such that its statistics only depend on the spatial difference $\v{x}_1 - \v{x}_2 = (x_1-x_2, y_1-y_2)$.

Define
	\begin{equation}
		\ti{W}(x_1,y_1,x_2,y_2) \defineas w'(x_1,y_1) w'(x_2,y_2)
		\label{appCE2:def_of_W}
	\end{equation}
(taken at the same time $t$).  Averaging $\ti{W}$ over $x_1$ holding $x_2$ fixed (or vice versa) gives zero, by definition.  Instead we define
	\begin{equation}
		\ti{W}(x_1,y_1,x_2,y_2) \defineas \ti{W}\left(x_1-x_2,y_1-y_2 \mid \tfrac{1}{2}(x_1+x_2) \right) = \ti{W}(x,y_1,y_2 \mid \ol{x}),
	\end{equation}
with the sum coordinate $\ol{x} = \frac12(x_1 + x_2)$ and the difference coordinate $x = x_1 - x_2$.  At a later point, we will also switch to sum and difference coordinates for $y$.  Now, define
	\begin{align}
		W(x,y_1,y_2) &\defineas \frac{1}{L_x} \int_0^{L_x} d\ol{x}|_x\, \ti{W}(x,y_1,y_2 \mid \ol{x}) \notag \\
			&= \frac{1}{L_x} \int_0^{L_x} d\ol{x}|_x\, w'(x_1,y_1)w'(x_2,y_2),
	\end{align}
where $L_x$ is some averaging length.  This averages the product $w'(x_1,y_1) w'(x_2,y_2)$ holding the separation $x_1-x_2$ fixed.  This zonal average presumably smooths rapidly fluctuating quantities (in space and time).  Similarly, we can define
	\begin{gather}
		\ti{\Psi}(x,y_1,y_2 \mid \xbar) \defineas \psi'(x_1,y_1) \psi'(x_2,y_2), \\
		\Psi(x,y_1,y_2) \defineas \frac{1}{L_x} \int_0^{L_x} d\ol{x}|_x\, \ti{\Psi}(x,y_1,y_2 \mid \ol{x}) = \frac{1}{L_x} \int_0^{L_x} d\ol{x}|_x\, \psi'(x_1,y_1)\psi'(x_2,y_2).
	\end{gather}
We can relate $W$ and $\Psi$.  Recall that $w'(x,y) = \nablabarsq \psi'(x,y) = (\nabla^2 - \LD) \psi'(x,y)$.  Then
	\begin{align}
		w'(x_1,y_1) &= [ \partial_{x_1}^2|_{x_2} + \partial_{y_1}^2 -\LD] \psi'(x_1,y_1), \\
		w'(x_2,y_2) &= [ \partial_{x_2}^2|_{x_1} + \partial_{y_2}^2 -\LD] \psi'(x_2,y_2).
	\end{align}
Now, use $\partial_{x_1} = \partial_x + \tfrac{1}{2} \partial_{\ol{x}}$, and $\partial_{x_2} = -\partial_x + \tfrac{1}{2} \partial_{\ol{x}}$.  Substituting these relations into \eref{appCE2:def_of_W}, we have
	\begin{align}
		W(x,y_1,y_2) = \frac{1}{L_x} \int_0^{L_x} &  d\ol{x}|_x \, \left[\partial_x^2 + \partial_{x\ol{x}} + \tfrac{1}{4} \partial_{\ol{x}}^2 + \partial_{y_1}^2 - \LD\right] \notag \\
				& \circ \left[ \partial_x^2 - \partial_{x\ol{x}} + \tfrac{1}{4} \partial_{\ol{x}}^2 + \partial_{y_1}^2 - \LD \right] \psi'(x_1,y_1) \psi'(x_2,y_2).
	\end{align}
By the assumed periodicity in $\ol{x}$ (or other assumption), the $\partial_{\ol{x}}$ terms vanish.  Define
	\begin{align}
		\nabla_j^2 &\defineas  \partial_x^2 + \partial_{y_j}^2, \\
		\nablabarsq_j &\defineas \nabla_j^2 - \LD,
	\end{align}
for $j=1,2$.  Then, we see that
	\begin{equation}
		W(x,y_1,y_2) = \nablabarsq_1 \nablabarsq_2 \Psi(x,y_1,y_2).
		\label{appCE2:WandC_y1y2}
	\end{equation}
In shorthand notation, we also write
	\begin{equation}
		W(x,y_1,y_2) = \ol{w'_1 w'_2},
		\label{appCE2:Wshorthand}
	\end{equation}
where $w'_j = w'(x_j, y_j)$ and the overbar means spatial average holding $x = x_1 - x_2$ fixed.  Similarly, for the velocity correlation tensor, one finds (with $u \defineas v_x$ and $v \defineas v_y$)
	\begin{equation}
		V_{ij}(x_1,x_2,y) \defineas \begin{pmatrix} \ol{u'_1 u'_2} & \ol{ u'_1 v'_2} \\ \ol{u'_2 v'_1} & \ol{v'_1 v'_2} \end{pmatrix} =
			\begin{pmatrix} \partial_{y_1}\partial_{y_2} & \partial_x \partial_{y_1} \\ -\partial_x \partial_{y_2} & -\partial_x^2 \end{pmatrix} \Psi(x,y_1,y_2).
		\label{appCE2:velocitytensor}
	\end{equation}
Because the choice of denoting one point as $\v{x}_1$ and the other as $\v{x}_2$ is arbitrary, all correlation functions have the exchange symmetry \cite{srinivasan:2012}
	\begin{equation}
		W(x,y_1,y_2) = W(-x,y_2,y_1).
	\end{equation}
	
Now, we derive an evolution equation for $W$.  From \eref{appCE2:Wshorthand} we have
	\begin{equation}
		\partial_t W = \ol{ (\partial_t w_1') w_2' } + \ol{w_1' (\partial_t w_2') }.
	\end{equation}
Substituting in from \eref{appCE2:QLwprime} and applying the averaging, one eventually finds
	\begin{align}
		\partial_t W + \left( \nablabarsq_2 L_1 - \nablabarsq_1 L_2 \right) \partial_x \Psi &= -2\m W - \n (-1)^h \left( \nabla_1^{2h} + \nabla_2^{2h} \right) W \notag \\
			& + \ol { \xi_1 w_2' + w_1' \xi_2 },
		\label{appCE2:Wevolution_y1y2}
	\end{align}
where
	\begin{gather}
		L_j \defineas U_j \nablabarsq_j + (\b - \ol{U}''_j), \\
		U_j \defineas U(y_j), \\
		\ol{U}''_j = \ol{\partial}_{y_j}^2 U(y_j),
	\end{gather}
for $j=1,2$.  For later use, notice that
	\begin{equation}
		\left( \nablabarsq_2 L_1 - \nablabarsq_1 L_2\right) \partial_x \Psi = (U_1 - U_2) \partial_x W + \left[ (\b - \ol{U}_1'') \nablabarsq_2 - (\b - \ol{U}_2'') \nablabarsq_1 \right] \partial_x \Psi.
	\end{equation}
	
Now, we switch to using sum and difference coordinates in $y$, with $y \defineas y_1 - y_2$ and $\ol{y} \defineas (y_1+y_2)/2$, with $\partial_{y_1} = \partial_y + \tfrac{1}{2} \partial_{\ol{y}}$ and $\partial_{y_2} = -\partial_y + \tfrac{1}{2}\partial_{\ol{y}}$.  We write
	\begin{equation}
		W(x,y_1,y_2) \defineas W(x,y \mid \ol{y}).
	\end{equation}
In terms of $y$ and $\ybar$, the Laplacians are
	\begin{align}
		\nabla_1^2 &= \nabla^2 + \partial_y \partial_\ybar + \tfrac{1}{4} \partial_\ybar^2, \\
		\nabla_2^2 &= \nabla^2 - \partial_y \partial_\ybar + \tfrac{1}{4} \partial_\ybar^2,
	\end{align}
where now $\nabla^2 = \partial_x^2 + \partial_y^2$ is the ``separation'' Laplacian.  We also define
	\begin{equation}
		\nablabarsq = \nabla^2 - \LD.
	\end{equation}
The symbols $\nabla^2$ and $\nablabarsq$ were used in slightly different context in the fluctuating amplitude equations, but now we reuse them purely in the averaged equations and the meaning should be clear.  From \eref{appCE2:WandC_y1y2}, we can relate $W$ and $\Psi$ in the new coordinates,
	\begin{equation}
		W(x,y\mid \ybar) = \biggl(\nablabarsq + \partial_y \partial_\ybar + \frac{1}{4}\partial_\ybar^2\biggr)\biggl(\nablabarsq - \partial_y \partial_\ybar + \frac{1}{4}\partial_\ybar^2\biggr) \Psi(x,y\mid \ybar). \label{appCE2:WandCcollective}
	\end{equation}
	
Using the sum and difference coordinates, the evolution equation for $W$, \eref{appCE2:Wevolution_y1y2}, becomes after some algebra
	\begin{align}
		\partial_t W(x,y &\mid \ybar) + (U_+ - U_-) \partial_x W - (\ol{U}_+'' - \ol{U}_-'') (\nablabarsq + \tfrac{1}{4} \partial_\ybar^2 ) \partial_x \Psi \notag \\
			& - [2\b - (\ol{U}_+'' + \ol{U}_-'')] \partial_\ybar \partial_y \partial_x \Psi = \langle \xi_1 w_2' + w_1' \xi_2 \rangle_x - 2\m W - 2 \n D_h W,
			\label{appCE2:CE2W_noforcing}
	\end{align}
where now $U_\pm = U\bigl(\ybar \pm \frac12 y\bigr)$, $\ol{U}_{\pm}''  = U''\bigl(\ybar \pm \tfrac12 y\bigr) - \hat{\a}_{ZF} \LD U \bigl(\ybar \pm \tfrac12 y \bigr)$, and$D_h$ is the hyperdiffusion operator, given by
	\begin{equation}
		D_h = (-1)^h \frac{1}{2} \left\{ \left[ \partial_x^2 + \left( \partial_y + \tfrac{1}{2}\partial_{\ol{y}}\right)^2\right]^h + \left[ \partial_x^2 + \left( \partial_y - \tfrac{1}{2}\partial_{\ol{y}}\right)^2\right]^h \right\}.
	\end{equation}

We must now compute the term resulting from the external stochastic forcing, $\ol{\xi_1 w_2' + w_1' \xi_2}$.  We make an ergodic assumption such that a zonal average is equivalent to an ensemble average over the realizations of the forcing, 
	\begin{equation}
		\ol{\xi_1 w_2' + w_1' \xi_2} = \avg{\xi_1 w_2' + w_1' \xi_2}.
	\end{equation}
With this assumption, the desired term can be calculated exactly.  The assumption that the forcing is white noise (delta-correlated in time) is also crucial.  From \eref{appCE2:QLwprime}, we can write
	\begin{equation}
		w'(x_2,y_2,t) = w_0(x_2,y_2,t_0) + \int_{t_0}^t dt'\, N(t') + \int_{t_0}^t dt'\, \xi(x_2,y_2,t'),
	\end{equation}
where $t_0 < t$ and $N$ contains all the appropriate terms.  The ensemble average $\avg{ \xi(x_1,y_1,t) w'(x_2,y_2,t) }$ becomes
	\begin{equation}
		\left\langle \left(w_0(x_2,y_2,t_0) + \int_{t_0}^t dt' N(t') \right) \xi(x_1,y_1,t) \right\rangle + \int_{t_0}^t dt'\, \langle \xi(x_1,y_1,t) \xi(x_2,y_2,t') \rangle.
	\end{equation}
The first average vanishes because the fields $w_0$ and $N$ at times prior to $t$ are uncorrelated with the random forcing at time $t$.  The second average is given by the definition of the forcing \eref{appCE2:forcingcovariance}.  One is left with the integral
	\begin{equation}
		\avg{ \xi_1 w'_2 } = F(x_1-x_2,y_1-y_2) \int_{t_0}^t dt'\, \de(t-t').
	\end{equation}
The integral over the delta function is somewhat subtle because $t'=t$ occurs exactly at the endpoint, but it gives exactly $\frac12$.  This can be seen intuitively because any physical correlation function must be nonsingular and symmetric about its time argument.  Thus half of the `weight' of the correlation function sits at $t<t'$ and the other half at $t>t'$.  If one considers white noise as the limit of some process with finite correlation time, then one must conclude that only half of the correlation function is integrated over, leading to the value of the integral as $\frac12$.  Similarly, it is not hard to check that $\avg{ w'_1 \xi_2}$ evaluates to the same result of $\frac12 F$.

Thus the evolution equation for $W$ becomes, finally,
	\begin{align}
		\partial_t W(x,y &\mid \ybar) + (U_+ - U_-) \partial_x W - \bigl(\ol{U}_+'' - \ol{U}_-'' \bigr) \biggl(\nablabarsq + \frac{1}{4} \partial_\ybar^2 \biggr) \partial_x \Psi \notag \\
			& - \bigl[2\b - (\ol{U}_+'' + \ol{U}_-'') \bigr] \partial_\ybar \partial_y \partial_x \Psi = F(x,y) - 2\m W - 2 \n D_h W.
			\label{appCE2:CE2W}
	\end{align}

The Reynolds stress term in the equation for the zonal flow can be written in terms of the eddy correlation function.  In sum and difference coordinates, mean-square quantities are obtained by evaluating correlation functions at zero separation, i.e., by setting $(x,y)=0$.  For example
	\begin{equation}
		\ol{ w'(x_1,y_1) w'(x_1,y_1) } = W(x,y_1,y_1) = W(0,0 \mid y_1)
	\end{equation}
(with $\ol{y} = y_1$).  From \eref{appCE2:velocitytensor}, we have
	\begin{equation}
	\ol{u_1' v_2'} + \ol{u'_2 v'_1} = 2 \partial_x \partial_y \Psi.
	\end{equation}
Evaluating at $x_2 = x_1$ and $y_2=y_1 = \ybar$, so that $x=0$, $y=0$, we have
	\begin{equation}
	\ol{u'v'}(\ybar) = \partial_x \partial_y \Psi(0,0 \mid \ybar ).
	\end{equation}
Thus, as a function of $\ybar$, the mean flow equation \eref{appCE2:QLU} can be written as
	\begin{equation}
		\bigl[ \partial_t + \m + \n (-1)^h \partial_\ybar^{2h} \bigr]\ol{I} U(\ybar) + \partial_\ybar \partial_x \partial_y \Psi(0,0 \mid \ybar ) = 0,
		\label{appCE2:CE2meanflow}
	\end{equation}
where $\ol{I} = 1 - \azf \LD \partial_\ybar^{-2}$.

Equations \eqref{appCE2:CE2W} and \eqref{appCE2:CE2meanflow} form a closed system called CE2 (second-order cumulant), along with \eref{appCE2:WandCcollective} relating $W(x,y \mid \ybar)$ and $\Psi(x,y \mid \ybar)$.

\chapter{Correlation Function Corresponding to a Wave}
\label{app:wave_correlation}
We consider in this section the one-time, two-point correlation function corresponding to a wave.  First we consider the general case of a superposition of waves.  Let
	\begin{equation}
		\psi'(x,y,t) = 2 \sum_\v{k} c_\v{k} \cos(k_x x + k_y y - \w_\v{k} t + \p_\v{k}).
	\end{equation}
Then, letting $\psi'_1 = \psi'(x_1,y_1,t)$ and $\psi'_2 = \psi'(x_2,y_2,t)$, we have
	\begin{align}
		\psi'_1 \psi'_2 = &\sum_{\v{k},\, \v{k}'} 2 c_\v{k} c_{\v{k}'} \bigl\{ \cos \bigl[ \tfrac12 (k_x + k_x')x + (k_x - k_x')\xbar + \tfrac12 (k_y + k_y') y + (k_y - k_y') \ybar - z_{\v{k} \v{k}'}^-\bigr] \notag \\
			& + \cos \bigl[ \tfrac12 (k_x - k_x')x + (k_x + k_x')\xbar + \tfrac12 (k_y - k_y') y + (k_y + k_y') \ybar - z_{\v{k} \v{k}'}^+ \bigr] \bigr\},
	\end{align}
where $x = x_1-x_2$, $\xbar = \frac12 (x_1 + x_2)$, and $z_{\v{k} \v{k}'}^\pm = (\w_\v{k} \pm \w_{\v{k}'})t - (\p_\v{k} \pm \p_{\v{k}'})$.  Using a zonal average, the correlation function is obtained by integrating over $\xbar$ with $x$ held fixed:
	\begin{equation}
		\Psi(x,y \mid \ybar) = \frac{1}{L_x} \int_0^{L_x} d\ol{x}|_x \psi'_1 \psi'_2.
	\end{equation}
The first cosine vanishes unless $k_x' = k_x$, while the second cosine vanishes unless $k_x' = -k_x$.  For simplicity assume all the $k_x, k_x' > 0$.  Then we are left with
	\begin{equation}
		\Psi(x,y \mid \ybar) = \sum_{\v{k},\, k_y'} 2 c_\v{k} c_{\v{k}'} \cos[ k_x x + \tfrac12 (k_y + k_y')y + (k_y - k_y') \ybar - (\w_\v{k} - \w_{\v{k}'})t + \p_\v{k} - \p_{\v{k}'} ].
	\end{equation}
If we separate out in the sum the terms for which $k_y' = k_y$, then we have
	\begin{align}
		\Psi(x,y \mid \ybar) &= \sum_\v{k} 2 c_\v{k}^2 \cos( k_x x + k_y y) + \sum_\v{k} \sum_{k_y' \neq k_y}  2 c_\v{k} c_{\v{k}'} \cos\bigl[k_x x \notag \\
		& \quad  + \tfrac12(k_y + k_y') y + (k_y - k_y')\ybar - (\w_\v{k} - \w_{\v{k}'})t + \p_\v{k} - \p_{\v{k}'} \bigr]. \label{app:W_manywaves}
	\end{align}
It can be verified by substitution that this is a solution to the unforced, undamped CE2 equations without zonal flow, $\partial_t W = 2 \b \partial_\ybar \partial_y \partial_x \Psi$ (and using $\w_\v{k} = -k_x \b / \kbsq$).  We see that the first term of \eref{app:W_manywaves}, which corresponds to the covariance of individual waves, is unchanging in time and homogeneous in space.  But in the second term, waves with different $k_y$ give rise to a correlation function that oscillates in time and has $\ybar$ dependence.  This is a manifestation of the coherent beating between waves.  There is no decorrelation mechanism present; that requires nonlinear physics.

One can imagine using another averaging procedure instead of the zonal average.  With the zonal average, the only coherent structures allowed are zonally symmetric.  One might also want to investigate zonally asymmetric structures, which precludes the use of a zonal average \cite{bakas:2013}.  To study these more general coherent structures, the correlation function can be defined using a coarse graining in time or space (this approach typically requires the mean field and fluctuations to obey a scale-separation assumption) or an ensemble average.
	
To illustrate an alternate derivation for a single wave, let
	\begin{equation}
		\psi'(\v{x}) = \psi_0 \left( e^{i\v{p}\cdot\v{x} - i\w t} + e^{-i \v{p}\cdot\v{x} + i\w t}\right).
	\end{equation}
Then
	\begin{align}
	 	\psi'_1 \psi'_2 &= \psi_0^2\left( e^{2i\v{p} \cdot \ol{\v{x}}} e^{-2i\w t} + e^{i\v{p}\cdot\v{x}} + e^{-i\v{p}\cdot\v{x}} + e^{-2i\v{p} \cdot \ol{\v{x}}} e^{2i\w t} \right).
	\end{align}
At this point, a coarse graining in time over an intermediate time between $\w^{-1}$ and the timescale of the coherent structure eliminates the oscillating terms.  Equivalently, one could perform a coarse graining in space over an intermediate scale between $p^{-1}$ and the size of the coherent structure.  Then, one obtains
	\begin{equation}
		\Psi = \psi_0^2 \left( e^{i\v{p}\cdot\v{x}} + e^{-i\v{p}\cdot\v{x}} \right).
	\end{equation}
This $\Psi$ is homogeneous (independent of $\ol{\v{x}}$).  Its Fourier transform is
	\begin{equation}
		\Psi_H(\v{k}) = (2\pi)^2 \psi_0^2 \left[ \de(\v{k} - \v{p}) + \de(\v{k} + \v{p}) \right].
	\end{equation}
The inclusion of the mode at $-\v{p}$ as well as the mode at $\v{p}$ is essential and arises from the reality condition.
\chapter{Derivation of the Amplitude Equation}
\label{app:glderivation}
Here we derive the amplitude equation \eref{gl:amplitudeequation} directly from the CE2 equations \eref{CE2} and verify the results numerically.  First, we review the procedure for the perturbation expansion \cite{cross:2009}.  Then we fill in the algebraic details.

\section{Review of the Perturbation Expansion}
We limit ourselves to quadratic nonlinearity.  Let $\phi$ be an abstract vector, $\Lambda$ be a linear operator, $N$ be a bilinear operator, and $F$ be external forcing.  Any of $\Lambda$, $N$, and $F$ may depend explicitly on the small parameter $\e$.  The basic equation is taken to be
	\begin{equation}
		0 = \Lambda \phi + N(\phi,\phi) + F.
	\end{equation}
Without loss of generality, $N$ can be assumed to be symmetric in its arguments (if it is not, a new symmetrized operator can be defined and used instead).  Given a nonzero equilibrium $\phi_e$, we change variables by letting $\phi = \phi_e + u$ to give
	\begin{equation}
		0 = L u + N(u,u),
		\label{gl:abstractequation}
	\end{equation}
where $Lu$ = $\Lambda u + 2N(\phi_e,u)$.

\begin{figure}
		\centering
		\includegraphics{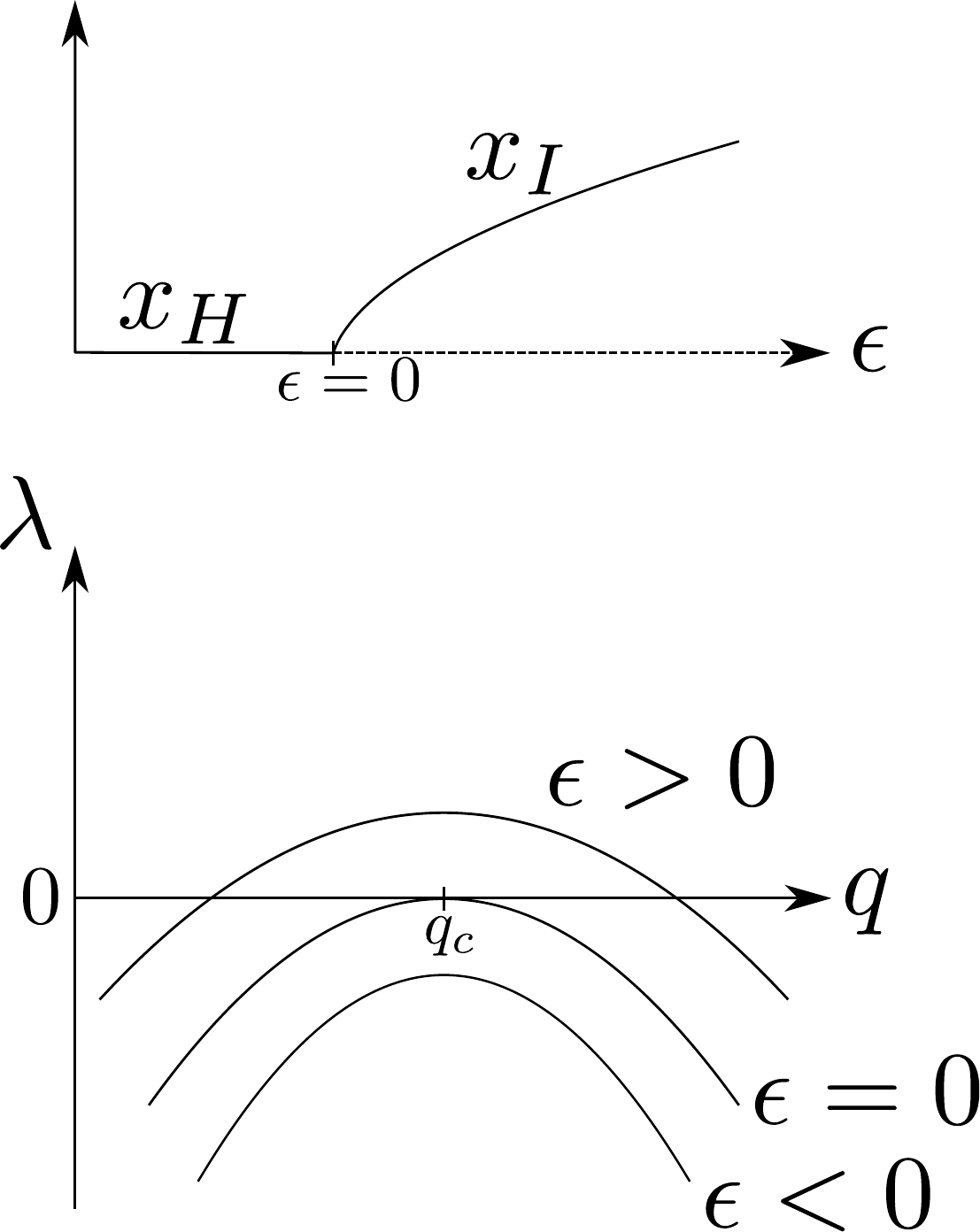}
		\caption{Schematic of bifurcation.  Top: The homogeneous equilibrium $x_H$ is stable (solid) for $\e<0$ and zonostrophically unstable (dashed) for $\e>0$.  At $\e=0$, a new set of inhomogeneous equilibria $x_I$ appears; some of these equilibria are stable and some are unstable.  Bottom: Growth rate $\l$ of perturbations about the homogeneous equilibrium $x_H$ as a function of the \zf{} wavenumber $q$.}
		\label{gl:bifschematic}
	\end{figure}

We take as given that at $\e=0$, the equilibrium $\phi_e$ transitions from stable to unstable due to a perturbation with wavenumber $q_c$.  This calculation is motivated by the discovery of the zonostrophic instability, described in \secref{sec:zi}.  Figure \ref{gl:bifschematic} depicts the schematic of the bifurcation.

In performing the perturbation procedure, we use a multiple-scale expansion with slowly varying space and time scales.  This is accomplished by introducing the slow scales $Y=\e^{1/2} \ybar$ and $T = \e t$, then letting $\partial_\ybar \to \partial_\ybar + \e^{1/2} \partial_Y$ and $\partial_t \to \partial_t + \e \partial_T$.  Using these, we expand $L = L_0 + \e^{1/2} L_1 + \e L_2 + \e^{3/2} L_3 + \cdots$, $N = N_0 + \e^{1/2} N_1 + \cdots$, and $u = \e^{1/2} u_1 + \e u_2 + \cdots$\,.  Expansion in $\e^{1/2}$ rather than in $\e$ arises due to generic behavior of supercritical bifurcations.  Collecting terms of the same order, we obtain the equations at $\O(\e^{1/2})$, $\O(\e)$, and $\O(\e^{3/2})$:
	\begin{align}
		\O\bigl(\e^{1/2}\bigr): \qquad 0 &= L_0 u_1, \\
		\O(\e): \qquad 0 &= L_0 u_2 + L_1 u_1 + N_0(u_1,u_1), \\
		\O\bigl(\e^{3/2}\bigr): \qquad 0 &= L_0 u_3 + L_1 u_2 + L_2 u_1 + 2 N_0(u_1,u_2) + N_1(u_1,u_1). \label{gl:eps32}
	\end{align}

At $\O(\e^{1/2})$, the condition $L_0 u_1=0$ states that $u_1$ is an eigenvector with a zero eigenvalue.  Then $u_1$ can be a linear combination of null eigenvectors with a to-be-determined amplitude.  The reality condition on $u$ restricts the form to be
	\begin{equation}
		u_1 = A(Y,T) r + A(Y,T)^* r^*,
		\label{gl:u1form}
	\end{equation}
where $r \sim e^{i q_c \ybar}$ (and its complex conjugate) are the right null eigenvectors.  These eigenvectors are periodic in $\ybar$ with critical wavenumber $q_c$, which is the first wavenumber to go unstable as $\e$ crosses zero.  Given an inner product $(\cdot,\cdot)$, then associated with the right null eigenvector is a left null eigenvector $l$ of $L_0$, such that $( l, L_0 u ) = 0$ for any $u$.  The $\ybar$ dependence of $l$ will also be $e^{i q_c \ybar}$.  The amplitude $A$ will be determined by nonlinearities occurring at higher order.

At $\O(\e)$, we first note that $L_1 u_1=0$ automatically.  This is because $q_c$ is marginally stable at the instability threshold: given a dispersion relation $\lambda(q,\e)$ as a function of wavenumber $q$ and control parameter $\e$, then both $\lambda(q_c,0)=0$ and $\partial \lambda / \partial q(q_c,0)=0$ (see \figref{gl:bifschematic}).  The former equality yields $L_0 u_1 = 0$ and the latter equality yields the condition $L_1 u_1=0$.  In order to ensure that a solution for $u_2$ exists, a solvability condition obtained by taking the inner product with the left null eigenvector must be satisfied.  This solvability condition is $(l, L_0 u_2 + N_0(u_1,u_1)) = (l, N_0(u_1,u_1)) = 0$.  Because $l \sim e^{i q_c \ybar}$ and $N_0(u_1,u_1) \sim 1$ or $e^{\pm 2 i q_c \ybar}$ due to the quadratic nonlinearity, this solvability condition is always satisfied.  Thus, given that a solution exists, one may write $u_2$ as a linear combination of homogeneous and particular solutions:
	\begin{equation}
		u_2 = u_{2h} + u_{2p},
	\end{equation}
	where
	\begin{align}
		u_{2h} &= A_2(Y,T) r  + A_2(Y,T)^* r^*, \\
		L_0 u_{2p} &= -N_0(u_1,u_1). \label{gl:u2pqen}
	\end{align}
Since we have not yet determined $A$, we must proceed to higher order.  Another unknown parameter $A_2$ has been introduced, but we will not need it in order to solve for $A$.

At $\O(\e^{3/2})$, note that $L_1 u_{2h} = 0$ for the same reason that $L_1 u_1=0$.  Upon writing the solvability condition from \eref{gl:eps32}, one finds that several terms vanish, leaving
	\begin{equation}
		0 = (l_1, L_2 u_1) + \bigl(l_1, 2 N_0(u_1,u_{2p}) \bigr).
		\label{gl:abstractsolvabilitycondition}
	\end{equation}
This is the desired partial differential equation which determines the amplitude $A$.  Note that one never explicitly needs $L_1$ or $N_1$.

\section{Details}
We now apply this procedure to \eref{CE2}.  For simplicity, we set the viscosity to zero, take infinite deformation radius, and cross the instability threshold by varying the strength of the forcing (rather than by varying the friction as in the main text); modification for other scenarios is obvious.  Let the forcing be given by $F(x,y) = (1+\e) F_0(x,y)$, where instability threshold is at $\e=0$.  We shift variables relative to the equilibrium at $W_{\rm eqb} = (1+\e) F_0 / 2\m$, $U_{\rm eqb}=0$.  Explicitly, the abstract vector $u$ consists of two components, $u = \{W, U\}$.  Then we have the basic structure of \eref{gl:abstractequation} with
	\begin{subequations}
	\begin{align}
		[L(u)]_W &= -\partial_t W(x,y,\ybar) + 2\b \partial_y \partial_x \partial_\ybar D^{-1} W(x,y,\ybar) - 2\m W(x,y,\ybar) \notag  \\
					&\quad - \frac{1+\e}{2\m} \bigg\{ [U(\ybar+\tfrac12 y)-U(\ybar- \tfrac12 y)] \partial_x F_0(x,y) \notag \\
					&\quad - \partial_\ybar^2 [U(\ybar+ \tfrac12 y) - U(\ybar- \tfrac12 y)] \frac{\partial_x}{\nabla^2} F_0(x,y) \bigg\}, \\
		[L(u)]_U &= -\partial_t U(\ybar) - \partial_y \partial_x \partial_\ybar D^{-1} W(0,0,\ybar) - \m U(\ybar), \label{gl:LuUbasic}
	\end{align}
	\end{subequations}
where $W$ in \eref{gl:LuUbasic} should be evaluated at $x=0$ and $y=0$ after performing the derivatives, $[\cdot]_W$ and $[\cdot]_U$ refer to the $W$ and $U$ components of the abstract vector, and
	\begin{equation}
		D = \nabla^4 + \frac{1}{2} \partial_x^2 \partial_\ybar^2 - \frac{1}{2} \partial_y^2 \partial_\ybar^2 + \frac{1}{16} \partial_\ybar^4.
	\end{equation}
Note that $D^{-1}$ commutes with $\partial_x$, $\partial_y$, and $\partial_\ybar$.

For the nonlinear operator $N(v,z)$, with $v=\{v_W,v_U\}$ and $z=\{z_W,z_U\}$, we have the unsymmetrized version $N^{\rm un}$, 
	\begin{subequations}
	\begin{align}
		[N^{\rm un}(v,z)]_W &= - [v_U(\ybar+ \tfrac12 y) - v_U(\ybar - \tfrac12 y)] \partial_x z_W \notag \\
		 	& \quad+ [\partial_\ybar^2 v_U(\ybar + \tfrac12 y) - \partial_\ybar^2 v_U(\ybar - \tfrac12 y)] (\nabla^2 + \tfrac14 \partial_\ybar^2) \partial_x D^{-1} z_W \notag \\
			&	\quad - [\partial_\ybar^2 v_U(\ybar + \tfrac12 y) + \partial_\ybar^2 v_U(\ybar - \tfrac12 y)] \partial_y \partial_x \partial_\ybar D^{-1} z_W, \\
			\left[N^{\rm un}(v,z)\right]_U &= 0.
	\end{align}
	\end{subequations}
The symmetrized operator is then given by
	\begin{equation}
		N(v,z) = \frac12 [ N^{\rm un}(v,z) + N^{\rm un}(z,v) ].
	\end{equation}
	
We now introduce the slow space and time scales.  One subtlety that was not mentioned in the general procedure described above is that the $u_i$ may need to be expanded in $\e$.  This occurs for two reasons.  First, because the $U(\ybar \pm \frac12 y)$ terms lead to
	\begin{equation}
		A\bigl(\e^{1/2}\bigl(\ybar \pm \tfrac12 y\bigr)\bigr) = A\bigl(Y \pm \tfrac12 \e^{1/2} y\bigr) = A(Y) + \frac{1}{2} y \e^{1/2} \partial_Y A + \frac{1}{8} y^2 \e \partial_Y^2 A + \cdots\,. 
	\end{equation}
Second, the right null eigenvector $r$ itself contains the differential operator $\partial_\ybar$ (i.e., it depends on $q$), which must be expanded in the multiple-scale procedure.  It is extremely convenient to introduce these expansions at the outset so as to keep the entire $\e$ expansion in a single place.  This procedure is even more motivated when we absorb these extra terms into $L_1$ and $L_2$, for these terms are necessary in order to satisfy $L_1 u_1=0$.  If instead we kept separate the $\e$ expansion of $u_1$, the result would be an awkward expression like $L_0 u_{1;1} + L_1 u_{1;0} = 0$.  To introduce our convenient shortcut, first recall that since $N_1$ is never needed, we only need to perform this within $L$.  Then, for the places where $U\bigl(\ybar \pm \tfrac12 \bigr)$ occurs within $L$, we substitute, keeping only to the order required,
	\begin{equation}
		U\bigl(\ybar \pm \tfrac12 y\bigr) \to \biggl(1 \pm \frac{1}{2}y \e^{1/2} \partial_Y + \frac{1}{8} y^2 \e \partial_Y^2\biggr) U\bigl(\ybar \pm \tfrac12 y\bigr),
	\end{equation}
and then later on we substitute the specific form of $u_1$, we substitute $A(Y)$ rather than $A\bigl(Y + \tfrac12 \e^{1/2}y\bigr)$.  The second place we introduce the expansion is that since $r$ depends on $\partial_\ybar$, we have
	\begin{equation}
		r(\partial_\ybar) \to r\bigl(\partial_\ybar + \e^{1/2} \partial_Y\bigr) = r(\partial_\ybar) + \e^{1/2} \partial_Y \pd{}{(\partial_\ybar)}r(\partial_\ybar) + \frac{1}{2} \e \partial_Y^2 \pdd{}{(\partial_\ybar)} r(\partial_\ybar).
	\end{equation}
Then, letting $\partial_\ybar \to iq$ (which we can do because we will only need to perform this expansion on a term $e^{iq_c \ybar}$ and not other harmonics), we see that
	\begin{equation}
	r(q) \to \biggl(1 - i\e^{1/2} \partial_Y \pd{}{q} - \frac{1}{2} \e \partial_Y^2 \pdd{}{q}\biggr) r(q).
	\end{equation}
To implement this, one can set, in $L$,
	\begin{equation}
		W(\ybar) \to \biggl(1 - i \e^{1/2} \partial_Y `` \partial_q " - \frac{1}{2} \e \partial_Y^2 `` \partial_q^2 "\biggr) W(\ybar),
	\end{equation}
where the $`` \partial_q "$ means that the derivative acts only on $r(q)$, not on the $e^{iq\ybar}$ part of $u_1$.  We need only make this replacement in $W$, not $U$, because the $U$ component of the right null eigenvector does not contain any derivatives $\partial_\ybar$.  One can verify that this shortcut gives the same results as if one proceeded more straightforwardly.

The problem is most conveniently expressed in terms of the Fourier transform of the difference variables $x,y$.  We use the convention
	\begin{equation}
		f(k_x,k_y) = \int dx\, dy\, e^{-ik_x x} e^{-ik_y y} f(x,y).
	\end{equation}
After Fourier transform,  the required linear operators are given by
	\begin{subequations}
	\begin{align}
		[L_0 u]_W &= \biggl(-\frac{2 \b k_x k_y \partial_\ybar}{g_0(k_x,k_y,\ybar)} - 2\m\biggr) W(k_x,k_y \mid \ybar) - \int dk_y'\, e^{i k_y' \ybar} h_0(k_x,k_y,k_y') \frac{\hat{U}(k_y')}{2\pi}, \\
		[L_0 u]_U &= \partial_\ybar \frac{1}{(2\pi)^2} \int dk_x dk_y \frac{k_x k_y}{g_0(k_x,k_y,\ybar)} W(k_x,k_y \mid \ybar) - \m U(\ybar),
	\end{align}
	\end{subequations}
	\begin{subequations}
	\begin{align}
		[L_2 u]_W &= -\partial_T W(k_x,k_y \mid \ybar) - \int dk_y'\, e^{ik_y' \ybar} h_0(k_x,k_y,k_y') \frac{\hat{U}(k_y')}{2\pi} \notag \\
				& \qquad- 2\beta \partial_Y^2 \biggl( s_0(k_x,k_y,\ybar) - \frac{i k_x k_y \partial_\ybar^2 g_1(k_x,k_y,\ybar)}{g_0(k_x,k_y,\ybar)} `` \partial_q " \notag \\
				& \qquad +  \frac{k_x k_y \partial_\ybar}{2 g_0(k_x,k_y,\ybar)} `` \partial_q^2 " \biggr) W(k_x,k_y \mid \ybar) + \partial_Y^2 \mu `` \partial_q^2 " W(k_x,k_y \mid \ybar), \\
		[L_2 u]_U & = -\partial_T U(\ybar) + \partial_Y^2 \frac{1}{(2\pi)^2} \int dk_x\, dk_y\, \biggl[ s_0(k_x,k_y,\ybar) \notag \\
		& \qquad - \frac{i k_x k_y \partial_\ybar^2 g_1(k_x,k_y,\ybar)}{g_0(k_x,k_y,\ybar)} `` \partial_q " +  \frac{k_x k_y \partial_\ybar}{2 g_0(k_x,k_y,\ybar)} `` \partial_q^2 " \biggr] W(k_x,k_y \mid \ybar).
	\end{align}
	\end{subequations}
And the nonlinear operator is given by
	\begin{subequations}
	\begin{align}
		[N_0^{\rm un}(v,z)]_W &= \int dk_y'\, e^{ik_y' \ybar} \Bigl[ s_1(k_x,k_y,k_y',\ybar) z_W\bigl(k_x,k_y- \tfrac{1}{2} k_y' \mid \ybar\bigr) \notag \\
			& \qquad - s_1(k_x,k_y,-k_y',-\ybar) z_W\bigl(k_x,k_y+\tfrac{1}{2} k_y' \mid \ybar\bigr)\Bigr] \frac{\hat{v}_U(k_y')}{2\pi}, \\
		[N_0^{\rm un}(v,z)]_U & = 0,
	\end{align}
	\end{subequations}
where again the symmetrized version is $N_0(v,z) = \frac12 [ N^{\rm un}_0(v,z) + N^{\rm un}_0(z,v) ]$.

Here, $\hat{U}$ is the Fourier transform of $U$.  Our expressions for $U(\ybar)$ will always consist of periodic exponentials, and so $\hat{U}$ contains delta functions and the convolution integral can be immediately performed.  We also have defined
	\begin{gather}
		k^2 = k_x^2 + k_y^2, \\
		g_0(k_x,k_y,\ybar) = k^4 - \frac{1}{2}k_x^2 \partial_\ybar^2 + \frac{1}{2} k_y^2 \partial_\ybar^2 + \frac{1}{16}\partial_\ybar^4, \\
		g_1(k_x,k_y,\ybar) = -k_x^2 + k_y^2 + \frac{1}{4} \partial_\ybar^2, \\
		g_2(k_x,k_y,\ybar) = -\frac{1}{2} k_x^2 + \frac{1}{2} k_y^2 + \frac{3}{8} \partial_\ybar^2, \\
		h_0(k_x,k_y,k_y') = \frac{ik_x}{2\m} \Biggl\{ \Biggl[ 1 - \frac{k_y'^2}{k_x^2 + (k_y - \tfrac{1}{2} k_y')^2} \Biggr] F_0\bigl(k_x, k_y- \tfrac12 k_y'\bigr)  \notag \\
				\qquad \qquad \qquad - \Biggl[ 1 - \frac{k_y'^2}{k_x^2 + (k_y +\tfrac12 k_y')^2} \Biggr] F_0\bigl(k_x, k_y+\tfrac12 k_y' \bigr) \Biggr\}, \\
		s_0(k_x,k_y,\ybar) = k_x k_y \partial_\ybar \left[ \frac{\partial_\ybar^2 g_1(k_x,k_y,\ybar)^2 - g_0(k_x,k_y,\ybar) g_2(k_x,k_y,\ybar) }{g_0(k_x,k_y,\ybar)^3} - \frac{g_1(k_x,k_y,\ybar)}{g_0(k_x,k_y,\ybar)^2} \right], \\
		s_1(k_x,k_y,k_y',\ybar) = -ik_x \Biggl(1 + \frac{k_y'^2 \bigl[-k_x^2- \bigl(k_y-\frac12 k_y' \bigr)^2 + \frac14 \partial_\ybar^2 \bigr] - i k_y'^2 \partial_\ybar \bigl(k_y - \frac12 k_y' \bigr)}{g_0(k_x,k_y-\frac12 k_y',\ybar)} \Biggr).
	\end{gather}
We also define $\hat{g}_0(k_x,k_y,k_\ybar)$ as  $g_0(k_x,k_y,\ybar)$ with $\partial_\ybar \to i k_\ybar$, and similarly for $\hat{g}_1$, $\hat{g}_2$, $\hat{s}_0$, and $\hat{s}_1$.

We define an inner product
	\begin{equation}
		(v,z) = \int d\ybar\, v_U^*(\ybar) z_U(\ybar) + \int d\ybar\, dk_x\, dk_y\, v_W^*(k_x,k_y,\ybar) z_W(k_x,k_y,\ybar).
	\end{equation}
	
At $\O(\e^{1/2})$, we find that $u_1$ is given by \eref{gl:u1form} where the right null eigenvector is given by $r = e^{i q_c \ybar} \{r_W(k_x,k_y,q), U_0\}$, where
	\begin{align}
		r_W(k_x,k_y,q) &= - \frac{h_0(k_x,k_y,q) U_0}{\hat{g}_3(k_x,k_y,q)}, \\
		\hat{g}_3(k_x,k_y,q) &= 2\m + \frac{2i \b k_x k_y q}{\hat{g}_0(k_x,k_y,q)},
	\end{align}
and $U_0$ is a constant with dimension of velocity, whose purpose is to help keep track of dimensional consistency.  For any computation it can be set to unity.  The complex conjugate of the left null eigenvector is found to be $l^* = e^{-i q_c \ybar} \{l_W^*(k_x,k_y), 1\}$, where
	\begin{equation}
		l_W^*(k_x,k_y) = \frac{i q k_x k_y}{(2\pi)^2 \hat{g}_0(k_x,k_y,q) \hat{g}_3(k_x,k_y,q)}.
	\end{equation}
The $q$ dependence of $r_W$ and $l_W$ is now suppressed except for where it matters in \eref{gl:c2}; they should be evaluated at $q=q_c$.

At $\O(\e)$, we need to solve the particular solution of \eref{gl:u2pqen}.  Take an ansatz
	\begin{subequations}
	\begin{align}
		u_{2pW} &= a_W(k_x,k_y) A^2 e^{i2q_c \ybar} + a_W(k_x,k_y)^* A^{*2} e^{-i2q_c \ybar} + b_W(k_x,k_y) AA^*, \\
		u_{2pU} &= a_U A^2 e^{i2q_c \ybar} + a_U^* A^{*2} e^{-i2q_c\ybar} + b_U AA^*.
	\end{align}
	\end{subequations}
After some algebra we find
	\begin{align}
		a_W(k_x,k_y) &= -\frac{h_0(k_x,k_y,2q_c)}{\hat{g}_3(k_x,k_y,2q_c)} a_U \notag \\
		& \qquad \qquad + U_0 \hat{g}_3(k_x,k_y,2q)^{-1} \bigl[\hat{s}_1(k_x,k_y,q_c,q_c)r_W(k_x,k_y-\tfrac12 q_c) \notag \\
		& \qquad \qquad  - \hat{s}_1(k_x,k_y,-q_c,-q_c) r_W(k_x,k_y+\tfrac12 q_c)\bigr], \\
		a_U &= \frac{N_a}{D_a}, \\
		N_a &= \frac{2iq_cU_0}{(2\pi)^2} \int dk_x dk_y\, \bigl[ \hat{g}_0(k_x,k_y,2q_c) \hat{g}_3(k_x,k_y,2q_c) \bigr]^{-1} \notag \\
		& \times [ \hat{s}_1(k_x,k_y,q_c,q_c) r_W(k_x,k_y- \tfrac12 q_c) - \hat{s}_1(k_x,k_y,-q_c,-q_c) r_W(k_x,k_y + \tfrac12 q_c)], \\
		D_a &=  \mu + \frac{2iq_c}{(2\pi)^2} \int dk_x dk_y\, \frac{k_x k_y h_0(k_x,k_y,2q_c)}{\hat{g}_0(k_x,k_y,2q_c) \hat{g}_3(k_x,k_y,2q_c)},
	\end{align}
and
	\begin{align}
		b_W(k_x,k_y) &= \frac{U_0}{2\m} \bigl[ \hat{s}_1(k_x,k_y,q_c,-q_c) r_W(k_x, k_y-\tfrac12 q_c)^* \notag \\
				& \quad - \hat{s}_1(k_x,k_y,-q_c,q_c) r_W(k_x,k_y+\tfrac12 q_c)^* -\hat{s}_1(k_x,k_y,q_c,-q_c) r_W(k_x,k_y-\tfrac12 q_c) \notag \\
				& \quad + \hat{s}_1(k_x,k_y,-q_c,q_c) r_W(k_x, k_y+\tfrac12 q_c) \bigr], \\
		b_U &= 0.
	\end{align}

At $\O(\e^{3/2})$, the solvability condition \eref{gl:abstractsolvabilitycondition} becomes
	\begin{equation}
		c_0 \partial_T A(\ybar,t) = c_1 A + c_2 \partial_Y^2 A - c_3 |A|^2 A,
	\end{equation}
where
	\begin{align}
		c_0 &= U_0 + \int dk_x\, dk_y\, l_W^*(k_x,k_y) r_W(k_x,k_y) \bigr|_{q=q_c}, \\
		c_1 &= -U_0 \int dk_x\, dk_y\, l_W^*(k_x,k_y) h_0(k_x,k_y) \bigr|_{q=q_c}, \\
		c_2 &= \left. \pdd{}{q} \right|_{q=q_c} \frac{1}{2} \int dk_x\, dk_y\, l_w^*(k_x,k_y,q) U_0 h_0(k_x,k_y,q), \label{gl:c2} \\
		c_3  &= -\int dk_x\, dk_y\,  l_W^*(k_x,k_y) \Bigl\{ U_0 \bigl[ \hat{s}_1(k_x,k_y,q,0) b_W(k_x,k_y - \tfrac12 q) \notag \\
		& \qquad - \hat{s}_1(k_x,k_y,-q,0) b_W(k_x,k_y + \tfrac12 q)\bigr] + U_0\bigl[ \hat{s}_1(k_x,k_y,-q,2q) a_W(k_x,k_y + \tfrac12 q) \notag \\
		& \qquad - \hat{s}_1(k_x,k_y,q,-2q) a_W(k_x,k_y - \tfrac12 q)\bigr] + a_U\bigl[ \hat{s}_1(k_x,k_y,2q,-q) r_W(k_x,k_y-q)^* \notag \\
		& \qquad - \hat{s}_1(k_x,k_y,-2q,q) r_W(k_x,k_y+q)^* \bigr] \Bigr\}\Bigr|_{q=q_c}.
	\end{align}
	
After returning to the unscaled variables by letting $T \to \e t$, $Y \to \e^{1/2} \ybar$, and $A \to A/\e$, we recover \eref{gl:amplitudeequation}.   The coefficients $c_i$ involve integrals over the forcing spectrum which is here presented in a form where the the wavenumbers are shifted, i.e., contain terms like $k_y - q/2$.  It is also possible to shift the integration variable so all integrals contain just the unshifted forcing $F_0(k_x,k_y)$, after which the $q$ derivatives in $c_2$ can be explicitly computed \cite{bakas:tbs}.

It is possible to obtain $c_0$, $c_1$, and $c_2$, which govern the linear behavior, via the alternate and much simpler route of using the analytic dispersion relation \eref{zi:dispersionrelation}.  The dispersion relation can be put into the form $D(\l,\e,q)=0$.  The conditions of the instability threshold require $D(0,0,q_c)=0$ and $\partial D/\partial q(0,0,q_c)=0$.  Thus, expanding $D$ to lowest order about $(0,0,q_c)$, we find
	\begin{equation}
		-\pd{D}{\l}(0,0,q_c)\, \l = \e \pd{D}{\e}(0,0,q_c) + \frac{1}{2} \pdd{D}{q}(0,0,q_c) (q-q_c)^2.
	\end{equation}
Then up to a constant of proportionality, we see that $c_0 = -\partial D/\partial \l(0,0,q_c)$, $c_1 = \partial D/\partial \e(0,0,q_c)$, and $c_2 = - \tfrac12 \partial^2 D/ \partial q^2 (0,0,q_c)$.  This was used to put $c_2$ above into a succinct form.  But this approach does not give $c_3$; for that one needs the full bifurcation calculation which includes nonlinear terms.

To verify these analytic expressions, we take an example forcing $F_0(k_x,k_y) = \mathcal{A} k \exp[-(k-k_f)^2/\sigma_k^2]$, with $k^2 = k_x^2 + k_y^2$, $\mathcal{A} = 4\sqrt{\pi} \varepsilon/\sigma_k$, $k_f=1$, and $\sigma_k=0.5$.  For the other parameters we use $\m=0.1$, $\b=1$.  Then the critical value of the control parameter is calculated to be $\varepsilon_c = 0.1297$ with a critical wavenumber $q_c = 0.676$.  We compute $c_0 = 1.10$, $c_1=0.10$, $c_2=0.0015$, and $c_3=2.28$.  Comparisons between analytic and numerically computed results are shown in Figure \ref{fig:gl:GL_coeffs} and are in excellent agreement.

\chapter{Projection for Ideal State Equilibrium}
\label{app:iscalc}
In this Appendix we perform the projection of \eref{iscalc:CE2steadystate} onto the Galerkin basis functions.  We find explicit formulas for the nonlinear algebraic equation, in a suitable form for numerical implementation.  The shorthand notations in \eref{iscalc:first_shorthand}--\eref{iscalc:last_shorthand} are used throughout.

\section{Eddy Equation}
Projection of the eddy equation entails operating on \eref{iscalc:CE2W} with
	\begin{equation}
		\left(\frac{2\pi}{a} \frac{2\pi}{b} \frac{2\pi}{q}\right)^{-1}  \int_{-\pi/a}^{\pi/a} dx   \int_{-\pi/b}^{\pi/b} dy 	\int_{-\pi/q}^{\pi/q} d\ybar\,  \p_{rst}^*,
	\end{equation}
where
	\begin{equation}
		\p_{mnp} = e^{imax} e^{inby} e^{ipq\ybar}.
	\end{equation}

\subsection*{First Term}
First term of \eref{iscalc:CE2W}:
	\begin{equation}
		-[U_+ - U_-] \partial_x W = -\bigl[ U\bigl(\ybar + \tfrac12 y) - U\bigl(\ybar - \tfrac12 y \bigr) \bigr] \partial_x W.
	\end{equation}
Note that
	\begin{equation}
		U \bigl(\ybar + \tfrac12 y \bigr) - U\bigl(\ybar - \tfrac12 y \bigr) = \sum_{p'=-P}^P U_{p'} e^{ip'q\ybar} \bigl(e^{ip'qy/2} - e^{-ip'qy/2} \bigr)
	\end{equation}
and
	\begin{equation}
		\partial_x W = \sum_{mnp} W_{mnp}\, ik_x e^{imax} e^{inby} e^{ipq\ybar}.
	\end{equation}

Then we have
	\begin{equation}
		-[U_+ - U_-] \partial_x W = -\sum_{p'mnp} U_{p'} W_{mnp}\, ik_x e^{ip'q\ybar} \bigl(e^{ip'qy/2} - e^{-ip'qy/2} \bigr) e^{imax} e^{inby} e^{ipq\ybar}.
	\end{equation}
Now project onto $\p_{rst}$, yielding
	\begin{align}
		-\sum_{p'mnp} U_{p'} W_{mnp}\,  ik_x \frac{1}{2\pi/a} & \int_{-\pi/a}^{\pi/a} dx\, e^{i(m-r)ax} \frac{1}{2\pi/b} \int_{-\pi/b}^{\pi/b} dy\, e^{iy(nb-sb)} \bigl(e^{ip'qy/2} - e^{-ip'qy/2} \bigr) \notag \\
		& \qquad \qquad \times \frac{1}{2\pi/q} \int_{-\pi/q}^{\pi/q} d\ybar\, e^{i(p'+p-t)q\ybar}.
	\end{align}
For the $y$ integral, use
	\begin{equation}
		\frac{1}{2\pi/b} \int_{-\pi/b}^{\pi/b} dy\, e^{i \a y} = \sinc\left( \frac{\a\pi}{b} \right).
	\end{equation}
Performing the integrals results in
	\begin{equation}
		I_{rstp'mnp}^{(1)} U_{p'} W_{mnp},
	\end{equation}
with
	\begin{equation}
		I_{rstp'mnp}^{(1)} = -ik_x  (\s_+ - \s_-) \de_{m,r} \de_{p'+p-t, 0}.
	\end{equation}
Now split into real and imaginary parts:
	\begin{equation}
		I_{rstp'mnp}^{(1)} U_{p'} W_{mnp} = J_{rstp'mnp}^{(1)} U_{p'} F_{mnp} + i K_{rstp'mnp}^{(1)} U_{p'} E_{mnp},
	\end{equation}
where
	\begin{equation}
		J_{rstp'mnp}^{(1)} = -K_{rstp'mnp}^{(1)} = k_x  (\s_+ - \s_-) \de_{m,r} \de_{p'+p-t,0}.
	\end{equation}

\subsection*{Second Term}
Second term of \eref{iscalc:CE2W}:
	\begin{equation}
		\Bigl(\ol{U}_+'' - \ol{U}_-'' \Bigr) \biggl(\nablabarsq + \frac{1}{4} \partial_\ybar^2 \biggr) \partial_x \Psi,
	\end{equation}
where
	\begin{equation}
		\ol{U}_\pm'' = U_\pm'' - \azf \LD U_\pm = -\sum_{p'} U_{p'} \kbsqybarU e^{ip'q\ybar} e^{\pm ip'qy/2}.
	\end{equation}
And
	\begin{equation}
		\biggl(\nablabarsq + \frac{1}{4} \partial_\ybar^2 \biggr) \partial_x \Psi = - \sum_{mnp} C_{mnp}\, ik_x \biggl( \kbsq + \frac{1}{4} \kybarW^2 \biggr) e^{imax} e^{inby} e^{ipq\ybar}.
	\end{equation}
Then
	\begin{align}
		\Bigl(\ol{U}_+'' - \ol{U}_-'' \Bigr) \biggl(\nablabarsq + \frac{1}{4} \partial_\ybar^2 \biggr) \partial_x \Psi = & \sum_{p'mnp} U_{p'} C_{mnp}\, ik_x \kbsqybarU \biggl( \kbsq + \frac{1}{4} \kybarW^2 \biggr) e^{ip'q\ybar} \notag \\
		& \qquad \times \bigl( e^{ip'qy/2} - e^{-ip'qy/2} \bigr) e^{imax} e^{inby} e^{ipq\ybar}.
	\end{align}
Project onto $\p_{rst}$ and obtain
	\begin{equation}
		I_{rstp'mnp}^{(2)} U_{p'} C_{mnp},
	\end{equation}
with
	\begin{equation}
		I_{rstp'mnp}^{(2)} = ik_x \kbsqybarU \biggl(\kbsq + \frac{1}{4} \kybarW^2 \biggr) (\s_+ - \s_-) \de_{m,r} \de_{p'+p-t,0}.
	\end{equation}
Now split into real and imaginary parts:
	\begin{equation}
		I_{rstp'mnp}^{(2)} U_{p'} C_{mnp} = J_{rstp'mnp}^{(2)} U_{p'} H_{mnp} + i K_{rstp'mnp}^{(2)} U_{p'} G_{mnp},
	\end{equation}
where
	\begin{equation}
		J_{rstp'mnp}^{(2)} = -K_{rstp'mnp}^{(2)} = -k_x \kbsqybarU \biggl(\kbsq + \frac{1}{4} \kybarW^2 \biggr) (\s_+ - \s_-) \de_{m,r} \de_{p'+p-t,0}.
	\end{equation}

\subsection*{Third Term}
Third term of \eref{iscalc:CE2W}:
	\begin{equation}
		2 \b \partial_\ybar \partial_y \partial_x \Psi.
	\end{equation}
The projection process should be clear.  Here we obtain
	\begin{equation}
		I_{rstmnp}^{(3)} C_{mnp},
	\end{equation}
with
	\begin{equation}
		I_{rstmnp}^{(3)} = -i 2\b k_x k_y \kybarW \de_{m,r} \de_{n,s} \de_{p,t} \\
	\end{equation}
In real and imaginary parts:
	\begin{equation}
		I_{rstmnp}^{(3)} C_{mnp} = J_{rstmnp}^{(3)} H_{mnp} + i K_{rstmnp}^{(3)} G_{mnp}
	\end{equation}
with
	\begin{equation}
		J_{rstmnp}^{(3)} = -K_{rstmnp}^{(3)} = 2\b k_x k_y \kybarW \de_{m,r} \de_{n,s} \de_{p,t}
	\end{equation}	
	
\subsection*{Fourth Term}
Fourth term of \eref{iscalc:CE2W}:
	\begin{equation}
		-\Bigl(\ol{U}_+'' + \ol{U}_-'' \Bigr) \partial_\ybar \partial_y \partial_x \Psi.
	\end{equation}
After projection onto $\p_{rst}$, we obtain
	\begin{equation}
		I_{rstp'mnp}^{(4)} U_{p'} C_{mnp},
	\end{equation}
with
	\begin{equation}
		I_{rstp'mnp}^{(4)} = -i \kbsqybarU k_x k_y \kybarW (\s_+ + \s_-) \de_{m,r} \de_{p'+p-t,0}.
	\end{equation}
Now split into real and imaginary parts:
	\begin{equation}
		I_{rstp'mnp}^{(4)} U_{p'} C_{mnp} = J_{rstp'mnp}^{(4)} U_{p'} H_{mnp} + i K_{rstp'mnp}^{(4)} U_{p'} G_{mnp},
	\end{equation}
where
	\begin{equation}
		J_{rstp'mnp}^{(4)} = -K_{rstp'mnp}^{(4)} = \kbsqybarU k_x k_y \kybarW (\s_+ + \s_-) \de_{m,r} \de_{p'+p-t,0}.
	\end{equation}

\subsection*{Fifth Term}
Fifth term of \eref{iscalc:CE2W}:
	\begin{equation}
		F(x,y).
	\end{equation}
This is the forcing term.  After projection onto $\p_{rst}$, we obtain
	\begin{equation}
		I^{(5)}_{rst} = J^{(5)}_{rst} = \de_{t,0} \biggl(\frac{2\pi}{a}\frac{2\pi}{b}\biggr)^{-1} F_{nb}(k_x,k_y).
	\end{equation}
where $F_{nb}$ is ``narrowband forcing'' as described by \citet{srinivasan:2012} in the continuous Fourier transform.  The prefactor of $(2\pi/a)^{-1} (2\pi/b)^{-1}$ essentially comes from the conversion factor from a Fourier transform amplitude to a Fourier series amplitude.  Specifying $F_{nb}(k_x,k_y)$ defines $F(x,y)$.  We use
	\begin{equation}
		F_{nb}(k_x,k_y) = \begin{cases} 2\pi\ve k_f / \de k & k_f - \de k < k < k_f + \de k \\ 0 & \text{otherwise} \end{cases},
	\end{equation} 
where $\ve$ is an equivalent energy input in the case of $\LD = 0$.  Note $I^{(5)}$ is pure real.

\subsection*{Sixth Term}
Sixth and final term of \eref{iscalc:CE2W}:	
	\begin{equation}
		-(2\m + 2\n D_h) W.
	\end{equation}
After projection onto $\p_{rst}$ we obtain
\begin{equation}
		I_{rstmnp}^{(6)} W_{mnp},
	\end{equation}
with
	\begin{equation}
		I_{rstmnp}^{(6)} = - \bigl[ 2\m + \n \bigl( h_+^{2h} + h_-^{2h} \bigr) \bigr] \de_{m,r} \de_{n,s} \de_{p,t}.
	\end{equation}
In real and imaginary parts:
	\begin{equation}
		I_{rstmnp}^{(6)} W_{mnp} = J_{rstmnp}^{(6)} E_{mnp} + i K_{rstmnp}^{(6)} F_{mnp},
	\end{equation}
with
	\begin{equation}
		J_{rstmnp}^{(6)} = K_{rstmnp}^{(6)} = -\bigl[ 2\m + \n \bigl( h_+^{2h} + h_-^{2h} \bigr) \bigr] \de_{m,r} \de_{n,s} \de_{p,t}.
	\end{equation}	

\section{Zonal Flow Equation}
For the zonal flow equation \eref{iscalc:CE2U}, we can project onto $e^{itq\ybar}$, or equivalently, just equate the $e^{ipq\ybar}$ coefficients.  Upon substitution of the Galerkin series, we have
	\begin{equation}
		-U_p \bigl( \m + \n k_\ybar^{2h} \bigr) \kbsqybarU / \kybarU^2 + \sum_{mn} i k_x k_y k_\ybar C_{mnp} = 0,
	\end{equation}
where here we use $\kybarU = pq$.  Noting the real part of $C$ cancels out after summation, this becomes
	\begin{equation}
		-U_p \bigl( \m + \n k_\ybar^{2h} \bigr) \kbsqybarU / \kybarU^2 - \sum_{mn} k_x k_y k_\ybar H_{mnp} = 0.
	\end{equation}
By symmetry, one could sum only over positive $m,n$ (and put in a factor of 4), though we do not need to do this.  The above equation can be written
	\begin{equation}
		I_{tp'}^{(7)} U_{p'} + I_{tmnp}^{(8)} H_{mnp} = 0,
	\end{equation}
where
	\begin{align}
		I_{tp'}^{(7)} &= - \bigl(\m + \n \kybarU^{2h} \bigr) \bigl(\kbsqybarU / \kybarU^2 \bigr) \de_{t,p'}, \\
		I_{tmnp}^{(8)} &= - k_x k_y \kybarW \de_{t,p}.
	\end{align}
\chapter{Projection for Ideal State Stability}
\label{app:isstability}
In this Appendix we perform the projection of the linearized system \eref{is:dCE2} onto the basis functions.  We find explicit formulas for the matrix equation, in a suitable form for numerical implementation.  The shorthand notations in \eref{isstability:first_shorthand}--\eref{isstability:last_shorthand} are used throughout.

\section{Eddy Equation}
Let
	\begin{equation}
		\ti{\p}_{mnp} = e^{imax} e^{inby} e^{i(Q+pq)\ybar}.
	\end{equation}
We will project \eref{is:dCE2W} onto $\ti{\p}_{rst}$ in the same way as for the ideal state calculation, by applying
	\begin{equation}
		\left(\frac{2\pi}{a} \frac{2\pi}{b} \frac{2\pi}{q} \right)^{-1} \int_{-\pi/a}^{\pi/a} dx   \int_{-\pi/b}^{\pi/b} dy 	\int_{-\pi/q}^{\pi/q} d\ybar\, \ti{\p}_{rst}^*
	\end{equation}
We suppress the $e^{\s t}$ dependence of the perturbations $\de W$ and $\de U$ from now on.  Projecting the LHS of \eref{is:dCE2W} is trivial; one merely obtains $\s \de W_{rst}$.  Now we project the RHS.  The matrix coefficients are closely related to those in the ideal state calculation.

\subsection*{First and second term}
The first term on the RHS is
	\begin{equation}
		-(\de U_+ - \de U_-) \partial_x W.
	\end{equation}
We have
	\begin{equation}
		\de U_\pm = \de U\bigl(\ybar \pm \tfrac 12 y\bigr) = \sum_{p'} \de U_{p'} e^{i(Q+p'q)\ybar} e^{\pm i(Q+p'q)y/2},
	\end{equation}	
so that
	\begin{equation}
		\de U_+ - \de U_-  = \sum_{p'} \de U_{p'} e^{i(Q+p'q)\ybar} \left( e^{i(Q+p'q)y/2} - e^{-i(Q+p'q)y/2} \right).
	\end{equation}
Also,
	\begin{equation}
		\partial_x W = \sum_{mnp} W_{mnp}\, ik_x e^{imax} e^{inby} e^{ipq\ybar}.
	\end{equation}
Thus,
	\begin{align}
		-(\de U_+ - \de U_-) \partial_x W &= -\sum_{p'mnp} \de U_{p'} W_{mnp}\,  ik_x e^{i(Q+p'q)\ybar} e^{imax} e^{inby} e^{ipq\ybar}  \notag \\
			& \qquad \qquad \times  \left( e^{i(Q+p'q)y/2} - e^{-i(Q+p'q)y/2} \right).
	\end{align}
Now project onto $\ti{\p}_{rst}$.  Obtain
	\begin{equation}
		\ti{I}_{rstp'mnp}^{(1)} \de U_{p'} W_{mnp},
	\end{equation}
where
	\begin{equation}
		\ti{I}_{rstp'mnp}^{(1)} = -ik_x (\sdU{+} - \sdU{-}) \de_{m,r} \de_{p'+p-t,0}.
	\end{equation}
	
The second term on the RHS of \eref{is:dCE2W} is
	\begin{equation}
		-(U_+-U_-) \partial_x \de W.
	\end{equation}
With the aid of our convenient notation, we can obtain the result after projection from the first term by making the replacements $\de U \to U$ and $W \to \de W$ (including in the subscripts of the coefficients).  We obtain
	\begin{equation}
		\ti{I}_{rstp'mnp}^{(2)} U_{p'} \de W_{mnp},
	\end{equation}
where
	\begin{equation}
		\ti{I}_{rstp'mnp}^{(2)} = -ik_x (\sU{+} - \sU{-}) \de_{m,r} \de_{p'+p-t,0}.
	\end{equation}
	
\subsection*{Third and Fourth Term}
Third term:
	\begin{equation}
		\Bigl(\de \ol{U}_+'' - \de \ol{U}_-'' \Bigr) \biggl(\nablabarsq + \frac14 \partial_\ybar^2 \biggr) \partial_x \Psi.
	\end{equation}
We obtain after projection
	\begin{equation}
		\ti{I}_{rstp'mnp}^{(3)} \de U_{p'} \Psi_{mnp},
	\end{equation}
where
	\begin{equation}
		\ti{I}_{rstp'mnp}^{(3)} = ik_x \kbsqybardU \biggl(\kbsq + \frac14 \kybarW^2 \biggr) (\sdU{+} - \sdU{-}) \de_{m,r} \de_{p'+p-t,0}.
	\end{equation}

The fourth term is obtained after the appropriate replacements:
	\begin{equation}
		\ti{I}_{rstp'mnp}^{(4)} U_{p'} \de \Psi_{mnp},
	\end{equation}
where
	\begin{equation}
		\ti{I}_{rstp'mnp}^{(4)} = ik_x \kbsqybarU \biggl(\kbsq + \frac14 \kybardW^2 \biggr) (\sU{+} - \sU{-}) \de_{m,r} \de_{p'+p-t,0}.
	\end{equation}
		
\subsection*{Fifth and Sixth Term}
Fifth term:
	\begin{equation}
		-\Bigl(\de \ol{U}_+'' + \de \ol{U}_-'' \Bigr) \partial_\ybar \partial_y \partial_x \Psi.
	\end{equation}
We obtain after projection
	\begin{equation}
		\ti{I}_{rstp'mnp}^{(5)} \de U_{p'} \Psi_{mnp},
	\end{equation}
where
	\begin{equation}
		\ti{I}_{rstp'mnp}^{(5)} = -i \kbsqybardU k_x k_y \kybarW (\sdU{+} + \sdU{-}) \de_{m,r} \de_{p'+p-t,0}.
	\end{equation}

The sixth term is obtained after the appropriate replacements:
	\begin{equation}
		\ti{I}_{rstp'mnp}^{(6)} U_{p'} \de \Psi_{mnp},
	\end{equation}
where
	\begin{equation}
		\ti{I}_{rstp'mnp}^{(6)} = -i \kbsqybarU k_x k_y \kybardW (\sU{+} + \sU{-}) \de_{m,r} \de_{p'+p-t,0}.
	\end{equation}

\subsection*{Seventh and Eighth Term}
Seventh term:
	\begin{equation}
		2 \b \partial_\ybar \partial_y \partial_x \de \Psi.
	\end{equation}
After projection, we obtain
	\begin{equation}
		\ti{I}_{rstmnp}^{(7)} \de \Psi_{mnp},
	\end{equation}
where
	\begin{equation}
		\ti{I}_{rstmnp}^{(7)} = -2i \b k_x k_y \kybardW \de_{m,r} \de_{n,s} \de_{p,t}.
	\end{equation}
	
Eighth term:
	\begin{equation}
		-(2\m + 2\n D_h) \de W.
	\end{equation}
After projection we obtain
	\begin{equation}
		\ti{I}_{rstmnp}^{(8)} \de W_{mnp},
	\end{equation}
where
	\begin{equation}
		\ti{I}_{rstmnp}^{(8)} = -\bigl[2\m + \n\bigl( h_{+,\de W}^{2h} + h_{-,\de W}^{2h} \bigr)\bigr] \de_{m,r} \de_{n,s} \de_{p,t}.
	\end{equation}

\section{Zonal Flow Equation}
Since the zonal flow equation \eref{is:dCE2U} is linear, projection is equivalent to matching the coefficients of the exponentials.  It is simple to find that at each $p$,
	\begin{equation}
		\s \de U_p = i \frac{\kybardU^2}{\kbsqybardU} \sum_{mn} k_x k_y \kybardW \de \Psi_{mnp} - \bigl(\m + \n \kybardU^{2h}\bigr) \de U_p,
	\end{equation}
where here we use $\kybardU = \kybardW = Q+pq$, $k_x=ma$, $k_y=nb$, and $\kbsqybardU = \kybardU^2 + \azf \LD$.

\chapter{Stability of Kraichnan-Spiegel Closure}
\label{app:kraichnanspiegel}
\section{The Kraichnan-Spiegel Model}
In the pedagogical closure described in \chref{ch:closure}, we proved that any nonzero steady-state solution is linearly stable.  Here, we prove the same result for the Kraichnan-Spiegel (KS) closure \cite{kraichnan:1962}.  The KS closure is a model for energy transfer in 3D, isotropic turbulence.  The significance of this result stems from the fact that stability of the equilibria of closures is an important topic, and this is one of the first definitive results.

A limit of the KS closure assuming local transfer gives the Leith diffusion model \cite{leith:1967}.  Hence, stability of the Leith model follows from stability of the KS model.

In general, an energy balance equation can be written
	\begin{equation}
		\pd{E(k)}{t} = 2 \g_k E(k) + T(k),
	\end{equation}
where $E(k)$ is the energy spectrum and $T(k)$ is the nonlinear transfer term.  The $\g_k$ term includes all linear terms, generalizing the original KS model by allowing not only for viscous damping but linear drive as well.  In Navier--Stokes, the quadratic nonlinearity means that in $k$-space the fundamental interactions are among three Fourier modes, or triads.  The KS approximation involves treating the fundamental nonlinear transfer as occurring only between two modes.  The KS closure takes a specific form for $T(k)$:
	\begin{equation}
		\pd{E(k)}{t} = 2 \g_k E(k) + \int_0^\infty dp\, \bigl[S_e(p \mid k) - S_e(k \mid p) \bigr],
	\end{equation}
where the ``emission'' term $S_e(k \mid p)$, corresponding to the energy emitted by mode $k$ and absorbed by mode $p$, is given by
	\begin{equation}
		S_e(k \mid p) = \eta \bigl[k^{-2} E(k) \bigr]^{-3/2} (kp)^{7/4} g(p/k),
	\end{equation}
and $S_e(p \mid k)$ is an absorption term at mode $k$ corresponding to emission from mode $p$.  Here, $\eta$ is a dimensionless numerical constant, $g(p/k) = g(k/p)$, and $g(x)$ decays quickly for $x \gg 1$ ($g$ enforces locality in wavenumber space).  We will not be concerned with the functional form of $g(k/p)$, only its symmetry, so we write $g(k/p) \to g_{kp}$ and note that it is symmetric in its indices.  The balance equation written explicitly is
	\begin{equation}
		\pd{E(k)}{t} = 2\g_k E(k) + \eta \int dp\, g_{kp}\, (kp)^{7/4} \bigl[ p^{-3} E(p)^{3/2} - k^{-3} E(k)^{3/2} \bigr].
		\label{ks:KSbalanceequation}
	\end{equation}
Now, assume that a steady state solution $E(k)$ exists which is nowhere zero.  Then one can write a ``steady-state condition'' which will be later used to eliminate $\g_k$:
	\begin{equation}
		2\g_k = \frac{\eta}{E(k)} \int dp\, g_{kp}\, (kp)^{7/4} \bigl[ k^{-3} E(k)^{3/2} - p^{-3} E(p)^{3/2} \bigr].
		\label{ks:steadystatecondition}
	\end{equation}
	
\section{Linear Stability}
We show that in the Kraichnan-Spiegel closure, any nonzero equilibrium is linearly stable.  We do this by providing a positive definite functional, quadratic in the perturbation, which decays in time.

\subsection*{Linearization}
Linearize about an equilibrium $E(k)$, assuming one exists, by letting $E(k,t) = E(k) + \de E(k,t)$.  Then \eref{ks:KSbalanceequation} becomes
	\begin{align}
		\pd{\de E(k,t)}{t} &= 2\g_k \de E(k,t) + \frac{3 \eta}{2} \int dp\, g_{kp}\, (kp)^{7/4} \notag \\
			& \qquad \qquad \qquad  \times \bigl[ p^{-3} E(p)^{1/2} \de E(p,t) - k^{-3} E(k)^{1/2} \de E(k,t) \bigr].
	\end{align}
It will be convenient to write this in terms of $w_k \defineas \de E(k,t) / E(k)$:
	\begin{equation}
		\pd{w_k}{t} = 2\g_k w_k + \frac{3 \eta}{2 E(k)} \int dp\, g_{kp}\, (kp)^{7/4} \bigl[ p^{-3} E(p)^{3/2} w_p - k^{-3} E(k)^{3/2} w_k \bigr].
	\end{equation}
Now, substitute the steady-state condition \eref{ks:steadystatecondition} to obtain
	\begin{align}
		\pd{w_k}{t} &= \frac{\eta}{E(k)} \int dp\, g_{kp}\, (kp)^{7/4} \bigl[ k^{-3} E(k)^{3/2} - p^{-3} E(p)^{3/2} \bigr] w_k \notag \\
			& \qquad + \frac{3 \eta}{2 E(k)} \int dp\, g_{kp}\, (kp)^{7/4} \bigl[ p^{-3} E(p)^{3/2} w_p - k^{-3} E(k)^{3/2} w_k  \bigr].
	\end{align}
The notation can be simplified by defining
	\begin{align}
		Q_k & \defineas k^{-3} E(k)^{3/2}, \\
		C_{kp} & \defineas g_{kp}\, (kp)^{7/4},
	\end{align}
where $C_{kp}$ is symmetric in its indices, to yield
	\begin{equation}
		\pd{w_k}{t} = -\frac{\eta}{2E(k)} \int dp\, C_{kp} \bigl( Q_k w_k + 2 Q_p w_k - 3Q_p w_p \bigr).
	\end{equation}

\subsection*{Quadratic Functional}
Consider the quadratic functional
	\begin{equation}
		W(t) = \frac{1}{2} \int_0^\infty dk\, Q_k E(k)\, w_k^2.
	\end{equation}
$W(t)$ is positive definite with respect to $w_k$.  The evolution of $W(t)$ is given by
	\begin{align}
		\d{W(t)}{t} &= \int dk\, Q_k E(k)\, w_k \pd{w_k}{t} \notag \\
			&= -\frac{\eta}{2} \int dk\, dp\, C_{kp} \left( Q_k^2 w_k^2 + 2 Q_k Q_p w_k^2 - 3 Q_k Q_p w_k w_p \right).
		\label{ks:Wevolution}
	\end{align}

We will now show that $dW/dt \le 0$, meaning that perturbations decay and the equilibrium is linearly stable.  Note that for terms inside the square brackets in \eref{ks:Wevolution}, we are free to swap the indices $k \leftrightarrow p$, since $C_{kp}$ is symmetric in $k,p$.  Then, through a series of manipulations using this fact (we use the equals sign as if the following took place under the integral),
	\begin{align}
		Q_k^2 w_k^2 &+ 2 Q_k Q_p w_k^2 - 3Q_k Q_p w_k w_p \notag \\
			& = Q_k^2 w_k^2 + Q_k Q_p w_k^2 + Q_k Q_p w_p^2 - 3Q_k Q_p w_k w_p \notag \\
			& = Q_k Q_p w_k^2 - 2 Q_k Q_p w_k w_p + Q_k Q_p w_p^2 + Q_k^2 w_k^2 - Q_k Q_p w_k w_p \notag \\
			& = Q_k Q_p (w_k^2 - 2w_k w_p + w_p^2) + Q_k^2 w_k^2 - Q_k Q_p w_k w_p \notag \\
			& = Q_k Q_p (w_k - w_p)^2 + Q_k^2 w_k^2 - Q_k Q_p w_k w_p.
	\end{align}
In the second line, we have let $2 Q_k Q_p w_k^2 \to Q_k Q_p w_k^2 + Q_k Q_p w_p^2$.  Now, observe
	\begin{align}
		Q_k^2 w_k^2 - Q_k Q_p w_k w_p &= \tfrac12 \bigl( 2Q_k^2 w_k^2 - 2 Q_k Q_p w_k w_p \bigr) \notag \\
			&= \tfrac12 \bigl(Q_k^2 w_k^2 - 2 Q_k Q_p w_k w_p + Q_p^2 w_p^2 \bigr) \notag \\
			&= \tfrac12 (Q_kw_k - Q_p w_p)^2.
	\end{align}
Putting this all together, we obtain
	\begin{equation}
		\d{W}{t} = -\frac{\eta}{2} \int dk\, dp\, C_{kp} \biggl[ Q_kQ_p (w_k - w_p)^2 + \frac{1}{2} \left( Q_k w_k - Q_p w_p \right)^2 \biggr] \le 0.
	\end{equation}
We have found that $dW/dt \le 0$, and vanishes only when $w_k = 0$.  We have not required any further conditions on $g(x)$.

\subsection*{With external forcing}
It is not difficult to add random (isotropic) forcing to the model, which becomes nonrandom, positive $F(k)$ at the energy balance equation.  The conclusion remains unchanged, as we now show.  The energy balance equation can be written
	\begin{equation}
		\pd{E(k)}{t} = F(k) + 2 \g_k E(k) + \eta \int \cdots,
	\end{equation}
where the $\cdots$ indicate terms that were previously present without forcing.  The steady state condition, assuming $E(k)$ is nonzero everywhere, is now
	\begin{equation}
		2\g_k = -\frac{F(k)}{E(k)} + \frac{\eta}{E(k)} \int \cdots.
	\end{equation}
In the linearization equation, the $F(k)$ vanishes, giving
	\begin{equation}
		\pd{w_k}{t} = 2\g_k w_k + \frac{3\eta}{2E(k)}\int \cdots.
	\end{equation}
Substituting the steady state condition gives
	\begin{equation}
		\pd{w_k}{t} = -\frac{F(k)}{E(k)} w_k - \frac{\eta}{2E(k)} \int \cdots.
	\end{equation}
Taking the same quadratic functional $W(t)$, we obtain the evolution equation
	\begin{equation}
		\d{W}{t} = - \int dk\, F(k) Q_k w_k^2 + \cdots.
	\end{equation}	
The new term involves the forcing $F(k)$.  In $dW/dt$, the forcing contributes a negative definite term, so the total $dW/dt$ is still negative definite.

\singlespacing
\bibliographystyle{thesisstyle_url}

\cleardoublepage
\ifdefined\phantomsection
  \phantomsection  
\else
\fi
\addcontentsline{toc}{chapter}{Bibliography}

\bibliography{thesis_bib_master}

\begin{thebibliography}{162}
\providecommand{\natexlab}[1]{#1}
\providecommand{\url}[1]{\texttt{#1}}
\providecommand{\urlprefix}{URL }

\bibitem[{Allgower and Georg(2003)}]{allgower:2003}
E.~L. Allgower and K.~Georg.
\newblock \emph{Introduction to Numerical Continuation Methods}.
\newblock Society for Industrial and Applied Mathematics, Philadelphia, PA,
  USA, 2003.

\bibitem[{Bakas and Ioannou()}]{bakas:tbs}
N.~A. Bakas and P.~J. Ioannou.
\newblock To be published.

\bibitem[{Bakas and Ioannou(2011)}]{bakas:2011}
N.~A. Bakas and P.~J. Ioannou.
\newblock Structural stability theory of two-dimensional fluid flow under
  stochastic forcing.
\newblock \emph{J. Fluid Mech.} 682 (2011), 332--361.
\newblock
  \urlprefix\url{http://journals.cambridge.org/article_S002211201100228X}.

\bibitem[{Bakas and Ioannou(2013{\natexlab{a}})}]{bakas:2013}
N.~A. Bakas and P.~J. Ioannou.
\newblock Emergence of large scale structure in barotropic $\beta$-plane
  turbulence.
\newblock \emph{Phys. Rev. Lett.} 110 (2013{\natexlab{a}}), 224501.
\newblock
  \urlprefix\url{http://link.aps.org/doi/10.1103/PhysRevLett.110.224501}.

\bibitem[{Bakas and Ioannou(2013{\natexlab{b}})}]{bakas:2013b}
N.~A. Bakas and P.~J. Ioannou.
\newblock On the mechanism underlying the spontaneous emergence of barotropic
  zonal jets.
\newblock \emph{J. Atmos.\ Sci.} 70 (2013{\natexlab{b}}), 2251--2271.
\newblock
  \urlprefix\url{http://journals.ametsoc.org/doi/abs/10.1175/JAS-D-12-0102.1}.

\bibitem[{Bakas and Ioannou(2013{\natexlab{c}})}]{bakas:2013c}
N.~A. Bakas and P.~J. Ioannou.
\newblock A theory for the emergence of coherent structures in beta-plane
  turbulence.
\newblock \emph{arXiv:1303.6435}  (2013{\natexlab{c}}).
\newblock \urlprefix\url{http://arxiv.org/abs/1303.6435v2}.

\bibitem[{Biglari \emph{et~al.}(1990)Biglari, Diamond, and
  Terry}]{biglari:1990}
H.~Biglari, P.~H. Diamond, and P.~W. Terry.
\newblock Influence of sheared poloidal rotation on edge turbulence.
\newblock \emph{Phys. Fluids B} 2 (1990), 1--4.
\newblock
  \urlprefix\url{http://scitation.aip.org/content/aip/journal/pofb/2/1/10.1063/1.859529}.

\bibitem[{Bouchet \emph{et~al.}(2013)Bouchet, Nardini, and
  Tangarife}]{bouchet:2013}
F.~Bouchet, C.~Nardini, and T.~Tangarife.
\newblock Kinetic theory of jet dynamics in the stochastic barotropic and 2{D}
  {N}avier-{S}tokes equations.
\newblock \emph{J. Stat. Phys.} 153 (2013), 572--625.
\newblock \urlprefix\url{http://dx.doi.org/10.1007/s10955-013-0828-3}.

\bibitem[{Bowman(1996)}]{bowman:1996}
J.~C. Bowman.
\newblock On inertial-range scaling laws.
\newblock \emph{J. Fluid Mech.} 306 (1996), 167--181.
\newblock
  \urlprefix\url{http://journals.cambridge.org/article_S0022112096001279}.

\bibitem[{Bowman and Krommes(1997)}]{bowman:1997}
J.~C. Bowman and J.~A. Krommes.
\newblock The realizable {M}arkovian closure and realizable test-field model.
  {II}. {A}pplication to anisotropic drift-wave dynamics.
\newblock \emph{Phys. Plasmas} 4 (1997), 3895--3909.
\newblock \urlprefix\url{http://link.aip.org/link/?PHP/4/3895/1}.

\bibitem[{Bowman \emph{et~al.}(1993)Bowman, Krommes, and
  Ottaviani}]{bowman:1993}
J.~C. Bowman, J.~A. Krommes, and M.~Ottaviani.
\newblock The realizable {M}arkovian closure. {I}. {G}eneral theory, with
  application to three-wave dynamics.
\newblock \emph{Phys. Fluids B} 5 (1993), 3558--3589.
\newblock \urlprefix\url{http://link.aip.org/link/?PFB/5/3558/1}.

\bibitem[{Boyd(2001)}]{boyd:2001}
J.~P. Boyd.
\newblock \emph{{Chebyshev and Fourier} Spectral Methods}.
\newblock Courier Dover Publications, 2001.

\bibitem[{Burrell(1999)}]{burrell:1999}
K.~H. Burrell.
\newblock Tests of causality: Experimental evidence that sheared $\mathbf{E}
  \times \mathbf{B}$ flow alters turbulence and transport in tokamaks.
\newblock \emph{Phys. Plasmas} 6 (1999), 4418--4435.
\newblock
  \urlprefix\url{http://scitation.aip.org/content/aip/journal/pop/6/12/10.1063/1.873728}.

\bibitem[{Busse(1967)}]{busse:1967}
F.~Busse.
\newblock The stability of finite amplitude cellular convection and its
  relation to an extremum principle.
\newblock \emph{J. Fluid Mech.} 30 (1967), 625--649.
\newblock
  \urlprefix\url{http://journals.cambridge.org/action/displayAbstract?fromPage=online&aid=382011}.

\bibitem[{Busse and Clever(1979)}]{busse:1979}
F.~Busse and R.~Clever.
\newblock Instabilities of convection rolls in a fluid of moderate {Prandtl}
  number.
\newblock \emph{J. Fluid Mech.} 91 (1979), 319--335.
\newblock
  \urlprefix\url{http://journals.cambridge.org/action/displayAbstract?fromPage=online&aid=374359}.

\bibitem[{Busse(1978)}]{busse:1978}
F.~H. Busse.
\newblock Non-linear properties of thermal convection.
\newblock \emph{Rep. Prog. Phys.} 41 (1978), 1929.
\newblock \urlprefix\url{http://stacks.iop.org/0034-4885/41/i=12/a=003}.

\bibitem[{Carnevale \emph{et~al.}(1981)Carnevale, Frisch, and
  Salmon}]{carnevale:1981}
G.~F. Carnevale, U.~Frisch, and R.~Salmon.
\newblock H theorems in statistical fluid dynamics.
\newblock \emph{J. Phys. A} 14 (1981), 1701.
\newblock \urlprefix\url{http://stacks.iop.org/0305-4470/14/i=7/a=026}.

\bibitem[{Carnevale and Martin(1982)}]{carnevale:1982}
G.~F. Carnevale and P.~C. Martin.
\newblock Field theoretical techniques in statistical fluid dynamics: With
  application to nonlinear wave dynamics.
\newblock \emph{Geophysical \& Astrophysical Fluid Dynamics} 20 (1982),
  131--163.
\newblock
  \urlprefix\url{http://www.tandfonline.com/doi/abs/10.1080/03091928208209002}.

\bibitem[{Carter and Maggs(2009)}]{carter:2009}
T.~A. Carter and J.~E. Maggs.
\newblock Modifications of turbulence and turbulent transport associated with a
  bias-induced confinement transition in the large plasma device.
\newblock \emph{Phys. Plasmas} 16 (2009), 012304.
\newblock
  \urlprefix\url{http://scitation.aip.org/content/aip/journal/pop/16/1/10.1063/1.3059410}.

\bibitem[{Clever and Busse(1974)}]{clever:1974}
R.~M. Clever and F.~H. Busse.
\newblock Transition to time-dependent convection.
\newblock \emph{J. Fluid Mech.} 65 (1974), 625--645.
\newblock \urlprefix\url{http://dx.doi.org/10.1017/S0022112074001571}.

\bibitem[{Colchin \emph{et~al.}(2002)Colchin, Schaffer, Carreras, McKee,
  Maingi, Carlstrom, Rudakov, Greenfield, Rhodes, Doyle, Brooks, and
  Austin}]{colchin:2002}
R.~J. Colchin, M.~J. Schaffer, B.~A. Carreras, G.~R. McKee, R.~Maingi, T.~N.
  Carlstrom, D.~L. Rudakov, C.~M. Greenfield, T.~L. Rhodes, E.~J. Doyle, N.~H.
  Brooks, and M.~E. Austin.
\newblock Slow {L-H} transitions in {DIII-D} plasmas.
\newblock \emph{Phys. Rev. Lett.} 88 (2002), 255002.
\newblock
  \urlprefix\url{http://link.aps.org/doi/10.1103/PhysRevLett.88.255002}.

\bibitem[{Connaughton \emph{et~al.}(2011)Connaughton, Nazarenko, and
  Quinn}]{connaughton:2011}
C.~Connaughton, S.~Nazarenko, and B.~Quinn.
\newblock Feedback of zonal flows on wave turbulence driven by small-scale
  instability in the {Charney-Hasegawa-Mima} model.
\newblock \emph{EPL} 96 (2011), 25001.
\newblock \urlprefix\url{http://stacks.iop.org/0295-5075/96/i=2/a=25001}.

\bibitem[{Connaughton \emph{et~al.}(2010)Connaughton, Nadiga, Nazarenko, and
  Quinn}]{connaughton:2010}
C.~P. Connaughton, B.~T. Nadiga, S.~V. Nazarenko, and B.~E. Quinn.
\newblock Modulational instability of {Rossby} and drift waves and generation
  of zonal jets.
\newblock \emph{J. Fluid Mech.} 654 (2010), 207--231.

\bibitem[{Constantinou \emph{et~al.}(2013)Constantinou, Farrell, and
  Ioannou}]{constantinou:2013}
N.~C. Constantinou, B.~F. Farrell, and P.~J. Ioannou.
\newblock Emergence and equilibration of jets in beta-plane turbulence:
  applications of stochastic structural stability theory.
\newblock \emph{J. Atmos. Sci.}  (2013), --.
\newblock \urlprefix\url{http://dx.doi.org/10.1175/JAS-D-13-076.1}.

\bibitem[{Cox and Matthews(2002)}]{cox:2002}
S.~Cox and P.~Matthews.
\newblock Exponential time differencing for stiff systems.
\newblock \emph{J. Comput. Phys.} 176 (2002), 430--455.
\newblock
  \urlprefix\url{http://www.sciencedirect.com/science/article/pii/S0021999102969950}.

\bibitem[{Cross and Greenside(2009)}]{cross:2009}
M.~Cross and H.~Greenside.
\newblock \emph{Pattern Formation and Dynamics in Nonequilibrium Systems}.
\newblock Cambridge University Press, 2009.

\bibitem[{Cross and Hohenberg(1993)}]{cross:1993}
M.~C. Cross and P.~C. Hohenberg.
\newblock Pattern formation outside of equilibrium.
\newblock \emph{Rev. Mod. Phys.} 65 (1993), 851--1112.
\newblock \urlprefix\url{http://link.aps.org/doi/10.1103/RevModPhys.65.851}.

\bibitem[{Crowley(1994)}]{crowley:1994}
T.~P. Crowley.
\newblock Rensselaer heavy ion beam probe diagnostic methods and techniques.
\newblock \emph{IEEE Trans. Plasma Sci.} 22 (1994), 291--309.

\bibitem[{Danilov and Gurarie(2004)}]{danilov:2004}
S.~Danilov and D.~Gurarie.
\newblock Scaling, spectra and zonal jets in beta-plane turbulence.
\newblock \emph{Phys. Fluids} 16 (2004), 2592--2603.
\newblock \urlprefix\url{http://link.aip.org/link/?PHF/16/2592/1}.

\bibitem[{Davidson(2004)}]{davidson:book}
P.~A. Davidson.
\newblock \emph{Turbulence: An Introduction for Scientists and Engineers}.
\newblock Oxford University Press, 2004.

\bibitem[{Diamond \emph{et~al.}(2005)Diamond, Itoh, Itoh, and
  Hahm}]{diamond:2005}
P.~H. Diamond, S.-I. Itoh, K.~Itoh, and T.~S. Hahm.
\newblock Zonal flows in plasma -- a review.
\newblock \emph{Plasma Phys. Control. Fusion} 47 (2005), R35.
\newblock \urlprefix\url{http://stacks.iop.org/0741-3335/47/i=5/a=R01}.

\bibitem[{Diamond \emph{et~al.}(1994)Diamond, Liang, Carreras, and
  Terry}]{diamond:1994}
P.~H. Diamond, Y.-M. Liang, B.~A. Carreras, and P.~W. Terry.
\newblock Self-regulating shear flow turbulence: A paradigm for the \textit{L}
  to \textit{H} transition.
\newblock \emph{Phys. Rev. Lett.} 72 (1994), 2565--2568.
\newblock \urlprefix\url{http://link.aps.org/doi/10.1103/PhysRevLett.72.2565}.

\bibitem[{Dimits \emph{et~al.}(2000)Dimits, Bateman, Beer, Cohen, Dorland,
  Hammett, Kim, Kinsey, Kotschenreuther, Kritz, Lao, Mandrekas, Nevins, Parker,
  Redd, Shumaker, Sydora, and Weiland}]{dimits:2000}
A.~M. Dimits, G.~Bateman, M.~A. Beer, B.~I. Cohen, W.~Dorland, G.~W. Hammett,
  C.~Kim, J.~E. Kinsey, M.~Kotschenreuther, A.~H. Kritz, L.~L. Lao,
  J.~Mandrekas, W.~M. Nevins, S.~E. Parker, A.~J. Redd, D.~E. Shumaker,
  R.~Sydora, and J.~Weiland.
\newblock Comparisons and physics basis of tokamak transport models and
  turbulence simulations.
\newblock \emph{Phys. Plasmas} 7 (2000), 969--983.
\newblock
  \urlprefix\url{http://scitation.aip.org/content/aip/journal/pop/7/3/10.1063/1.873896}.

\bibitem[{Dritschel and McIntyre(2008)}]{dritschel:2008}
D.~G. Dritschel and M.~E. McIntyre.
\newblock Multiple jets as {PV} staircases: The {P}hillips effect and the
  resilience of eddy-transport barriers.
\newblock \emph{J. Atmos. Sci.} 65 (2008), 855--874.
\newblock \urlprefix\url{http://dx.doi.org/10.1175/2007JAS2227.1}.

\bibitem[{Drummond and Pines(1962)}]{drummond:1962}
W.~E. Drummond and D.~Pines.
\newblock Non-linear stability of plasma oscillations.
\newblock In \emph{Proceedings of the Conference on Plasma Physics and
  Controlled Nuclear Fusion Research (Salzburg, 1961) [Nucl.\ Fusion Suppl.\
  Pt.~3]}, 1049--1057. International Atomic Energy Agency, Vienna, 1962.

\bibitem[{Dubrulle and Nazarenko(1997)}]{dubrulle:1997}
B.~Dubrulle and S.~Nazarenko.
\newblock Interaction of turbulence and large-scale vortices in incompressible
  {2D} fluids.
\newblock \emph{Physica D} 110 (1997), 123.
\newblock
  \urlprefix\url{http://www.sciencedirect.com/science/article/pii/S0167278997001206}.

\bibitem[{Dyachenko \emph{et~al.}(1992)Dyachenko, Nazarenko, and
  Zakharov}]{dyachenko:1992}
A.~Dyachenko, S.~Nazarenko, and V.~Zakharov.
\newblock Wave-vortex dynamics in drift and $\b$-plane turbulence.
\newblock \emph{Physics Letters A} 165 (1992), 330--334.
\newblock
  \urlprefix\url{http://www.sciencedirect.com/science/article/pii/037596019290503E}.

\bibitem[{Eckhaus(1965)}]{eckhaus:1965}
W.~Eckhaus.
\newblock \emph{Studies in non-linear stability theory}.
\newblock Springer, 1965.

\bibitem[{Estrada()}]{estrada:2014}
T.~Estrada.
\newblock Zonal flows in magnetically confined plasmas: Experiments,
  \textit{Zonal Jets}.
\newblock Cambridge University Press. Edited by Boris Galperin and Peter Read.
  To be published 2015.

\bibitem[{Estrada \emph{et~al.}(2012)Estrada, Ascas\'{i}­bar, Blanco, Cappa,
  Diamond, Happel, Hidalgo, Liniers, van Milligen, Pastor, Tafalla, and the
  TJ-II~Team}]{estrada:2012}
T.~Estrada, E.~Ascas\'{i}­bar, E.~Blanco, A.~Cappa, P.~H. Diamond, T.~Happel,
  C.~Hidalgo, M.~Liniers, B.~P. van Milligen, I.~Pastor, D.~Tafalla, and the
  TJ-II~Team.
\newblock Spatial, temporal and spectral structure of the turbulence-flow
  interaction at the {L--H} transition.
\newblock \emph{Plasma Phys. Control. Fusion} 54 (2012), 124024.
\newblock \urlprefix\url{http://stacks.iop.org/0741-3335/54/i=12/a=124024}.

\bibitem[{Estrada \emph{et~al.}(2009)Estrada, Happel, Eliseev, L\'{o}pez-Bruna,
  Ascas\'{i}­bar, Blanco, Cupido, Fontdecaba, Hidalgo, Jim\'{e}nez-G\'{o}mez,
  Krupnik, Liniers, Manso, McCarthy, Medina, Melnikov, van Milligen, Ochando,
  Pastor, Pedrosa, Tabar\'{e}s, Tafalla, and Team}]{estrada:2009}
T.~Estrada, T.~Happel, L.~Eliseev, D.~L\'{o}pez-Bruna, E.~Ascas\'{i}­bar,
  E.~Blanco, L.~Cupido, J.~M. Fontdecaba, C.~Hidalgo, R.~Jim\'{e}nez-G\'{o}mez,
  L.~Krupnik, M.~Liniers, M.~E. Manso, K.~J. McCarthy, F.~Medina, A.~Melnikov,
  B.~van Milligen, M.~A. Ochando, I.~Pastor, M.~A. Pedrosa, F.~L. Tabar\'{e}s,
  D.~Tafalla, and T.-I. Team.
\newblock Sheared flows and transition to improved confinement regime in the
  {TJ-II} stellarator.
\newblock \emph{Plasma Phys. Control. Fusion} 51 (2009), 124015.
\newblock \urlprefix\url{http://stacks.iop.org/0741-3335/51/i=12/a=124015}.

\bibitem[{Estrada \emph{et~al.}(2010)Estrada, Happel, Hidalgo, Ascas\'{i}bar,
  and Blanco}]{estrada:2010}
T.~Estrada, T.~Happel, C.~Hidalgo, E.~Ascas\'{i}bar, and E.~Blanco.
\newblock Experimental observation of coupling between turbulence and sheared
  flows during {L-H} transitions in a toroidal plasma.
\newblock \emph{EPL} 92 (2010), 35001.
\newblock \urlprefix\url{http://stacks.iop.org/0295-5075/92/i=3/a=35001}.

\bibitem[{Estrada \emph{et~al.}(2011)Estrada, Hidalgo, Happel, and
  Diamond}]{estrada:2011}
T.~Estrada, C.~Hidalgo, T.~Happel, and P.~H. Diamond.
\newblock Spatiotemporal structure of the interaction between turbulence and
  flows at the {L-H} transition in a toroidal plasma.
\newblock \emph{Phys. Rev. Lett.} 107 (2011), 245004.
\newblock
  \urlprefix\url{http://link.aps.org/doi/10.1103/PhysRevLett.107.245004}.

\bibitem[{Farrell and Ioannou(2003)}]{farrell:2003}
B.~F. Farrell and P.~J. Ioannou.
\newblock Structural stability of turbulent jets.
\newblock \emph{J. Atmos. Sci.} 60 (2003), 2101--2118.
\newblock
  \urlprefix\url{http://dx.doi.org/10.1175/1520-0469(2003)060<2101:SSOTJ>2.0.CO;2}.

\bibitem[{Farrell and Ioannou(2007)}]{farrell:2007}
B.~F. Farrell and P.~J. Ioannou.
\newblock Structure and spacing of jets in barotropic turbulence.
\newblock \emph{J. Atmos. Sci.} 64 (2007), 3652--3665.
\newblock \urlprefix\url{http://dx.doi.org/10.1175/JAS4016.1}.

\bibitem[{Fj{\o}rtoft(1953)}]{fjortoft:1953}
R.~Fj{\o}rtoft.
\newblock On the changes in the spectral distribution of kinetic energy for
  twodimensional, nondivergent flow.
\newblock \emph{Tellus} 5 (1953), 225--230.
\newblock \urlprefix\url{http://dx.doi.org/10.1111/j.2153-3490.1953.tb01051.x}.

\bibitem[{Fonck \emph{et~al.}(1990)Fonck, Duperrex, and Paul}]{fonck:1990}
R.~J. Fonck, P.~A. Duperrex, and S.~F. Paul.
\newblock Plasma fluctuation measurements in tokamaks using beam-plasma
  interactions.
\newblock \emph{Rev. Sci. Instrum.} 61 (1990), 3487--3495.
\newblock
  \urlprefix\url{http://scitation.aip.org/content/aip/journal/rsi/61/11/10.1063/1.1141556}.

\bibitem[{Frederiksen(1999)}]{frederiksen:1999}
J.~S. Frederiksen.
\newblock Subgrid-scale parameterizations of eddy-topographic force, eddy
  viscosity, and stochastic backscatter for flow over topography.
\newblock \emph{J. Atmos. Sci.} 56 (1999), 1481--1494.
\newblock
  \urlprefix\url{http://dx.doi.org/10.1175/1520-0469(1999)056<1481:SSPOET>2.0.CO;2}.

\bibitem[{Frisch(1995)}]{frisch:1995}
U.~Frisch.
\newblock \emph{Turbulence}.
\newblock Cambridge University Press, Cambridge, 1995.

\bibitem[{Fujisawa(2009)}]{fujisawa:2009}
A.~Fujisawa.
\newblock A review of zonal flow experiments.
\newblock \emph{Nucl. Fusion} 49 (2009), 013001.
\newblock \urlprefix\url{http://stacks.iop.org/0029-5515/49/i=1/a=013001}.

\bibitem[{Fujisawa \emph{et~al.}(2004)Fujisawa, Itoh, Iguchi, Matsuoka,
  Okamura, Shimizu, Minami, Yoshimura, Nagaoka, Takahashi, Kojima, Nakano,
  Ohsima, Nishimura, Isobe, Suzuki, Akiyama, Ida, Toi, Itoh, and
  Diamond}]{fujisawa:2004}
A.~Fujisawa, K.~Itoh, H.~Iguchi, K.~Matsuoka, S.~Okamura, A.~Shimizu,
  T.~Minami, Y.~Yoshimura, K.~Nagaoka, C.~Takahashi, M.~Kojima, H.~Nakano,
  S.~Ohsima, S.~Nishimura, M.~Isobe, C.~Suzuki, T.~Akiyama, K.~Ida, K.~Toi,
  S.-I. Itoh, and P.~H. Diamond.
\newblock Identification of zonal flows in a toroidal plasma.
\newblock \emph{Phys. Rev. Lett.} 93 (2004), 165002.

\bibitem[{Gallagher \emph{et~al.}(2012)Gallagher, Hnat, Connaughton, Nazarenko,
  and Rowlands}]{gallagher:2012}
S.~Gallagher, B.~Hnat, C.~Connaughton, S.~Nazarenko, and G.~Rowlands.
\newblock The modulational instability in the extended {Hasegawa-Mima} equation
  with a finite {Larmor} radius.
\newblock \emph{Phys. Plasmas} 19 (2012), 122115.

\bibitem[{Galperin \emph{et~al.}(2010)Galperin, Sukoriansky, and
  Dikovskaya}]{galperin:2010}
B.~Galperin, S.~Sukoriansky, and N.~Dikovskaya.
\newblock Geophysical flows with anisotropic turbulence and dispersive waves:
  flows with a $\beta$-effect.
\newblock \emph{Ocean Dyn.} 60 (2010), 427--441.
\newblock \urlprefix\url{http://dx.doi.org/10.1007/s10236-010-0278-2}.

\bibitem[{Gill(1974)}]{gill:1974}
A.~Gill.
\newblock The stability of planetary waves on an infinite beta--plane.
\newblock \emph{Geophys. Fluid Dyn.} 6 (1974), 29--47.

\bibitem[{Golubitsky \emph{et~al.}(1988)Golubitsky, Stewart, and
  Schaeffer}]{golubitsky:book}
M.~Golubitsky, I.~Stewart, and D.~Schaeffer.
\newblock \emph{Singularities and Groups in Bifurcation Theory, Volume II}.
\newblock Applied Mathematical Sciences. Springer, 1988.

\bibitem[{Gupta \emph{et~al.}(2006)Gupta, Fonck, McKee, Schlossberg, and
  Shafer}]{gupta:2006}
D.~K. Gupta, R.~J. Fonck, G.~R. McKee, D.~J. Schlossberg, and M.~W. Shafer.
\newblock Detection of zero-mean-frequency zonal flows in the core of a
  high-temperature tokamak plasma.
\newblock \emph{Phys. Rev. Lett.} 97 (2006), 125002.

\bibitem[{Hall \emph{et~al.}(2002)Hall, Lisak, Anderson, Fedele, and
  Semenov}]{hall:2002}
B.~Hall, M.~Lisak, D.~Anderson, R.~Fedele, and V.~E. Semenov.
\newblock Statistical theory for incoherent light propagation in nonlinear
  media.
\newblock \emph{Phys. Rev. E} 65 (2002), 035602.
\newblock \urlprefix\url{http://link.aps.org/doi/10.1103/PhysRevE.65.035602}.

\bibitem[{Hammett \emph{et~al.}(1993)Hammett, Beer, Dorland, Cowley, and
  Smith}]{hammett:1993}
G.~W. Hammett, M.~A. Beer, W.~Dorland, S.~C. Cowley, and S.~A. Smith.
\newblock Developments in the gyrofluid approach to tokamak turbulence
  simulations.
\newblock \emph{Plasma Phys. Control. Fusion} 35 (1993), 973.
\newblock \urlprefix\url{http://stacks.iop.org/0741-3335/35/i=8/a=006}.

\bibitem[{Hasegawa and Mima(1978)}]{hasegawa:1978}
A.~Hasegawa and K.~Mima.
\newblock Pseudo-three-dimensional turbulence in magnetized nonuniform plasma.
\newblock \emph{Phys. Fluids} 21 (1978), 87--92.
\newblock \urlprefix\url{http://link.aip.org/link/?PFL/21/87/1}.

\bibitem[{Hasegawa and Wakatani(1983)}]{hasegawa:1983}
A.~Hasegawa and M.~Wakatani.
\newblock Plasma edge turbulence.
\newblock \emph{Phys. Rev. Lett.} 50 (1983), 682--686.
\newblock \urlprefix\url{http://link.aps.org/doi/10.1103/PhysRevLett.50.682}.

\bibitem[{Hasegawa and Wakatani(1987)}]{hasegawa:1987}
A.~Hasegawa and M.~Wakatani.
\newblock Self-organization of electrostatic turbulence in a cylindrical
  plasma.
\newblock \emph{Phys. Rev. Lett.} 59 (1987), 1581--1584.
\newblock \urlprefix\url{http://link.aps.org/doi/10.1103/PhysRevLett.59.1581}.

\bibitem[{Hatch()}]{pc:hatch:2013}
D.~R. Hatch.
\newblock Private communication (2013).

\bibitem[{Hatch \emph{et~al.}(2011{\natexlab{a}})Hatch, Terry, Jenko, Merz, and
  Nevins}]{hatch:2011a}
D.~R. Hatch, P.~W. Terry, F.~Jenko, F.~Merz, and W.~M. Nevins.
\newblock Saturation of gyrokinetic turbulence through damped eigenmodes.
\newblock \emph{Phys. Rev. Lett.} 106 (2011{\natexlab{a}}), 115003.
\newblock
  \urlprefix\url{http://link.aps.org/doi/10.1103/PhysRevLett.106.115003}.

\bibitem[{Hatch \emph{et~al.}(2011{\natexlab{b}})Hatch, Terry, Jenko, Merz,
  Pueschel, Nevins, and Wang}]{hatch:2011b}
D.~R. Hatch, P.~W. Terry, F.~Jenko, F.~Merz, M.~J. Pueschel, W.~M. Nevins, and
  E.~Wang.
\newblock Role of subdominant stable modes in plasma microturbulence.
\newblock \emph{Phys. Plasmas} 18 (2011{\natexlab{b}}), 055706.
\newblock
  \urlprefix\url{http://scitation.aip.org/content/aip/journal/pop/18/5/10.1063/1.3563536}.

\bibitem[{Herring(1963)}]{herring:1963}
J.~R. Herring.
\newblock Investigation of problems in thermal convection.
\newblock \emph{J. Atmos. Sci.} 20 (1963), 325--338.
\newblock
  \urlprefix\url{http://dx.doi.org/10.1175/1520-0469(1963)020<0325:IOPITC>2.0.CO;2}.

\bibitem[{Hirsch \emph{et~al.}(2001)Hirsch, Holzhauer, Baldzuhn, and
  Kurzan}]{hirsch:2001}
M.~Hirsch, E.~Holzhauer, J.~Baldzuhn, and B.~Kurzan.
\newblock Doppler reflectometry for the investigation of propagating density
  perturbations.
\newblock \emph{Rev. Sci. Instrum.} 72 (2001), 324--327.
\newblock
  \urlprefix\url{http://scitation.aip.org/content/aip/journal/rsi/72/1/10.1063/1.1308998}.

\bibitem[{Holland \emph{et~al.}(2006)Holland, Yu, James, Nishijima, Shimada,
  Taheri, and Tynan}]{holland:2006}
C.~Holland, J.~H. Yu, A.~James, D.~Nishijima, M.~Shimada, N.~Taheri, and G.~R.
  Tynan.
\newblock Observation of turbulent-driven shear flow in a cylindrical
  laboratory plasma device.
\newblock \emph{Phys. Rev. Lett.} 96 (2006), 195002.
\newblock
  \urlprefix\url{http://link.aps.org/doi/10.1103/PhysRevLett.96.195002}.

\bibitem[{Horton and Hasegawa(1994)}]{horton:1994}
W.~Horton and A.~Hasegawa.
\newblock Quasi-two-dimensional dynamics of plasmas and fluids.
\newblock \emph{Chaos} 4 (1994), 227--251.
\newblock \urlprefix\url{http://link.aip.org/link/?CHA/4/227/1}.

\bibitem[{Hoyle(2006)}]{hoyle:2006}
R.~Hoyle.
\newblock \emph{Pattern Formation: An Introduction to Methods}.
\newblock Cambridge University Press, 2006.

\bibitem[{Huang and Robinson(1998)}]{huang:1998}
H.-P. Huang and W.~A. Robinson.
\newblock Two-dimensional turbulence and persistent zonal jets in a global
  barotropic model.
\newblock \emph{J. Atmos. Sci.} 55 (1998), 611--632.
\newblock
  \urlprefix\url{http://dx.doi.org/10.1175/1520-0469(1998)055<0611:TDTAPZ>2.0.CO;2}.

\bibitem[{Hutchinson(2005)}]{hutchinson:book}
I.~H. Hutchinson.
\newblock \emph{Principles of Plasma Diagnostics}.
\newblock Cambridge Univ Press, 2005.

\bibitem[{Ido \emph{et~al.}(2002)Ido, Kamiya, Miura, Hamada, Nishizawa, and
  Kawasumi}]{ido:2002}
T.~Ido, K.~Kamiya, Y.~Miura, Y.~Hamada, A.~Nishizawa, and Y.~Kawasumi.
\newblock Observation of the fast potential change at {L-H} transition by a
  heavy-ion-beam probe on {JFT-2M}.
\newblock \emph{Phys. Rev. Lett.} 88 (2002), 055006.
\newblock
  \urlprefix\url{http://link.aps.org/doi/10.1103/PhysRevLett.88.055006}.

\bibitem[{Itoh \emph{et~al.}(2005)Itoh, Hallatschek, and Itoh}]{itoh:2005}
K.~Itoh, K.~Hallatschek, and S.-I. Itoh.
\newblock Excitation of geodesic acoustic mode in toroidal plasmas.
\newblock \emph{Plasma Phys. Control. Fusion} 47 (2005), 451.
\newblock \urlprefix\url{http://stacks.iop.org/0741-3335/47/i=3/a=004}.

\bibitem[{Johansen \emph{et~al.}(2009)Johansen, Youdin, and
  Klahr}]{johansen:2009}
A.~Johansen, A.~Youdin, and H.~Klahr.
\newblock Zonal flows and long-lived axisymmetric pressure bumps in
  magnetorotational turbulence.
\newblock \emph{Astrophys. J.} 697 (2009), 1269.
\newblock \urlprefix\url{http://stacks.iop.org/0004-637X/697/i=2/a=1269}.

\bibitem[{Kassam and Trefethen(2005)}]{kassam:2005}
A.~Kassam and L.~N. Trefethen.
\newblock Fourth-order time stepping for stiff {PDEs}.
\newblock \emph{SIAM J. Sci. Comput.} 26 (2005), 1214--1233.
\newblock
  \urlprefix\url{http://epubs.siam.org/doi/abs/10.1137/S1064827502410633}.

\bibitem[{Kelley(2003)}]{kelley:2003}
C.~T. Kelley.
\newblock \emph{Solving Nonlinear Equations with Newton's Method}.
\newblock Society for Industrial and Applied Mathematics, 2003.

\bibitem[{Kim and Diamond(2003)}]{kim:2003}
E.-j. Kim and P.~H. Diamond.
\newblock Zonal flows and transient dynamics of the {L--H} transition.
\newblock \emph{Phys. Rev. Lett.} 90 (2003), 185006.
\newblock
  \urlprefix\url{http://link.aps.org/doi/10.1103/PhysRevLett.90.185006}.

\bibitem[{Kolesnikov and Krommes(2005{\natexlab{a}})}]{kolesnikov:2005b}
R.~A. Kolesnikov and J.~A. Krommes.
\newblock Bifurcation theory of the transition to collisionless
  ion-temperature-gradient-driven plasma turbulence.
\newblock \emph{Phys. Plasmas} 12 (2005{\natexlab{a}}), 122302.
\newblock
  \urlprefix\url{http://scitation.aip.org/content/aip/journal/pop/12/12/10.1063/1.2116887}.

\bibitem[{Kolesnikov and Krommes(2005{\natexlab{b}})}]{kolesnikov:2005}
R.~A. Kolesnikov and J.~A. Krommes.
\newblock Transition to collisionless ion-temperature-gradient-driven plasma
  turbulence: A dynamical systems approach.
\newblock \emph{Phys. Rev. Lett.} 94 (2005{\natexlab{b}}), 235002.
\newblock
  \urlprefix\url{http://link.aps.org/doi/10.1103/PhysRevLett.94.235002}.

\bibitem[{Kraichnan(1959)}]{kraichnan:1959}
R.~H. Kraichnan.
\newblock The structure of isotropic turbulence at very high {R}eynolds
  numbers.
\newblock \emph{J. Fluid Mech.} 5 (1959), 497--543.
\newblock
  \urlprefix\url{http://journals.cambridge.org/article_S0022112059000362}.

\bibitem[{Kraichnan(1964{\natexlab{a}})}]{kraichnan:1964b}
R.~H. Kraichnan.
\newblock Diagonalizing approximation for inhomogeneous turbulence.
\newblock \emph{Phys. Fluids} 7 (1964{\natexlab{a}}), 1169--1177.
\newblock \urlprefix\url{http://link.aip.org/link/?PFL/7/1169/1}.

\bibitem[{Kraichnan(1964{\natexlab{b}})}]{kraichnan:1964a}
R.~H. Kraichnan.
\newblock Direct-interaction approximation for shear and thermally driven
  turbulence.
\newblock \emph{Phys. Fluids} 7 (1964{\natexlab{b}}), 1048--1062.
\newblock \urlprefix\url{http://link.aip.org/link/?PFL/7/1048/1}.

\bibitem[{Kraichnan(1967)}]{kraichnan:1967}
R.~H. Kraichnan.
\newblock Inertial ranges in two-dimensional turbulence.
\newblock \emph{Phys. Fluids} 10 (1967), 1417--1423.
\newblock
  \urlprefix\url{http://scitation.aip.org/content/aip/journal/pof1/10/7/10.1063/1.1762301}.

\bibitem[{Kraichnan(1971)}]{kraichnan:1971}
R.~H. Kraichnan.
\newblock Inertial-range transfer in two- and three-dimensional turbulence.
\newblock \emph{J. Fluid Mech.} 47 (1971), 525--535.
\newblock
  \urlprefix\url{http://journals.cambridge.org/article_S0022112071001216}.

\bibitem[{Kraichnan(1972)}]{kraichnan:1972}
R.~H. Kraichnan.
\newblock Test-field model for inhomogeneous turbulence.
\newblock \emph{J. Fluid Mech.} 56 (1972), 287--304.
\newblock
  \urlprefix\url{http://journals.cambridge.org/action/displayAbstract?fromPage=online&aid=372667}.

\bibitem[{Kraichnan and Spiegel(1962)}]{kraichnan:1962}
R.~H. Kraichnan and E.~A. Spiegel.
\newblock Model for energy transfer in isotropic turbulence.
\newblock \emph{Phys. Fluids} 5 (1962), 583--588.
\newblock
  \urlprefix\url{http://scitation.aip.org/content/aip/journal/pof1/5/5/10.1063/1.1706660}.

\bibitem[{Krall and Trivelpiece(1973)}]{krall:1973}
N.~Krall and A.~Trivelpiece.
\newblock \emph{Principles of plasma physics}.
\newblock Number v. 0-911351 in International series in pure and applied
  physics. McGraw-Hill, 1973.
\newblock \urlprefix\url{http://books.google.com/books?id=b0BRAAAAMAAJ}.

\bibitem[{Kramer and Zimmermann(1985)}]{kramer:1985}
L.~Kramer and W.~Zimmermann.
\newblock On the {E}ckhaus instability for spatially periodic patterns.
\newblock \emph{Physica D} 16 (1985), 221--232.
\newblock
  \urlprefix\url{http://www.sciencedirect.com/science/article/pii/0167278985900594}.

\bibitem[{Krommes(2002)}]{krommes:2002}
J.~A. Krommes.
\newblock Fundamental statistical descriptions of plasma turbulence in magnetic
  fields.
\newblock \emph{Phys. Rep.} 360 (2002), 1--352.
\newblock
  \urlprefix\url{http://www.sciencedirect.com/science/article/pii/S0370157301000667}.

\bibitem[{Krommes(2006)}]{krommes:australia}
J.~A. Krommes.
\newblock Analytical descriptions of plasma turbulence.
\newblock In \emph{Turbulence and Coherent Structures in Fluids, Plasmas and
  Nonlinear Media}. World Scientific, 2006.

\bibitem[{Krommes and Kim(2000)}]{krommes:2000}
J.~A. Krommes and C.-B. Kim.
\newblock Interactions of disparate scales in drift-wave turbulence.
\newblock \emph{Phys. Rev. E} 62 (2000), 8508--8539.
\newblock \urlprefix\url{http://link.aps.org/doi/10.1103/PhysRevE.62.8508}.

\bibitem[{Krommes and Parker()}]{krommes:book}
J.~A. Krommes and J.~B. Parker.
\newblock Genesis and maintenance of zonal jets: Turbulence and instabilities,
  \textit{Zonal Jets}.
\newblock Cambridge University Press. Edited by Boris Galperin and Peter Read.
  To be published 2015.

\bibitem[{Krommes and Smith(1987)}]{krommes:1987}
J.~A. Krommes and R.~A. Smith.
\newblock Rigorous upper bounds for transport due to passive advection by
  inhomogeneous turbulence.
\newblock \emph{Ann. Phys.} 177 (1987), 246--329.
\newblock
  \urlprefix\url{http://www.sciencedirect.com/science/article/pii/0003491687901229}.

\bibitem[{Kunz and Lesur(2013)}]{kunz:2013}
M.~W. Kunz and G.~Lesur.
\newblock Magnetic self-organization in {H}all-dominated magnetorotational
  turbulence.
\newblock \emph{Monthly Notices of the Royal Astronomical Society} 434 (2013),
  2295--2312.
\newblock
  \urlprefix\url{http://mnras.oxfordjournals.org/content/434/3/2295.abstract}.

\bibitem[{Lee(1952)}]{lee:1952}
T.~Lee.
\newblock On some statistical properties of hydrodynamical and
  magneto-hydrodynamical fields.
\newblock \emph{Q. Appl. Math.} 10 (1952), 69.

\bibitem[{Leith(1967)}]{leith:1967}
C.~E. Leith.
\newblock Diffusion approximation to inertial energy transfer in isotropic
  turbulence.
\newblock \emph{Phys. Fluids} 10 (1967), 1409--1416.
\newblock
  \urlprefix\url{http://scitation.aip.org/content/aip/journal/pof1/10/7/10.1063/1.1762300}.

\bibitem[{Lin \emph{et~al.}(1998)Lin, Hahm, Lee, Tang, and White}]{lin:1998}
Z.~Lin, T.~S. Hahm, W.~W. Lee, W.~M. Tang, and R.~B. White.
\newblock Turbulent transport reduction by zonal flows: Massively parallel
  simulations.
\newblock \emph{Science} 281 (1998), 1835--1837.
\newblock
  \urlprefix\url{http://www.sciencemag.org/content/281/5384/1835.abstract}.

\bibitem[{Lorenz(1972)}]{lorenz:1972}
E.~N. Lorenz.
\newblock Barotropic instability of {Rossby} wave motion.
\newblock \emph{J. Atmos. Sci.} 29 (1972), 258--265.

\bibitem[{Majda and Wang(2006)}]{majda:2006}
A.~Majda and X.~Wang.
\newblock \emph{Nonlinear dynamics and statistical theories for basic
  geophysical flows}.
\newblock Cambridge University Press, 2006.

\bibitem[{Makwana \emph{et~al.}(2014)Makwana, Terry, Pueschel, and
  Hatch}]{makwana:2014}
D.~Makwana, K.\, W.~Terry, P.\, J.~Pueschel, M.\, and R.~Hatch, D.\.
\newblock Subdominant modes in zonal-flow-regulated turbulence.
\newblock \emph{Phys. Rev. Lett.} 112 (2014), 095002.
\newblock
  \urlprefix\url{http://link.aps.org/doi/10.1103/PhysRevLett.112.095002}.

\bibitem[{Makwana \emph{et~al.}(2012)Makwana, Terry, and Kim}]{makwana:2012}
K.~D. Makwana, P.~W. Terry, and J.-H. Kim.
\newblock Role of stable modes in zonal flow regulated turbulence.
\newblock \emph{Phys. Plasmas} 19 (2012), 062310.
\newblock
  \urlprefix\url{http://scitation.aip.org/content/aip/journal/pop/19/6/10.1063/1.4729906}.

\bibitem[{Makwana \emph{et~al.}(2011)Makwana, Terry, Kim, and
  Hatch}]{makwana:2011}
K.~D. Makwana, P.~W. Terry, J.-H. Kim, and D.~R. Hatch.
\newblock Damped eigenmode saturation in plasma fluid turbulence.
\newblock \emph{Phys. Plasmas} 18 (2011), 012302.
\newblock
  \urlprefix\url{http://scitation.aip.org/content/aip/journal/pop/18/1/10.1063/1.3530186}.

\bibitem[{Manfroi and Young(1999)}]{manfroi:1999}
A.~J. Manfroi and W.~R. Young.
\newblock Slow evolution of zonal jets on the beta plane.
\newblock \emph{J. Atmos. Sci.} 56 (1999), 784--800.
\newblock
  \urlprefix\url{http://dx.doi.org/10.1175/1520-0469(1999)056<0784:SEOZJO>2.0.CO;2}.

\bibitem[{Manin and Nazarenko(1994)}]{manin:1994}
D.~Y. Manin and S.~V. Nazarenko.
\newblock Nonlinear interaction of small-scale {Rossby} waves with an intense
  large-scale zonal flow.
\newblock \emph{Phys. Fluids} 6 (1994), 1158--1167.
\newblock
  \urlprefix\url{http://scitation.aip.org/content/aip/journal/pof2/6/3/10.1063/1.868286}.

\bibitem[{Marston \emph{et~al.}(2008)Marston, Conover, and
  Schneider}]{marston:2008}
J.~B. Marston, E.~Conover, and T.~Schneider.
\newblock Statistics of an unstable barotropic jet from a cumulant expansion.
\newblock \emph{J. Atmos. Sci.} 65 (2008), 1955--1966.
\newblock \urlprefix\url{http://dx.doi.org/10.1175/2007JAS2510.1}.

\bibitem[{McIntyre(2008)}]{mcintyre:2008}
M.~E. McIntyre.
\newblock Potential-vorticity inversion and the wave-turbulence jigsaw: some
  recent clarifications.
\newblock \emph{Adv. Geosci.} 15 (2008), 47--56.
\newblock \urlprefix\url{http://www.adv-geosci.net/15/47/2008/}.

\bibitem[{Mendon\c{c}a \emph{et~al.}(2014)Mendon\c{c}a, {a}o, and
  Smolyakov}]{mendonca:2014}
J.~T. Mendon\c{c}a, R.~M. O.~G. {a}o, and A.~I. Smolyakov.
\newblock Nonlinear evolution of a single coherent mode in a turbulent plasma.
\newblock \emph{Plasma Phys. Control. Fusion} 56 (2014), 055004.
\newblock \urlprefix\url{http://stacks.iop.org/0741-3335/56/i=5/a=055004}.

\bibitem[{Mendon\c{c}a and Benkadda(2012)}]{mendonca:2012}
J.~T. Mendon\c{c}a and S.~Benkadda.
\newblock Nonlinear instability saturation due to quasi-particle trapping in a
  turbulent plasma.
\newblock \emph{Phys. Plasmas} 19 (2012), 082316.
\newblock
  \urlprefix\url{http://scitation.aip.org/content/aip/journal/pop/19/8/10.1063/1.4747531}.

\bibitem[{Mendon\c{c}a and Hizanidis(2011)}]{mendonca:2011}
J.~T. Mendon\c{c}a and K.~Hizanidis.
\newblock Improved model of quasi-particle turbulence (with applications to
  {A}lfv\'{e}n and drift wave turbulence).
\newblock \emph{Phys. Plasmas} 18 (2011), 112306.
\newblock
  \urlprefix\url{http://scitation.aip.org/content/aip/journal/pop/18/11/10.1063/1.3656956}.

\bibitem[{Meyer \emph{et~al.}(2011)Meyer, Bock, Conway, Freethy, Gibson,
  Hiratsuka, Kirk, Michael, Morgan, Scannell, Naylor, Saarelma, Saveliev,
  Shevchenko, Suttrop, Temple, Vann, the MAST, and teams}]{meyer:2011}
H.~Meyer, M.~D. Bock, N.~Conway, S.~Freethy, K.~Gibson, J.~Hiratsuka, A.~Kirk,
  C.~Michael, T.~Morgan, R.~Scannell, G.~Naylor, S.~Saarelma, A.~Saveliev,
  V.~Shevchenko, W.~Suttrop, D.~Temple, R.~Vann, the MAST, and N.~teams.
\newblock {L--H transition and pedestal studies on MAST}.
\newblock \emph{Nucl. Fusion} 51 (2011), 113011.
\newblock \urlprefix\url{http://stacks.iop.org/0029-5515/51/i=11/a=113011}.

\bibitem[{Miki \emph{et~al.}(2012)Miki, Diamond, G\"{u}rcan, Tynan, Estrada,
  Schmitz, and Xu}]{miki:2012}
K.~Miki, P.~H. Diamond, O.~D. G\"{u}rcan, G.~R. Tynan, T.~Estrada, L.~Schmitz,
  and G.~S. Xu.
\newblock Spatio-temporal evolution of the {L}$\to${I}$\to${H} transition.
\newblock \emph{Phys. Plasmas} 19 (2012), 092306.
\newblock
  \urlprefix\url{http://scitation.aip.org/content/aip/journal/pop/19/9/10.1063/1.4753931}.

\bibitem[{Mikkelsen and Dorland(2008)}]{mikkelsen:2008}
D.~R. Mikkelsen and W.~Dorland.
\newblock Dimits shift in realistic gyrokinetic plasma-turbulence simulations.
\newblock \emph{Phys. Rev. Lett.} 101 (2008), 135003.
\newblock
  \urlprefix\url{http://link.aps.org/doi/10.1103/PhysRevLett.101.135003}.

\bibitem[{Moyer \emph{et~al.}(1995)Moyer, Burrell, Carlstrom, Coda, Conn,
  Doyle, Gohil, Groebner, Kim, Lehmer, Peebles, Porkolab, Rettig, Rhodes,
  Seraydarian, Stockdale, Thomas, Tynan, and Watkins}]{moyer:1995}
R.~A. Moyer, K.~H. Burrell, T.~N. Carlstrom, S.~Coda, R.~W. Conn, E.~J. Doyle,
  P.~Gohil, R.~J. Groebner, J.~Kim, R.~Lehmer, W.~A. Peebles, M.~Porkolab,
  C.~L. Rettig, T.~L. Rhodes, R.~P. Seraydarian, R.~Stockdale, D.~M. Thomas,
  G.~R. Tynan, and J.~G. Watkins.
\newblock Beyond paradigm: Turbulence, transport, and the origin of the radial
  electric field in low to high confinement mode transitions in the {DIII-D}
  tokamak.
\newblock \emph{Phys. Plasmas} 2 (1995), 2397--2407.
\newblock
  \urlprefix\url{http://scitation.aip.org/content/aip/journal/pop/2/6/10.1063/1.871263}.

\bibitem[{Nakata \emph{et~al.}(2012)Nakata, Watanabe, and Sugama}]{nakata:2012}
M.~Nakata, T.-H. Watanabe, and H.~Sugama.
\newblock Nonlinear entropy transfer via zonal flows in gyrokinetic plasma
  turbulence.
\newblock \emph{Phys. Plasmas} 19 (2012), 022303.
\newblock
  \urlprefix\url{http://scitation.aip.org/content/aip/journal/pop/19/2/10.1063/1.3675855}.

\bibitem[{Newell \emph{et~al.}(1990)Newell, Passot, and Souli}]{newell:1990}
A.~C. Newell, T.~Passot, and M.~Souli.
\newblock The phase diffusion and mean drift equations for convection at finite
  {R}ayleigh numbers in large containers.
\newblock \emph{J. Fluid Mech.} 220 (1990), 187--252.
\newblock \urlprefix\url{http://dx.doi.org/10.1017/S0022112090003238}.

\bibitem[{Nozawa and Yoden(1997)}]{nozawa:1997}
T.~Nozawa and S.~Yoden.
\newblock Formation of zonal band structure in forced two-dimensional
  turbulence on a rotating sphere.
\newblock \emph{Phys. Fluids} 9 (1997), 2081--2093.
\newblock \urlprefix\url{http://link.aip.org/link/?PHF/9/2081/1}.

\bibitem[{Numata \emph{et~al.}(2007)Numata, Ball, and Dewar}]{numata:2007}
R.~Numata, R.~Ball, and R.~L. Dewar.
\newblock Bifurcation in electrostatic resistive drift wave turbulence.
\newblock \emph{Phys. Plasmas} 14 (2007), 102312.
\newblock \urlprefix\url{http://link.aip.org/link/?PHP/14/102312/1}.

\bibitem[{Ogura(1962{\natexlab{a}})}]{ogura:1962a}
Y.~Ogura.
\newblock Energy transfer in a normally distributed and isotropic turbulent
  velocity field in two dimensions.
\newblock \emph{Phys.\ Fluids} 5 (1962{\natexlab{a}}), 395--401.
\newblock
  \urlprefix\url{http://scitation.aip.org/content/aip/journal/pof1/5/4/10.1063/1.1706631}.

\bibitem[{Ogura(1962{\natexlab{b}})}]{ogura:1962b}
Y.~Ogura.
\newblock Energy transfer in an isotropic turbulent flow.
\newblock \emph{J. Geophys.\ Res.} 67 (1962{\natexlab{b}}), 3143--3149.
\newblock
  \urlprefix\url{http://onlinelibrary.wiley.com/doi/10.1029/JZ067i008p03143/abstract}.

\bibitem[{O'Kane and Frederiksen(2004)}]{okane:2004}
T.~J. O'Kane and J.~S. Frederiksen.
\newblock The {QDIA} and regularized {QDIA} closures for inhomogeneous
  turbulence over topography.
\newblock \emph{J. Fluid Mech.} 504 (2004), 133--165.
\newblock
  \urlprefix\url{http://journals.cambridge.org/article_S0022112004007980}.

\bibitem[{Orszag(1969)}]{orszag:1969}
S.~A. Orszag.
\newblock Numerical methods for the simulation of turbulence.
\newblock \emph{Phys. Fluids} 12 (1969), II--250.
\newblock
  \urlprefix\url{http://scitation.aip.org/content/aip/journal/pof1/12/12/10.1063/1.1692445}.

\bibitem[{Orszag(1971)}]{orszag:1971:dealias}
S.~A. Orszag.
\newblock On the elimination of aliasing in finite-difference schemes by
  filtering high-wavenumber components.
\newblock \emph{J. Atmos. Sci.} 28 (1971), 1074--1074.
\newblock
  \urlprefix\url{http://dx.doi.org/10.1175/1520-0469(1971)028<1074:OTEOAI>2.0.CO;2}.

\bibitem[{Orszag(1977)}]{orszag:1977}
S.~A. Orszag.
\newblock Lectures on the statistical theory of turbulence.
\newblock In \emph{Fluid Dynamics}. Gordon and Breach, 1977.

\bibitem[{Parker and Krommes()}]{parker:book}
J.~B. Parker and J.~A. Krommes.
\newblock Zonal flow as pattern formation, \textit{Zonal Jets}.
\newblock Cambridge University Press. Edited by Boris Galperin and Peter Read.
  To be published 2015.

\bibitem[{Parker and Krommes(2013)}]{parker:2013}
J.~B. Parker and J.~A. Krommes.
\newblock Zonal flow as pattern formation.
\newblock \emph{Phys. Plasmas} 20 (2013), 100703.
\newblock
  \urlprefix\url{http://scitation.aip.org/content/aip/journal/pop/20/10/10.1063/1.4828717}.

\bibitem[{Parker and Krommes(2014)}]{parker:2014}
J.~B. Parker and J.~A. Krommes.
\newblock Generation of zonal flows through symmetry breaking of statistical
  homogeneity.
\newblock \emph{New J. Phys.} 16 (2014), 035006.
\newblock \urlprefix\url{http://iopscience.iop.org/1367-2630/16/3/035006}.

\bibitem[{Pedlosky(1987)}]{pedlosky:book}
J.~Pedlosky.
\newblock \emph{Geophysical {F}luid {D}ynamics}.
\newblock Springer-Verlag, 1987.

\bibitem[{Plunk(2007)}]{plunk:2007}
G.~Plunk.
\newblock Gyrokinetic secondary instability theory for electron and ion
  temperature gradient driven turbulence.
\newblock \emph{Physics of Plasmas (1994-present)} 14 (2007), 112308.

\bibitem[{Pueschel \emph{et~al.}(2013)Pueschel, G\:orler, Jenko, Hatch, and
  Cianciara}]{pueschel:2013}
M.~J. Pueschel, T.~G\:orler, F.~Jenko, D.~R. Hatch, and A.~J. Cianciara.
\newblock Second and tertiary instability in electromagnetic plasma
  microturbulence.
\newblock \emph{Phys.\ Plasmas} 20 (2013), 102308.

\bibitem[{Pushkarev \emph{et~al.}(2013)Pushkarev, Bos, and
  Nazarenko}]{pushkarev:2013}
A.~V. Pushkarev, W.~J.~T. Bos, and S.~V. Nazarenko.
\newblock Zonal flow generation and its feedback on turbulence production in
  drift wave turbulence.
\newblock \emph{Phys. Plasmas} 20 (2013), 042304.
\newblock
  \urlprefix\url{http://scitation.aip.org/content/aip/journal/pop/20/4/10.1063/1.4802187}.

\bibitem[{Rhines(1975)}]{rhines:1975}
P.~B. Rhines.
\newblock Waves and turbulence on a beta-plane.
\newblock \emph{J. Fluid Mech.} 69 (1975), 417--443.
\newblock
  \urlprefix\url{http://journals.cambridge.org/action/displayAbstract?fromPage=online&aid=373502}.

\bibitem[{Rogers \emph{et~al.}(2000)Rogers, Dorland, and
  Kotschenreuther}]{rogers:2000}
B.~N. Rogers, W.~Dorland, and M.~Kotschenreuther.
\newblock Generation and stability of zonal flows in ion-temperature-gradient
  mode turbulence.
\newblock \emph{Phys. Rev. Lett.} 85 (2000), 5336--5339.
\newblock \urlprefix\url{http://link.aps.org/doi/10.1103/PhysRevLett.85.5336}.

\bibitem[{Schmitz \emph{et~al.}(2012)Schmitz, Zeng, Rhodes, Hillesheim, Doyle,
  Groebner, Peebles, Burrell, and Wang}]{schmitz:2012}
L.~Schmitz, L.~Zeng, T.~L. Rhodes, J.~C. Hillesheim, E.~J. Doyle, R.~J.
  Groebner, W.~A. Peebles, K.~H. Burrell, and G.~Wang.
\newblock Role of zonal flow predator-prey oscillations in triggering the
  transition to {H}-mode confinement.
\newblock \emph{Phys. Rev. Lett.} 108 (2012), 155002.
\newblock
  \urlprefix\url{http://link.aps.org/doi/10.1103/PhysRevLett.108.155002}.

\bibitem[{Schober \emph{et~al.}(1986)Schober, Allroth, Schroeder, and
  M\"uller-Krumbhaar}]{schober:1986}
H.~R. Schober, E.~Allroth, K.~Schroeder, and H.~M\"uller-Krumbhaar.
\newblock Dynamics of periodic pattern formation.
\newblock \emph{Phys. Rev. A} 33 (1986), 567--575.
\newblock \urlprefix\url{http://link.aps.org/doi/10.1103/PhysRevA.33.567}.

\bibitem[{Scott and Dritschel(2012)}]{scott:2012}
R.~K. Scott and D.~G. Dritschel.
\newblock The structure of zonal jets in geostrophic turbulence.
\newblock \emph{J. Fluid Mech.} 711 (2012), 576--598.
\newblock
  \urlprefix\url{http://journals.cambridge.org/article_S0022112012004107}.

\bibitem[{Smolyakov \emph{et~al.}(2000{\natexlab{a}})Smolyakov, Diamond, and
  Malkov}]{smolyakov:2000}
A.~I. Smolyakov, P.~H. Diamond, and M.~Malkov.
\newblock Coherent structure phenomena in drift wave--zonal flow turbulence.
\newblock \emph{Phys. Rev. Lett.} 84 (2000{\natexlab{a}}), 491--494.
\newblock \urlprefix\url{http://link.aps.org/doi/10.1103/PhysRevLett.84.491}.

\bibitem[{Smolyakov \emph{et~al.}(2000{\natexlab{b}})Smolyakov, Diamond, and
  Shevchenko}]{smolyakov:2000b}
A.~I. Smolyakov, P.~H. Diamond, and V.~I. Shevchenko.
\newblock Zonal flow generation by parametric instability in magnetized plasmas
  and geostrophic fluids.
\newblock \emph{Phys. Plasmas} 7 (2000{\natexlab{b}}), 1349--1351.
\newblock
  \urlprefix\url{http://scitation.aip.org/content/aip/journal/pop/7/5/10.1063/1.873950}.

\bibitem[{Srinivasan and Young(2012)}]{srinivasan:2012}
K.~Srinivasan and W.~R. Young.
\newblock Zonostrophic instability.
\newblock \emph{J. Atmos. Sci.} 69 (2012), 1633--1656.
\newblock \urlprefix\url{http://dx.doi.org/10.1175/JAS-D-11-0200.1}.

\bibitem[{Terry(2000)}]{terry:2000}
P.~W. Terry.
\newblock Suppression of turbulence and transport by sheared flow.
\newblock \emph{Rev. Mod. Phys.} 72 (2000), 109--165.
\newblock \urlprefix\url{http://link.aps.org/doi/10.1103/RevModPhys.72.109}.

\bibitem[{Terry \emph{et~al.}(2006)Terry, Baver, and Gupta}]{terry:2006}
P.~W. Terry, D.~A. Baver, and S.~Gupta.
\newblock Role of stable eigenmodes in saturated local plasma turbulence.
\newblock \emph{Phys. Plasmas} 13 (2006), 022307.
\newblock
  \urlprefix\url{http://scitation.aip.org/content/aip/journal/pop/13/2/10.1063/1.2168453}.

\bibitem[{Tobias \emph{et~al.}(2011)Tobias, Dagon, and Marston}]{tobias:2011}
S.~M. Tobias, K.~Dagon, and J.~B. Marston.
\newblock Astrophysical fluid dynamics via direct statistical simulation.
\newblock \emph{Astrophys. J.} 727 (2011), 127.
\newblock \urlprefix\url{http://stacks.iop.org/0004-637X/727/i=2/a=127}.

\bibitem[{Tobias and Marston(2013)}]{tobias:2013}
S.~M. Tobias and J.~B. Marston.
\newblock Direct statistical simulation of out-of-equilibrium jets.
\newblock \emph{Phys. Rev. Lett.} 110 (2013), 104502.
\newblock
  \urlprefix\url{http://link.aps.org/doi/10.1103/PhysRevLett.110.104502}.

\bibitem[{Trefethen(2000)}]{trefethen:2000}
L.~N. Trefethen.
\newblock \emph{Spectral Methods in Matlab}.
\newblock Society for Industrial and Applied Mathematics, Philadelphia, PA,
  USA, 2000.

\bibitem[{Trines \emph{et~al.}(2010)Trines, Bingham, Silva, Mendon\c{c}a,
  Shukla, Murphy, Dunlop, Davies, Bamford, Vaivads, and Norreys}]{trines:2010}
R.~M. G.~M. Trines, R.~Bingham, L.~O. Silva, J.~T. Mendon\c{c}a, P.~K. Shukla,
  C.~D. Murphy, M.~W. Dunlop, J.~A. Davies, R.~Bamford, A.~Vaivads, and P.~A.
  Norreys.
\newblock Applications of the wave kinetic approach: from laser wakefields to
  drift wave turbulence.
\newblock \emph{J. Plasma Phys.} 76 (2010), 903--914.
\newblock
  \urlprefix\url{http://journals.cambridge.org/article_S0022377810000449}.

\bibitem[{Tuckerman and Barkley(1990)}]{tuckerman:1990}
L.~S. Tuckerman and D.~Barkley.
\newblock Bifurcation analysis of the {E}ckhaus instability.
\newblock \emph{Physica D} 46 (1990), 57--86.
\newblock
  \urlprefix\url{http://www.sciencedirect.com/science/article/pii/0167278990901134}.

\bibitem[{Tynan \emph{et~al.}(2006)Tynan, Holland, Yu, James, Nishijima,
  Shimada, and Taheri}]{tynan:2006}
G.~R. Tynan, C.~Holland, J.~H. Yu, A.~James, D.~Nishijima, M.~Shimada, and
  N.~Taheri.
\newblock Observation of turbulent-driven shear flow in a cylindrical
  laboratory plasma device.
\newblock \emph{Plasma Phys. Control. Fusion} 48 (2006), S51.
\newblock \urlprefix\url{http://stacks.iop.org/0741-3335/48/i=4/a=S05}.

\bibitem[{Vallis(2006)}]{vallis:2006}
G.~K. Vallis.
\newblock \emph{Atmospheric and oceanic fluid dynamics}.
\newblock Cambridge University Press. Cambridge, 2006.

\bibitem[{Vallis and Maltrud(1993)}]{vallis:1993}
G.~K. Vallis and M.~E. Maltrud.
\newblock Generation of mean flows and jets on a beta plane and over
  topography.
\newblock \emph{J. Phys. Oceanogr.} 23 (1993), 1346--1362.
\newblock
  \urlprefix\url{http://dx.doi.org/10.1175/1520-0485(1993)023<1346:GOMFAJ>2.0.CO;2}.

\bibitem[{Vasavada and Showman(2005)}]{vasavada:2005}
A.~R. Vasavada and A.~P. Showman.
\newblock Jovian atmospheric dynamics: an update after \textit{Galileo} and
  \textit{Cassini}.
\newblock \emph{Rep. Prog. Phys.} 68 (2005), 1935.
\newblock \urlprefix\url{http://stacks.iop.org/0034-4885/68/i=8/a=R06}.

\bibitem[{Vedenov \emph{et~al.}(1962)Vedenov, Velikhov, and
  Sagdeev}]{vedenov:1962}
A.~Vedenov, E.~Velikhov, and R.~Sagdeev.
\newblock The quasi-linear theory of plasma oscillations.
\newblock In \emph{Proceedings of the Conference on Plasma Physics and
  Controlled Nuclear Fusion Research (Salzburg, 1961) [Nucl.\ Fusion Suppl.\
  Pt.~2]}, 465--475. International Atomic Energy Agency, Vienna, 1962.
\newblock Translated in U.S.A.E.C. Division of Technical Information document
  AEC--tr--5589 (1963), pp.~204--37.

\bibitem[{Wagner \emph{et~al.}(1982)Wagner, Becker, Behringer, Campbell,
  Eberhagen, Engelhardt, Fussmann, Gehre, Gernhardt, Gierke, Haas, Huang,
  Karger, Keilhacker, Kl\"uber, Kornherr, Lackner, Lisitano, Lister, Mayer,
  Meisel, M\"uller, Murmann, Niedermeyer, Poschenrieder, Rapp, R\"ohr,
  Schneider, Siller, Speth, St\"abler, Steuer, Venus, Vollmer, and
  Y\"u}]{wagner:1982}
F.~Wagner, G.~Becker, K.~Behringer, D.~Campbell, A.~Eberhagen, W.~Engelhardt,
  G.~Fussmann, O.~Gehre, J.~Gernhardt, G.~v. Gierke, G.~Haas, M.~Huang,
  F.~Karger, M.~Keilhacker, O.~Kl\"uber, M.~Kornherr, K.~Lackner, G.~Lisitano,
  G.~G. Lister, H.~M. Mayer, D.~Meisel, E.~R. M\"uller, H.~Murmann,
  H.~Niedermeyer, W.~Poschenrieder, H.~Rapp, H.~R\"ohr, F.~Schneider,
  G.~Siller, E.~Speth, A.~St\"abler, K.~H. Steuer, G.~Venus, O.~Vollmer, and
  Z.~Y\"u.
\newblock Regime of improved confinement and high beta in neutral-beam-heated
  divertor discharges of the {ASDEX} tokamak.
\newblock \emph{Phys. Rev. Lett.} 49 (1982), 1408--1412.
\newblock \urlprefix\url{http://link.aps.org/doi/10.1103/PhysRevLett.49.1408}.

\bibitem[{Waltz and Holland(2008)}]{waltz:2008}
R.~E. Waltz and C.~Holland.
\newblock Numerical experiments on the drift wave-zonal flow paradigm for
  nonlinear saturation.
\newblock \emph{Phys. Plasmas} 15 (2008), 122503.
\newblock
  \urlprefix\url{http://scitation.aip.org/content/aip/journal/pop/15/12/10.1063/1.3033206}.

\bibitem[{Whyte \emph{et~al.}(2010)Whyte, Hubbard, Hughes, Lipschultz, Rice,
  Marmar, Greenwald, Cziegler, Dominguez, Golfinopoulos, Howard, Lin,
  McDermott, Porkolab, Reinke, Terry, Tsujii, Wolfe, Wukitch, Lin, and the
  Alcator C-Mod~Team}]{whyte:2010}
D.~Whyte, A.~Hubbard, J.~Hughes, B.~Lipschultz, J.~Rice, E.~Marmar,
  M.~Greenwald, I.~Cziegler, A.~Dominguez, T.~Golfinopoulos, N.~Howard, L.~Lin,
  R.~McDermott, M.~Porkolab, M.~Reinke, J.~Terry, N.~Tsujii, S.~Wolfe,
  S.~Wukitch, Y.~Lin, and the Alcator C-Mod~Team.
\newblock I-mode: an {H}-mode energy confinement regime with {L}-mode particle
  transport in {Alcator C-Mod}.
\newblock \emph{Nucl. Fusion} 50 (2010), 105005.
\newblock \urlprefix\url{http://stacks.iop.org/0029-5515/50/i=10/a=105005}.

\bibitem[{Wilkinson(1994)}]{wilkinson:1994}
J.~H. Wilkinson.
\newblock \emph{Rounding Errors in Algebraic Processes}.
\newblock Dover Publications, 1994.

\bibitem[{Winsor \emph{et~al.}(1968)Winsor, Johnson, and Dawson}]{winsor:1968}
N.~Winsor, J.~L. Johnson, and J.~M. Dawson.
\newblock Geodesic acoustic waves in hydromagnetic systems.
\newblock \emph{Phys. Fluids} 11 (1968), 2448--2450.
\newblock
  \urlprefix\url{http://scitation.aip.org/content/aip/journal/pof1/11/11/10.1063/1.1691835}.

\bibitem[{Wordsworth(2009)}]{wordsworth:2009}
R.~D. Wordsworth.
\newblock A phase-space study of jet formation in planetary-scale fluids.
\newblock \emph{Phys. Fluids} 21 (2009), 056602.
\newblock
  \urlprefix\url{http://scitation.aip.org/content/aip/journal/pof2/21/5/10.1063/1.3140002}.

\bibitem[{Xanthopoulos \emph{et~al.}(2011)Xanthopoulos, Mischchenko, Helander,
  Sugama, and Watanabe}]{xanthopoulos:2011}
P.~Xanthopoulos, A.~Mischchenko, P.~Helander, H.~Sugama, and T.-H. Watanabe.
\newblock Zonal flow dynamics and control of turbulent transport in
  stellarators.
\newblock \emph{Phys. Rev. Lett.} 107 (2011), 245002.
\newblock
  \urlprefix\url{http://link.aps.org/doi/10.1103/PhysRevLett.107.245002}.

\bibitem[{Xu \emph{et~al.}(2011)Xu, Wan, Wang, Guo, Zhao, Liu, Naulin, Diamond,
  Tynan, Xu, Chen, Jiang, Liu, Yan, Zhang, Wang, Liu, and Ding}]{xu:2011}
G.~S. Xu, B.~N. Wan, H.~Q. Wang, H.~Y. Guo, H.~L. Zhao, A.~D. Liu, V.~Naulin,
  P.~H. Diamond, G.~R. Tynan, M.~Xu, R.~Chen, M.~Jiang, P.~Liu, N.~Yan,
  W.~Zhang, L.~Wang, S.~C. Liu, and S.~Y. Ding.
\newblock First evidence of the role of zonal flows for the {L--H} transition
  at marginal input power in the {EAST} tokamak.
\newblock \emph{Phys. Rev. Lett.} 107 (2011), 125001.
\newblock
  \urlprefix\url{http://link.aps.org/doi/10.1103/PhysRevLett.107.125001}.

\bibitem[{Yan \emph{et~al.}(2010{\natexlab{a}})Yan, Tynan, Holland, Xu, Muller,
  and Yu}]{yan:2010b}
Z.~Yan, G.~R. Tynan, C.~Holland, M.~Xu, S.~H. Muller, and J.~H. Yu.
\newblock Scaling properties of turbulence driven shear flow.
\newblock \emph{Phys. Plasmas} 17 (2010{\natexlab{a}}), 012302.
\newblock
  \urlprefix\url{http://scitation.aip.org/content/aip/journal/pop/17/1/10.1063/1.3276521}.

\bibitem[{Yan \emph{et~al.}(2010{\natexlab{b}})Yan, Tynan, Holland, Xu, Muller,
  and Yu}]{yan:2010}
Z.~Yan, G.~R. Tynan, C.~Holland, M.~Xu, S.~H. Muller, and J.~H. Yu.
\newblock Shear flow and drift wave turbulence dynamics in a cylindrical plasma
  device.
\newblock \emph{Phys. Plasmas} 17 (2010{\natexlab{b}}), 032302.
\newblock
  \urlprefix\url{http://scitation.aip.org/content/aip/journal/pop/17/3/10.1063/1.3322823}.

\bibitem[{Zhou \emph{et~al.}(2012)Zhou, Heidbrink, Boehmer, McWilliams, Carter,
  Vincena, Friedman, and Schaffner}]{zhou:2012}
S.~Zhou, W.~W. Heidbrink, H.~Boehmer, R.~McWilliams, T.~A. Carter, S.~Vincena,
  B.~Friedman, and D.~Schaffner.
\newblock Sheared-flow induced confinement transition in a linear magnetized
  plasma.
\newblock \emph{Phys. Plasmas} 19 (2012), 012116.
\newblock
  \urlprefix\url{http://scitation.aip.org/content/aip/journal/pop/19/1/10.1063/1.3677361}.

\bibitem[{Zhu and Hammett(2010)}]{zhu:2010}
J.-Z. Zhu and G.~W. Hammett.
\newblock Gyrokinetic statistical absolute equilibrium and turbulence.
\newblock \emph{Physics of Plasmas} 17 (2010), 122307.
\newblock \urlprefix\url{http://link.aip.org/link/?PHP/17/122307/1}.

\end{thebibliography}

\end{document}